\documentclass[twocolumn,aps,prd,amsmath,amssymb,preprintnumbers,longbibliography]{revtex4-1}
\usepackage{amsmath} \usepackage{amsfonts} \usepackage{amssymb}
\usepackage{bbm}
\usepackage{epsfig}
\usepackage{graphics}
\usepackage{graphicx}
\usepackage{titlesec}
\usepackage{mathtools}
\usepackage{environ}
\usepackage{dsfont}

\usepackage[colorlinks=true]{hyperref}
\hypersetup{urlcolor=black,linkcolor=black,citecolor=black}
\usepackage[toc,page]{appendix}

\makeatletter
\renewcommand*\env@matrix[1][c]{\hskip -\arraycolsep
  \let\@ifnextchar\new@ifnextchar
  \array{*\c@MaxMatrixCols #1}}
\makeatother

\newcommand{\ee}{\mathrm{e}}
\newcommand{\ba}{\begin{eqnarray}}
\newcommand{\ea}{\end{eqnarray}}
\newcommand{\nn}{\nonumber}

\titleformat{\subsection}[block]{\normalfont\bfseries}{\thesubsection.}{1ex}{}
\titlespacing{\subsection}{0pt}{10pt}{1pt}[0pt]
\titleformat*{\section}{\large\bfseries}

\renewcommand{\thesubsection}{\arabic{section}.\arabic{subsection}}

\makeatletter
\def\p@subsection{}
\makeatother
\makeatletter
\def\p@subsubsection{}
\makeatother

\numberwithin{equation}{section}

\usepackage[utf8]{inputenc}
\usepackage[T1]{fontenc}
\usepackage[svgnames]{xcolor}
\usepackage{microtype,enumitem,lmodern,dsfont}

\usepackage{mathtools,amssymb,tensor}

\usepackage{cleveref}
\Crefname{equation}{Eq.}{Eqs.}
\crefname{section}{sec.}{secs.}
\crefname{appendix}{app.}{apps.}
\crefname{cond}{condition}{conditions}
\AtBeginDocument{}
\newcounter{condition}[equation]

\usepackage{todonotes}
\usepackage{braket}

\newcommand{\tr}{\mathrm{tr}}
\newcommand{\raw}{\rightarrow}

\newcommand{\exval}[1]{\langle #1 \rangle}
\newcommand{\eq}[1]{eq.~\eqref{eq:#1}}
\newcommand{\eqs}[2]{eqs.~\eqref{eq:#1}, \eqref{eq:#2}}

\newcommand{\com}{\, ,\quad}
\newcommand{\mn}{\mu\nu}

\newcommand{\cD}{\mathcal{D}}
\newcommand{\cL}{\mathcal{L}}
\newcommand{\dif}{\mathrm{d}}

\newcommand{\p}{\partial}

\begin{document}
\unitlength=1mm 

\title[ ]{Quantum scale symmetry}

\author{C. Wetterich}

\affiliation{\href{eqn:mailto:c.wetterich@thphys.uni-heidelberg.de}{c.wetterich@thphys.uni-heidelberg.de}\\
{Universität Heidelberg, Institut für Theoretische Physik, Philosophenweg 16, D-69120 Heidelberg}}

\begin{abstract}

Quantum scale symmetry is the realization of scale invariance in a quantum field theory. No parameters with dimension of length or mass are present in the quantum effective action. Quantum scale symmetry is generated by quantum fluctuations via the presence of fixed points for running couplings. As for any global symmetry, the ground state or cosmological state may be scale invariant or not. Spontaneous breaking of scale symmetry leads to massive particles and predicts a massless Goldstone boson. A massless particle spectrum follows from scale symmetry of the effective action only if the ground state is scale symmetric. Approximate scale symmetry close to a fixed point leads to important predictions for observations in various areas of fundamental physics.

We review consequences of scale symmetry for particle physics, quantum gravity and cosmology. For particle physics, scale symmetry is closely linked to the tiny ratio between the Fermi scale of weak interactions and the Planck scale for gravity. For quantum gravity, scale symmetry is associated to the ultraviolet fixed point which allows for a non-perturbatively renormalizable quantum field theory for all known interactions. The interplay between gravity and particle physics at this fixed point permits to predict couplings of the standard model or other ``effective low energy models'' for momenta below the Planck mass. In particular, quantum gravity determines the ratio of Higgs boson mass and top quark mass. In cosmology, approximate scale symmetry explains the almost scale-invariant primordial fluctuation spectrum which is at the origin of all structures in the universe. The pseudo-Goldstone boson of spontaneously broken approximate scale symmetry may be responsible for dynamical dark energy and a solution of the cosmological constant problem.

%\vspace{1cm}

%\noindent\makebox[\textwidth]{
%\centering
%\includegraphics[scale=0.6]{Flowplot1pdfblack.pdf}

%\vspace{1cm}

\vspace{15cm}

\end{abstract}

%\onecolumngrid
\maketitle 

%\enlargethispage{-3.2cm}

%\twocolumngrid

%
%\hspace{-3cm}
%\vspace{-2cm}

%\newpage

%

\hypertarget{TOC}{\tableofcontents}

\newpage

\section{Introduction}
\label{sec:Introduction}

Scale symmetry is associated to the absence of intrinsic mass or length scales. Do we see scale symmetry in particle physics and cosmology? There are two well known fields where approximate scale symmetry plays a central role. The first is the approximately scale invariant power spectrum of the primordial cosmic fluctuations. These fluctuations are the seed for the structure in the Universe, and therefore of our existence. Their approximate scale invariance hints to a role of scale symmetry in the very early inflationary cosmology \cite{STAR,GUTH,MUCHI,LIN1,ALST,LIN2,SHACWCI} or other variants of primordial cosmology. 

The second is the approximate scale invariance of particle physics at high momenta if the standard model of particle physics \cite{SMGLA,SMWEI,SMAL} can be extrapolated sufficiently above the Fermi scale $\varphi_0=174.1$ GeV, where the electroweak gauge symmetry is spontaneously broken \cite{HIG1,HIG2,EB}. With the absence of new physics in the present findings of the large hadron collider (LHC), the possibility of such an extrapolation is accepted more and more widely. In a momentum range above $1$ TeV the dimensionless couplings in the standard model are all small. Effects of their running are at least quadratic in the couplings, often even cubic or quartic, and therefore even smaller. The role of a possible intrinsic mass scale associated to the Fermi scale, or to the characteristic ``confinement scale'' $\Lambda_{\rm QCD}$ for quantum chromodynamics (QCD), is tiny for characteristic momenta $p$ above $1$ TeV, being typically suppressed by powers of $\varphi^2_0/p^2$ or $\Lambda^2_{\rm QCD}/p^2$. Scale symmetry holds in this high momentum range with rather high accuracy. 

Scale symmetry, often also called dilatation symmetry, and sometimes somewhat inaccurately conformal symmetry, has played an important role in the theoretical discussion of several fundamental questions in particle physics, cosmology and quantum gravity. Approximate scale symmetry has been advocated \cite{CWFT,CWMHB,BAR} for ensuring the naturalness of the observed small value of the Fermi scale as compared to the Planck mass $M=2.436\cdot 10^{18}$GeV, namely $\varphi_{0}/M\approx 7\cdot 10^{-17}$, the gauge hierarchy \cite{GIL,WEIGH}. Furthermore, approximate scale symmetry has been employed for a possible solution \cite{CWQ} of the cosmological constant problem \cite{WEICC}, namely explaining why the value of the cosmological ``constant'' $\lambda$ is many orders of magnitude smaller than a characteristic value given by the Planck mass, $\lambda/M^{4}\approx 10^{-120}$. For this explanation $\lambda$ becomes a dynamical quantity, changing with time in cosmology. The corresponding association of $\lambda$ with the decreasing potential energy of a scalar field has predicted \cite{CWQ} a form of dynamical dark energy several years before the observational discovery of dark energy \cite{DEOB,DEOB2,DEOB3}. Thus, scale symmetry may be related to both observed tiny ratios in fundamental physics, the gauge hierarchy $\varphi_{0}/M$ and the cosmological constant $\lambda/M^{4}$. 

Quantum gravity is a missing corner stone in the building of a unified picture of theoretical fundamental physics. One of the simplest possibilities is that quantum gravity is a renormalizable quantum field theory with diffeomorphisms (general coordinate transformations) as a gauge symmetry. Quantum gravity is not perturbatively renormalizable. Renormalizability of quantum gravity has therefore to be non-perturbative, associated to an ultraviolet (UV) fixed point with non-vanishing interactions. The presence of such a fixed point is called ``asymptotic safety'' \cite{WEIAS}. While established first only in the vicinity of two dimensions, non-perturbative computations have become available for four dimensions by the modern form of functional renormalization \cite{CWFR}, \cite{MRCW}. Since the pioneering work of Reuter on functional renormalization for gravity \cite{MRQG}, substantial evidence in favor of the existence of an UV-fixed point for four-dimensional gravity -- the Reuter fixed point -- has accumulated \cite{DP,SOU,RS,LAUREU1}.

As for any fixed point, the UV-fixed point for quantum gravity realizes exact scale symmetry. Close to a fixed point one should observe approximate scale symmetry. It is tempting to associate the approximate scale symmetry of the primordial cosmic fluctuations with the vicinity of the UV-fixed point in quantum gravity \cite{CWIQ}.

The aim of the present work is a discussion of the relations between different facets of scale symmetry in particle physics, cosmology and quantum gravity, and to put them into a common context. A central theme of this work is the presence of several types of scale symmetry in particle physics coupled to quantum gravity. They are related to the presence of different (approximate) fixed points. The most prominent ones are: (1) The ``gravity scale symmetry'' associated to the UV-fixed point used for the definition of a renormalizable quantum field theory for gravity. (2) The ``particle scale symmetry'' of the standard model (SM) without gravity, seen as an effective low energy theory below the Planck mass. This is related to the (almost) second order character of the electroweak phase transition, and reflected by the presence of a SM-fixed point. (3) The ``cosmic scale symmetry'', related to the infrared (IR)-fixed point. This matters for the properties of the Universe at the largest distances and concerns the issue of the cosmological constant.

\begin{figure}[t!]
\includegraphics[scale=0.4]{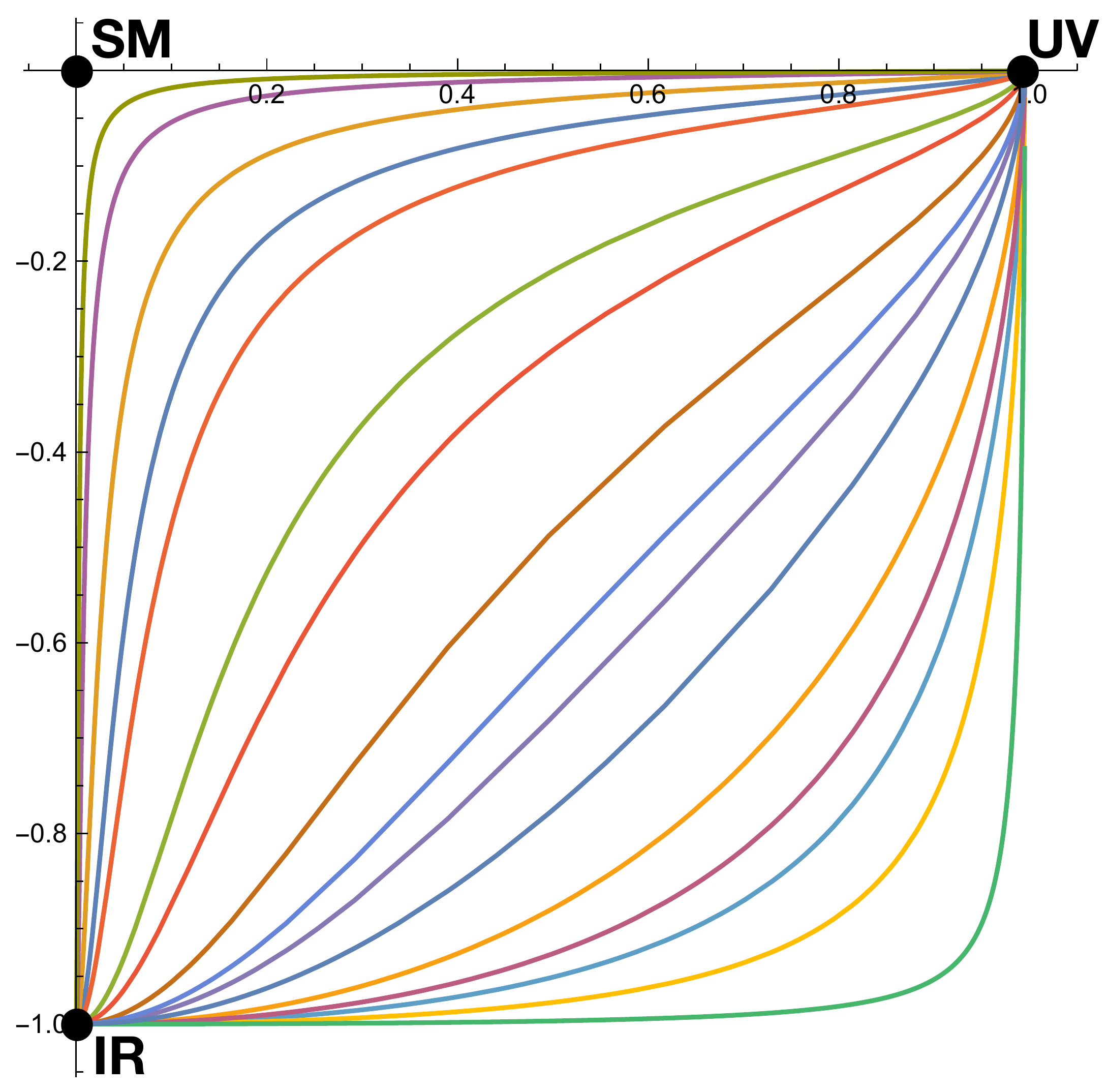}
\caption{Crossover flow trajectories. We plot $1/\sqrt{1+w^{2}}$ ($x$-axis) and $\gamma/\sqrt{1+\gamma^{2}}$ ($y$-axis), with $w^{-1}$ measuring the strength of gravity and $\gamma$ the distance from the electroweak phase transition, both in units of the renormalization scale $k$ or $\mu$. Trajectories start for $k\raw\infty$ in the upper right corner at the UV-fixed point where $\gamma=0$ and $w=w_{*}>0$. As $k$ is lowered, the trajectories sufficiently close to the electroweak phase transition approach first the standard model fixed point at $\gamma=0$, $w=0$ (upper left corner). Subsequently, they are attracted for $k\raw 0$ by the infrared fixed point at $\gamma\raw -\infty$, $w=0$. Parameters are $b=c=0.1$, and initial conditions given by $\gamma_{0}=\gamma(k=M)$, range from $\gamma_{0}=-0.001$ (upper left) to $\gamma_{0}=-6.0$ (lower right), see sect.~\ref{sec:Flow_Diagram_PP_QG}. As $\gamma_{0}$ approaches zero, the trajectory becomes a sequence of two straight lines connecting the three fixed points. The observed standard model corresponds to very small $\gamma_{0}$. For $b>2c$ the only possible trajectory has $\gamma_{0}=0$.}\label{fig:AA} 
\end{figure}

The overall situation can be summarized by the flow diagram Fig.~\ref{fig:AA}. It exhibits the three fixed points, with UV-fixed point in the upper right corner, SM-fixed point in the upper left corner, and IR-fixed point in the lower left corner. Realistic trajectories approach the SM-fixed point very closely. The renormalization flow connects the different facets of scale symmetry which are related to the three different fixed points. For cosmology, a typical time flow follows the renormalization group (RG-)trajectory, from the UV-fixed point in the past to the IR-fixed point in the future. Trajectories depend on the renormalization scale $k$ or $\mu$ via a dimensionless ratio $k/\tilde{\mu}$. There are different possibilities how cosmology maps to the RG-trajectories. First, $\tilde{\mu}$ may be given by scalar field $\chi$. If $\chi$ increases with time, $k/\tilde{\mu}$ moves from the UV to the IR as time increases. Second, $k^{-1}$ may be associated with some geometrical length scale which increases for increasing time. We will see for explicit solutions of field equations for cosmological models that this behavior is indeed realized. This connects the approximate scale invariance in early epochs of cosmology to the vicinity of the UV-fixed point.

Classical scale symmetry is realized if the classical or microscopic action does not involve a parameter with dimension of mass or length. Scale invariant classical gravity has been discussed since a long time \cite{HW,JOR,RDI}. A formulation of classical scale symmetry for the standard model coupled to gravity introduces in the classical action an additional scalar singlet field beyond the metric and the fields of the standard model \cite{FU1,ETG,MIN,ZEE,SLA,FU2}. The Planck mass is proportional to the singlet scalar field. For a scale invariant potential for the singlet and doublet scalars, the Fermi scale is no longer a parameter of the classical action, being proportional to the expectation value of a scalar field as well.

In a theory with fluctuations, as quantum field theory or statistical physics, classical or microscopic scale symmetry can be violated by fluctuations \cite{COL1}. This is rooted in the violation of scale symmetry by the measure in the functional integral or partition function. The results are running dimensionless couplings, e.g., $g(\mu)$ for the strong gauge coupling. Here, $\mu$ sets typically the scale for momenta of particles involved in a scattering process. Indeed, the overall momentum dependence of vertices is typically described by a momentum dependent dimensionless coupling $g$ (besides a structural momentum dependence of certain vertices). Since $g$ is dimensionless, any momentum dependence of $g$ requires the presence of an intrinsic scale $\bar{\mu}$ such that $g(\mu/\bar{\mu})$ can be a dimensionless function of a dimensionless quantity. The scale $\bar{\mu}$ may be an ultraviolet cutoff $\Lambda_{UV}$, or some other intrinsic scale where the coupling takes a given value $\bar{g}$, e.g. $g(\mu=\bar{\mu})=\bar{g}$. 

This dependence on an intrinsic scale often translates to explicit mass scales, a phenomenon called ``dimensional transmutation''. A well known example is the confinement scale $\Lambda_{QCD}$ associated with asymptotic freedom in QCD \cite{GW,POL}. The scale $\Lambda_{QCD}$ determines the mass of protons and neutrons. Another example is the Coleman-Weinberg mechanism for radiatively induced spontaneous symmetry breaking \cite{COLW}. These examples demonstrate clearly how quantum fluctuations can break scale symmetry. Several models assume classical scale symmetry for the standard model \cite{HEM,MENI1,MENI2,FKV}. The running dimensionless couplings of these models induce an intrinsic scale $\bar{\mu}$ that depends on the precise setting. This amounts to an explicit breaking of scale symmetry.

On the other hand, the running of dimensionless couplings or mass ratios may have fixed points. At a fixed point, the running of couplings is absent -- $g(\mu/\bar{\mu})$ is replaced by the fixed point value $g_{*}$. No intrinsic mass scale is present at the fixed point. At a fixed point, scale symmetry is exact. We call the scale symmetry associated to a fixed point ``quantum scale symmetry''. This is motivated by two properties. First, the scale symmetry associated to a fixed point is an exact symmetry in the presence of quantum fluctuations. Second, the quantum fluctuations are responsible for the presence of the fixed point, such that scale symmetry is actually induced by quantum fluctuations, rather than being destroyed by quantum fluctuations. Quantum scale symmetry does not need classical scale symmetry on the microscopic level. The microscopic or classical action may involve intrinsic mass of length scales. At a fixed point, the memory of these microscopic scales is wiped out for $\mu$ sufficiently below the microscopic scale. Well known examples are fixed points in statistical physics, as the Wilson-Fisher fixed point for scalar theories in three dimensions. It can be realized, for example, on a lattice, with intrinsic scale given by the lattice distance $a$. For critical phenomena at the phase transition, characterized by the fixed point, no memory of the lattice distance survives in the limit $\mu a\raw 0$. This holds provided one uses suitable renormalized fields.

The presence or absence of scale symmetry is best discussed in terms of the quantum effective action $\Gamma$. The effective action is a functional of fields, similar to the classical action. It includes, however, all fluctuation effects. Quantum scale invariance is realized if the effective action does not involve any intrinsic mass or length scale. The effective action contains all the necessary information about full propagators and vertices - the one-particle-irreducible vertices can be found by simple functional differentiation of $\Gamma$. Since $\Gamma$ contains all effects of fluctuations, the running of dimensionless couplings as well as their fixed points are directly encoded in $\Gamma$. The first derivative of $\Gamma$ generates the exact quantum field equations, while the second functional derivative constitutes the inverse propagator. Once $\Gamma$ is known or assumed, the remaining issues can be dealt with ``classical field theory''. Thus ``classical field equations'' are valid for the full quantum field theory. Only the central object from which they are derived is the quantum effective action $\Gamma$ rather than the classical action $S$.

The central role of the effective action for quantum scale symmetry becomes visible if one realizes that a fixed point is not a property of a single coupling or a finite number of couplings. (Those may be sufficient for partial fixed points.) A theory contains infinitely many effective vertices, each with a momentum dependence to be specified. Exact scale symmetry requires that none of these vertices involves an intrinsic mass scale $\bar{\mu}$. This is directly encoded in the effective action $\Gamma$. Scale symmetry requires that $\Gamma$ is a functional of suitably renormalized fields that does not involve any intrinsic scale $\bar{\mu}$.

The notion of renormalized fields is important for scale symmetry. Typically, the relation between the renormalized fields and the microscopic fields involves the microscopic length scale. Therefore, an intrinsic scale $\bar{\mu}$ appears in this relation. If one expresses $\Gamma$ in terms of microscopic fields, it will no longer be independent of $\bar{\mu}$. Thus a more precise characterization of quantum scale invariance states that there exists a suitable choice of renormalized fields $\varphi_{R}$ such that $\Gamma[\varphi_{R}]$ does not involve any parameters with dimension of mass or length.

It is possible to discuss scale symmetry in terms of the Noether construction with the associated dilatation currents, for an example see refs.~\cite{PSW,FEHIRO,FHR2}. The quantum effective action contains the necessary information for this construction in the quantum field theory context. In the present work we will directly use the quantum effective action for our discussion of scale symmetry. Presence of scale symmetry is signaled by the absence of a characteristic scale $\bar{\mu}$, and approximate scale symmetry close to a fixed point can be discussed in terms of a weak dependence of $\Gamma$ on $\bar{\mu}$.

The association between scale symmetry and fixed points is known since a long time in the theory of critical phenomena in statistical physics \cite{FISH,WILK}. The status of quantum scale symmetry in particle physics and cosmology is precisely the same as the one for critical phenomena. For both statistical physics and quantum field theory for particle physics, gravity and cosmology, fluctuations are the central ingredient for the emergence of scale symmetry. This is why we can take over the concepts for critical phenomena. The Wilsonian approach \cite{WIL,WILK,WEGN} is common for the issue of scale symmetry for all fluctuating systems. The present paper employs these concepts for a systematic view of the various facets of scale symmetry in particle physics, quantum gravity and cosmology. Our wording of quantum scale symmetry emphasizes that for these fields quantum fluctuations are the essential agents for the understanding of fixed points, as well as for their vicinity, and that the quantum effective action is the essential tool for our investigation. It also permits an easy distinction from classical scale symmetry.

In a complete theory of particle physics and gravity quantum scale symmetry can be an exact symmetry \cite{CWQ,BS1}, or it can be explicitly broken by an intrinsic mass scale $\bar{\mu}$ \cite{CWQ}. Exact scale symmetry is typically realized if the theory is defined exactly on a fixed point. The route of an exact scale symmetry in particle physics and gravity has been followed in refs.~\cite{GRSZ,FHR1}. If the theory is defined in the vicinity of an UV-fixed point, scale symmetry becomes exact only in the limit of infinite momentum. The flow of the relevant parameters away from the fixed point induces intrinsic mass scales. This explicit scale symmetry breaking is often named ``dilatation anomaly'' or ``scale anomaly''. Models with a small intrinsic mass $\bar{\mu}$ show approximate scale symmetry. This line of research has been followed in refs.~\cite{CWQ,CWVG,STR1,CWIQ}. In the range of momenta much larger than $\bar{\mu}$ there is practically no difference between exact scale symmetry and approximate scale symmetry. This typically extends to field expectation values much larger than $\bar{\mu}$. The limit $\bar{\mu}\raw 0$ defines a theory closer and closer to the fixed point and connects to exact scale symmetry in a smooth way.

Quantum scale symmetry is not necessarily associated with a massless particle spectrum. As for any other global continuous symmetry a scale invariant effective action permits two regimes. The solution of the quantum field equations may or may not break scale symmetry. For solutions which break scale symmetry one encounters ``spontaneously broken scale symmetry''. This happens, for example, if some scalar field takes a non-zero value $\chi$. In this ``SSB regime'' particles can be massive, with mass given by $m=h\chi$ and $h$ a dimensionless coupling. A spontaneously broken global symmetry implies the presence of a massless Goldstone boson. For spontaneously broken scale symmetry this is the dilaton. The SSB-regime is therefore characterized by a (partially) massive particle spectrum and a massless dilaton. On the other hand, if the solution respects scale symmetry as well, quantum scale symmetry is not spontaneously broken. In this symmetric regime (SYM-regime) all particles are massless. Indeed, for $m=h\chi$ and $\chi=0$ one has $m=0$. The SYM-regime is therefore characterized by a massless particle spectrum.

Only in the SYM-regime the world would look the same on all scales. This property, often associated with scale symmetry, therefore needs an additional ingredient, namely the absence of spontaneous breaking of scale symmetry. Fixed points with their associated scale symmetry are compatible with the appearance of effective mass and length scales associated to spontaneous breaking of scale symmetry. Spontaneous breaking of scale symmetry is usually not realized for critical phenomena in statistical physics. At a second order phase transition the world looks indeed the same on all scales. In contrast, cosmological solutions of field equations derived from a scale invariant effective action often involve a non-vanishing scalar field. They realize scale symmetry in the SSB regime. In our present world we see explicit mass and length scales as the mass of the electron or the size of an atom, or else the gravitational constant. We will see that this is perfectly consistent with scale symmetry. All observed masses are then proportional to a scalar field $\chi$.

One of the main distinctions between exact and approximate scale symmetry concerns the properties of the (pseudo) Goldstone boson in case of spontaneous scale symmetry breaking by a scalar field $\chi$. In case of exact scale symmetry a massless Goldstone boson is predicted, which can only have derivative couplings. For approximate scale symmetry the Goldstone boson acquires a small mass $\sim\bar{\mu}^{2}/\chi$ and is called a ``pseudo-Goldstone boson''. The couplings of the pseudo-Goldstone boson no longer need to be pure derivative couplings. The exchange of this pseudo-Goldstone boson, often called ``cosmon'', mediates a ``fifth force'' \cite{PSW}. In the limit $\bar{\mu}/\chi\raw 0$ the properties of the pseudo-Goldstone boson smoothly connect to the exact Goldstone boson or dilaton. A pseudo-Goldstone boson or cosmon with a tiny, typically time-dependent mass can play an important cosmological role in the present universe, being responsible for dynamical dark energy \cite{CWQ}.

This review is organized as follows: In sect.~\ref{sec:Fixed_points_and_scales} we describe the basics of quantum scale symmetry, associated to fixed points. The difference between classical and quantum scale symmetry resides in the role of quantum fluctuations. We discuss the effective action for a quantum scale invariant standard model coupled to gravity. This extends to approximate scale symmetry due to the presence of relevant couplings at the fixed point, and the associated generation of intrinsic mass scales. In sect.~\ref{sec:Networks_FP_Crossover} we discuss networks of fixed points and the crossover between them. They are responsible for the presence of two different scale symmetries, one for quantum gravity and the other for the low energy effective theory below the Planck mass. The discussion of this section is directly connected to the flow diagram Fig.~\ref{fig:AA}. We discuss the predictivity for observable parameters due to couplings becoming irrelevant at the fixed point or due to partial fixed points. In sects.~\ref{sec:Fixed_points_and_scales}, \ref{sec:Networks_FP_Crossover} the scale for flowing couplings is set by a momentum scale $\mu$ characteristic for the investigated vertices. In sect.~\ref{sec:Flow_In_Field_Space} we discuss the situation where the scale is set by the value of a scalar field $\chi$. Any ground state or cosmological solution with non-vanishing $\chi$ induces a spontaneous breaking of scale symmetry. We discuss the interplay between spontaneous breaking by $\chi$ and explicit breaking by a possible intrinsic mass scale $\bar{\mu}$. We also address the relations between the flow with $\mu$ and the flow with $\chi$.

In sect.~\ref{sec:Particle_Scale_Sym} we turn to particle scale symmetry. It is due to the SM-fixed point in Fig.~\ref{fig:AA}. We discuss its origin in the (almost) second order character of the electroweak phase transition. We investigate the role of particle scale symmetry for the naturalness of the gauge hierarchy. We describe the quantum scale invariant standard model as well as a possible explicit symmetry breaking by a mass parameter $\bar{\mu}$. We argue that the mass of the pseudo-Goldstone boson of spontaneously broken dilatation symmetry, $m\sim\bar{\mu}^{2}/\chi$, cannot be determined within the low energy effective theory. It needs a specification of the UV-completion, for example by the UV-fixed point for quantum gravity.

Sect.~\ref{sec:V} is devoted to gravity scale symmetry. Gravity scale symmetry is the scale symmetry for the combined theory of particle physics and quantum gravity. It is associated to the UV-fixed point for asymptotic safety. In Fig.~\ref{fig:AA} all trajectories start from this UV-fixed point for $k\raw\infty$. We briefly introduce quantum Einstein gravity and dilaton quantum gravity. For the latter the Planck mass is proportional to a scalar field $\chi$. For dilaton quantum gravity a scale invariant effective action can be found easily. One can transform the scale invariant frame to the more familiar Einstein frame with fixed Planck mass $M$ by a field transformation of the metric. This Weyl scaling introduces $M$ as an explicit mass appearing in the field transformation, thus hiding in a certain way the presence of scale symmetry. However, $M$ is not an intrinsic scale in this case, being not visible in the original scale invariant frame. We describe several versions of a scale invariant effective action for gravity. The vicinity of an UV-fixed point involves flowing couplings. If relevant, they lead to undetermined parameters that have to be specified for the definition of the model. 

We discuss in detail quantum gravity effects for the effective potential $U$ for scalars. The interplay of the dependence of $U$ on the renormalization scale $k$ and the scalar field $\chi$ leads to an extended notion of a fixed point. The latter is characterized by a scaling form of the potential, where $U/k^{4}$ only depends on the ratio $y=\chi^{2}/k^{2}$. This is a whole function, replacing fixed point values for a finite number of couplings. 

Besides the UV-fixed point, quantum gravity also exhibits an IR-fixed point with possible high relevance for the cosmological constant problem. All flow trajectories in Fig.~\ref{fig:AA} end in this IR-fixed point for $k\raw 0$. We discuss the crossover from the UV- to the IR-fixed point. We finally review the quantum gravity predictions for the effective potential of the Higgs scalar and the successful prediction for the mass ratio between the Higgs boson and the top quark.

In sect.~\ref{sec:Scale_Sym_Cosmo} we discuss the important consequences of scale symmetry for cosmology. This concerns all epochs, from the almost scale invariant spectrum of primordial cosmic fluctuations in inflationary cosmology, to cosmological scaling solutions with early dark energy during radiation and matter domination, and to dynamical dark energy in the present cosmological epoch. The dynamics of a single scalar field may be responsible for modifications of Einstein gravity for all epochs. Its potential energy may both drive inflation and be responsible for the present accelerated expansion. 

After a short discussion of the quantitative relation between the cosmic fluctuation spectrum and scale invariance we investigate Starobinski inflation, cosmon inflation and Higgs inflation with the perspective of scale symmetry, and also describe inflationary models with exact scale symmetry. After inflation cosmology makes a transition to cosmological scaling solutions for the radiation and matter dominated universe. These scaling solutions can be understood in terms of scale symmetry and its explicit or spontaneous breaking. If intrinsic scales $\bar{\mu}$ are small enough, the scaling solutions typically involve a small constant fraction of early dark energy. Exact scale symmetry predicts the absence of a time variation of couplings or apparent violations of the equivalence principle. Small violations of scale symmetry could lead to tiny residual effects which may be observable. 

Realistic models of quintessence need an exit from the scaling solution a couple of billion years ago. The associated stop of the evolution of the cosmon is due to intrinsic scales generated by the flow of relevant couplings. We discuss two cases, one with a crossover in the potential as observed in dilaton quantum gravity (crossover quintessence), the other resulting in the flow of beyond standard model couplings within growing neutrino quintessence. It is amazing to see for crossover quintessence how a quantum gravity computation of the scaling form of the effective potential translates directly to predictions for observable quantities in the present cosmological epoch. We summarize our findings in sect.~\ref{sec:Conclusion}.

\section{Fixed points and scales}\label{sec:Fixed_points_and_scales}

The present and next section mainly takes over the concepts and language of critical phenomena to particle physics, quantum gravity and cosmology. The conceptual part is not meant to be new, but rather puts various well known phenomena in particle physics in a common framework with respect to scale symmetry. Our discussion of the effective action has direct correspondence in statistical physics. The effective action is a generalization of the Gibbs free energy, now defined for arbitrary inhomogeneous fields. In statistical physics, various ``Landau theories'' can be seen as approximations to the effective action.

We stress the natural emergence of quantum scale symmetry in every continuous complete renormalizable quantum field theory. It is related to the absence of intrinsic mass or length scales in the quantum effective action at a fixed point. We establish the appearance of particle masses in a quantum scale invariant theory, due to spontaneous symmetry breaking. In particular, we discuss the quantum scale invariant standard model.

\subsection{Classical scale symmetry}

We start our discussion with classical scale symmetry. It is realized in a classical field theory if the classical action contains no parameter with dimension mass. We use $\hbar=1$, $c=1$ such that length and time have dimension $\textit{mass}^{-1}$. We often call all parameters with dimension $\textit{mass}^{P}$, $P\neq 0$, ``parameters with dimension mass'', with no restriction to $P=1$. Dimensionless parameters have $P=0$. Instead of classical field theory we may also neglect fluctuation effects in a quantum field theory (classical approximation). In this case the classical action $S$ can be associated with a microscopic action, defined at some high momentum scale. 

We employ a euclidean language with
\begin{equation}\label{eq:1} 
S=\int_{x}\, \sqrt{g}\cL=\int\dif^{4}x\, \sqrt{g}\cL
\end{equation}
and $g=\det(g_{\mn})$ the determinant of the metric $g_{\mn}$. The metric signature is $(-,+,+,+)$, such that $g$ is negative and $\sqrt{g}$ imaginary. For flat Minkowski space one has $\sqrt{g}=i$. The weight factor $e^{-S}$ reads in flat Minkowski space $\exp(-i\int_{x}\cL)$. Our conventions for $\cL$ are such that the potential is positive, while a stable kinetic term for a scalar field amounts to a negative sign of the term with two time derivatives due to $g^{00}<0$. (The sign of $\cL$ is opposite to the usual Lagrange density.) These conventions permit for a simple continuation to Euclidean signature. We often use $\tilde{\cL}=\sqrt{g}\cL$. Our conventions for the curvature tensor are such that the curvature scalar $R$ is positive for de Sitter space, and positive for the sphere for euclidean geometry with signature $(+,+,+,+)$.

Our first example is electrodynamics with
\begin{equation}\label{eq:2} 
\cL=\dfrac{1}{4e^{2}}F_{\mn}F^{\mn}+i\bar{\psi}\gamma^{\mu}D_{\mu}\psi+m_{e}\bar{\psi}\gamma^{5}\psi .
\end{equation}
Here we use a normalization of the gauge field $A_{\mu}$ such that the covariant derivative acting on the field $\psi$ for the electron reads
\begin{equation}\label{eq:3} 
D_{\mu}=\p_{\mu}-i A_{\mu }.
\end{equation}
The electromagnetic coupling $e$ is related to the fine structure constant $\alpha=e^{2}/4\pi$, and the electromagnetic field strength is given as usual by
\begin{equation}\label{eq:4} 
F_{\mn}=\p_{\mu}A_{\nu}-\p_{\nu}A_{\mu}\com F^{\mn}=g^{\mu\rho}g^{\nu\sigma}F_{\rho\sigma}.
\end{equation}
The matrices $\gamma^{\mu}$ are the euclidean Dirac matrices multiplied with the inverse vierbein $e_{m}^{\mu}$
\begin{align}\label{eq:5} 
&\gamma^{\mu}=e_{m}^{\mu}\gamma^{m}\com\lbrace\gamma^{m},\gamma^{n}\rbrace=2\delta^{mn}\com g_{\mn}=e^{m}_{\mu}e_{\nu}^{n}\delta_{mn}\, ,\nn\\
&e^{m}_{\mu}e^{\mu}_{n}=\delta^{m}_{n}\com \sqrt{g}=e=\det(e^{m}_{\mu}).
\end{align}
In flat Minkowski space one has $e_{k}^{m}=\delta_{k}^{m}$, $e_{0}^{m}=i\delta^{m}_{0}$, with $\mu=(0,k)$, $k=1,2,3$. We use conventions \cite{CWEF} for $\psi$ where $\bar{\psi}\gamma^{5}\psi$ is a scalar, $\bar{\psi}=\psi^{\dagger}\gamma^{0}$, and $m_{e}$ is the electron mass. In these conventions analytic continuation to euclidean space only amounts to a change of the phase for the vierbein $e^{0}_{\mu}$ \cite{CWEF}.

Scale transformations perform a multiplicative rescaling of all fields, including the vierbein and the metric,
\begin{align}\label{eq:6} 
&A^{\prime}_{\mu}=A_{\mu}\com\psi^{\prime}=\alpha^{3/2}\psi\, ,\nn\\
&e_{\mu}^{m}\,^{\prime}=\alpha^{-1}e_{\mu}^{m}\com g^{\prime}_{\mn}=\alpha^{-2}g_{\mn}\com \sqrt{g^{\prime}}=\alpha^{-4}\sqrt{g}.
\end{align}
Scale invariance of $S$ is therefore realized if $\cL^{\prime}=\alpha^{4}\cL$. This is indeed the case for the first two ``kinetic'' terms in eq.~\eqref{eq:2}. The last ``mass term'' scales as $\cL^{\prime}_{m}=\alpha^{3}\cL_{m}$ and therefore violates scale symmetry, as expected since it involves the intrinsic scale $m_{e}$. In flat space there exists another version of scale symmetry where one rescales coordinates instead of the metric and the vierbein. The formulation with fixed coordinates and scaling of the geometric fields is simpler, however. It is easily extended to curved space in quantum gravity, and also allows for a straightforward Noether construction. In eq.~\eqref{eq:6} the scalar and fermion fields transform according to their canonical dimension, gauge fields are invariant, and the scaling dimension of the metric is length squared, as appropriate for unscaled coordinates. In the absence of electrons the last two terms in eq.~\eqref{eq:2} can be omitted. The first term yields Maxwell's equations in vacuum, and one recovers the well known scale invariance of these equations. Approximate scale invariance is realized for high momenta $p$ where $m^{2}_{e}/p^{2}$ can be neglected, such that only the first two terms in eq.~\eqref{eq:2} contribute.

Scale invariant electrodynamics is easily constructed. One replaces $m_{e}=h\chi$, with scalar field $\chi$ scaling as
\begin{equation}\label{eq:7} 
\chi^{\prime}=\alpha\chi\, ,
\end{equation}
and adding a scalar kinetic term and quartic potential
\begin{equation}\label{eq:8} 
\cL_{\chi}=\dfrac{1}{2}g^{\mn}\p_{\mu}\chi\p_{\nu}\chi+\tilde{\lambda}\chi^{4}.
\end{equation}
For $\tilde{\lambda}>0$ the only homogeneous solution for $\chi$ is $\chi=0$ and scale symmetry is in the SYM-regime, with a massless electron and scalar field. For $\tilde{\lambda}=0$ one has solutions with arbitrary $\chi=\chi_{0}$. For $\chi_{0}\neq 0$ the electron is massive, with effective electron mass $m_{e}=h\chi_{0}$, while $\chi-\chi_{0}$ is associated to the massless Goldstone boson of spontaneously broken scale symmetry. (The field for the Goldstone boson that has only derivative couplings obtains by a non-linear field redefinition. It exists only for $\chi_{0}\neq 0$.) The action \eqref{eq:8} admits both solutions with spontaneous scale symmetry breaking (SSB-regime) and with unbroken scale symmetry (SYM-regime). For scale symmetry the transition between the SSB- and SYM-regimes is somewhat special, since in both regimes a massless scalar is present. For usual second order phase transitions involving an order parameter breaking a global continuous symmetry a massless scalar occurs only in the SSB-regime and at the phase transition. In case of scale symmetry the scalar remains massless in the SYM-regime as well, since all particles are massless in this regime.

The generalization to QCD is straightforward. There are now eight gluon fields $A_{\mu}^{z}$, with non-abelian field strength
\begin{equation}\label{eq:9} 
F_{\mn}^{z}=\p_{\mu}A_{\nu}^{z}-\p_{\nu}A_{\mu}^{z}+f^{z}\,_{vw}A_{\mu}^{v}A_{\nu}^{w}\, ,
\end{equation}
and $f_{zvw}$ the totally antisymmetric structure constants of $SU(3)$. For the first term in eq.~\eqref{eq:2} one replaces $F_{\mn}F^{\mn}$ by $F_{\mn}^{z}F^{\mn}_{z}$ and $e$ is replaced by the strong gauge coupling $g$. (Lowering and raising the index $z$ is only for visual purposes, $f_{zvw}=\delta_{zy}f^{y}\,_{vw}$ etc.). The electron is replaced by several quark fields in the triplet representation of $SU(3)$ and the two last terms in eq.~\eqref{eq:2} involve now sums over the quark flavors. The covariant derivative involves the generators $T_{z}$ of $SU(3)$ in the fundamental representation
\begin{equation}\label{eq:10} 
D_{\mu}=\p_{\mu}-i A_{\mu}^{z}T_{z}.
\end{equation}
In the limit of vanishing quark masses classical QCD is scale invariant.

For the standard model of particle physics one adds to QCD the gauge fields of $SU(2)\times U(1)$, extends the covariant derivatives correspondingly, and adds the leptons. Left handed and right handed fermions transform differently with respect to the standard model gauge symmetry $SU(3)\times SU(2)\times U(1)$, such that no field independent mass term is consistent with the symmetries. This part of the standard model is scale invariant.

Masses of the $W$- and $Z$-bosons, as well as for the quarks and charged leptons, are induced by spontaneous breaking of $SU(2)\times U(1)$ through the vacuum expectation value $\varphi_{0}$ of the Higgs scalar field, which sets the Fermi scale. The action for the complex doublet $h$ is given by
\begin{equation}\label{eq:11} 
\cL_{H}=\dfrac{1}{2}(D_{\mu}h)^{\dagger}D^{\mu}h+\dfrac{\lambda_{H}}{2}\left (h^{\dagger}h-\varphi_{0}^{2}\right )^{2}+\cL_{Y}
\end{equation}
with
\begin{equation}\label{eq:12} 
\cL_{Y}=\sum_{i,j}\, y_{ij}\bar{\psi}_{R}^{i}\gamma^{5}h^{\dagger}\psi_{L}^{j}+h.c.\, .
\end{equation}
Here $\psi_{L}$ and $\psi_{R}$ are left handed and right handed fermion fields
\begin{equation}\label{eq:13} 
\gamma^{5}\psi_{L}=\psi_{L}\com \gamma^{5}\psi_{R}=-\psi_{R}\, ,
\end{equation}
and $i,j$ denote the various species of fermion fields. The dimensionless Yukawa couplings $y_{ij}$ need to be compatible with $SU(3)\times SU(2)\times U(1)$ symmetry. For $\exval{h}=\varphi_{0}$ the $W,Z$-bosons acquire a mass through the Higgs mechanism \cite{HIG1,HIG2,EB}, and quark and lepton mass matrices are given by
\begin{equation}\label{eq:14} 
m_{ij}=y_{ij}\varphi_{0}.
\end{equation}
The neutrinos remain massless since the neutrino doublets have no neutrino singlets as counterparts. The only parameter violating scale symmetry is $\varphi_{0}$ in eq.~\eqref{eq:11}. For $\varphi_{0}=0$ the standard model exhibits classical scale symmetry. This may be easily verified by extending the scale transformations \eqref{eq:6}, \eqref{eq:7} to all fields.

\subsection{Classically scale invariant standard model}\label{sec:Scale_Inv_SM}

Similar to our discussion of scale invariant QED one can formulate a classically scale invariant standard model by replacing the parameter $\varphi_{0}$ by a field,
\begin{equation}\label{eq:15} 
\varphi_{0}^{2}=\varepsilon\chi^{2}\, ,
\end{equation}
with dimensionless $\varepsilon$ and $\chi$ a scalar singlet for which we add the term $\cL_{\chi}$ in eq.~\eqref{eq:8}. This can be extended to a scale invariant standard model coupled to gravity \cite{FU1,FU2} by replacing $\cL_{\chi}$ by $\cL_{\chi R}$,
\begin{equation}\label{eq:16} 
\cL_{\chi R}=-\dfrac{1}{2}\chi^{2}R+\dfrac{1}{2}\left (B-6\right )\p^{\mu}\chi\p_{\mu}\chi+\tilde{\lambda}\chi^{4}.
\end{equation}
Here $R$ is the curvature scalar formed with the metric $g_{\mn}$. We have normalized the scalar $\chi$ such that the variable Planck mass is given by $\chi$. The model is stable for all $B>0$, despite the negative sign of the kinetic term for $B<6$. 

The field equations derived from the action with classical scale symmetry admit solutions with a Robertson-Walker metric and a time dependent homogeneous scalar field $\chi(t)$. After a rather short while the scalar field settles at some constant value $\chi_{0}$ which only depends on the initial conditions. This type of solution realizes spontaneous breaking of the scale symmetry. The Fermi scale obeys $\varphi_{0}^{2}=\varepsilon_{\chi}\chi_{0}^{2}$, such that the particle spectrum is massive. The particle masses can be identified with the observed masses for suitable values of the gauge and Yukawa couplings, as well as the quartic scalar coupling $\lambda_{H}$. The units of $\chi_{0}$ are arbitrary. We choose units such that $\chi_{0}$ coincides with the Planck mass in present cosmology, $\chi_{0}=M=2.435 \cdot 10^{18}$ GeV. 

Realistic phenomenology requires the dimensionless quartic coupling $\varepsilon$ to be tiny,
\begin{equation}\label{eq:17} 
\varepsilon=\dfrac{\varphi_{0}^{2}}{M^{2}}=5 \cdot 10^{-33}\, ,
\end{equation}
reflecting the gauge hierarchy. The present cosmological constant is given by
\begin{equation}\label{eq:18} 
\lambda=\tilde{\lambda}\chi_{0}^{4}\, ,
\end{equation}
such that $\tilde{\lambda}$ is again a tiny dimensionless quartic coupling, reflecting the cosmological constant problem
\begin{equation}\label{eq:19} 
\tilde{\lambda}=\dfrac{\lambda}{M^{4}}=7\cdot 10^{-121}.
\end{equation}
Accepting these tiny values the cosmology of Fujii's scale symmetric model \cite{FU1,FU2} is standard cosmology, supplemented by the dynamics of the Goldstone boson $\chi-\chi_{0}$. Since $\chi$ settles to $\chi_{0}$ very rapidly \cite{CWQ}, dark energy is given by a static cosmological constant $\lambda$. Dynamical dark energy is only realized at some very early stage in cosmology before $\chi$ settles to its constant value.

Besides the coupling of the singlet scalar $\chi$ to gravity there is also a scale invariant coupling of the Higgs doublet to gravity \cite{CWQ,BS1}
\begin{equation}\label{eq:19AA} 
\cL_{\varphi R}=-\dfrac{1}{2}\,\xi_{H}\, h^{\dagger}h\, R\, .
\end{equation}
For a sufficiently large value of the dimensionless coupling $\xi_{H}$ this type of model leads to scale invariant Higgs inflation \cite{GRSZ}.

\subsection{Quantum scale symmetry}

Quantum fluctuations violate scale symmetry through running dimensionless couplings. In a functional integral formulation of quantum field theory this is due to the lack of scale invariance of the functional measure. In other words, the functional integral has to be regularized, and this procedure introduces a scale. (We will comment in sect.~\ref{sec:Exact_Approx_Quantum_Scale_Sym} on the possibility to ensure scale invariance by a field dependent regulator scale \cite{CWQ,SHAZEN}.) In the context of classical scale invariance of the standard model coupled to gravity, the running of the gauge couplings, Yukawa couplings, $\lambda_{H}$, $\varepsilon$, $B$ and $\tilde{\lambda}$ may induce intrinsic scales. Intrinsic scales are absent, however, if all couplings are at fixed points where they do not run. In this case quantum scale symmetry is realized.

We will discuss the issue of scale symmetry breaking by quantum fluctuations in terms of the properties of the quantum effective action $\Gamma$. The presence of quantum fluctuations implies that the quantum effective action no longer retains a simple polynomial form, even if one starts with a polynomial classical action. We first describe the example of QCD and turn to other models later.

For the example of perturbative QCD the gluon kinetic term takes the form \cite{CWGF}
\begin{equation}\label{eq:20} 
\cL_{F}=-\dfrac{1}{4}F_{z}^{\mn}(Z_{F})^{z\rho}_{\nu y}\, F^{y}_{\rho\mu}.
\end{equation}
The ``wave function renormalization'' $Z_{F}$ is a dimensionless matrix-function of a covariant differential operator $D$,
\begin{equation}\label{eq:21} 
Z_{F}=Z_{F}\left (D/\Lambda^{2}_{UV}\right ).
\end{equation}
The matrix elements $D^{z\nu}_{\mu y}$ of $D$ are given by
\begin{equation}\label{eq:22} 
\cD_{\mu}\,^{\nu}=-D^{2}\delta^{\nu}_{\mu}+D_{\mu}D^{\nu}+2i F_{\mu}\,^{\nu}.
\end{equation}
In eq.~\eqref{eq:22} $\cD_{\mu}\,^{\nu}$ are matrices in color space, with elements $(\cD_{\mu}\,^{\nu})^{z}\,_{y}=\cD^{z\nu}_{\mu y}$, where we employ the notation
\begin{align}\label{eq:23} 
&(A_{\mu})_{yz}=A_{\mu}^{w}\, (T_{w})_{yz}\com (F_{\mn})_{yz}=F_{\mn}^{w}\, (T_{w})_{yz}\, ,\nn\\
&(T_{w})_{yz}=-i f_{wyz}\, ,
\end{align}
and
\begin{equation}\label{eq:24} 
D_{\mu}=\p_{\mu}-i A_{\mu}\com D^{2}=D^{\mu}D_{\mu}=g^{\mn}D_{\mu}D_{\nu}.
\end{equation}
(The operator $\cD$ can be seen as the projection of $-D^{2}$ on the transversal gauge field fluctuations \cite{CWGF}.) For
\begin{equation}\label{eq:25} 
(Z_{F})^{z\rho}_{\nu y}=\dfrac{1}{g^{2}}\delta_{y}^{z}\delta_{\nu}^{\rho}
\end{equation}
one recovers the action of classical QCD. Whenever $Z_{F}$ depends on $\cD$, it also depends on some ultraviolet regularization scale $\Lambda_{UV}$. (For the example of lattice gauge theories, $\Lambda_{UV}$ can be identified with the inverse lattice distance.) This follows from the simple observation that $Z_{F}$ is dimensionless and therefore can only depend on dimensionless ratios.

We define the renormalized gauge coupling $g(\mu/\Lambda_{UV})$ by properties of $\Gamma$ for gauge field configurations for which $\cD$ takes values of the order $\mu^{2}$. For example, we may define $g(\mu)$ by the four gluon vertex corresponding to the fourth functional derivative of $\Gamma$ evaluated at $A_{\mu}=0$. This vertex depends on momenta, and we chose a configuration of momenta of the order $\mu$ for which they act as an IR-cutoff. In terms of the renormalized gauge coupling we write
\begin{equation}\label{eq:26} 
Z_{F}=g^{-2}\left (\dfrac{\mu }{\Lambda_{UV}}\right )\,\tilde{Z}_{F}\, .
\end{equation}
(If no confusion with the inverse function is possible, we sometimes employ for the purpose of notational simplicity $g^{-1}(x)$ for $1/g(x)$ or $(g(x))^{-1}$ etc.) For a renormalizable theory and $\mu\ll \Lambda_{UV}$, the dimensionless function $\tilde{Z}_{F}$ depends on $\cD/\mu^{2}$ and $g(\mu/\Lambda_{UV})$, but contains no longer any explicit dependence on the UV-regularization scale $\Lambda_{UV}$.

As an example, in one loop perturbation theory for a pure $SU(N)$-gauge theory one finds
\begin{equation}\label{eq:27} 
Z_{F}=g^{-2}\left (\dfrac{\mu }{\Lambda_{UV}}\right )+\dfrac{11N}{48\pi^{2}}\ln\left (\dfrac{\cD}{\mu^{2}}\right )\, ,
\end{equation}
with running renormalized gauge coupling
\begin{equation}\label{eq:28} 
g^{-2}\left (\dfrac{\mu }{\Lambda_{UV}}\right )=c_{UV}-\dfrac{11N}{48\pi^{2}}\ln\left (\dfrac{\Lambda_{UV}^{2}}{\mu^{2}}\right ).
\end{equation}
We can identify the integration constant $c_{UV}=g^{-2}(\mu=\Lambda_{UV})$. As it should be, $Z_{F}$ does not depend on $\mu$ - the $\mu$-dependence of the two terms in eq.~\eqref{eq:27} cancels. The scale $\mu$ has only been introduced for the definition of $g(\mu)$ and is not present in the effective action. With eq.~\eqref{eq:26} one obtains
\begin{equation}\label{eq:29} 
\tilde{Z}_{F}=1+\dfrac{11N g^{2}(\mu)}{48\pi^{2}}\ln\left (\dfrac{\cD}{\mu^{2}}\right ).
\end{equation}
This quantity depends no longer explicitly on $\Lambda_{UV}$, only implicitly via $g^{2}(\mu)$. For $D$ close to $\mu^{2}$ one finds $\tilde{Z}_{F}$ close to one, with small corrections that may be neglected to a good approximation.

The issue of quantum scale symmetry is directly related to the running of $g(\mu)$. With eq.~\eqref{eq:20} the only intrinsic scale in the effective action arises from $\Lambda_{UV}$ in eq.~\eqref{eq:21}. For a renormalizable theory this dependence on $\Lambda_{UV}$ is only present through the dependence of $g(\mu/\Lambda_{UV})$ on $\Lambda_{UV}$. Since $g$ is dimensionless it can depend on $\Lambda_{UV}$ only if it depends on $\mu$. For a fixed point $g$ no longer depends on $\mu$ and is therefore also independent of $\Lambda_{UV}$. The quantum effective action no longer involves an intrinsic mass scale and therefore is quantum scale invariant. The crucial point of this argument is that $\Gamma$ depends on $\Lambda_{UV}$ only via the renormalized gauge coupling $g(\mu/\Lambda_{UV})$. We will understand below this key property of renormalizable quantum field theories by the discussion of relevant parameters near a fixed point. The gauge coupling of pure Yang Mills theories is a special case of a single relevant parameter. In general, a renormalizable field theory will have several distinct renormalizable couplings, but their number has to be finite.

The role of the ``renormalization scale'' $\mu$ is twofold. If one changes $\mu$ with a simultaneous change of $g(\mu)$, nothing depends on $\mu$ since $\mu$ is not present in the effective action \eqref{eq:20}. On the other hand, for a fixed value of $g(\mu=\bar{\mu})=\bar{g}$ the scale $\bar{\mu}$ becomes an intrinsic mass scale if $g(\mu)$ depends on $\mu$. Only for a particular $\mu=\bar{\mu}$ the running coupling $g(\mu)$ equals $\bar{g}$. The presence of such an intrinsic mass scale indicates the violation of scale symmetry. If dimensionless couplings do not flow with $\mu$, no such intrinsic mass scale is introduced by the running couplings. If no other parameter with dimension mass is present, the theory is scale invariant.

We need the flow equation or renormalization group equation for the running renormalized gauge coupling. For small couplings this can be computed in perturbation theory. For QCD one finds in two loop order \cite{CAS}
\begin{equation}\label{eq:30} 
\p_{t}g^{2}=\mu\p_{\mu}g^{2}=\beta_{g^{2}}=-b_{0}g^{4}-b_{1}g^{6}\, ,
\end{equation}
with $t=\ln(\mu/\Lambda_{UV})$ and
\begin{equation}\label{eq:31} 
b_{0}=\dfrac{1}{16\pi^{2}}\left (22-\dfrac{4N_{f}}{3}\right )\com b_{1}=\dfrac{1}{(16\pi^{2})^{2}}\left (204-\dfrac{76 N_{f}}{3}\right )\, ,
\end{equation}
and $N_{f}$ the number of quark flavors. Fixed points occur for zeros of the $\beta$-function $\beta_{g^{2}}$. If the $\beta$-function vanishes, the flow of $g$ vanishes and $g$ becomes $\mu$-independent. Eq.~\eqref{eq:30} has a fixed point at $g^{2}=0$. For $b_{0}>0$ the theory is asymptotically free. This means that the fixed point $g^{2}=0$ is reached in the $UV$ for $\mu\raw \infty$. Indeed, as $\mu$ increases, $g^{2}$ decreases until it reaches zero. In the ultraviolet the interactions vanish and the theory becomes a free theory. On the other hand, towards the IR the coupling increases. In general we always consider the flow towards the IR, unless stated otherwise explicitly. For a region of positive $\beta$ the coupling decreases, while for negative $\beta$ it increases.

For an understanding of the qualitative behavior one may consider
\begin{equation}\label{eq:32} 
\p_{t} g^{-2}=-g^{-4}\p_{t}g^{2}=b_{0}+b_{1}g^{2}.
\end{equation}
For positive $b_{0}$ and $b_{1}$ this implies that $g^{-2}$ always decreases until it reaches zero at some scale $\mu=\Lambda_{QCD}$. At this scale $g^{2}(\mu)$ diverges. Perturbation theory remains no longer valid for large $g^{2}$. Neglecting the term $\sim b_{1}$, eq.~\eqref{eq:32} is easily solved
\begin{equation}\label{eq:33} 
g^{-2}(\mu)=g^{-2}(\Lambda_{UV})-\dfrac{b_{0}}{2}\ln\left (\dfrac{\Lambda_{UV}^{2}}{\mu^{2}}\right ).
\end{equation}
One recovers eq.~\eqref{eq:28} with $N$ replaced by $3-{2N_{f}}/{11}$. The scale $\Lambda_{QCD}$ corresponds to the zero of the r.h.s. of eq.~\eqref{eq:33}, namely
\begin{equation}\label{eq:34} 
\Lambda_{QCD}=\Lambda_{UV}\exp\left (-\dfrac{1}{b_{0}g^{2}(\Lambda_{UV})}\right ).
\end{equation}
The running gauge coupling generates the scale $\Lambda_{QCD}$ by dimensional transmutation.

On the other hand, if both $b_{0}$ and $b_{1}$ are negative, the fixed point at $g^{2}=0$ is reached in the IR. Such theories are called trivial since no long distance interactions survive for finite $g^{2}(\Lambda_{UV})$ and $\Lambda_{UV}\raw\infty$. Towards the UV one encounters a ``Landau pole'' where $g^{2}$ diverges. It is given by eq.~\eqref{eq:34}, now with the negative sign of $b_{0}$ or, expressed in terms of some nonzero $g(\mu)$,
\begin{equation}\label{eq:35} 
\Lambda_{LP}=\mu\exp\left (\dfrac{1}{|b_{0}| g^{2}(\mu)}\right ).
\end{equation}
Trivial theories can be continued towards short distances only for momenta $p^{2}<\Lambda_{LP}^{2}$. In a strict sense they are not renormalizable. They may be, however, effectively renormalizable if one considers them as low energy effective theories below a certain scale, say $M$, embedded into a more complete renormalizable theory. The standard model is of this type since the hypercharge-gauge coupling has a Landau pole far above the Planck mass $M$.

Finally, if $b_{0}$ and $b_{1}$ have opposite signs the flow equation \eqref{eq:30} exhibits two fixed points, one at $g^{2}=0$, and the other for
\begin{equation}\label{eq:36} 
g^{2}_{*}=-\dfrac{b_{0}}{b_{1}}.
\end{equation}
In contrast to asymptotic freedom the interactions do not vanish for the fixed point \eqref{eq:36}. Defining a renormalizable theory by reaching the ``interacting fixed point'' \eqref{eq:36} in the UV-limit realizes asymptotic safety. Precisely at the fixed point such a theory is scale invariant. It is also ``finite'' in the following sense. If $g^{2}$ is independent of $\mu$, also $\tilde{Z}_{F}$ in eq.~\eqref{eq:26} is independent of $\mu$, since the $\mu$-dependence of the two factors on the r.h.s. of eq.~\eqref{eq:26} has to cancel. Furthermore, there is no dependence on $\Lambda_{UV}$. If the sum of all contributions to a vertex diverges with $\Lambda_{UV}\raw \infty$, this would contradict the observation that the vertex cannot depend on $\Lambda_{UV}$. Finiteness does not hold, however, for arbitrary values of $g^{2}$, but only for $g^{2}=g^{2}_{*}$. Single diagrams in perturbation theory are not finite. We also observe that the normalization of the gauge field by the way it appears in the covariant derivative \eqref{eq:24} implies that $A_{\mu}$ is already a renormalized field.

The behavior away from the interacting fixed point for $g^{2}(\Lambda_{UV})\neq g^{2}_{*}$ depends on the sign of $b_{0}$. For $b_{0}<0$ and $g^{2}(\Lambda_{UV})<g^{2}_{*}$ the theory is trivial. A renormalized coupling $g^{2}(\mu)$ in the range $0\leq g^{2}(\mu)\leq g^{2}_{*}$ flows between the UV-fixed point $g_{*}^{2}$ and the IR-fixed point $g^{2}=0$. There is no Landau pole and the theory is renormalizable. For a given scale $\mu$ any arbitrary value of $g^{2}(\mu)$ in the range $0<g^{2}(\mu)<g^{2}_{*}$ can be realized. Thus $g$ is a renormalizable coupling. For $g^{2}(\mu)>g^{2}_{*}$ the flow is between the UV-fixed point $g^{2}_{*}$ and a QCD-like behavior with $g^{2}(\mu\raw \tilde{\Lambda})\raw\infty$ as one flows towards the IR. Again this corresponds to a renormalizable theory with renormalized coupling $g^{2}(\mu)>g^{2}_{*}$.

For $b_{0}>0$ the interacting fixed point becomes infrared stable. Arbitrary $g^{2}(\mu)$ flow towards $g^{2}_{*}$ for $\mu/\Lambda_{UV}\raw 0$. For $0<g^{2}(\mu)<g^{2}_{*}$ the theory is asymptotically free with UV-fixed point at $g^{2}=0$. Values of $g^{2}(\mu)>g^{2}_{*}$ can only be reached for finite $\Lambda_{UV}$, i.e., for an effectively renormalizable ``low energy theory''. A renormalizable theory predicts $g^{2}(\mu)$ in the range
\begin{equation}
0\leq g^{2}(\mu)\leq g^{2}_{*}\, .
\end{equation}
For $b_{0}>0$, $b_{1}<0$ the flow \eqref{eq:30} is an example of self-organized criticality in particle physics \cite{BORW}. The fixed point $g^{2}_{*}$ is reached in the IR for arbitrary $g^{2}(\mu)>0$. Independently of the parameters of the model the long distance theory is scale invariant for any $g^{2}(\Lambda_{UV})>0$. 

For QCD with $N_{f}$ flavors the IR-fixed point \eqref{eq:36} is realized for $8<N_{f}<17$. This is the Banks-Zaks fixed point \cite{CAS,BZA}. The flow equation \eqref{eq:30} is based on a perturbative two loop calculation. A reliable fixed point should lie in the region of validity of this approximation. Somewhat generously one may assume that this holds for $g^{2}_{*}< 4\pi$, corresponding to a number of flavors $N_{f}>11$. Functional renormalization investigations confirm this fixed point \cite{BRGI,BGF1,BGJR}. A systematic study for perturbatively realized asymptotic safety for a whole class of gauge theories coupled to fermions and scalars has been initiated by Litim and Sannino \cite{LS} and yields a rather complete picture \cite{BOLI}. At this stage we have encountered first examples of quantum scale symmetry with non-vanishing interactions. For running couplings with a fixed point at $g_{*}^{2}\neq 0$ one may define a specific model by choosing $g^{2}(\mu)=g^{2}_{*}$. The quantum effective action for this model does not contain an intrinsic scale. It is quantum scale invariant.

\subsection{Quantum scale invariant standard model with gravity}

An interacting fixed point of the type \eqref{eq:36} is not the only possible way to realize quantum scale symmetry in QCD. An alternative is the replacement of $\Lambda_{UV}$ in eq.~\eqref{eq:21} by a scalar field $\chi$ \cite{CWQ}. This may be supplemented by scale invariant kinetic and potential terms for $\chi$, as well as a coupling to gravity according to eq.~\eqref{eq:16}. The effective action no longer contains any intrinsic scale. In particular, $\Lambda_{QCD}$ is proportional to $\chi$, as exemplified by the one loop relation
\begin{equation}\label{eq:37} 
\Lambda_{QCD}=\chi\,\exp\left (-\dfrac{1}{b_{0}g^{2}(\chi)}\right )=\chi\,\exp\left (-\dfrac{1}{b_{0}\bar{g}^{2}}\right )\, .
\end{equation}
In eq.~\eqref{eq:37} $g^{2}(\chi)$, as defined by a vertex with momenta $p^{2}\sim \chi^{2}$, does not depend on $\chi$, e.g. $g^{2}(\chi)=\bar{g}^{2}$. (Otherwise this would imply a dependence on some intrinsic scale $\Lambda_{UV}$, that we have removed by the identification $\Lambda_{UV}=\chi$.)

The $\chi$-independence of $g^{2}(\chi)$ does not mean that the gauge coupling is not running if we compute it for momenta $p^{2}=\mu^{2}$, $\mu\neq \chi$. Indeed, with fixed $g^{2}(\chi)$ the running of the gauge coupling depends on the ratio $\mu/\chi$. To one loop order eq.~\eqref{eq:33} reads
\begin{equation}\label{eq:38} 
g^{-2}(\mu)=g^{-2}(\chi)-\dfrac{b_{0}}{2}\ln\left (\dfrac{\chi^{2}}{\mu^{2}}\right ).
\end{equation}
In case of spontaneous breaking of scale symmetry by a non-vanishing value of $\chi$ one may define an effective field theory for momenta $p^{2}<\chi^{2}$. The running \eqref{eq:38} accounts for the running of the gauge coupling in this effective field theory with $\chi$ kept fixed. The fact that only the ratio $\chi^{2}/\mu^{2}$ appears in the logarithm \eqref{eq:38} has a simple reason. If we lower the momenta $p^{2}$ in some $n$-point function from $p^{2}=\chi^{2}$ to $p^{2}=\mu^{2}$, we include the effects of additional fluctuations in the range between $\mu$ and $\chi$. For these contributions $\chi$ acts as an UV-cutoff, and $\mu$ as an IR-cutoff. Thus the ratio $\mu/\chi$ appears in the difference between $g^{2}(\mu)$ and $g^{2}(\chi)$.

The scale invariant version of QCD arises from a fixed point in the running of $g^{2}(\chi)$ with $\chi$,
\begin{equation}\label{eq:39} 
\chi\p_{\chi}g^{2}(\chi)=\hat{\beta}(g^{2}(\chi))\com \hat{\beta}(\bar{g}^{2})=0\, ,
\end{equation}
with $g^{2}(\chi)$ in eq.~\eqref{eq:37} identified with the fixed point value $\bar{g}^{2}$. Thus quantum scale symmetry is again connected to a fixed point. We will discuss in sect.~\ref{sec:Mass_Pseudo_Goldstone} how this fixed point could be associated to a fixed point in quantum gravity. There we also discuss the conditions under which the quantum scale invariant standard model is realized.

We emphasize that the function $\hat{\beta}$ in eq.~\eqref{eq:39} differs from the perturbative beta function $\beta_{g^{2}}$ in eq.~\eqref{eq:30}. The flow of the gauge coupling is determined by two different $\beta$-functions, $\beta_{g^{2}}$ and $\hat{\beta}$. While $\hat{\beta}$ accounts for the change of the gauge coupling at momenta $p^{2}=\chi^{2}$ with $\chi$, the function $\beta_{g^{2}}$ describes the response of the renormalizable gauge coupling to a change in the ratio $p^{2}/\chi^{2}$. Instead of $p^{2}=\chi^{2}$ we can take any fixed ratio $\mu^{2}/\chi^{2}=c^{2}$. Keeping $c$ fixed, e.g., by varying the momentum scale $\mu$ together with $\chi$, the response of the gauge coupling is given by $\hat{\beta}$, as seen in eq.~\eqref{eq:38}
\begin{equation}\label{eq:40} 
g^{-2}(c\chi)=g^{-2}(\chi)+b_{0}\ln(c).
\end{equation}
Thus $\hat{\beta}$ accounts for the change under a simultaneous variation of $\mu$ and $\chi$, while $\beta_{g^{2}}$ describes the relative variation of $\mu/\chi$, e.g. the change with $c$. This holds for arbitrarily small $c$. In consequence, quantum scale symmetry is perfectly compatible with a non-vanishing $\beta$-function $\beta_{g^{2}}$. The fixed point condition that leads to quantum scale symmetry is $\hat{\beta}=0$. We will discuss the relation between the different $\beta$-functions in detail in sect.~\ref{sec:Flow_In_Field_Space}.

The extension to a quantum scale invariant standard model coupled to gravity \cite{CWQ,SHAZEN} is straightforward. All dimensionless couplings, as gauge couplings, Yukawa couplings or quartic scalar couplings, take $\chi$-independent values if defined at $p^{2}=\mu^{2}=\chi^{2}$. This generalizes the fixed point \eqref{eq:39}. The running of these couplings with a change of the ratio $\mu^{2}/\chi^{2}$ is accounted for by the well known perturbative $\beta$-functions that do not vanish.

No intrinsic mass scale is present in the scalar sector as well. In particular, the effective potential for the Higgs doublet takes the approximate form
\begin{equation}\label{eq:41} 
U_{H}=\dfrac{1}{2}\lambda_{H}\left (\dfrac{\tilde{\mu}}{\chi}\right )\left (h^{\dagger}h-\varepsilon\left (\dfrac{\tilde{\mu}}{\chi}\right )\chi^{2}\right )^{2}\, ,
\end{equation}
neglecting higher order couplings. Here $\tilde{\mu}=c\chi$ is some appropriately fixed physical IR-scale proportional to $\chi$, such that $U_{H}$ indeed does not involve any intrinsic mass scale. A rather natural choice is an identification of $\tilde{\mu}$ with the Fermi scale, as given by the implicit relation
\begin{equation}\label{eq:42} 
c^{2}=\varepsilon(c).
\end{equation}
The running of $\lambda_{H}$ and $\varepsilon$ with a change of the ratio $\mu/\chi$ is given by the usual perturbative $\beta$-functions. In particular, the running of $\varepsilon$ with $\mu/\chi$ involves for $\mu/\chi\ll 1$ the perturbative anomalous mass dimension of the Higgs doublet \cite{CWPFP,CWFT}, given up to small corrections by
\begin{equation}\label{eq:43} 
\mu\, \p_{\mu}\varepsilon\biggl |_{\chi}=A\varepsilon\com A=\dfrac{1}{16\pi^{2}}\left (6\lambda_{H}+6y_{t}^{2}-\dfrac{9}{2}g_{2}^{2}-\dfrac{9}{10}g_{1}^{2}\right ).
\end{equation}
Here $g_{2}$ and $g_{1}$ are the gauge couplings of $SU(2)$ and $U(1)$, respectively, and $y_{t}$ is the Yukawa coupling of the top quark. We will later discuss the important property that the flow of $\varepsilon$ has a partial fixed point for $\varepsilon=0$.

With this setting the effective action for the standard model does not contain any parameter with dimension of mass or length. This is the quantum scale invariant standard model introduced in ref.~\cite{CWQ}. Its coupling to gravity contains in addition to the part \eqref{eq:16} a coupling of the Higgs boson to gravity \eqref{eq:19AA}, as discussed in ref.~\cite{SMWEI}. This term is important for Higgs inflation \cite{BS1}, as we will discuss in sect.~\ref{sec:Higgs_Infl}, or scale invariant Higgs inflation in sect.~\ref{sec:Scale_Inv_Higgs_Infl}. The quantum scale invariant standard model corresponds to a model defined precisely at a fixed point.

\subsection{Spontaneous scale symmetry breaking}\label{sec:SponScalSymBre} 

The scale invariant standard model can be consistent with observation only for $\chi_{0}\neq 0$, implying a spontaneous breaking of scale symmetry. Inserting $\tilde{\mu}=c\chi$ one has in eq.~\eqref{eq:41} constant $\lambda_{H}$ and $\varepsilon$ that we assume both positive. The most general scale invariant potential including the term from eq.~\eqref{eq:16} reads
\begin{equation}\label{eq:2.44A} 
U=\dfrac{\lambda_{H}}{2}\left (h^{\dagger}h-\varepsilon\chi^{2}\right )^{2}+\tilde{\lambda}\chi^{4}\, .
\end{equation}
For $\tilde{\lambda}>0$ the unique minimum of $U$ occurs for $\chi=0$, $h=0$. This ``symmetric'' state preserves scale symmetry. All particles are massless for this state. For $\tilde{\lambda}=0$ the potential becomes degenerate. There is a valley of minima for arbitrary $\chi=\chi_{0}$, with $h_{0}^{\dagger}h_{0}=\varepsilon\chi_{0}^{2}$. For any $\chi_{0}\neq 0$ scale symmetry is spontaneously broken. The particle spectrum is massive, as required by observation. As for any spontaneously broken symmetry there is a massless Goldstone boson that only has derivative couplings -- the dilaton.

In the absence of scale invariant gravity the ground state is given by the minimum of the effective potential. Without further input a value $\tilde{\lambda}=0$ may seem artificial. Then spontaneous scale symmetry would appear as a rather accidental event only for the particular choice $\tilde{\lambda}=0$. This is the reason why for the statistical physics of critical phenomena scale symmetry is an exact symmetry at a second order phase transition which is usually not spontaneously broken. For scale invariant gravity the situation changes profoundly. In the presence of the scale invariant couplings to the curvature scalar \eqref{eq:16}, \eqref{eq:19} the solution of the field equations is a de Sitter space with constant $R$ given by a cosmological constant $\tilde{\lambda}\chi^{4}_{0}$. The value of $\chi_{0}$ is arbitrary for every $\tilde{\lambda}\geq 0$. Thus spontaneous breaking of scale symmetry by the cosmological solution is a generic phenomenon, in contrast to flat space where it only occurs for $\tilde{\lambda}=0$.

The value of $\chi_{0}$ setting the scale for de Sitter space is arbitrary, including $\chi_{0}=0$ for which geometry is flat space. This situation may seem somewhat paradoxical since the massive particle spectrum for $\chi_{0}\neq 0$ looks rather different from the massless particle spectrum for $\chi_{0}=0$. For any $\chi_{0}\neq 0$ the value $\chi_{0}$ is not observable. Only dimensionless ratios can be observed as, for example, the ratio between the momentum of a particle and $\chi_{0}$. For any nonzero, even arbitrarily small value of $\chi_{0}$ it makes sense to speak about particles with momenta smaller or larger than its mass which is proportional to $\chi_{0}$. This differs from the scale symmetric state with $\chi_{0}=0$, for which there is no momentum which is smaller than the mass. These issues will play a role for our discussion of scale symmetry in cosmology in sect.~\ref{sec:Scale_Sym_Cosmo}. The presence of a massless Goldstone boson for any $\chi_{0}\neq 0$, even in case of $\tilde{\lambda}>0$ where the potential has no flat direction, will become visible there. Actually, the non-linear field corresponding to the Goldstone boson is most easily seen after a Weyl scaling to the Einstein frame as discussed in sect.~\ref{sec:Weyl_Scaling}.

The scalar field $\chi$ appearing in the quantum scale invariant standard model or scale invariant gravity \eqref{eq:16} may be called ``metron'', since it sets in a scale invariant theory all scales of mass or length used for measurements. For $\chi_{0}\neq 0$ it is closely related to the dilaton, the Goldstone boson of spontaneously broken scale symmetry. In an appropriate normalization the dilaton is given by
\begin{equation}\label{eq:DIA} 
\tau=\ln\left (\chi/\chi_{0}\right )\, .
\end{equation}
Under scale transformations the metron transforms multiplicatively, while the dilaton is shifted additively,
\begin{equation}\label{eq:DIB} 
\chi^{\prime}=\alpha\chi\com \tau^{\prime}=\tau+\ln(\alpha)\, .
\end{equation}
It is this shift symmetry that forbids all non-derivative couplings of the dilaton $\tau$ in a field basis where the other fields are invariant. This field basis is the Einstein frame discussed in sect.~\ref{sec:Weyl_Scaling}. Die to this close relation the metron is also often called somewhat unprecisely dilaton. A precise wording should distinguish between the metron and the dilaton. The metron is well defined for arbitrary values, while the dilaton exists only as an excitation around a non-zero $\chi_{0}$.

While the value of $\chi_{0}$ is not observable for $\chi_{0}>0$, the value of the dimensionless coupling $\tilde{\lambda}$ is. The effective squared Planck mass for a given de-Sitter solution is
\begin{equation}\label{eq:2.44B} 
M^{2}=\chi_{0}^{2}+\xi_{H}h_{0}^{\dagger}h_{0}=\chi_{0}^{2}\left (1+\xi_{H}\varepsilon\right )\, .
\end{equation}
The observable dimensionless ratio between the cosmological constant $\lambda=\tilde{\lambda}\chi_{0}^{4}$ and the fourth power of $M$ is given by
\begin{equation}\label{eq:2.44C} 
\dfrac{\lambda}{M^{4}}=\dfrac{\tilde{\lambda}}{(1+\xi_{H}\varepsilon)^{2}}\, .
\end{equation}
In the present universe $\tilde{\lambda}$ must be a very small quantity, $\tilde{\lambda}\lesssim 10^{-120}$. Thus $\tilde{\lambda}=0$ is not needed for spontaneous scale symmetry breaking, but a very small value of $\tilde{\lambda}$ is required for realistic cosmology. We will address this issue later. We will see in sect.~\ref{sec:Infrared_QG} that a non-zero value of $\tilde{\lambda}$ is not compatible with a fixed point of asymptotically safe quantum gravity.

This issue generalizes to other scale invariant models. After solving the field equations for possible additional scalar fields one ends with a single scalar field $\chi$, where quantum scale invariance implies in four dimensions
\begin{equation}\label{eq:2.46A} 
U=\tilde{\lambda}\chi^{4}\, .
\end{equation}
As has been noted in refs.~\cite{AMRAB,RASABA,BERRAB}, the scalar effective potential at its minimum always vanishes, independently if in flat space scale symmetry is spontaneously broken (for $\tilde{\lambda}=0$) or not (for $\tilde{\lambda}\geq 0$). This does not help for the cosmological constant problem which concerns the coupling to gravity. The reason is that the solutions to the field equations with constant $\chi_{0}$ do not require $\chi_{0}$ to be at a minimum of the effective potential, as we have discussed above. In a scale invariant theory of gravity the squared Planck mass is replaced by $M^{2}(\chi)=\xi\chi^{2}$. The observable dimensionless ratio $U(\chi)/M^{4}(\chi)$ is given by $\tilde{\lambda}/\xi^{2}$
\begin{equation}\label{eq:2.46B} 
\dfrac{U}{M^{4}(\chi)}=\dfrac{\tilde{\lambda}}{\xi^{2}}\, .
\end{equation}
Observation requires a present value $\tilde{\lambda}/\xi^{2}\approx10^{-120}$ that needs explanation. We will see that in quantum gravity $\tilde{\lambda}/\xi^{2}$ depends on the field $\chi$ and vanishes in the infrared limit for $\chi\raw\infty$.

\subsection{Quantum scale symmetry and quantum effective action}\label{sec:QScaleSymAndQuqnatumEffAction}

Our discussion of quantum scale symmetry is based on the quantum effective action $\Gamma$. For the example of a scalar field theory, $\Gamma$ obtains from the classical action $S$ by the implicit definition
\begin{equation}\label{eq:DEF1} 
\exp\left (-\Gamma[\varphi]\right )=\int\,\cD\chi\exp\left \{-S[\varphi+\chi]+\int_{x}\, \dfrac{\p\Gamma}{\p\varphi}\chi\right \}\, .
\end{equation}
Here $\varphi(x)$ is the macroscopic field and $\chi(x)$ denotes the fluctuating field, with $\varphi+\chi$ the microscopic field. The functional integral over $\chi$ accounts for all quantum fluctuations. The exact quantum field equation
\begin{equation}\label{eq:DEF2} 
\dfrac{\p\Gamma}{\p\varphi(x)}=J(x)
\end{equation}
involves the source term $J(x)$. The effective action is the generating functional for the one-particle-irreducible Green's functions. It obtains from the generating function for the connected Green's function $W[J]$ by a Legendre transform
\begin{equation}\label{eq:DEF3} 
\Gamma[\varphi]=-W[J]+\int_{x}\,  J\varphi\com \varphi=\dfrac{\p W}{\p J}\, .
\end{equation}
For local gauge theories one may proceed to gauge fixing. For a physical gauge fixing one can extract a gauge invariant effective action \cite{CWGF,CWGIF}.

Quantum scale symmetry is realized whenever one can find suitable renormalized fields, such that the quantum effective action, expressed as a functional of these renormalized fields, contains no intrinsic mass scale. If the renormalized fields have canonical kinetic terms, the scale transformations acting on the renormalized fields are given by eqs.~\eqref{eq:6}, \eqref{eq:7}, e.g. according to the canonical scaling dimensions of fermions, scalars, gauge fields, metric and vierbein. (Recall that the canonical scaling dimension does not equal the canonical mass dimension. Gauge fields have mass dimension one and scaling dimension zero.) If no intrinsic mass scale is present, the effective action $\Gamma$ is invariant under this scale transformation. This follows from dimensional analysis, as we show in the following.

Diffeomorphism symmetry, or Lorentz symmetry for flat space, implies that every factor $\p_{\mu}$ or $A_{\mu}$ with a lower Lorentz index has to be contracted by an inverse vierbein $e_{m}^{\mu}$, or pairs of such lower indices are contracted with $g^{\mn}$. We assume that in the particle sector the renormalized fields can be chosen such that their kinetic term, e.g. the term with the lowest nonzero number of derivatives, is canonical. The canonical scaling dimensions of the renormalized fields are determined by scale invariance of the canonical kinetic terms. The invariance of the gauge fields, together with the invariant coordinates, guarantees the invariance of covariant derivatives $D_{\mu}$, and therefore extends scale invariance to covariant kinetic terms. This also holds if the metric connection is included in the covariant derivatives.

For the non-derivative terms, as the scalar effective potential or masses or Yukawa couplings for fermions, we observe that omission of a factor $\gamma^{m}e_{m}^{\mu}D_{\mu}$ or $g^{\mn}D_{\mu}D_{\nu}$ in the canonical kinetic terms for fermions or scalars produces expressions with canonical mass dimension $m^{3}$ or $m^{2}$. Thus fermion bilinears as $\bar{\psi}\psi$ or scalar bilinears as $h^{\dagger}h$ have to be multiplied by a quantity with dimension $m$ or $m^{2}$, respectively. If this quantity is field independent such non-derivative terms introduce intrinsic masses. In this case they are also not scale invariant, since a factor $\alpha$ or $\alpha^{2}$ is missing due to the missing inverse vierbein or inverse metric. The only possibilities for avoiding the introduction of an intrinsic mass, and which are consistent with diffeomorphism symmetry, is the multiplication of a fermion bilinear by a scalar, and the multiplication of a scalar bilinear by another scalar bilinear. Such terms are scale invariant since the missing factors of $\alpha$ or $\alpha^{2}$ are precisely compensated by the scalar fields. If we do not insist on polynomial interactions the scalars can be replaced by $(\bar{\psi}\psi)^{\frac{1}{3}}$. This does not change the argument.

For higher derivative terms we observe that each additional derivative produces an additional factor of $\alpha$ since it has to be contracted with an inverse vierbein. If the introduction of an intrinsic mass is avoided, each derivative has to be multiplied with the inverse of a scalar field. Thus the basic building blocks for higher derivative terms without intrinsic mass scales are $\chi^{-1} e_{m}^{\mu}D_{\mu}$. These building blocks are invariant under the canonical scale transformation \eqref{eq:6}, \eqref{eq:7}. For flat space and cartesian coordinates the reader may find it easier to switch to an equivalent scale transformation where $\p_{\mu}$ and $A_{\mu}$ scale $\sim\alpha$, while $e_{\mu}^{m}$ and $g_{\mn}$ are invariant. There is a direct match between canonical mass dimension and canonical scaling dimension in this case, such that the argument based on dimensional analysis becomes straightforward.

If $\Gamma$ is invariant under the scale transformations \eqref{eq:6}, \eqref{eq:7} we can directly perform the Noether construction for conserved currents \cite{CWQ,PSW}. One uses for this purpose that the quantum field equations obtained from the first functional derivative of $\Gamma$ are exact. This contrasts with the classical action. We discuss the quantum dilatation current and its conservation explicitly in appendix~\ref{app:A}. One concludes that the absence of an intrinsic mass in $\Gamma$ guarantees exact scale symmetry on the quantum level. We have already seen how the absence of an intrinsic mass scale arises from the presence of a fixed point for the running couplings. Therefore, a fixed point guarantees quantum scale symmetry, in accordance with the expectations. What is not needed for this argument is classical scale invariance. The classical or microscopic action may involve intrinsic mass scales. At a fixed point the memory of such microscopic intrinsic mass scales is erased by the quantum fluctuations. At a fixed point, quantum fluctuations generate scale symmetry, rather than destroying it.

We finally stress again that the canonical scaling applies to the renormalized fields. These renormalized fields may be linear or even non-linear expressions of the microscopic fields used to formulate the functional integral. Translating the canonical scaling of the renormalized fields back to the microscopic fields induces ``anomalous dimensions'' for the microscopic fields.

\subsection{Renormalizable quantum field theories and scale symmetry}

We next argue that quantum scale symmetry is always realized for the UV-behavior of continuum renormalizable quantum field theories. This argument is based on the necessary presence of an UV-fixed point for continuum renormalizable quantum field theories. Quantum scale symmetry is therefore not an exceptional property for particular cases. It is a central general ingredient for renormalizable quantum field theories.

Before presenting the argument for the above statement we should specify our definition of a renormalizable continuum quantum field theory. First, the notion of continuum field theory requires that the theory can be formulated for arbitrarily small distance in continuum space in terms of a finite number of fields. (This includes a finite number of fields in higher dimensions which appear as infinitely many four-dimensional fields after dimensional reduction. In contrast, string theories or discrete lattice theories are not in this class.) Second, by ``renormalizable'' we understand that the theory can indeed be continued to arbitrarily short distances. This goes beyond perturbative renormalizability. In our sense renormalizability is not a technical statement, as often used, but rather concerns the validity of the theory at arbitrarily small distances. We may call this a ``complete renormalizable theory''. For example, the standard model may not be renormalizable in our sense due to the presence of a Landau pole in the abelian (hypercharge) gauge coupling. It is perturbatively renormalizable, however, which guarantees that is is a valid effective low energy theory.

The presence of a fixed point renders a quantum field theory renormalizable, in the sense that it can be used consistently up to infinitely short distances or infinite momenta. If the theory is defined precisely at the fixed point it is scale invariant. Since no intrinsic momentum scale $\bar{\mu}$ is present, the theory is valid for arbitrary momenta and can be continued to $p^{2}\raw\infty$. More generally, a renormalizable theory is realized if the flow of couplings reaches a fixed point for $p^{2}\raw\infty$. Thus a fixed point is sufficient for defining a renormalizable quantum field theory.

For a quantum field theory that can be extrapolated to infinitely short distances a fixed point is also necessary. If the dimensionless couplings do not reach fixed values for $\mu\raw\infty$ they have to diverge, or first run to a ``forbidden region'' where the theory becomes inconsistent for other reasons. This signals the breakdown of a given model at some finite UV-scale $\Lambda_{UV}$. This is perfectly admissible for an effective low energy theory, but not for a theory that remains valid for $p^{2}\raw\infty$. (To be more precise, one has to specify which dimensionless couplings need to remain finite for a valid description. Only those need to reach fixed points in the UV. Other couplings, for example related to particular derivatives with respect to momenta or fields, may diverge for $p^{2}\raw\infty$. This happens if the UV-theory shows non-analytic behavior.)

Continuum quantum field theories as QCD are sometimes regularized by other, perhaps discrete, types of models as lattice gauge theories. Similar discrete formulations are proposed for quantum gravity. Such formulations may exhibit an intrinsic regularization scale $\Lambda_{UV}$ given, for example, by the inverse lattice distance. If such regularizations admit a continuum limit where the effective physics can be described in terms of continuous spacetime, and if for given ``physical momenta'' $p$ the regularization scale $\Lambda_{UV}$ can be moved to infinity, this continuum limit has again to correspond to a fixed point. (In practice, one often uses fixed points associated to a second order phase transition in parameter space.)

If a continuum quantum field theory remains valid for $p^{2}\raw\infty$ the existence of a fixed point for its running dimensionless couplings is mandatory. The existence of a fixed point does not exclude the presence of an intrinsic mass scale $\bar{\mu}$. Nevertheless, for $p^{2}/\bar{\mu}^{2}\raw\infty$ the effect of the intrinsic mass scale $\bar{\mu}$ becomes negligible. Then scale symmetry is realized in the UV-limit for this case too. We conclude that scale symmetry in the UV is the generic case for all continuum quantum field theories that remain valid for arbitrarily small distances. This is not the only possibility to formulate consistent theories - for example, there could be some intrinsic smallest distance or largest momentum. Even in this case (approximate) scale symmetry becomes relevant if the intrinsic momentum scale is far above the intrinsic mass scales and relevant physical momenta.

There exist two types of UV-fixed points that can be used to define renormalizable quantum field theories. For the first, the fixed point occurs for vanishing couplings, more precisely for a free quantum field theory without interactions. This is asymptotic freedom \cite{GW,POL}. Such theories are perturbatively renormalizable. One can perform a loop expansion, corresponding to a power series in (arbitrarily) small couplings close to the fixed point. Renormalization is possible on a diagrammatic level, with the usual subtraction of counterterms order by order in the expansion. For the other possibility, called asymptotic safety \cite{WEIAS}, the fixed point occurs for nonzero couplings. The theory is renormalizable, but typically not in a perturbative approach. Most models for critical phenomena in statistical physics are of the asymptotic safety type. A typical example is the Wilson-Fisher fixed point for three-dimensional scalar models with $O(N)$-symmetry. Asymptotic safety does, however, not exclude perturbative renormalizability. A fixed point with interactions does not contradict the perturbative assumption of small couplings. Perturbatively renormalizable models with asymptotic safety have been investigated rather systematically in ref.~\cite{LS}. Quantum gravity is presumably not a perturbatively renormalizable quantum field theory. A consistent quantum field theory for the metric therefore needs the existence of an UV-fixed point of the asymptotic safety type. Non-vanishing interactions at the fixed point seem rather natural since gravitational interactions cannot be ``turned off''. As for any other UV-fixed point, asymptotically safe quantum gravity is scale invariant in the UV-limit.

\subsection{Relevant parameters}\label{sec:RelevantParameters}

We next discuss the flow of couplings close to a fixed point. We consider a certain number of dimensionless couplings $g_{i}$, with fixed point values $g_{i*}$. In principle, the number of couplings in a quantum field theory is infinite, associated, for example, to coefficients of an expansion in powers of fields and momenta. For simplicity we only take a finite number, which becomes a good approximation if the number is sufficiently large and the couplings cover the relevant aspects of the model. The flow equation for the couplings is specified by their $\beta$-functions
\begin{equation}\label{eq:44} 
\p_{t}g_{i}=\mu\p_{\mu} g_{i}=\beta_{i}(g_{j})\, .
\end{equation}
A fixed point with couplings $g_{*}=\lbrace g_{j*}\rbrace$ occurs if all $\beta$-functions vanish
\begin{equation}\label{eq:2.45AA} 
\beta_{i}(g_{*})=0\, .
\end{equation}
The linearization near the fixed point is characterized by the stability matrix $T$
\begin{equation}\label{eq:45} 
\p_{t}(g_{i}-g_{i*})=\dfrac{\p\beta_{i}}{\p g_{j}}\biggl |_{g_{*}}(g_{j}-g_{j*})=-T_{ij}(g_{j}-g_{j*}).
\end{equation}

The eigenvalues of $T$ are the critical exponents $\theta_{i}$. We can diagonalize $T$ and denote the eigenvectors by $v_{i}=D_{ij}(g_{j}-g_{j*})$,
\begin{equation}\label{eq:46} 
\p_{t}v_{i}=-\theta_{i}v_{i}.
\end{equation}
Eigenvectors belonging to positive $\theta_{i}$ are called relevant parameters, those for negative are irrelevant, and those with $\theta_{i}=0$ marginal. The solutions of eq.~\eqref{eq:46}, with arbitrary $\Lambda$,
\begin{equation}\label{eq:47} 
v_{i}(\mu)=\left (\dfrac{\Lambda}{\mu}\right )^{\theta_{i}}v_{i}(\Lambda)\, ,
\end{equation}
show an increase of the relevant parameters as $\mu$ is lowered, while irrelevant parameters decrease. For marginal couplings higher orders in the expansion are needed for a decision if they increase (marginally relevant) or decrease (marginally irrelevant).

If we assume finite $v_{i}(\Lambda)$ and take the limit $\Lambda/\mu\raw \infty$, the irrelevant couplings vanish. This is the central ingredient for the predictivity of quantum field theories. Any microscopic information about irrelevant parameters is lost, such that only the information about relevant or marginal parameters remains available. If the number of relevant parameters is finite, all observable quantities only depend on a finite number of parameters. (We include the marginally relevant parameters in the set of ``relevant parameters''.)

For the relevant parameters we can take arbitrary values $v_{i}(\mu)$ if $v_{i}$ remains sufficiently small such that eq.~\eqref{eq:45} remains valid. The relevant parameters vanish in the UV-limit $\mu\raw\infty$, e.g., they approach the fixed point values arbitrarily closely. This is expected for the UV-limit of a model defined at a given UV-fixed point. In the UV-limit the fixed point is approached arbitrarily closely and scale symmetry is realized.

All available information about a model is contained in the values of the relevant parameters $v_{i}(\mu)$ at some given scale $\bar{\mu}$. We assume their number $N_{r}$ to be finite. As the flow moves away from the fixed point, the linearized flow equation \eqref{eq:45} typically remains no longer valid. Nevertheless, no new information is added. As a result, at any scale $\mu$ the description of the model can only involve $N_{r}$ independent dimensionless quantities. They can be identified with the values of $N_{r}$ renormalizable couplings at the scale $\mu$. This is the reason why a renormalizable quantum field theory with a given content of fields is fully specified by the values of $N_{r}$ renormalizable couplings at some scale $\mu$. All memory about physics at some much higher scale, e.g., $\Lambda_{UV}$ in eq.~\eqref{eq:21}, can only be transported by the values of the renormalizable couplings. In the particular case of QCD without quarks there is only one (marginally) relevant parameter. It can be identified with the gauge coupling. This is the reason why the dependence of $Z_{F}$ in eq.~\eqref{eq:21} on $\Lambda_{UV}$ can only arise via the renormalized gauge coupling $g(\mu/\Lambda_{UV})$.

The number $N_{r}$ constitutes the maximal number of independent parameters for a renormalizable quantum field theory. The number of observables far exceeds $N_{r}$. Therefore many observables or relations between observables become predictable in renormalizable quantum field theories. This predictivity extends to a wide class of microscopic theories beyond renormalizable quantum field theories. If there is a large range of scales for which a continuum quantum field theory is valid, many of the possible couplings will show a flow-behavior similar to irrelevant parameters. Information about such couplings is lost in the long distance limit.

The number of of renormalizable couplings in the standard model may be larger than the number $N_{r}$ of relevant theories in the full quantum field theory including gravity. Then part of the renormalizable couplings of the standard model, or relations between these couplings, become predictable. Couplings that behave as relevant or marginal couplings from the point of view of the standard model may be irrelevant at the UV-fixed point which necessarily includes the fluctuations of gravitational degrees of freedom. An example is the quartic Higgs-boson coupling $\lambda_{H}$. Due to a substantial gravity-induced positive anomalous dimension this coupling is irrelevant at the UV-fixed point of asymptotically safe quantum gravity \cite{CWGFE}. Anticipating this behavior the mass of the Higgs scalar has been predicted from asymptotically safe gravity to be $126$GeV, with a few GeV uncertainty \cite{SW}. This prediction has been confirmed by the measurement of the Higgs boson mass at the LHC, finding a mass of $125$GeV \cite{AAD1,CHAT1}. We discuss this issue in detail in sect.~\ref{sec:6.17}.

\subsection{Relevant parameters and mass scales}\label{sec:RelevantParametersAndMassScales}

We have discussed before how running renormalizable dimensionless couplings can induce mass scales by dimensional transmutation. A first example is the ``confinement scale'' $\Lambda_{QCD}$ in QCD \eqref{eq:34}, which is dominantly responsible for the nucleon masses. In general, a given model will have more than one relevant parameter. As a consequence, rather different mass scales can be generated by the running couplings.

A second example is the standard model. The running of the strong gauge coupling $g_{3}$ generates the confinement scale $\Lambda_{QCD}$. One the other hand, in the electroweak sector we may consider the dimensionless ratio
\begin{equation}\label{eq:48} 
\gamma=\dfrac{\delta}{\mu^{2}}\, ,
\end{equation}
where $\delta$ is a parameter with dimension mass squared which measures the distance from the critical surface of the almost second order vacuum electroweak phase transition \cite{CWPFP,CWFT,CWQR,CWMY,AOISO}. We can treat $\gamma$ as one of the running dimensionless couplings. In the effective low energy theory, where graviton fluctuations play no role, the running of $\gamma$ obeys
\begin{equation}\label{eq:49} 
\p_{t} \gamma=(-2+A)\gamma\, ,
\end{equation}
with $A$ given by eq.~\eqref{eq:43}. In this range $\gamma$ behaves as a relevant parameter with critical exponent $\theta=2-A$. Since $\theta$ is large, a very small value of $\gamma(\Lambda_{UV})$ is needed in order to obtain $\gamma(\mu)\approx 1$ for $\mu$ near $\varphi_{0}$ - this is the gauge hierarchy problem. A relevant parameter $\gamma$ generates the Fermi scale $\varphi_{0}$ as the scale where it becomes of order one. The Fermi scale is only a factor $10^{3}$ larger than $\Lambda_{QCD}$, but theoretically it could be many orders of magnitude larger. Two relevant parameters $g_{3}$ and $\gamma$ can induce largely different mass scales. We discuss in sect.~\ref{sec:6.17} the question if $\gamma$ remains a relevant parameter in a more complete theory including gravity.

A third example for a hierarchy of scales generated by the flow of different relevant parameters is the standard model coupled to quantum gravity. In quantum gravity (without additional scalar singlet fields) the Planck mass corresponds to a relevant parameter. For a scale dependent Planck mass $M(\mu)$ we may define a dimensionless coupling similar to eq.~\eqref{eq:48}
\begin{equation}\label{eq:50} 
w(\mu)=\dfrac{M^{2}(\mu)}{2\mu^{2}}.
\end{equation}
In the momentum regime where the fluctuations of gravitational degrees of freedom play an important role the running of $M^{2}(\mu)$ may be substantial. For a fixed point one expects a constant value $w(\mu)=w_{*}$. If the flow of $w(\mu)$ away from the fixed point is characterized by positive $\theta$ the Planck mass correspond to relevant parameter. If $w(\mu)$ scales for small $\mu$ proportional to $\mu^{-2}$ its running defines the Planck mass $M=M(\mu\raw 0)$ as an intrinsic scale.

At this place we do not give a precise definition of $M^{2}(\mu)$, nor do we present a computation of its evolution equation. We will be satisfied with a simple illustration how the dimension of $M^{2}$ turns $w$ into a relevant parameter. Qualitative properties of the quantum gravity fixed point may be illustrated by assuming an effective action of the form
\begin{equation}\label{eq:51} 
\cL_{R}=-\dfrac{1}{2}f(R).
\end{equation}
We define
\begin{equation}\label{eq:52} 
M^{2}(R)=\dfrac{\p f}{\p R}\, ,
\end{equation}
and associate
\begin{equation}\label{eq:53} 
\mu^{2}=R\, .
\end{equation}
This can be motivated by the fact that a curved geometry can act as an IR-cutoff, such that $R$ can replace external momenta for the definition of running couplings. As a result, $w(\mu)$ is identified with $w(R)$,
\begin{equation}\label{eq:54} 
w(R)=\dfrac{1}{2R}\dfrac{\p f}{\p R}\, .
\end{equation}
There are several reasons why the associations \eqref{eq:53}, \eqref{eq:54} are only partly valid. A better treatment of a flowing Planck mass will be discussed within functional renormalization in sect.~\ref{sec:V}. Nevertheless, some key features become already visible in this setting.

Let us assume a flow equation for $w$ of the form
\begin{equation}\label{eq:55} 
\p_{t}w=\beta(w)=-2w+2c-\dfrac{dc^{2}}{2w}\, ,
\end{equation}
with positive constants $c$ and $d$. This is motivated by the canonical scaling for $\mu^{2}/M^{2}\raw 0$, $w\raw\infty$, where graviton fluctuations decouple and $M^{2}$ becomes independent of $\mu$, such that $\beta(w\raw\infty)\raw -2w$. On the other hand, negative $w$ correspond to instabilities, such that the flow of $w$ should not cross  a barrier at $w=0$. We assume $d<1$ such that the $\beta$-function has two fixed points,
\begin{equation}\label{eq:56} 
w_{*}=\dfrac{c}{2}\left (1\pm\sqrt{1-d}\right ).
\end{equation}
The UV-fixed point for quantum gravity corresponds to the plus sign in eq.~\eqref{eq:56}. For small $d$ the critical exponent is close to two
\begin{equation}\label{eq:57} 
\theta=-\dfrac{\p\beta}{\p w}\biggl |_{w_{*}}=2-\dfrac{2d}{(1+\sqrt{1-d})^{2}}.
\end{equation}

For $w>w_{*}$ the relevant parameter $w-w_{*}$ increases as $\mu$ is lowered, reaching for large $w$ an asymptotic behavior $w\sim \mu^{-2}$, such that $M^{2}(\mu)=2\mu^{2}w(\mu)$ indeed settles to a constant value. With $M^{2}(R)$ independent of $R$ eqs.~\eqref{eq:51}, \eqref{eq:52} yield in the IR limit the standard Einstein-Hilbert action for gravity, $\cL_{R}=-M^{2}R/2+\ldots$. In the UV-limit the effective action becomes scale invariant, $\cL_{R}=-w_{*}R^{2}/2$. The qualitative behavior of $f$ can be approximated by
\begin{equation}\label{eq:58} 
\cL_{R}=-\dfrac{1}{2}\left (w_{*}R^{2}+M^{2}R\right ).
\end{equation}
For small $d$ this approximation becomes even quantitatively rather close to the solution of the flow equation \eqref{eq:55}. With eq.~\eqref{eq:58} describing the limits $R\raw\infty$ and $R\raw 0$, an expansion around the UV-fixed point for $R\raw\infty$ yields
\begin{equation}\label{eq:2.59AB} 
f(R)=w_{*}R^{2}+cR^{2-\frac{\theta}{2}}\, .
\end{equation}
The lessons from this simple illustration are:
\begin{enumerate}
\item For large $R$ (mimicking the UV-limit) the effective action $\cL\sim R^{2}$ becomes scale invariant.
\item A relevant parameter induces for small $R$ the Planck mass by dimensional transmutation.
\item There is no reason why the UV-limit of the effective action should be analytic -- in our example this happens only if $\theta$ equals two precisely.
\end{enumerate}

There is no reason why the mass scales $M$ and $\Lambda_{QCD}$ should be close to each other. This demonstrates in a simple way that largely hierarchical mass ratios as $\Lambda_{QCD}/M$ can be generated in a very natural way if several relevant parameters are present. This holds, in particular, for marginally relevant parameters or parameters with critical exponents close to zero. For QCD it is sufficient that the gauge coupling $g^{2}(M)$ is small at the scale $M$ generated by the flow in the gravitational sector.

In the linear approximation \eqref{eq:45} all relevant parameters $v_{i}$ grow independently of each other, each one according to its own critical exponent $\theta_{i}$. This could suggest that an independent intrinsic mass scale can be associated to every relevant parameter, more or less associated to the scale $\mu_{i}$ where $v_{i}(\mu_{i})\approx 1$. The independent running of the relevant parameters does not hold, however, beyond the linear regime. Once one of the relevant parameters has grown large enough (typically of the order one) it can influence the flow of the other relevant parameters. An example is the interplay between the scales $\Lambda_{QCD}$ and $\varphi_{0}$. It provides a lower bound for $\varphi_{0}/\Lambda_{QCD}$. Indeed, the vacuum electroweak phase transition is not precisely second order, but rather weakly first order or a crossover. Quark-antiquark condensates induced by the strong interactions give a mass to the $W$- and $Z$-bosons, similar to the Higgs-mechanism. There is therefore a minimal mass for these bosons that we may associate with a minimal value of the generalized Fermi constant. A minimal value of the order parameter of spontaneous symmetry breaking is not compatible with a second order phase transition. Nevertheless, the critical phenomena associated to a second order phase transition will be a very good approximation for all scales $\mu$ sufficiently above the minimal value of the order parameter.

\subsection{Exact and approximate quantum scale\\
 symmetry}\label{sec:Exact_Approx_Quantum_Scale_Sym} 

At this point we can compare models with exact and approximate quantum scale symmetry. Exact scale symmetry is realized if the model is defined precisely on a fixed point. This includes  the case of finite theories. For finite theories dimensionless couplings are scale independent in the UV. If there are free dimensionless couplings, the $\beta$-functions have formally a degenerate zero on the hypersurface of these couplings. 

It has been proposed to obtain a scale invariant quantum field theory by replacing a fixed mass scale in the regularization of the theory or in the renormalization condition by a scalar field $\chi$ \cite{CWQ,ETG,SHAZEN,SHATKA,MMSV,RPERS,COPR,MORPER}. We may consider this issue in the light of Wilsonian renormalization flow. A replacement of some ultraviolet cutoff $\Lambda_{UV}$ by $\chi$ is valid for an effective ``low energy theory'' for momenta smaller than $\chi$. For momenta larger than $\chi$, however, the scalar field can no longer act as an UV-cutoff. If couplings flow in the momentum range above $\chi$ a fixed point is needed. Only for UV-finite theories $\chi$ can act as an UV-regulator. Finiteness may be obtained if the range of possible momenta is restricted, as for lattice theories \cite{WILLAT}. In this case there is no continuum field theory for momenta of the order of $\chi$. For example, ref.~\cite{SHAZEN} proposes to replace a fixed lattice distance by the inverse of a scalar field $\chi^{-1}$, thereby realizing a scale invariant regularization.

Scale symmetry may also be obtained if one choses for the renormalization condition a dynamical scalar $\chi$ instead of a fixed momentum scale $\mu$. We will discuss in sect.~\ref{sec:Flow_In_Field_Space} that dimensionless couplings $g$ typically depend both on $\chi$ and on $\mu$. Such a normalization condition therefore reads more precisely $g(\mu=\chi)=\bar{g}$. A fixed value of $\bar{g}$ is possible if $g$ depends only on the ratio $\chi/\mu$, e.g. $g(\mu,\chi)=g(\chi/\mu)$. In this case scale symmetry is indeed exact. In contrast, if $\bar{g}$ depends on $\chi$ the model contains again an intrinsic mass scale, now connected to the flow of $\bar{g}(\chi)$ with $\chi$. We will argue in sect.~\ref{sec:Flow_In_Field_Space} that a $\chi$-independent $\bar{g}$ obtains precisely if $\bar{g}=g_{*}$ is the value of $g$ at some UV-fixed point. Either $g-g_{*}$ is an irrelevant parameter, or it is relevant but the theory is defined precisely on the fixed point. We present a detailed computation for this issue in sect.~\ref{sec:Mass_Pseudo_Goldstone} and give quantitative cosmological bounds on the $\chi$-dependence of $\bar{g}$ in sect.~\ref{sec:Approach_SM_FP}. (For these arguments $g$ may be considered as a multi-component dimensionless coupling.)

For a renormalizable continuum quantum field in our strict sense it should be possible to follow the flow of $g(\mu,\chi)$ for fixed $\chi$ to $\mu\raw\infty$. This is indeed the case if $g(\mu\raw\infty,\chi)=g_{*}^{\prime}$ is given by a fixed point value. For a scale invariant model defined precisely on a fixed point the ``scaling function'' $g(\chi/\mu)$ does not involve an intrinsic mass scale. The fixed point values $g_{*}=g(\chi/\mu=1)$, $g_{*}^{\prime}=g(\chi/\mu=0)$ or $g_{*}^{\prime\prime}=g(\chi/\mu\raw\infty)$ are different facets of this scaling function. They all do not depend on $\chi$. 

Models with approximate quantum scale symmetry are defined in the ``vicinity of an UV-fixed point''. This means that the fixed point is reached only for infinite momenta, while for any finite momenta at least one of the relevant parameters differs from its fixed point value. In this case the flow of relevant couplings away from the fixed point induces an intrinsic scale $\bar{\mu}$ which explicitly breaks scale symmetry. One can choose $\bar{\mu}$ as the largest intrinsic mass scale, and express all other mass scales in terms of $\bar{\mu}$ and dimensionless couplings. The value of $\bar{\mu}$ is arbitrary - it only sets the units of mass and is not measurable.

Consider now physics in a typical range of momenta $\mu$. As long as $\mu\gg \bar{\mu}$ the explicit scale symmetry breaking by $\bar{\mu}$ is negligible and the physics of an exactly scale invariant model is not distinguished from approximate scale symmetry. Observables can only depend on the ratio $\bar{\mu}/\mu$. For fixed $\mu$ one can take the limit $\bar{\mu}\raw 0$. This limit is smooth and realizes exact scale symmetry. Decreasing $\bar{\mu}$ towards zero simply means that the model is defined closer and closer to the fixed point. Alternatively, one may keep $\bar{\mu}$ fixed and approach the UV-fixed point for $\mu\raw\infty$. As we will discuss in sect.~\ref{sec:Flow_In_Field_Space}, the momentum scale $\mu$ may be replaced by a field expectation value $\chi$.

\section{Networks of fixed points and crossovers}\label{sec:Networks_FP_Crossover}

In this section we discuss the presence of several fixed points and the associated crossover between them. This is a central ingredient for the understanding of different versions of scale symmetry, e.g. one for particle physics coupled to gravity, and another one for particle physics without gravity. We include in our discussion approximate fixed points for which all couplings run very slowly instead of being precisely constant. This section substantiates the flow diagram Fig.~\ref{fig:AA}. We specify the SM-fixed point and display simple flow equations for the couplings $w$ and $\gamma$ whose trajectories are shown in Fig.~\ref{fig:AA}. In the present section we treat the Planck mass $M$ as an intrinsic scale. In the following sections we will extend the discussion to the case where $M$ is given by a scalar field and reflects spontaneous scale symmetry breaking.

The simple general properties discussed in this section are basic for what seems to be realized in Nature, namely the important role of different (approximate) fixed points. Such fixed points are necessarily connected by crossover trajectories.

\subsection{UV- and IR-fixed points}

An ultraviolet fixed point is reached as momenta grow and length scales shrink. In the opposite, an infrared fixed point is approached for decreasing momenta. We associate momenta here with the renormalization scale $\mu$. A renormalizable quantum field theory is defined at the UV-fixed point. In case of the existence of several fixed points in a system of flowing couplings we define as the UV-fixed point the one on which the quantum field theory is defined. Thus flow trajectories towards the IR (lowering $\mu$) always start at the UV-fixed point. If a flow trajectory encounters several approximate fixed points, the order of the fixed points from the UV to the IR is defined by the direction of this flow. The relative ordering is not free since trajectories have to exist between these fixed points, and their flow direction orders the sequence. If a trajectory comes close to an exact fixed point and departs from its vicinity for lower scales, we subsume this situation under a trajectory passing an approximate fixed point. Near an exact fixed point all couplings run indeed very slowly.

For an UV-fixed point with several relevant parameters a sequence of approximate fixed points is a rather general case. Different relevant parameters are often associated to different intrinsic mass scales. Assume that at the largest intrinsic mass scale a certain number of degrees of freedom decouples from the other particles, either because their masses exceed $\mu$, or their interactions get very small. The remaining massless particles form a subsystem, which is characterized by an approximate fixed point if the interactions within this subsystem still change only slowly. If several distinct intrinsic mass scales are present, this pattern can be repeated several times.

\subsection{Particle scale symmetry}\label{sec:ParticleScaleSym}

This type of sequence happens in quantum gravity coupled to the standard model. The highest intrinsic mass scale can be associated with the Planck mass $M$. For $\mu\ll M$ the contributions of the gravitational degrees of freedom to the flow of the standard model couplings become tiny. They involve inverse powers of $M$ and are therefore suppressed by powers of $\mu / M$. The gravitational degrees of freedom decouple. For $\mu\ll M$ the flow of the standard model couplings is determined by the perturbative $\beta$-functions of the standard model without gravity.

For $\mu\ll M$ the running of the standard model couplings is slow. At the scale $M$ the dimensionless gauge, Yukawa and quartic scalar couplings are still close to their ``quantum gravity fixed points'', determined by the UV-fixed point, and the slow running away from it. For $\mu <M$ the further flow of these couplings follows the perturbative renormalization group equations. Since for $\mu=M$ all these couplings are small, their running is very slow. The dimensionless parameter $\gamma$ measuring the distance to the (almost) second order electroweak phase transition has to be tiny at the scale $\mu=M$. Following eq.~\eqref{eq:49} it remains very small for the whole range $\mu\gg\varphi_{0}$. This behavior defines the approximate fixed point for the standard model without gravity. It is valid for scales $\varphi_{0}\ll\mu\ll M$. For $\mu\approx \varphi_{0}$ the relevant parameter $\gamma$ becomes of order one. This ends the range of validity of the standard model approximate fixed point. 

For the whole range $\varphi_{0}\ll\mu\ll M$ scale symmetry is a good approximation. We call the scale symmetry associated to the standard model fixed point ``particle scale symmetry'', in distinction to ``gravity scale symmetry'' which includes the gravitational degrees of freedom and becomes relevant for the UV-fixed point of quantum gravity coupled to particle physics. The essential ingredient for the existence of particle scale symmetry is the second order character of the vacuum electroweak phase transition. The critical behavior for a second order phase transition is always characterized by an exact fixed point and the associated exact quantum scale symmetry.

There is an exact fixed point that can be associated to the slow running of couplings in the standard model. This is the SM-fixed point in Fig.~\ref{fig:AA}. The SM-fixed point is a trivial fixed point with all gauge couplings and Yukawa couplings vanishing, as well as $\gamma=0$, $\lambda_{H}=0$ in the Higgs sector. For vanishing gauge and Yukawa couplings the vacuum electroweak phase transition is indeed of second order. We can add a small nonzero $\lambda_{H}$ as a marginally irrelevant coupling. It runs very slowly, $\beta_{\lambda}\sim\lambda_{H}^{2}$, such it can be taken effectively as constant. For vanishing non-abelian gauge couplings $g_{2}$ and $g_{3}$ also the Yukawa couplings and the hypercharge gauge coupling $g_{1}$ are marginally irrelevant and run very slowly towards zero. For vanishing gauge couplings $g_{3}=g_{2}=g_{1}=0$ the parameter $\gamma$ is the only relevant parameter. For the flow between the UV-fixed point and the SM-fixed point, it is the SM-fixed point that plays the role of an IR-fixed point. As characteristic for an IR-fixed point the flow towards the UV is unstable. Every nonzero Yukawa coupling, as well as $\lambda_{H}$, has a Landau pole. Since the validity of this fixed point is restricted to $\mu< M$, these Landau poles far above $M$ do not matter in practice.

The gauge couplings $g_{3}$ and $g_{2}$ are marginally relevant. As long as they remain small, they do not disturb the second order character of the vacuum electroweak phase transition and therefore the existence of the fixed point. Once one of these couplings gets large, however, it will typically induce a minimal mass for the $W$- and $Z$-bosons. The vacuum electroweak phase transition does not stay second order, but rather becomes first order or a crossover. For the standard model with $\gamma\raw 0$ the first coupling growing large is the strong gauge coupling $g_{3}$. This induces a minimal $W$, $Z$-mass of the order of $\Lambda_{QCD}^{(6)}$, e.g., the confinement scale obtained for six flavors of massless quarks. (For $\gamma=0$ all quarks presumably remain massless for $\mu\gg\Lambda_{QCD}^{(6)}$.) For a given $g_{3}(M)$ one finds $\Lambda_{QCD}^{(6)}$ substantially smaller than the confinement scale in the presence of the physical Fermi scale, far below $100$ MeV. Non-zero $g_{2}$ and $g_{1}$ can drive $\lambda_{H}$ towards negative values and induce a first order phase transition by the Coleman-Weinberg mechanism \cite{COLW}. For realistic values of the top-Yukawa coupling this effect can be neglected for $\mu>\Lambda_{QCD}^{(6)}$. 

We conclude that for the relevant range $\mu>\varphi_{0}$ one can neglect all effects related to the small minimal mass of the $W$, $Z$-bosons and treat the vacuum electroweak phase transition effectively as second order. Nevertheless, the presence of non-zero asymptotically free gauge couplings $g_{2}$ and $g_{3}$ implies that the SM-fixed point cannot be reached exactly. Incorporating this in Fig.~\ref{fig:AA} means that even the trajectory with $\gamma_{0}=0$ does not reach the SM-fixed point but finally has to bend towards the IR-fixed point.

There are two possible views on the particle scale symmetry associated to the approximate standard model fixed point. The first considers it as a consequence of the vicinity of the exact standard model fixed point described above. The transition at $\mu\approx M$ is as a crossover from the UV-fixed point of quantum gravity to the SM-fixed point for the standard model. Since the SM-fixed point is a genuine fixed point, particle scale symmetry is a necessary consequence of this fixed point. This holds independently of what happens in the short distance theory. It does not involve properties of the UV-fixed point and would hold even in models without an UV-fixed point. In this view particle scale symmetry emerges as the SM-fixed point is approached.

The second point of view considers particle scale symmetry as a remnant of the scale symmetry of the UV-fixed point. The scale symmetry at the UV-fixed point is explicitly broken if the Planck mass is an intrinsic scale associated to a relevant parameter. Nevertheless, this breaking only concerns the gravitational sector. The scale symmetry for the particle subsector with fixed metric for flat space is not affected. In this sense particle scale symmetry may be considered as originating from the scale symmetry at the UV-fixed point in a situation where explicit scale symmetry breaking only concerns degrees of freedom that decouple from the SM. Indeed, the second order character of the vacuum electroweak phase transition is already a property of the UV-fixed point. It implies a running $\p_{t}\gamma\sim \gamma$ for all scales. It dictates properties of the flow of couplings already for the UV-fixed point. The line $\gamma=0$ in Fig.~\ref{fig:AA} is protected from crossing by any trajectory at all scales. The connection to the UV-fixed point is smooth. At the crossover scale $M$ the change of the couplings is small. Only their $\beta$-functions undergo a rapid change.

\subsection{Sequence of particle physics fixed points}\label{sec:Sequence_Part_Phys_Fixed_Point}

Particle physics needs ingredients beyond the standard model. The observed neutrino oscillations imply a minimal value for the mass of the heaviest neutrino $\bar{m}_{\nu}$. In the standard model neutrino-masses cannot be described by renormalized couplings. A neutrino mass term for the left-handed neutrinos, which belong to the same doublet as the left-handed charged leptons, has to be quadratic in the expectation value $\varphi_{0}$ of the Higgs doublet
\begin{equation}\label{eq:59} 
m_{\nu}= H_{\nu}\dfrac{\varphi_{0}^{2}}{M_{B-L}}.
\end{equation}
Here $M_{B-L}$ is a heavy mass scale, often (but not necessarily) associated to spontaneous symmetry breaking of a local $B-L$-symmetry. Indeed, mass terms involving only the three known neutrinos violate the difference between baryon number $B$ and lepton number $L$ by two units. The parameter $H_{\nu}$ is dimensionless and reflects dimensionless combinations of couplings in some beyond-standard model sector that is responsible for the mass \eqref{eq:59}.

Typically, the parameter $H_{\nu}$ cannot be much larger than one. For $H_{\nu}<1$ and $\bar{m}_{\nu}>0.05$ eV this yields an upper bound for $M_{B-L}$
\begin{equation}\label{eq:60} 
M_{B-L}<6\cdot 10^{14}\,\mathrm{GeV}.
\end{equation}
This is substantially smaller than the Planck mass $M$. One concludes that the observed neutrino properties require an intermediate scale \cite{CWNBL}. A typical value may be $M_{B-L}=10^{12}$ GeV, with much smaller values requiring very small $H_{\nu}$. For $\mu>M_{B-L}$ additional particles beyond the standard model are needed and contribute to the flow equations. These could be singlet neutrinos (sterile or right-handed neutrinos) \cite{MINNEU,YAN,GMRS} or a $SU(2)$-triplet of massive scalars \cite{MACW,LSW,MOSE,SCHEVA}. The running in the range $M_{B-L}<\mu <M$ will correspond to an extended standard model fixed point, with $\beta$-functions including contributions of the additional particles. If the number of additional particles is not very high, this is quantitatively a small effect.

There may be additional characteristic mass scales between $M$ and $\varphi_{0}$, or even below $\varphi_{0}$. For example, grand unified models require spontaneous symmetry breaking at a scale $M_{GUT}$ not too far below the Planck scale. A typical sequence of scales in particle physics is depicted in Fig.~\ref{fig:1}. All these scale have ultimately to be connected with separate relevant parameters at the UV-fixed point, or some of their ratios have to become predictable. We conclude that sequences of largely different scales can be a rather natural consequence of properties of an UV-fixed point.

\begin{widetext}
\onecolumngrid
\begin{figure}[h]
\includegraphics[scale=0.45]{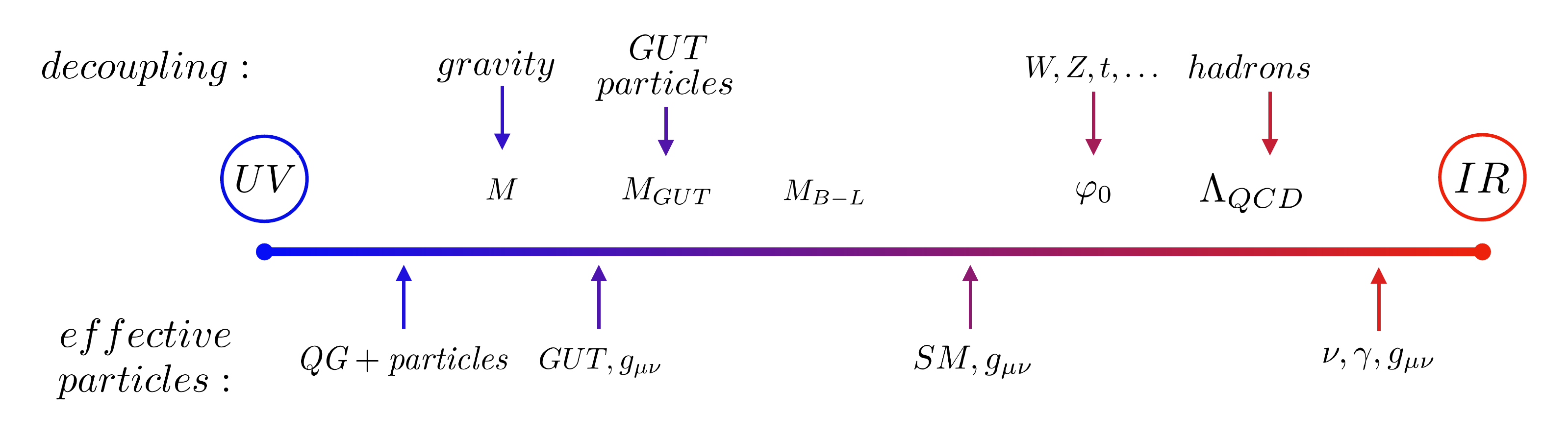}
\caption{Fixed points, including approximate ones, and crossover scales. For a minimal extension of the standard model the GUT-fixed point with associated crossover scale $M_{GUT}$ is not present.}\label{fig:1} 
\end{figure}
\end{widetext}
\twocolumngrid

\subsection{Predictivity in presence of several fixed points or partial fixed points}\label{sec:Predictivity_Sev_Fixed_Point}

In a renormalizable quantum field theory the predictivity for observables at some scale $\mu$, say $\mu=1$ GeV, depends on properties of all the fixed points encountered by the relevant flow trajectory in the space of couplings. First of all, the standard model fixed point implies that to a high accuracy all observables depend only on the renormalizable couplings of the standard model. Those can be associated to the relevant and marginal parameters for the standard model fixed point. Corrections involve non-renormalizable operators that are suppressed by typical mass scales where the validity of the standard model ends. An example are the neutrino masses \eqref{eq:59}.

Second, the number of free parameters cannot exceed the number of relevant parameters for the UV-fixed point. If this number is smaller than the number of renormalizable couplings in the standard model, some relations between standard model couplings can be predicted. An example is the quartic coupling of the Higgs scalar $\lambda_{H}$. In the standard model this is a renormalizable coupling, corresponding to a marginal parameter. For the UV-fixed point there are strong indications that $\lambda_{H}$ is an irrelevant parameter. The graviton fluctuations induce an anomalous dimension $A_{\lambda}>0$ which is of the order one \cite{PEPE,NAPER,CWGFE,EHLY}. Combining this with contributions from gauge bosons, fermions and possible other scalars yields
\begin{equation}\label{eq:61} 
\p_{t}\lambda_{H}=A_{H}\lambda_{H}-C_{H}\, ,
\end{equation}
with
\begin{equation}\label{eq:62} 
C_{H}=\dfrac{1}{16\pi^{2}}\left (b_{h}y^{4}-b_{g}g^{4}-b_{s}\lambda_{s}^{2}\right ).
\end{equation}
The coefficient $b_{h}$ depends on the fermions present for $\mu>M$ and their Yukawa couplings, and similarly for $b_{g}$ for the gauge bosons and $b_{s}$ for the scalars. (For more details see sect.~\ref{sec:6.17} where also the omitted terms from the scalar anomalous dimension and non-minimal gravitational couplings are added.) The couplings $y$, $g$ and $\lambda_{s}$ (for the scalars) are small and run very slowly for $\mu >M$, such that $C_{H}$ is small. One infers for the UV-fixed point a small value
\begin{equation}\label{eq:63} 
\lambda_{H*}=\dfrac{C_{H}}{A_{H}}.
\end{equation}
Here we neglect in $C_{H}$ the terms $\sim\lambda_{H}^{2}$ which only give small corrections. The critical exponent associated to $\lambda_{H}$ reads $\theta_{H}\approx -A_{H}$. It is negative, such that $\lambda_{H}$ is indeed an irrelevant parameter. In practice, a flow equation \eqref{eq:63} leads to the value $\lambda_{H}(M)\approx 0$. The increase of $\lambda_{H}$ for $\mu<M$ can be computed perturbatively. Including only standard model particles and taking for definiteness $C_{H}$ as obtained from the fermions and gauge bosons of the standard model has led to the prediction \cite{SW} of a Higgs boson mass of $126$ GeV with a few GeV uncertainty.

A third source of predictivity concerns partial fixed points for the flow of couplings in the standard model. Partial fixed points are related to the non-linear flow equations for marginal couplings, for which the $\beta$-function \eqref{eq:45} vanishes in the linear approximation. One example is the partial IR-fixed point \cite{CWPFP,CWMHB} in the ratio $\lambda_{H}/y_{t}^{2}$ which restricts the mass of the Higgs scalar as a function of the mass of the top quark. Since dimensionless couplings in the standard model run only slowly, the partial fixed point is not yet reached for $\mu=\varphi_{0}$. For the predictivity of $\lambda_{H}$ this replaces the partial fixed point value by an infrared interval \cite{CWMHB}, which determines upper and lower bounds for the mass of the Higgs scalar \cite{MPR,CMP}. A similar partial fixed point exists for the ratio $g_{3}^{2}/y_{t}^{2}$ \cite{PR,CHI,CWPFP,CWMHB}.

\subsection{Crossover}

Let us assume that the flow of couplings admits both an UV- and an IR-fixed point, with a ``crossover-trajectory'' leading from the first to the second. Unless a theory is exactly defined on one of the fixed points, the crossover situation defines an intrinsic mass scale $\bar{\mu}$. It is determined by fixing the value of the couplings at a given renormalization scale $\mu=\bar{\mu}$. For example, one may define $\bar{\mu}$ such that couplings are in some sense ``half way'' between the UV- and IR-fixed points. Comparing characteristic momenta $p$ of a given process with $\bar{\mu}$ permits an assessment where the situation can be placed on the critical trajectory. For $p\gg\bar{\mu}$ one is close to the UV-fixed point, whereas for $p\ll\bar{\mu}$ the situation is close to the IR-fixed point. If there is only one intrinsic scale $\bar{\mu}$ this is not a measurable quantity. It only sets units of mass and length. What is measurable are ratios as $p^{2}/\bar{\mu}^{2}$. They express the measured momenta in units of the intrinsic scale $\bar{\mu}$.

For a concrete example we consider the flow of a single dimensionless coupling $\lambda$. The flow equations are assumed to exhibit an UV-fixed point at $\lambda_{UV}$, and an IR-fixed point at $\lambda_{IR}$,
\begin{equation}\label{eq:64} 
\p_{t}\lambda=\beta_{\lambda}=c(\lambda-\lambda_{UV})(\lambda-\lambda_{IR}).
\end{equation}
We take $c>0$ with $\lambda_{UV}<\lambda_{IR}$. The general solution reads
\begin{equation}\label{eq:65} 
\dfrac{\lambda_{IR}-\lambda}{\lambda-\lambda_{UV}}=\left (\dfrac{\mu}{\bar{\mu}}\right )^{c(\lambda_{IR}-\lambda_{UV})}.
\end{equation}
As $\mu$ decreases from infinity to zero this describes a crossover from $\lambda(\mu\raw\infty)=\lambda_{UV}$ to $\lambda(\mu\raw 0)=\lambda_{IR}$. We have chosen the intrinsic scale $\bar{\mu}$ such that
\begin{equation}\label{eq:66} 
\lambda(\bar{\mu})=\dfrac{1}{2}\left (\lambda_{IR}-\lambda_{UV}\right ).
\end{equation}

For $\mu\ll\bar{\mu}$ one has approximately
\begin{equation}\label{eq:67} 
\lambda(\mu)=\lambda_{IR}-(\lambda_{IR}-\lambda_{UV})\left (\dfrac{\mu}{\bar{\mu}}\right )^{c(\lambda_{IR}-\lambda_{UV})}.
\end{equation}
The relative change of $\lambda$ becomes tiny
\begin{equation}\label{eq:68} 
\dfrac{1}{\lambda}\p_{t}\lambda=-c\dfrac{(\lambda_{IR}-\lambda_{UV})^{2}}{\lambda_{IR}}\left (\dfrac{\mu}{\bar{\mu}}\right )^{c(\lambda_{IR}-\lambda_{UV})}.
\end{equation}
Similarly, for $\mu\raw\infty$ and $\lambda$ close to $\lambda_{UV}$ the flow is very slow. We encounter a situation where scale symmetry is not an overall property of the theory. It is realized approximately only in specific momentum regions, namely the UV-region $p^{2}\gg \bar{\mu}^{2}$ and the IR-region $p^{2}\ll \bar{\mu}^{2}$. In the crossover region for $p^{2}\approx\bar{\mu}^{2}$ scale symmetry can be strongly violated.

For a sequence with several fixed points and several crossover regions scale symmetry is realized in the momentum ranges corresponding to $\mu$ in the vicinity of the fixed points, while substantial running and strong scale symmetry violations can be expected in the crossover regions. We note that only one of the intrinsic crossover scales sets the units. Ratios between two different crossover scales, as the gauge hierarchy $\varphi_{0}/M$, are measurable quantities. The overall conclusion of this section states that quantum scale symmetry is typically not an overall property of a theory. It is rather realized effectively in certain regions of momentum of field space, that can be associated to approximate fixed points.

\subsection{Flow diagram for particle physics and quantum gravity}\label{sec:Flow_Diagram_PP_QG} 

Having established the general properties of fixed points and the crossover between them, we can now address the flow diagram Fig.~\ref{fig:AA}. It accounts for three fixed points (UV,SM,IR), and the crossover between them. We restrict the discussion to two parameters, a running squared Planck mass $M^{2}(k)$ and a running deviation from the electroweak phase transition $\delta(k)$. Typically, $\delta(k)$ is a type of mass term for the Higgs boson with dimension mass squared -- for more details see sect.~\ref{sec:Particle_Scale_Sym}. The value $\delta=0$ corresponds to the vacuum electroweak phase transition. We use here $k$ as a generalized renormalization scale, not necessarily related to nonvanishing momentum.

We assume simple flow equations whose characteristic features are given by dimensional analysis and general properties. The flow equation for the running Planck mass, $\p_{t}=k\p_{k}$,
\begin{equation}\label{eq:FD1} 
\p_{t}M^{2}=4ck^{2}
\end{equation}
follows from dimensional analysis and we assume constant $c$. For $\delta$ we only include the contribution from gravitational fluctuations
\begin{equation}\label{eq:FD2} 
\p_{t}\delta=\dfrac{2bk^{2}}{M^{2}}\delta\, .
\end{equation}
The right hand side has to vanish for $\delta=0$ since the critical surface of a second order phase transition cannot be crossed. Gravity decouples for $k^{2}/M^{2}\raw 0$, explaining the factor $k^{2}/M^{2}$. We assume constant $b$. It is related to the gravitational contribution to $A=b/w$ in eq.~\eqref{eq:49}. Direct computations for quantum gravity in sect.~\ref{sec:V} indeed yield eqs.~\eqref{eq:FD1}, \eqref{eq:FD2} as valid approximations. A typical value, cf. eq.~\eqref{eq:6.278A}, is $b\approx0.1$, while $c$ is of a similar order of magnitude, depending on the precise particle content of the model.

Employing dimensionless quantities,
\begin{equation}\label{eq:FD3} 
\gamma=\dfrac{\delta}{k^{2}}\com w=\dfrac{M^{2}}{2k^{2}}\, ,
\end{equation}
the flow equations read
\begin{align}\label{eq:FD4} 
\p_{t}w&=-2w+2c\, ,\nn\\
\p_{t}\gamma&=\left (-2+\dfrac{b}{w}\right )\gamma\, .
\end{align}
The first equation corresponds to $d\raw 0$ in eq.~\eqref{eq:55}. The general solution of eq.~\eqref{eq:FD4} is given by
\begin{align}\label{eq:FD5} 
w&=c+\dfrac{\bar{M}^{2}}{2k^{2}}\, ,\nn\\
\gamma&=\gamma_{0}(1+2c)^{-\frac{b}{2c}}\dfrac{\bar{M}^{2}}{k^{2}}\left (1+2c\dfrac{k^{2}}{\bar{M}^{2}}\right )^{\frac{b}{2c}}\, .
\end{align}
It involves two integration constants. One is a scale $\bar{M}$ that we may associate with the observed (constant) value of the Planck mass. The other is
\begin{equation}\label{eq:FD6} 
\gamma_{0}=\gamma(k=\bar{M})\, .
\end{equation}

Following the flow for $k\raw 0$ one arrives at
\begin{equation}\label{eq:FD6A} 
M^{2}(k=0)=\bar{M}^{2}\com \delta(k=0)=\gamma_{0}(1+2c)^{-\frac{b}{2c}}\,\bar{M}^{2}\, .
\end{equation}
The use of a dimensionless flow parameter
\begin{equation}\label{eq:FD7} 
x=\ln\left (\dfrac{k^{2}}{\bar{M}^{2}}\right )
\end{equation}
absorbs $\bar{M}^{2}$,
\begin{align}\label{eq:FD8} 
w&=c+\dfrac{1}{2}\ee^{-x}\, ,\nn\\
\gamma&=\gamma_{0}(1+2c)^{-\frac{b}{2c}}\,\ee^{-x}\left (1+2c\ee^{x}\right )^{\frac{b}{2c}}\, .
\end{align}

The flow trajectories run from the UV-fixed point for $x\raw\infty$ to the IR-fixed point for $x\raw -\infty$. Different trajectories in Fig.~\ref{fig:AA} are characterized by different values of $\gamma_{0}<0$ that correspond to spontaneous breaking of the electroweak symmetry. Positive $\gamma_{0}$ result in unbroken electroweak symmetry. We have not shown them in Fig.~\ref{fig:AA} since the diagram is symmetric under a reflection at the axis $\gamma=0$. With our simple ansatz the trajectories can be given explicitly
\begin{equation}\label{eq:EFT} 
\gamma=2\gamma_{0}(w-c)\left (\dfrac{w}{(1+2c)(w-c)}\right )^{\frac{b}{2c}}\, .
\end{equation}
For easier visualization we actually use for the $x$-axis the function $f_{w}=1/\sqrt{1+w^{2}}$ which approaches $w^{-1}$ for large $w$ (weak gravity) and maps $w\raw 0$ (strong gravity) to one. Similarly, we take for the $y$-axis the function $f_{\gamma}=\gamma/\sqrt{1+\gamma^{2}}$. It maps $\gamma\raw\pm\infty$ to $f_{\gamma}=\pm 1$.

The UV-fixed point is given by
\begin{equation}\label{eq:FD9} 
\text{UV:}\quad w_{*}=c\, ,\; f_{w*}=\dfrac{1}{\sqrt{1+c^{2}}}\, ,\; \gamma_{*}=0\, ,\; f_{\gamma*}=0\, .
\end{equation}
It is in the upper right corner of Fig.~\ref{fig:AA}, with $f_{w*}$ only slightly smaller than one. The IR-fixed point corresponds to
\begin{equation}\label{eq:FD10} 
\text{IR:}\quad w_{*}^{-1}=0\, ,\; f_{w*}=0\, ,\; \gamma^{-1}=0\, ,\; f_{\gamma*}=-1\, .
\end{equation}
It is the lower left corner of the flow diagram. The standard model fixed point is
\begin{equation}\label{eq:FD11} 
\text{SM:}\quad w_{*}^{-1}=0\, ,\; f_{w*}=0\, ,\; \gamma_{*}=0\, ,\; f_{\gamma*}=0\, .
\end{equation}
It is in the upper left corner of Fig.~\ref{fig:AA}. For small values of $|\gamma_{0}|$ one observes first a crossover from the UV-fixed point to the vicinity of the SM-fixed point. This is followed by a second crossover to the IR-fixed point. For $|\gamma_{0}|\raw 0$ the SM-fixed point is approached arbitrarily closely.

For Fig.~\ref{fig:AA} we have employed the parameters $b=0.1$, $c=0.1$. The overall picture of flow trajectories is similar as long as $b<2c$. For $b>2$ the parameter $\gamma$ becomes irrelevant and one encounters self organized criticality for the electroweak phase transition. Only $\gamma_{0}=0$ is possible. We discuss this issue in sect.~\ref{sec:Pred_Gauge_Hier}. There are also trajectories leaving the UV-fixed point in Fig.~\ref{fig:AA} towards the right. They correspond to negative $\bar{M}^{2}$ in eq.~\eqref{eq:FD5}. Before $M^{2}(k)=0$ is reached, gravity becomes very strong and the approximation \eqref{eq:FD1} is no longer applicable.

A more complete picture of the three fixed points and the crossover between them includes other couplings beyond $w$ and $\gamma$. Typically, $b$ and $c$ depend on such additional couplings. A quantity important for cosmology is $\tilde{\lambda}=U_{0}/M^{4}$, with $U_{0}$ the value of the scalar effective potential at the minimum of $U/M^{4}$. We discuss the flow of $\tilde{\lambda}$ in detail in sect.~\ref{sec:V}. We find that $\tilde{\lambda}$ flows to zero as one approaches the IR-fixed point. We find additional steps of crossover in the flow of $\tilde{\lambda}$ from its value at the UV-fixed point to the value at the IR-fixed point. As we have discussed in sect.~\ref{sec:Sequence_Part_Phys_Fixed_Point} also the crossover in the particle physics sector proceeds in several steps. Nevertheless, the simple flow equations \eqref{eq:FD4} provide already for a rather reliable overview of the overall situation in the interplay between particle physics and quantum gravity.

\section{Flow in field space}\label{sec:Flow_In_Field_Space}

So far we have discussed the flow as a function of characteristic momenta, associated to the renormalization scale $\mu$. Characteristic momenta are often not the only scales in a problem. There are also characteristic expectation values of fields, that we denote generically by $\chi$. If the scales $\mu$ and $\chi$ are different, one may also have a nontrivial flow in dependence of the ratio $\mu/\chi$. Since $\chi$ can be associated with spontaneous scale symmetry breaking, we will find new types of scale symmetry that exist in the spontaneously broken regime. The scale symmetric standard model discussed in sect.~\ref{sec:Fixed_points_and_scales} is of this type. Throughout this section $\chi$ is taken as a renormalized field.

The central role of a possible $\chi$-dependence of couplings is apparent in the simplified flow of the couplings $w$ and $\gamma$ in eq.~\eqref{eq:FD4}. These flow equations hold for fixed $\chi$, with $w$ and $\gamma$ being functions of $\chi$. In the general solution \eqref{eq:FD5} we can replace $\bar{M}^{2}$ by a function $F(\chi)$. Following the flow to $k=0$ therefore also replaces in eq.~\eqref{eq:FD6A} $\bar{M}^{2}$ by $F(\chi)$. For $F=\chi^{2}$ this realizes the scale invariant standard model, with
\begin{equation}\label{eq:4.0A} 
\varepsilon_{H}\sim -\gamma_{0}(1+2c)^{-\frac{b}{2c}}\, .
\end{equation}

The present section provides the basic setting for a discussion which version of the standard model arises from some given microscopic theory. In particular, we explain how the observed momentum dependence of the gauge coupling in QCD is compatible with exact quantum scale symmetry.

\subsection{Field dependent couplings}

A classical example for field-dependent couplings is the discussion of the effective potential for the Higgs field (or the analogue in an abelian $U(1)$-gauge theory) by Coleman and Weinberg \cite{COLW}. By virtue of the gauge symmetry the quantum effective potential $U(\rho)$ can only depend on $\rho=h^{\dagger}h$. One may define the field dependent quartic coupling $\lambda_{H}(\rho)$ by
\begin{equation}\label{eq:69} 
\lambda_{H}(\rho)=\dfrac{\p^{2}U}{\p\rho^{2}}\com \rho=h^{\dagger}h\, .
\end{equation}
Also the quadratic ``radial mass term'' is a field-dependent function,
\begin{equation}\label{eq:70} 
m_{H}^{2}(\rho)=\dfrac{\p U}{\p \rho}+2\lambda_{H}(\rho)\rho\, ,
\end{equation}
with observable mass of the Higgs boson given by $m_{H}(\rho_{0})$ at the location $\rho_{0}=\varphi_{0}^{2}$ of the minimum of $U(\rho)$. In a quartic approximation (quadratic in $\rho$) of the effective potential one has
\begin{equation}\label{eq:71} 
m^{2}_{H}(\rho)=\lambda_{H}(\rho)(3\rho-\rho_{0}).
\end{equation}

Let us first omit the gauge bosons and consider a single real scalar field, with discrete $\mathbb{Z}_{2}$-symmetry $\varphi\raw -\varphi$, $\rho=\varphi^{2}/2$. If $m^{2}_{H}(\rho)$ exceeds all relevant momenta, it acts as an infrared cutoff for the scalar fluctuations. Only momenta larger than $m_{H}(\rho)$ contribute to the running of the dimensionless coupling $\lambda_{H}$, such that the renormalized coupling depends now on the ratio $\Lambda_{UV}/m_{H}(\rho)$ instead of $\Lambda_{UV}/\mu$. For $\mu^{2}\ll m^{2}_{H}(\rho)$ the coupling $\lambda_{H}$ depends no longer on the momentum scale $\mu$.

We may define a quartic coupling $\lambda_{H}(\mu,\rho)$ that depends both on a characteristic momentum scale $\mu$ and the field expectation value $\rho$. Since $\lambda_{H}$ is dimensionless, the dependence on the scales involved is of the form
\begin{equation}\label{eq:72} 
\lambda_{H}(\mu,\rho)=\lambda_{H}\left (\dfrac{\mu}{\Lambda_{UV}},\dfrac{\mu^{2}}{\rho}\right ).
\end{equation}
The running takes the qualitative form
\begin{align}\label{eq:73} 
\mu\p_{\mu}\lambda_{H}&=\beta_{\lambda}(\lambda_{H})\,\theta(\mu-m_{H}(\rho))\, ,\nn\\
\rho\p_{\rho}\lambda_{H}&=\dfrac{1}{2}\beta_{\lambda}(\lambda_{H})\,\theta(m_{H}(\rho)-\mu)\, ,
\end{align}
where the step function $\theta$ may be replaced by an appropriate smooth threshold function for a more accurate computation. On a trajectory with a fixed ratio $\mu^{2}/\rho$ one recovers the $\beta$-function as computed from the variation of a single scale or the variation of $\Lambda_{UV}$,
\begin{equation}\label{eq:74} 
\left (\mu\p_{\mu}+2\rho\p_{\rho}\right )\lambda_{H}=-\Lambda_{UV}\dfrac{\p}{\p\Lambda_{UV}}\lambda_{H}=\beta_{\lambda}(\lambda_{H}).
\end{equation}
For $\rho\gg\rho_{0}$ and $m_{H}(\rho)\gg\mu$ we can approximate
\begin{equation}\label{eq:75} 
\rho\p_{\rho}\lambda_{H}=\dfrac{1}{2}\beta_{\lambda}(\lambda_{H}).
\end{equation}
The running of $\lambda_{H}$ with $\sqrt{\rho}$ is in this range the same as the running with $\mu$ in the range $\mu\gg m_{H}(\rho)$.

The inclusion of the $SU(2)$-gauge bosons and the extension to a complex doublet scalar field does not change the qualitative picture. The gauge bosons acquire an effective mass $\sim g\sqrt{\rho}$ which again acts as an effective infrared cutoff for $\rho\neq 0$. The form of eqs.~\eqref{eq:73}-\eqref{eq:75} remains the same, with threshold function reflecting both $\rho$ dependent masses, e.g. $m_{H}(\rho)$ and $g\sqrt{\rho}$. Further extension by including the $U(1)$-hypercharge gauge boson of the standard model shows similar features. There is now a mass split between the $W$- and $Z$-boson, while the photon remains massless for all values of $\rho$. The effect of the photon-fluctuations is independent of $\rho$ and does not contribute to the flow of $\lambda_{H}$.

\subsection{Spontaneous and explicit scale symmetry breaking}\label{sec:SponAndExpSymBr}

Let us associate the scalar singlet field or metron $\chi$ with the characteristic quantity for the spontaneous breaking of scale symmetry - typically it will be a variable Planck mass. We consider a dimensionless coupling $\lambda$ that depends both on the field $\chi$ and the characteristic momentum $\mu$ of a suitable vertex defining $\lambda$. In case of several dimensionless couplings $\lambda_{i}$ we may take $\lambda$ as a vector with components $\lambda_{i}$. If $\lambda$ depends only on the ratio $\mu/\chi$, quantum scale symmetry is realized. Since there is no trace of some regularization scale $\Lambda_{UV}$, no intrinsic scale is generated even if $\lambda$ runs as a function of $\mu/\chi$. The scale symmetric standard model discussed in sect.~\ref{sec:Fixed_points_and_scales} realizes quantum scale symmetry in this way. Still, scale symmetry is spontaneously broken by any nonzero value of $\chi$. In this case a precisely massless Goldstone boson will be present.

Explicit breaking of scale symmetry occurs if $\lambda$ depends, in addition, on some UV-regularization scale via the dimensionless ratios $\mu/\Lambda_{UV}$ or $\chi/\Lambda_{UV}$. Now the flow of $\lambda$ generates an intrinsic scale $\bar{\mu}$, breaking the scale symmetry explicitly. No precisely massless Goldstone boson will be found. For small values of $\bar{\mu}/\chi$ there will be, however, an almost massless pseudo-Goldstone boson. 

We may first discuss the running of $\lambda$ with $\chi$ at zero momentum, $\mu=0$. Then $\lambda$ can only depend on $\chi/\Lambda_{UV}$. The flow of $\lambda$ with $\chi$ at $\mu=0$ is described by
\begin{equation}\label{eq:76} 
\chi\p_{\chi}\lambda=\hat{\beta}_{\lambda}(\lambda).
\end{equation}
Any non-zero flow generator $\hat{\beta}_{\lambda}$ induces an intrinsic mass scale $\bar{\mu}$. On the other hand, a zero of $\hat{\beta}_{\lambda}$ generates at fixed point for the $\chi$-dependence of $\lambda$ at $\mu=0$. At the fixed point there is no longer any dependence on $\Lambda_{UV}$. No intrinsic scale is present and quantum scale symmetry is realized. The issue of scale symmetry depends therefore on the zeros of $\hat{\beta}_{\lambda}$, rather than on the dependence of $\lambda$ on the ratio $\mu/\chi$. Correspondingly, intrinsic scales $\bar{\mu}$ correspond to properties of $\hat{\beta}_{\lambda}$.

For the simple Coleman-Weinberg type scenario \eqref{eq:73} there is no difference between $\hat{\beta}_{\lambda}$ and $\beta_{\lambda}$. One could extract the dependence of $\lambda$ on $\Lambda_{UV}$ equally well from the running with $\mu$ for $\chi=0$. This is, however, not the most general situation. Let us assume that for momenta smaller than $\chi$ some particles decouple because of a mass $\sim\chi$ or because of interactions inversely proportional to $\chi$. This happens for gravity if we associate $\chi$ with a variable Planck mass. The gravitational degrees of freedom in the metric field decouple. Another example are grand unified theories with $\chi$ the unification mass. For $\mu\ll\chi$ one is left with an effective theory for light particles, while the heavy gauge bosons and scalars decouple. The expectation value $\chi$ separates the UV-theory for $\mu\gg\chi$ from the low energy effective theory for $\mu\ll\chi$. From the point of view of the UV-theory $\chi$ acts as a partial IR-cutoff, while from the point of view of the effective low energy theory it is an UV-cutoff beyond which the effective theory looses its validity. Varying $\mu$ at a fixed scale $\chi$ defines the flow generator $\beta_{\lambda}$,
\begin{equation}\label{eq:77} 
\mu\p_{\mu}\lambda_{|\chi}=\beta_{\lambda}(\lambda).
\end{equation}
Typically, $\beta_{\lambda}$ depends on the ratio $\mu/\chi$. For $\mu\gg \chi$ all particles contribute to the flow, while for $\mu\ll\chi$ only the light particles of the effective low energy theory contribute,
\begin{equation}\label{eq:4.9A} 
\beta_{\lambda}=\begin{cases}
\beta_{\lambda}^{(UV)}&\textit{for }\mu\gg\chi\\
\beta_{\lambda}^{(IR)}&\textit{for }\mu\ll\chi\, .
\end{cases}
\end{equation}

For $\mu\ll\chi$ the variation with $\mu$ at fixed $\chi$ induces a dependence of $\lambda$ on the ratio $\mu/\chi$, driven by the fluctuations of the light particles with $\beta_{\lambda}=\beta_{\lambda}^{(IR)}$. For our examples of the standard model coupled to gravity or embedded in a grand unified theory the function, $\beta_{\lambda}^{(IR)}$ is given by the perturbative $\beta$-function of the standard model.

In contrast, the function $\hat{\beta}_{\lambda}$ accounts for the variation of $\lambda$ with fixed ratio $\mu/\chi$, generalizing eq.~\eqref{eq:76},
\begin{equation}\label{eq:77A} 
\chi\p_{\chi}\lambda_{|\mu/\chi}=\hat{\beta}_{\lambda}(\lambda).
\end{equation}
(The case $\mu=0$ is a particular case with $\mu/\chi=0$.) It involves the effects of fluctuations of all particles, including the ``heavy particles'' in a grand unified theory, or including the fluctuations of the gravitational degrees of freedom. The functions $\beta_{\lambda}$ and $\hat{\beta}_{\lambda}$ differ. They both depend on the ratio $\mu/\chi$ and on the renormalizable couplings of the model. The properties of $\hat{\beta}_{\lambda}$ decide on the realization of quantum scale symmetry and on the existence or not of an exactly massless Goldstone boson. Independently of the form of $\beta_{\lambda}$ quantum scale symmetry is realized, and spontaneously broken for $\chi\neq 0$, if $\hat{\beta}_{\lambda}$ vanishes at a fixed point. The Goldstone boson of spontaneously broken scale symmetry is present despite the non-trivial dependence of $\lambda$ on $\mu/\chi$, as described for $\mu\ll\chi$ by non-vanishing $\beta_{\lambda}^{(IR)}$. For the standard model coupled to gravity the UV-fixed point corresponds to vanishing $\hat{\beta}_{\lambda}$.

The running of couplings with $\mu$ for fixed $\chi$ may again induce scales, as $\Lambda_{QCD}$ for the strong interactions. For $\hat{\beta}_{\lambda}=0$ such scales are all proportional to $\chi$ as in eq.~\eqref{eq:37}. Only for $\hat{\beta}_{\lambda}\neq 0$ scales as $\Lambda_{QCD}$ also depend on $\Lambda_{UV}$, reflecting explicit scale symmetry breaking by the presence of intrinsic mass scales. For $\hat{\beta}_{\lambda}\neq 0$ the coupling $\bar{g}$ in eq.~\eqref{eq:37} is no longer $\chi$-independent, but rather depends on $\chi/\Lambda_{UV}$ according to the solution of eq.~\eqref{eq:76}.

We may summarize the relations between the different flow generators by admitting a general dependence of the dimensionless coupling $\lambda(\mu ,\chi ,\Lambda)=\lambda(r,s)$ on the dimensionless ratios $r=\mu/\chi$ and $s=\chi/\Lambda$, with $\Lambda=\Lambda_{UV}$ some intrinsic UV-regularization scale. With
\begin{align}\label{eq:4.10A} 
\beta_{\lambda}&=\mu\p_{\mu}\lambda_{|\chi,\Lambda}=(\mu\p_{\mu} r)\dfrac{\p\lambda}{\p r}\, ,
\end{align}
and
\begin{align}\label{eq:4.10B}
\hat{\beta}_{\lambda}&=\chi\p_{\chi}\lambda_{|\mu/\chi,\Lambda}=(\chi\p_{\chi} s)\dfrac{\p\lambda}{\p s}\, ,
\end{align}
the $\chi$-dependence at fixed $\mu$ and $\Lambda$ involves both the dependence on $r$ and $s$,
\begin{align}
\label{eq:77B}\chi\p_{\chi}\lambda_{|\mu,\Lambda}&=(\chi\p_{\chi} r)\dfrac{\p\lambda}{\p r}+(\chi\p_{\chi} s)\dfrac{\p\lambda}{\p s}\, .
\end{align}
Employing $\chi\p_{\chi}r=-\mu\p_{\mu}r$, one has
\begin{equation}\label{eq:77C} 
\chi\p_{\chi}\lambda_{\mu,\Lambda}=\hat{\beta}_{\lambda}-\beta_{\lambda}\, ,
\end{equation}
or
\begin{equation}\label{eq:77D} 
\hat{\beta}_{\lambda}=\beta_{\lambda}+\chi\p_{\chi}\lambda |_{\mu,\Lambda}\, .
\end{equation}
We observe that the dependence on $\Lambda$, and therefore the possible existence of an intrinsic scale $\bar{\mu}$, arises only through the dependence on $s$,
\begin{equation}\label{eq:4.13B} 
\Lambda\p_{\Lambda}\lambda=(\Lambda\p_{\Lambda}s)\dfrac{\p\lambda}{\p s}=-(\chi\p_{\chi}s)\dfrac{\p\lambda}{\p s}=-\hat{\beta}_{\lambda}\, .
\end{equation}

Similar to $\lambda$, the derivatives $\beta_{\lambda}$ and $\hat{\beta}_{\lambda}$ are functions of $r$ and $s$. The dependence on $s$ arises only indirectly through the dependence of renormalized couplings on $s$. The dependence on $r$ is induced both by $r$-dependent renormalized couplings and possible ``threshold functions'' depending explicitly on $r$. We may investigate limiting cases. For $\mu\gg\chi$ the presence of $\chi$ plays no role for the running of $\lambda$ with $\mu$, such that $\lambda$ only depends on $\mu/\Lambda$ in this range. For $\mu\gg\chi$ one therefore has
\begin{equation}\label{eq:4.13A} 
\mu\p_{\mu}\lambda=-\Lambda\p_{\Lambda}\lambda=\beta_{\lambda}^{(UV)}\, .
\end{equation}
This identifies for $r\raw\infty$, $s\raw 0$
\begin{equation}\label{eq:4.13C} 
\hat{\beta}_{\lambda}=\beta_{\lambda}^{(UV)}\, .
\end{equation}
As stated before, the fluctuations of all particles contribute to $\hat{\beta}_{\lambda}$. The relation \eqref{eq:4.13C} holds for $r\raw\infty$, while for the evaluation of $\hat{\beta}_{\lambda}$ at $\mu=0$ one needs the opposite limit $r\raw 0$. In general, the relation between $\hat{\beta}_{\lambda}(r\raw\infty)$ and $\hat{\beta}_{\lambda}(r\raw 0)$ depends on the detailed flow of couplings. A simple property is easy to understand, however. If $\hat{\beta}_{\lambda}$ vanishes for a range of large finite $r$, there is no dependence on the UV-scale $\Lambda$. As a result, $\lambda$ cannot depend on $s$ and $\hat{\beta}_{\lambda}=0$ holds for all $r$ and $s$. Intrinsic mass scales and the corresponding explicit breaking of scale symmetry are always related to a nonzero value of $\beta_{\lambda}^{(UV)}$.

For $\beta_{\lambda}^{(IR)}\neq 0$ the flow of $\lambda$ with the momentum scale $\mu$ continues for fixed $\chi$ and $\mu\ll\chi$. There are several ways how this flow can end. First, there could be a new scale $\tilde{\mu}<\chi$ such that the flow stops for $\mu<\tilde{\mu}$. An example is the Fermi scale in the standard model. If $\tilde{\mu}$ is proportional to $\chi$ the stop of the flow by spontaneous scale symmetry breaking occurs in two steps at $\chi$ and at $\tilde{\mu}\sim\chi$, $\tilde{\mu}\ll \chi$. For $\beta_{\lambda}^{(UV)}=0$ no violation of scale symmetry is present. Second, the coupling $\lambda$ may grow large at some scale $\tilde{\mu}$, procuring dynamics that stops the flow. An example for $\tilde{\mu}$ is the confinement scale $\Lambda_{QCD}$ of strong interactions. For $\beta_{\lambda}^{(UV)}\neq 0$ the scale $\tilde{\mu}$ will depend both on $\Lambda_{UV}$ and $\chi$, while $\beta_{\lambda}^{(UV)}=0$ implies $\tilde{\mu}\sim \chi$. Finally, the running of $\lambda$ with $\mu$ for fixed $\chi$ may approach an IR-fixed point where it stops.

\subsection{Fixed points in field space}

Let us next consider the case of explicit scale symmetry breaking by an intrinsic scale $\bar{\mu}$. We concentrate on vanishing momenta, $\mu=0$. The coupling $\lambda$ is now a function of $\bar{\mu}/\chi$, with $\Lambda_{UV}$ eliminated in favor of $\bar{\mu}$. It corresponds to a solution of the flow equation \eqref{eq:76}. Assume that $\hat{\beta}_{\lambda}$ has a zero corresponding to a fixed point. At the fixed point $\bar{\mu}$ vanishes. Close to the fixed point $\bar{\mu}$ is small and the fixed point is reached for $\bar{\mu}/\chi\raw 0$. For fixed $\bar{\mu}$ one can reach the fixed point in the limit $\chi\raw\infty$. This leads to the interesting situation that scale symmetry is (approximately) realized for certain regions in field space, in our case for $\chi/\bar{\mu}\gg 1$. In other regions of field space, e.g. for $\chi\approx\bar{\mu}$, scale symmetry is broken explicitly. 

Scale symmetry is no longer an overall property of such a model, holding for arbitrary values of the scalar field. The relevance of scale symmetry is restricted to particular regions in field space. Only exactly on a fixed point, e.g. for $\bar{\mu}=0$, scale symmetry holds for arbitrary $\chi$. This situation is the analogue of a well known feature in momentum space. If the theory is defined by the vicinity of an UV-fixed point, scale symmetry is realized effectively only for high enough momenta, e.g. $\mu\gg\bar{\mu}$. For ``low momenta'' $\mu\approx\bar{\mu}$ it is broken explicitly. This situation is taken over by replacing $\mu$ with $\chi$. As a direct consequence of the fixed point behavior for $\chi\raw\infty$ the mass of the pseudo-Goldstone resulting from spontaneous scale symmetry breaking will be field dependent. It vanishes for $\chi/\bar{\mu}\raw\infty$.

\subsection{Crossover in field space}\label{sec:Crossover_Field_Space} 

There may be situations with two different fixed points in field space, and an associated crossover between them. As an example we investigate the flow (at $\mu=0$)
\begin{equation}\label{eq:78} 
\chi\p_{\chi}\lambda=\hat{\beta}_{\lambda}=-a\lambda(c-\lambda)
\end{equation}
with positive $a$ and $c$. It has fixed points for
\begin{equation}\label{eq:79} 
\lambda_{*}=0\com\lambda_{*}=c.
\end{equation}
The solution reads, with $\rho=\chi^{2}/2$,
\begin{equation}\label{eq:80} 
\lambda=\dfrac{c}{1+\left (\dfrac{\chi}{\sqrt{2}\bar{\mu}}\right )^{ac}}=c\left [1+\left (\frac{\rho}{\bar{\mu}^{2}}\right )^{\frac{ac}{2}}\right ]^{-1}.
\end{equation}
For $\chi\raw\infty$ one reaches the fixed point $\lambda_{*}=0$, while in the opposite limit $\chi\raw 0$ the other fixed point $\lambda_{*}=c$ is attained. The intrinsic scale $\bar{\mu}$ can be used for a specification at what point on the crossover trajectory a given value of $\chi$ can be situated.

If we associate $\lambda=\p^{2}U/\p\rho^{2}$ we find for the flow of $\p U/\p\rho=U^{\prime}$ with $\rho$
\begin{equation}\label{eq:81} 
\rho\p_{\rho}U^{\prime}=\lambda\rho=\dfrac{c\rho}{1+\left (\frac{\rho}{\bar{\mu}^{2}}\right )^{\frac{ac}{2}}}.
\end{equation}
The solution for $U^{\prime}(\rho)$ involves an integration constant $m^{2}$ that can play the role of a second intrinsic mass scale violating scale symmetry. The same holds for the solution of the flow equation for $U$. If a fixed point is reached for $\rho\raw 0$ these integration constants have to be set such that for $\rho\raw 0$ one has
\begin{equation}\label{eq:82} 
U(\rho\raw 0)=\dfrac{c}{2}\rho^{2}\left [1-\dfrac{2}{\bigl (1+\frac{ac}{2}\bigl )\bigl (2+\frac{ac}{2}\bigl )}\left (\frac{\rho}{\bar{\mu}^{2}}\right )^{\frac{ac}{2}}\right ].
\end{equation}
Indeed, one finds for $\rho=0$ a massless scalar, $(U^{\prime}+2\rho U^{\prime\prime})(\rho=0)=0$.

In the opposite limit for $\rho\raw\infty$ the fixed integration constants imply
\begin{equation}\label{eq:83} 
U(\rho\raw \infty)=\dfrac{c\bar{\mu}^{4}}{\bigl (1-\frac{ac}{2}\bigl )\bigl (2-\frac{ac}{2}\bigl )}\left (\frac{\rho}{\bar{\mu}^{2}}\right )^{2-\frac{ac}{2}}+h_{1}\bar{\mu}^{2}\rho+h_{2}\bar{\mu}^{4}\, ,
\end{equation}
where the dimensionless coefficients $h_{1}$ and $h_{2}$ may be computed by a numerical solution of eq.~\eqref{eq:81}. We concentrate on
\begin{equation}\label{eq:84} 
ac<2\, ,
\end{equation}
such that $U$ grows monotonically with $\rho$ and the first term in eq.~\eqref{eq:83} is leading. In the limit $ac=2$ one finds
\begin{equation}\label{eq:85} 
\dfrac{\p U}{\p \rho}=c\bar{\mu}^{2}\ln\left (1+\dfrac{\rho}{\bar{\mu}^{2}}\right ).
\end{equation}

One observes that for $\rho\raw\infty$ the squared scalar mass
\begin{equation}\label{eq:86} 
m^{2}(\rho)=\dfrac{\p U}{\p\rho}+2\rho\lambda(\rho)=\dfrac{2c(3-ac)}{2-ac}\bar{\mu}^{2}\left (\frac{\rho}{\bar{\mu}^{2}}\right )^{1-\frac{ac}{2}}
\end{equation}
does not vanish. This seems to contradict the expectation of a massless Goldstone boson associated to the fixed point for $\rho\raw\infty$. The reason is that the mass term \eqref{eq:86} corresponds to an expansion around some homogeneous field $\chi$. Such a homogeneous field is, however, not a solution of the field equation for a pure scalar theory. We will see in sect.~\ref{sec:Scale_Sym_Cosmo} that a scale invariant coupling to gravity indeed allows for a homogeneous asymptotic solution with $\rho\raw\infty$. The relevant mass term for the physical scalar excitation is proportional to the dimensionless ratio $m^{2}(\rho)/\rho$, which indeed vanishes for $\rho\raw\infty$. For finite $\rho$ one finds a pseudo-Goldstone boson, with a mass vanishing indeed in the limit $\rho\raw\infty$. We also observe that for fixed $\rho$ the mass term \eqref{eq:86} vanishes for $\bar{\mu}\raw 0$.

\subsection{Flow with momentum for fixed field}\label{sec:Flow_Mom_Fixed_Field}

Let us next discuss the momentum dependence of $\lambda$ by investigating its dependence on $\mu$ at fixed $\chi$. We first consider a scale invariant theory defined precisely on a fixed point. In this case $\hat{\beta}_{\lambda}$ vanishes for all values of $\mu/\chi$. Therefore $\lambda$ can depend only on the ratio $\mu/\chi$, implying
\begin{equation}\label{eq:87} 
\mu\p_{\mu}\lambda_{|\chi}=\beta_{\lambda}=-\chi\p_{\chi}\lambda_{|\mu}.
\end{equation}
The coupling $\lambda(\mu,\chi)$ depends only on the ratio $\mu/\chi$, as described by a ``scaling function'' $\lambda_{*}(\mu/\chi)$ for the whole range of $\mu$. In momentum space it covers the UV-range for $\mu\raw\infty$ as well as the IR-range for $\mu\raw 0$. Since $\lambda$ depends only on the dimensionless ratio $\mu/\chi$, the IR-limit $\mu\raw 0$ is also realized for fixed $\mu$ by the limit $\chi\raw\infty$. In this sense the fixed point of eq.~\eqref{eq:78} for $\chi\raw\infty$ could be associated to an infrared fixed point for vanishing momenta. In particular, if $\lambda$ becomes independent of $\mu/\chi$ in the limit $\mu/\chi\raw 0$ one has
\begin{equation}\label{eq:88} 
\beta_{\lambda}=\mu\p_{\mu}\lambda=0\com\chi\p_{\chi}\lambda_{|\mu}=0\, ,
\end{equation}
in accordance with vanishing $\hat{\beta}_{\lambda}$ by virtue of eq.~\eqref{eq:77D}.

The UV-limit is realized for fixed $\chi$ by $\mu\raw\infty$ or for fixed $\mu$ by $\chi\raw 0$. If $\lambda$ admits a finite limit for vanishing $\chi$, e.g. $\mu/\chi\raw\infty$, then both $\beta_{\lambda}$ and $\chi\p_{\chi}\lambda_{|\mu}$ vanish in this limit. This limit corresponds to an UV-fixed point without spontaneous symmetry breaking, $\chi=0$. We emphasize that a scale invariant theory defined precisely on a fixed point involves only one fixed point, covering the whole momentum range of arbitrary $\mu$. This contrasts with the crossover situation described previously that involves two different fixed points and a location on the crossover trajectory specified by an intrinsic scale $\bar{\mu}$. For a scale invariant theory one still may observe a type of crossover as a function of $\mu/\chi$. This does, however, not induce scale symmetry breaking. An exactly massless Goldstone boson is expected for all nonzero values of $\chi$.

Despite the preservation of scale symmetry and the presence of a unique ``overall fixed point'' characterized by scaling functions as $\lambda_{*}(\mu/\chi)$, the flow with $\mu$ at fixed $\chi$ may be associated in practice to different fixed points, e.g. the UV-fixed point for $\mu\raw\infty$ and the IR-fixed point for $\mu\raw 0$. The physical properties, as the spectrum of massless particles, may be rather different for $\mu\raw\infty$ and $\mu\raw 0$. This happens for the quantum scale invariant standard model coupled to gravity. The UV-fixed point in Fig.~\ref{fig:AA} corresponds in this case to $\mu\gg\chi$, while the SM-fixed point is realized for $\mu\ll\chi$. The flow trajectory linking the two fixed points corresponds to the scaling solution. We will often adopt the language of different fixed points even for situations where they are connected by a scaling solution for a common ``overall fixed point''.

We associate the overall fixed point with its scaling solution to the UV-fixed point at which the renormalizable quantum field theory is defined. We next discuss small deviations from the UV-fixed point, corresponding to relevant parameters. They will induce an intrinsic scale $\bar{\mu}$, or several such scales. A particularly simple situation arises if the flow with $\mu$ stops for momenta smaller than the intrinsic scale $\bar{\mu}$ generated by some relevant coupling,
\begin{equation}\label{eq:89} 
\beta_{\lambda}(\mu<\bar{\mu})=0\, ,
\end{equation}
while for $\mu>\bar{\mu}$ the flow according to the scaling solution is not affected much by $\bar{\mu}$. (Such situations are rather common, as exemplified by quantum electrodynamics where $\bar{\mu}$ is identified with the electron mass.) In this case we may employ the scaling function $\lambda_{*}(\mu/\chi)$ for all $\mu>\bar{\mu}$, and replace $\mu$ by $\bar{\mu}$ for $\mu<\bar{\mu}$. In particular, for $\mu=0$ we have $\lambda(\chi)=\lambda_{*}(\bar{\mu}/\chi)$. The $\chi$-dependence of $\lambda$ obeys
\begin{equation}\label{eq:90} 
\hat{\beta}_{\lambda}=\chi\p_{\chi}\lambda_{|\mu=0}=-\beta_{*\lambda}(\bar{\mu}/\chi)\, ,
\end{equation}
with $\beta_{*\lambda}$ the flow generator for the $\mu$-dependence of the scaling function at fixed $\chi$. For non-zero $\beta_{*\lambda}(\bar{\mu}/\chi)$ also $\hat{\beta}_{\lambda}$ differs from zero, indicating explicit scale symmetry breaking.

If the scaling solution $\lambda_{*}$ becomes independent of $\mu/\chi$ for $\mu/\chi\raw 0$, one has $\beta_{*\lambda}(\mu/\chi\raw 0)=0$, cf. eq.~\eqref{eq:88}. Accordingly, also $\hat{\beta}_{\lambda}$ vanishes for $\chi\raw\infty$, such that a fixed point with quantum scale symmetry is reached in the limit $\bar{\mu}/\chi\raw 0$. We next consider for arbitrary $\chi$ the momentum range $\mu\gg\bar{\mu}$. The range $\mu\gg\bar{\mu}$ corresponds by our assumption to the scaling regime associated to the UV-fixed point, with $\lambda(\mu,\chi)=\lambda_{*}(\mu/\chi)$ given by the scaling function. One infers that $\hat{\beta}_{\lambda}$ vanishes also for $\bar{\mu}/\mu\raw 0$. 

We have depicted the scaling regimes for $\lambda(\chi,\mu)$ closed to fixed points in Fig.~\ref{fig:2}(a). Exact scale symmetry is realized for the limits $\mu\raw\infty$ and $\chi\raw\infty$, indicated by thick lines at the boundary of the shown region. Approximate scale symmetry holds in the neighborhood of these lines. Qualitatively, the $\mu$- and $\chi$-dependence of $\lambda(\mu,\chi)$ is well described if we replace in the scaling function $\lambda_{*}(\mu/\chi)$ the scale $\mu$ by $\sqrt{\mu^{2}+\bar{\mu}^{2}}$,
\begin{equation}\label{eq:90A} 
\lambda(\mu,\chi)=\lambda_{*}\left (\dfrac{\sqrt{\mu^{2}+\bar{\mu}^{2}}}{\chi}\right ).
\end{equation}
The intrinsic scale $\bar{\mu}$ becomes unimportant both for $\bar{\mu}\ll\mu$ and $\bar{\mu}\ll \chi$, such that approximate scaling is realized for these limits.

These types of simple scenarios extend to scaling functions for which $\beta_{*\lambda}$ has two zeros. For example, we may consider the analogue of eq.~\eqref{eq:78}
\begin{equation}\label{eq:90B} 
\beta_{*\lambda}=a\lambda(c-\lambda).
\end{equation}
The scaling solution is given by eq.~\eqref{eq:80}, with $\bar{\mu}$ replaced by $\mu$. The qualitative behavior of $\lambda(\mu,\chi)$ is given by $(\rho=\chi^{2}/2)$
\begin{equation}\label{eq:90C} 
\lambda(\mu,\chi)=c\left [1+\left (\dfrac{\rho}{\mu^{2}+\bar{\mu}^{2}}\right )^{\frac{ac}{2}}\right ]^{-1}.
\end{equation}
We have depicted the scaling regimes of this scenario in fig.~\ref{fig:2}(b).Exact scale symmetry occurs now for $\chi\raw\infty$ and $\chi\raw 0$, as well as for $\mu\raw\infty$ (thick lines). At $\mu=0$ we observe now a crossover from a fixed point for $\chi\raw 0$ to another fixed point for $\chi\raw\infty$. Both fixed points are inherited from the common scaling function for the fixed point used for the definition of the renormalizable quantum theory. By analogy with the momentum dependence of the scaling solution we call the fixed point in field space for $\chi\raw\infty$ the IR-fixed point, and the one for $\chi\raw 0$ the UV-fixed point. In cosmology the value of a scalar field $\chi$ can change in time. We will discuss in sect.~\ref{sec:Scale_Sym_Cosmo} crossover cosmologies where $\chi$ changes from the UV-fixed point ($\chi\raw 0$) in the infinite past to the IR-fixed point ($\chi\raw\infty$) in the infinite future.

%\onecolumngrid
\begin{widetext}
\onecolumngrid
\begin{figure}[h]
\includegraphics[scale=0.4]{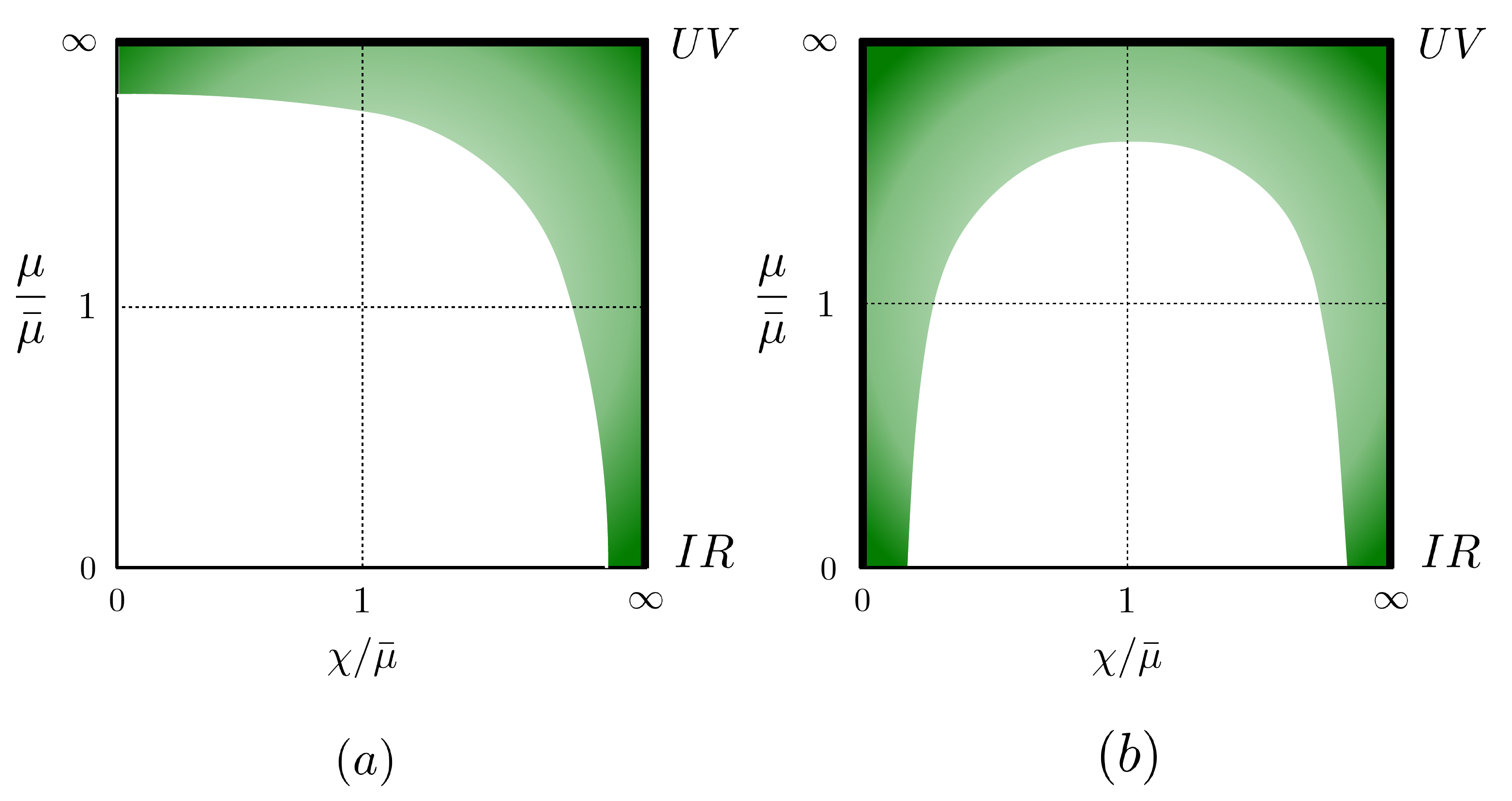}
\caption{Scaling regions for $\lambda(\mu,\chi)$. Thick lines are fixed points with exact quantum scale symmetry. The shaded regions denote qualitatively the vicinity of the fixed points. (a) The fixed point for $\chi/\bar{\mu}\raw\infty$ is induced by a zero of $\beta_{*\lambda}$ for $\chi/\mu\raw\infty$. (b) The fixed points at $\chi/\bar{\mu}\raw\infty$ and $\chi/\bar{\mu}\raw0$ correspond to a crossover with two zeros of $\beta_{*\lambda}$ for $\chi/{\mu}\raw\infty$ and $\chi/{\mu}\raw0$.}\label{fig:2} 
\end{figure}
\end{widetext}
\twocolumngrid
%\twocolumngrid

More complicated situations can arise if the stop of the flow at $\bar{\mu}$ is only partial. A subsector corresponding to an effective low momentum theory for $\mu<\bar{\mu}$ may still induce a non-vanishing $\beta_{\lambda}$, and a different fixed point could be reached for $\mu\raw 0$. This would realize the crossover to an IR-fixed point in a more complicated way.

Our discussion of the different scenarios may perhaps appear somewhat lengthy. A precise understanding is, however, required for the issue of the Goldstone boson of spontaneously broken scale symmetry for $\chi\neq 0$. It makes a difference if a quantum field theory is defined precisely on a fixed point, or in the vicinity of a fixed point in the sense that not all relevant couplings vanish. Precisely on the fixed point quantum scale symmetry is exact and the Goldstone boson is massless. In the vicinity of the fixed point scale symmetry is explicitly broken by the intrinsic scale $\bar{\mu}$. For $\chi\gg\bar{\mu}$ a characteristic pseudo-Goldstone boson mass is given by
\begin{equation}\label{eq:91} 
m\sim\dfrac{\bar{\mu}^{2}}{\chi}.
\end{equation}
It vanishes for $\chi\raw\infty$.

\section{Particle scale symmetry}\label{sec:Particle_Scale_Sym} 

This section discusses scale symmetry for the standard model of particles physics, considered as an effective low energy theory for momenta smaller than a scale $M$. Typically, $M$ may be associated to the Planck mass. We will investigate two versions. The first considers $M$ as an intrinsic scale that acts like an UV-cutoff $\Lambda_{UV}$ for the effective low energy theory. The second associates $M$ with the vacuum expectation value of a scalar field $\chi$ which spontaneously breaks scale symmetry. If the standard model does not extend to the Planck mass, the scale $M$ may be taken as the scale for grand unification $M_{GUT}$, or the scale of B--L - symmetry breaking $M_{B-L}$. For definiteness, we assume here $M$ to be the Planck mass.

In a wider picture including gravity, particle scale symmetry is associated to the SM-fixed point in Fig.~\ref{fig:AA}. As we have mentioned already, particle scale symmetry can be understood only based on the properties of this fixed point. No assumption about the detailed microphysics is needed for the discussion of the consequences of particle scale symmetry for the naturalness of the Fermi scale and related issues. We argue that particle scale symmetry is the symmetry that ``protects'' a small Fermi scale. We also address the scale invariant standard model for which two types of scale symmetry are present: gravity scale symmetry and particle scale symmetry. 

The fixed point associated to exact particle scale symmetry is a trivial fixed point with vanishing gauge and Yukawa couplings. In view of the observed non-vanishing couplings particle scale symmetry is only an approximate symmetry. The running couplings induce an explicit breaking of particle scale symmetry by inducing the Fermi scale and the confinement scale. There is no Goldstone boson associated to particle scale symmetry. In contrast, gravity scale symmetry for the scale symmetric standard model can be exact. It has to be spontaneously broken, implying the presence of a Goldstone boson, or a pseudo-Goldstone boson if an intrinsic mass scale exists. We argue that the issue of the presence of an intrinsic mass scale depends on the properties of the UV-fixed point. As a consequence, the mass of the pseudo-Goldstone boson cannot be computed within the standard model. It needs the properties of the UV-completion of the standard model. The quantum scale invariant standard model is realized if the gauge and Yukawa couplings at the Planck scale correspond to fixed point values at the UV-fixed point. In this case they are, in principle, predictable.

\subsection{Scale symmetry and vacuum electroweak phase transition}\label{sec:Scale_Sym_Vac_Electroweak_PT} 

The largest mass scale in the standard model is the Fermi scale $\varphi_{0}$, given by the vacuum expectation value of the Higgs doublet. The dimensionless parameter
\begin{equation}\label{eq:PS1} 
\varepsilon=\dfrac{\varphi_{0}^{2}}{M^{2}}=5\cdot 10^{-33}
\end{equation}
is tiny. The gauge hierarchy problem \cite{GIL,WEIGH} concerns different aspects of the understanding of this small parameter. We first discuss the issue of naturalness of a small parameter $\varepsilon$ in the light of particle scale symmetry \cite{CWFT,BAR,CWMHB}.

Particle scale symmetry is associated to the second order electroweak phase transition in the limit of vanishing gauge couplings. In the language of statistical physics this is a quantum phase transition at zero temperature. As for any second order phase transition, the critical surface in coupling constant space corresponds to an exact fixed point. We have discussed this fixed point for the standard model in sect.~\ref{sec:ParticleScaleSym}. The scale symmetry associated to this fixed point is particle scale symmetry. 

The transformation properties with respect to particle scale symmetry are simple: The Higgs scalar and the metric transform as
\begin{equation}\label{eq:PS1A} 
h\raw\alpha_{p}h\com g_{\mn}\raw\alpha^{-2}_{p}g_{\mn}\, ,
\end{equation}
while other scalars as $\chi$ or some grand unified scalars are invariant. The light fermions transform as $\psi\raw\alpha_{p}^{3/2}\psi$ and the gauge bosons are invariant. Precisely on the second order phase transition ($\varepsilon=0$) particle scale symmetry becomes an exact symmetry of the low energy effective action. This effective action does not include dynamical gravity. Indeed, a term $\tilde{\cL}\sim\sqrt{g}M^{2}R$ violates particle scale symmetry, independently if $M$ is a constant or $M\sim\chi$. 

Away from the critical surface of the phase transition particle scale symmetry is violated by terms in the low energy effective action. Small deviations from the critical surface of the phase transition are protected by particle scale symmetry in the sense that their flow is very slow. This follows by continuity from the basic property of a critical surface that for all couplings on the surface the flow vanishes. For small non-zero gauge couplings particle scale symmetry is only approximate. As we have discussed in sect.~\ref{sec:Networks_FP_Crossover} the intrinsic mass scales associated to these running couplings are much smaller than $\varphi_{0}$. We can neglect the effect of these intrinsic mass scales for the discussion of the gauge hierarchy and treat the vacuum electroweak phase transition as a second order phase transition even in the presence of non-vanishing gauge and Yukawa couplings.

A non-vanishing parameter $\varepsilon$ expresses the deviation from the critical surface of the electroweak phase transition. As well known from statistical physics, any valid expansion method should reflect this property. One such method is renormalization group improved perturbation theory. In this approach the generators ($\beta$-functions) for running dimensionless couplings are computed in a loop expansion. This is the method employed for the computation of the running of the gauge couplings, Yukawa couplings and quartic scalar coupling. It can also be employed for the flow of the deviation from the critical surface of the electroweak phase transition \cite{CWFT,CWMHB,CWQR}. For small enough couplings low order perturbation theory converges well. This contrasts with a bare loop expansion which computes directly the couplings of interest in a perturbative expansion. In bare perturbation theory a computation of $\varepsilon$ does not converge, requiring severe cancellations or ``fine tuning'' for obtaining a small $\varepsilon$.

The reason for the different convergence properties arises from the fact that possible fixed points are visible in renormalization group improved perturbation theory, but not in bare perturbation theory. At a fixed point, many different contributions of bare perturbation theory have to cancel in order to obtain the fixed point values of couplings. The presence of fixed points is related to universality. Details of the microscopic theory (e.g. the regularization or UV-completion) become unimportant - the ``memory'' of the microscopic physics is partly lost. This important property is not directly visible in bare perturbation theory. Typically, the low orders (one and two loop) of the $\beta$-functions are universal. They neither depend on the UV-regularization, nor on the precise definition of the running renormalized couplings. This universality is absent for bare perturbation theory. We conclude that bare perturbation theory is not an appropriate method for addressing the gauge hierarchy problem.

The location of the critical surface of the electroweak phase transition in coupling constant space is not a universal quantity. It depends on the precise regularization and on the precise definition of the individual couplings as coordinates in coupling constant space. Let us define some type of effective potential $U_{k}(\rho)$ for the Higgs doublet, $\rho=h^{\dagger}h$, for which quantum fluctuations with momenta $q^{2}\gtrsim k^{2}$ have already been included, while the integration over fluctuations with momenta $q^{2}\lesssim k^{2}$ still needs to be performed. For the computation of the fluctuation effects from the low momentum modes $k$ acts as a variable UV-cutoff. (Within functional renormalization the potential $U_{k}(\rho)$ can be associated with the effective average potential \cite{CWFR}, but the concept used here is wider.) A flowing mass term may be defined as
\begin{equation}\label{eq:PS2} 
\bar{m}^{2}(k)=\dfrac{\p U_{k}(\rho)}{\p\rho}\bigl |_{\rho}=0\, .
\end{equation}
This definition yields the lowest order in a polynomial expansion of $U_{k}$ around the origin,
\begin{equation}\label{eq:PS3} 
U_{k}(\rho)=\bar{U}_{k}+\bar{m}^{2}_{k}\, h^{\dagger}h+\ldots\; .
\end{equation}
It is particularly well adapted if the minimum of $U_{k}(\rho)$ occurs for $\rho=0$. For a minimum of $U_{k}(\rho)$ at $\rho_{0}(k)\neq 0$ a mass term may be defined by a polynomial expansion in the radial direction,
\begin{equation}\label{eq:PS4} 
m_{H}^{2}(k)=\left (\dfrac{\p U_{k}}{\p\rho}+2\rho\dfrac{\p^{2} U_{k}}{\p\rho^{2}}\right )\bigl |_{\rho=\rho_{0}(k)}\, .
\end{equation}
Whatever precise definition of the mass term we choose, it will depend on the precise way how the separation into high momenta and low momenta (as compared to $k$) is defined.

For $k=0$ all fluctuations are included and $U_{k=0}(\rho)$ is the (quantum) effective potential, with $\rho_{0}(k=0)=\varphi_{0}^{2}$ and $m_{H}^{2}(k=0)$ the squared Higgs boson mass. The relation between $m_{H}^{2}(k)$ and $m_{H}^{2}(0)$ typically involves an additive ``quadratic term'' $\sim k^{2}$,
\begin{equation}\label{eq:PS5} 
m_{H}^{2}(0)=m_{H}^{2}(k)+ck^{2}\, ,
\end{equation}
with coefficient $c$ involving the couplings of the theory. Not surprisingly, the coefficient $c$ depends strongly on the way how the separation of fluctuation modes is done. A similar quadratic term $\sim c^{\prime}k^{2}$ relates $\bar{m}^{2}(0)$ and $\bar{m}^{2}(k)$.

The quadratic term $\sim ck^{2}$ is the essence of the so-called quadratic divergence of the mass of the Higgs scalar. Indeed, in a theory with a physical UV-cutoff $\Lambda_{UV}$ we can start with $k=\Lambda_{UV}$. The term $c\Lambda_{UV}^{2}$ diverges for $\Lambda_{UV}\raw\infty$. Now $c$ depends on the way the cutoff is implemented or the regularization scheme \cite{CWFT,CWMHB,BAR}. It will differ between a sharp cutoff and a smooth cutoff in momentum space. For a smooth cutoff the values of $c$ will depend on the precise form of the cutoff function. In short, the quadratic term reflects the detailed microphysics. It is not universal.

What is common to all regularizations and all definitions of the mass term is the existence of a ``critical trajectory'' $m_{H*}^{2}(k)$ with the property
\begin{equation}\label{eq:PS6} 
m_{H*}^{2}(k=0)=0\, .
\end{equation}
The same holds for $\bar{m}^{2}_{*}(k)$. The existence of the critical trajectory is directly linked to the second order character of the vacuum electroweak phase transition. Precisely on the phase transition the scalar mass term vanishes - the correlation length $\xi=m^{-1}_{H}$ diverges. If the minimum of $U_{k}(\rho)$ occurs for $\rho_{0}(k)\neq 0$, there exists also a critical trajectory $\rho_{0*}(k)$ with
\begin{equation}\label{eq:PS7} 
\rho_{0*}(k=0)=0\, .
\end{equation}
At the phase transition the order parameter $\rho_{0}$ vanishes. The critical trajectories $m_{H*}^{2}(k)$, $\bar{m}_{*}^{2}(k)$, $\rho_{0*}(k)$ are related, reflecting different facets of the phase transition. The critical trajectory separates trajectories that correspond to the symmetric or disordered phase with $\bar{m}^{2}(k=0)>0$, $\rho_{0}(k=0)=0$ from trajectories that correspond to the ordered phase with spontaneous symmetry breaking, where $\rho_{0}(k=0)>0$. No trajectory can cross the critical trajectory. More precisely, in case of several flowing couplings the critical trajectory is generalized to a critical hypersurface in coupling constant space. This hypersurface separates the disordered and ordered phase. Every point on the critical hypersurface flows for $k\raw 0$ to $\bar{m}^{2}=m_{H}^{2}=\rho_{0}=0$, as characteristic for a second order phase transition. All trajectories which are on the critical hypersurface at some scale $\bar{k}$ are critical trajectories that stay on the critical hypersurface for all $k<\bar{k}$. The critical hypersurface cannot be crossed by any trajectory. If a trajectory hits the critical hypersurface, it would remain on this hypersurface until $k=0$. (We recall that strictly speaking our discussion of the properties of a second order phase transition only holds for vanishing gauge couplings. The small running gauge couplings do not affect the qualitative properties of the flow in the relevant range of scales. Also the fact that local gauge symmetries are not spontaneously broken in a strict sense does not alter the results.)

This general properties can be seen by a functional renormalization group investigation \cite{CWQR}. The flow equation for the mass term takes the form \cite{CWQR,BTW,HONOWE,GIGS,GISON,EGJP,BOGISO,Gies:2017zwf,Sondenheimer:2017jin,Reichert:2017puo,Held:2018cxd}
\begin{equation}\label{eq:5.9} 
\p_{t}m^{2}=Am^{2}+Bk^{2}\, ,
\end{equation}
where $A$ and $B$ depend on other couplings that typically may flow mildly with $k$. We derive this type of flow equation in sect.~\ref{sec:6.17.4}, with
\begin{equation}\label{eq:5.10} 
B=\dfrac{3y_{t}^{2}}{8\pi^{2}}+\ldots\, ,
\end{equation}
where $y_{t}$ is the top-quark Yukawa coupling. Let us pursue the approximation of constant $B$ and $A$. The general solution for $A<2$,
\begin{equation}\label{eq:5.11} 
m^{2}(k)=\dfrac{B}{2-A}k^{2}+dk^{A}\, ,
\end{equation}
involves the integration constant $d$ which specifies the particular flow trajectory. The critical trajectory corresponds to $d=0$,
\begin{equation}\label{eq:5.12} 
m_{*}^{2}(k)=\dfrac{B}{2-A}k^{2}\, ,
\end{equation}
and therefore to a fixed point of the dimensionless ratio
\begin{equation}\label{eq:5.12A} 
\dfrac{m_{*}^{2}(k)}{k^{2}}=\dfrac{B}{2-A}\, .
\end{equation}
On the fixed point trajectory the mass term does not vanish for $k\neq 0$. It only vanishes in the limit $k\raw 0$. As advocated, the other flow trajectories do not cross the critical trajectory. For $d>0$ one has $m^{2}(k)>m_{*}^{2}(k)$ for all $k$, while $d<0$ corresponds for all $k$ to $m^{2}<m_{*}^{2}(k)$. For small $A$ the constant $c$ in eq.~\eqref{eq:PS5} is given approximately by $c=-B/(2-A)$. It depends on the particular choice of the IR-regulator function via the regulator dependence of $B$.

\subsection{Anomalous mass dimension}\label{sec:Anomalous_Mass_Dim} 

The crucial quantity for the understanding of the scale of electroweak symmetry breaking is the distance from the critical trajectory \cite{CWFT,CWMHB,CWQR,AOISO}. In general, the distance from a critical trajectory $\delta$ obeys a universal flow equation. For a flow with $\rho_{0}(k)\neq 0$ we may define
\begin{equation}\label{eq:PS8} 
\tilde{\delta}(k)=\rho_{0*}(k)-\rho_{0}(k)\, ,
\end{equation}
while for $\rho_{0}(k)=0$ one can take
\begin{equation}\label{eq:PS9} 
\delta(k)=\bar{m}^{2}(k)-\bar{m}^{2}_{*}(k)\, .
\end{equation}
Trajectories with positive $\delta$ or $\tilde{\delta}$ belong to the symmetric phase, while negative $\delta$ or $\tilde{\delta}$ leads to spontaneous symmetry breaking with $\rho_{0}(k=0)>0$. The critical surface corresponds to $\delta=\tilde{\delta}=0$. The $\beta$-function for $\delta$ or $\tilde{\delta}$ has to vanish for $\delta=0$ or $\tilde{\delta}=0$ since the critical surface at $\delta=0$ ($\tilde{\delta}=0$) cannot be crossed. For small $\delta$ it is given by an anomalous mass dimension
\begin{equation}\label{eq:PS10} 
k\p_{k}\delta=A\delta\, .
\end{equation}
This behavior is explicit for the functional flow equation \eqref{eq:5.9}. For constant $A$ this leads to a slow powerlike decrease of $\delta$ with decreasing $k$,
\begin{equation}\label{eq:5.16A} 
\delta=\delta_{0}\left (\dfrac{k}{k_{0}}\right )^{A}\, .
\end{equation}

The anomalous mass dimension $A$ can be computed in perturbation theory. For the standard model it is given by eq.~\eqref{eq:43} \cite{CWPFP,CWFT}. This quantity is universal and does not depend on the regularization or the precise definition of $\delta$. (It holds as well for $\tilde{\delta}$.) In contrast to the flow of $\bar{m}_{*}^{2}$, $m_{H*}^{2}$ or $\rho_{0*}$, which only concerns the precise location of the critical hypersurface in coupling constant space and therefore depends on the ``microscopic physics'' or regularization, the flow of the deviation from the critical surface $\delta$ is only a property of the effective low energy theory. From the point of view  of the fixed point associated to the second order vacuum electroweak phase transition the deviation from the fixed point corresponds to a relevant dimensionless parameter $\gamma=\delta/k^{2}$, with flow given by eq.~\eqref{eq:49}. From this point of view it is obvious that the critical hypersurface at $\gamma=0$ cannot be crossed. The same picture arises from functional renormalization where
\begin{equation}\label{eq:MF5} 
A=\dfrac{\p}{\p m^{2}}\beta_{m}^{2}\bigl |_{m_{*}^{2}}\, .
\end{equation}
In lowest order one finds indeed \cite{CWQR} the perturbative anomalous dimension \eqref{eq:43}. For the general solution \eqref{eq:5.11}, \eqref{eq:5.12} one has
\begin{equation}\label{eq:5.16AA} 
\gamma=d_{\gamma}k^{A-2}\, ,
\end{equation}
such that for $A<2$ one finds an increase of $|\gamma|$ for decreasing $k$, as appropriate for a relevant coupling.

We summarize that a perturbative computation of quantities as $\bar{m}^{2}$ or $\rho_{0}$, with the associated ``quadratic divergences'' and the necessity of fine tuning order by order in the expansion for obtaining $\bar{m}^{2}=\rho_{0}=0$, only concerns the expression for the location of the critical hypersurface in terms of microphysical or ``bare'' couplings. It depends on the precise choice of the bare couplings and the precise ultraviolet regularization. In contrast, flow equations only involve renormalized couplings. The existence of a critical surface ($d=0$ in eq.~\eqref{eq:5.11}) follows immediately from the structure of the flow equation in the limit where a second order phase transition is a good approximation. The relation to the microscopic ``bare couplings'' is given by the solution of the flow equation with bare couplings specifying the initial values for some large $k$, say $k=M$. The precise embedding of the critical hypersurface in the space of microphysical couplings remains complicated. The precise location of the critical hypersurface in the space of bare couplings is, however, not important for the computation of physical quantities at low momenta which only depend on renormalized couplings, not on bare couplings.

The renormalized scalar mass term appears in the propagator or Greens function $G(p^{2})$ for the Higgs boson. The dependence of the inverse propagator on the squared momentum $p^{2}$ takes the characteristic form (in a euclidean setting)
\begin{equation}\label{eq:PS11} 
G^{-1}(p^{2})=Z_{\varphi}(p^{2})\left (p^{2}+m^{2}(p^{2})\right )\, ,
\end{equation}
with slowly varying wave function renormalization $Z_{\varphi}(p^{2})$. The critical hypersurface corresponds to $m^{2}(p^{2})=0$. We can take $m^{2}(p^{2})$ as a version for the deviation $\delta$ from the critical hypersurface. Associating the renormalization scale $\mu^{2}=p^{2}$ one finds the flow equation
\begin{equation}\label{eq:PS12} 
\mu\p_{\mu}m^{2}=Am^{2}\, .
\end{equation}
Only the universal anomalous mass dimension \eqref{eq:43} appears in the physical quantities. No influence of ``quadratic divergencies'' or other features concerning the expression of the critical surface in terms of bare couplings is present. Renormalization group improved perturbation has no problems with convergence (at least not in low orders). The two loop contribution to $A$ contains terms $\sim g^{4}$, $y_{t}^{4}$, $\lambda_{H}^{2}$ and is small as compared to the one loop expression.

These features of the vacuum electroweak phase transition fit perfectly into the general framework for the understanding of critical phenomena in statistical physics. The precise location of the critical surface in the space of microscopical couplings is often rather complicated and not known in practice. Very often no reliable methods for its computation are known. This location is not needed, either. All universal physics for critical phenomena concern only the deviations from the critical hypersurface, as parameterized by the relevant parameters.

\subsection{Naturalness of the Fermi scale}

The understanding of a small parameter as $\varepsilon$ involves two issues. The first asks if there are particular properties that protect the value $\varepsilon=0$ in the sense that it is not changed if the microphysics undergoes small changes. This is the issue of naturalness. A small parameter is then protected against large corrections. A typical case is a symmetry that implies $\varepsilon=0$. In this case a small value of $\varepsilon$ amounts to a small breaking of this symmetry. The second issue concerns an explanation why the parameter is small. We investigate here the question of naturalness of a small $\varepsilon$. In sect.~\ref{sec:6.17} we will turn to a possible explanation of its small value.

An example for a natural small parameter is the ratio of electron mass over Planck mass, $m_{e}/M$, in quantum electrodynamics coupled to gravity. The additional chiral symmetry for $m_{e}/M=0$ protects the small electron mass from large corrections in all calculational schemes that respect chiral symmetry. We will see that the gauge hierarchy $\varphi_{0}/M$ is natural in the same sense, the relevant symmetry being scale symmetry. Naturalness of a small quantity is not yet an explanation why the quantity is small. In this respect there is no difference between a small $m_{e}/M$ in QED or a small $\varphi_{0}/M$ in the standard model.

In a renormalizable quantum field theory the dependence on the microphysics arises only through the values of the renormalized couplings or relevant parameters. A violation of naturalness of a small parameter is associated to a large relative change of this parameter if some of the relevant parameters are subject to a small relative change. For a general discussion we may denote the investigated small dimensionless parameter by $\sigma$ - for the problem of naturalness of the gauge hierarchy $\sigma$ will correspond to $\gamma$. We are interested in the relative change of $\sigma$ under a relative change of relevant couplings $g_{i}$, taken at some scale $k$ which will be associated here to high momenta or small distances. (In a model with physical UV-cutoff $\Lambda_{UV}$, as a lattice model, we take $k$ somewhat smaller than $\Lambda_{UV}$ such that irrelevant couplings no longer matter.) We parametrize the relative change of $\sigma$ by
\begin{equation}\label{eq:PS13} 
\dfrac{\Delta \sigma}{\sigma}=\sum_{i}\, S_{i}\dfrac{\Delta g_{i}(k)}{g_{i}(k)}\, .
\end{equation}
For ``sensitivity coefficients'' $S_{i}$ of the order one or smaller, a small parameter $\sigma$ is natural, for large $S_{i}\gg 1$ it is not natural. For $\sigma\raw 0$ naturalness requires that $\Delta\sigma$ is proportional to $\sigma$.

Any observable parameter $\sigma$ is a function of the renormalized couplings $g_{i}$. Such a relation holds for all $k$ and we may therefore consider a $k$-dependent quantity $\sigma(k)=\sigma(g_{i}(k))$. The observable small parameter may be associated with $\sigma=\sigma(k=0)$ or $\sigma=\sigma(k=\mu)$ for some particular appropriate $\mu$. We can expand the flow equation for small $\sigma(k)$
\begin{equation}\label{eq:PS14} 
k\p_{k}\sigma=B_{\sigma}(g_{i})+A_{\sigma}(g_{i})\sigma+\ldots\; .
\end{equation}
For $B_{\sigma}\neq 0$ a small parameter $\sigma$ is not natural, since the coefficient $B_{\sigma}$ drives $\sigma(k)$ through zero. Typically, $\sigma(k)$ may reach a value $\sigma(\bar{k})=0$ at some particular scale $\bar{k}$. Then it differs from zero for $k>\bar{k}$ and $k<\bar{k}$. The precise value $\bar{k}$ where $\sigma(\bar{k})=0$ depends on the precise values of the renormalized couplings $g_{i}(k)$ and changes under a small relative change $\Delta g_{i}(k)/g_{i}(k)$. Obtaining $\bar{k}=0$ or $\bar{k}=\mu$ requires fine-tuning of the microphysical renormalizable couplings. On the other hand, for $B_{\sigma}=0$ a small value of $\sigma$ is natural. If $\sigma(k)=0$ at some scale $k$, it stays zero for all $k$.

From the point of view of flow equations a small parameter $\sigma$ is natural if $\sigma=0$ corresponds to a partial fixed point of the flow. Whenever $\sigma=0$ corresponds to an enhanced symmetry there is an associated partial fixed point of the flow, provided that the flow is consistent with this symmetry. The association of naturalness with a partial fixed point therefore recovers the case where naturalness is associated to an enhanced symmetry. It is more general, however, since partial fixed points are not always associated to enhanced symmetries.

A partial fixed point at $\sigma=0$ does not yet explain a small parameter $\sigma=\sigma(k=0)$. It only guarantees naturalness in the sense that a small value of $\sigma$ at some microphysical scale $\bar{k}$ guarantees a small $\sigma$ at all $k<\bar{k}$, including $k=0$, and that this property does not require a fine tuning of the renormalizable couplings $g_{i}(k)$. Most importantly, it leaves open the scale at which a small value of $\sigma$ may find an explanation. This can be a very high momentum scale.

On the basis of these properties it has been argued that a small value of the gauge hierarchy parameter $\varepsilon$ is natural \cite{CWFT}. A second order phase transition indeed corresponds to a partial fixed point \cite{CWPFP}, since the critical surface cannot be crossed by the flow. The flow of $\gamma$ vanishes for $\gamma=0$, independently of the values of the other relevant parameters or renormalized couplings. As has been noted in refs.~\cite{CWFT,CWMHB}, this partial fixed point is associated to an enhanced symmetry, namely scale symmetry. In this sense scale symmetry protects $\varepsilon$ and renders a small value natural. If a symmetry plays a central role, it is a good idea to choose a regularization consistent with this symmetry. The naturalness of a small value of $\varepsilon$ by a scale invariant regularization has been emphasized by Bardeen \cite{BAR} and advocated in later work \cite{HEM,MENI1,MENI2,FKV,GKM,MNP,AOISO}. Many conceptual and technical aspects of this proposal for naturalness have been clarified \cite{FGHW,THAM,CTH1,HLM,SHASHI}.

The partial fixed point due to the second order phase transition is a rather robust property. Even for a large class of grand unified theories, with a complicated sector of many scalar fields, the vacuum electroweak phase transition remains typically of second order. The critical surface may now be a rather complicated hypersurface in the space of couplings of the grand unified model. Nevertheless, the critical surface is never crossed such that the flow of $\delta$ is proportional to $\delta$. The anomalous mass dimension $A$ is modified in a grand unified theory. As long as $A<2$ the dimensionless ratio $\gamma$ remains a relevant coupling. The issue that the Higgs doublet is only part of a larger representation of some GUT symmetry (``doublet-triplet splitting''), or even a mixture of different representations, does not change the naturalness of a small parameter $\varepsilon$ due to the scale symmetry for a second order phase transition. As a consequence, it remains open at which scale an explanation of the small value of $\varepsilon$ has to be found \cite{CWWTL}. It may be near the Fermi scale, or above the Planck scale. We will discuss in sect.~\ref{sec:6.17} the proposal that fluctuations of the gravitational degrees of freedom may lead to an explanation of the small value of $\varepsilon$ \cite{CWMY}.

\subsection{Scale invariant standard model with singlet scalar}

So far we have discussed a scenario where the Planck mass $M$, or some other physical scale $\Lambda_{UV}$ where the standard model has to be extended, is an intrinsic mass scale. The particle scale symmetry associated to the vacuum electroweak phase transition is then the only scale symmetry relevant for momenta below $M$. It is explicitly broken by the non-zero value of the relevant parameter $\gamma$ or, equivalently, by the expectation value $\varphi_{0}$ of the Higgs doublet. No Goldstone boson is present in this scenario, since scale symmetry is not broken spontaneously. If one embeds the standard model into a more complete renormalizable quantum field theory including gravity, the scales $M$ or $M_{GUT}$ correspond in this scenario to explicit scale symmetry breaking by the values of parameters that are relevant with respect to the UV-fixed point.

A different scenario embeds the standard model into a renormalizable theory where $M$ is given by a scalar field $\chi$. The scale symmetry associated to the UV-fixed point is spontaneously broken by $\chi$. The ultraviolet completion of the standard model does not involve intrinsic mass scales. The effective reduction of the short distance theory to the standard model as an effective model for large distances occurs by spontaneous symmetry breaking. If there is no intrinsic mass scale above the Fermi scale, the momentum range $\mu\gg\varphi_{0}$ is characterized by two different scale symmetries. The first is the ``gravity scale symmetry'' associated to the UV-fixed point of the full theory including gravity. As we have discussed in sect.~\ref{sec:QScaleSymAndQuqnatumEffAction}, this scale symmetry is exact if we define the model precisely on the UV-fixed point. In this case the function $\hat{\beta}_{\lambda}$ in sect.~\ref{sec:SponAndExpSymBr} vanishes. The second scale symmetry is particle scale symmetry, as discussed in sect.~\ref{sec:ParticleScaleSym}. Particle scale symmetry is present in both scenarios, being only related to the second order vacuum electroweak phase transition. The issues related to particle scale symmetry discussed in the present section are the same if $M$ is an intrinsic scale or if $M=\chi$ reflects spontaneous symmetry breaking. In the second case the running of couplings relevant for particle scale symmetry concerns their dependence on $\mu/\chi$, as discussed in sect.~\ref{sec:Flow_In_Field_Space}.

For a Planck mass given by a scalar field, $M=\chi$, the two different scale symmetries correspond to different field transformations. Gravity scale symmetry is characterized by an invariance under the scalings \eqref{eq:6}, \eqref{eq:7}, with $h$ scaling as $\chi$, $h^{\prime}=\alpha h$. In contrast, for particle scale symmetry $\chi$ is kept fixed. The part $\cL_{\chi R}$ in eq.~\eqref{eq:16} violates particle scale symmetry. It is discarded for the low energy effective theory because the gravitational and $\chi$-mediated interactions are very small for the momentum range far below $M$. In case of grand unification, all scalar fields except for the Higgs doublet are invariant under particle scale symmetry. Again their kinetic and potential terms violate particle scale symmetry and are discarded from the effective low energy theory. With respect to particle scale symmetry $\chi$ and other scalars beyond the Higgs doublet are treated as fixed parameters. 

It is particle symmetry that is responsible for the naturalness of a small Fermi scale. Indeed, exact particle scale symmetry requires $\varepsilon=0$, such that small values of $\varepsilon$ are protected by this symmetry. Gravity scale symmetry is unrelated to this issue. Indeed, arbitrary values of $\varepsilon$ are compatible with gravity scale symmetry. Gravity scale symmetry for the standard model should not be invoked for an understanding of the naturalness problem.

The main observational difference between the two scenarios concerns the dynamics of the scalar field $\chi$. If we define the theory exactly on the UV-fixed point, gravity scale symmetry is exact. Its spontaneous breaking by a non-zero cosmological value of $\chi$ results in a massless Goldstone boson. We will see this explicitly in sect.~\ref{sec:Scale_Sym_Cosmo} where we discuss cosmological solutions. In contrast, defining the theory in the vicinity of the UV-fixed point, in the sense that some relevant parameter differs from zero and the fixed point is reached only for $\mu\raw \infty$, necessarily introduces an intrinsic scale $\bar{\mu}$ due to the running of the relevant couplings. We take here $\bar{\mu}$ as the largest such scale. For $\bar{\mu}\ll\chi$ one predicts a pseudo-Goldstone boson with mass $\sim\bar{\mu}^{2}/\chi$. In the limit $\bar{\mu}/\chi\raw 0$ this becomes the massless Goldstone boson, the dilaton.

\subsection{Mass of the pseudo-Goldstone boson}\label{sec:Mass_Pseudo_Goldstone} 

The pseudo-Goldstone boson of spontaneously broken approximate scale symmetry has been named ``cosmon'' \cite{PSW,CWQ} because of its possible role for the solution of the cosmological constant problem, c.f. sects.~\ref{sec:V}, \ref{sec:Scale_Sym_Cosmo}. For simplicity we employ here this name, even though the presence of the pseudo-Goldstone boson is a general property and not necessarily related to the cosmological constant problem or an important role in cosmology. The mass of the cosmon
\begin{equation}\label{eq:PS15} 
m_{c}\sim \dfrac{\bar{\mu}^{2}}{\chi}
\end{equation}
depends on the value of the scalar field $\chi$. It will vary for cosmological solutions for which $\chi$ is time-dependent. In the limit $\chi/\bar{\mu}\raw\infty$ the cosmon mass vanishes. If we associate the present value of $\chi$ with the present value of the Planck mass, $\chi(t_{0})=M$, the cosmon mass can be very small for $\bar{\mu}\ll M$. In the limit $\bar{\mu}\raw 0$ the cosmon has all properties of a Goldstone boson. In particular, all its couplings are derivative couplings. This implies that non-derivative couplings are suppressed by $\bar{\mu}/\chi$ and vanish in the limit $\bar{\mu}/\chi\raw 0$. This has important consequences for the issues of time varying fundamental constants and apparent violation of the equivalence principle, since the cosmon mediates a type of ``fifth force'' \cite{PSW}.

The present cosmon mass depends on the value of the intrinsic mass scale $\bar{\mu}$ (in units in which $\chi(t_{0})=2.4\cdot 10^{18}$GeV). A few examples are:
\begin{align}\label{eq:PS16} 
\bar{\mu}&=\Lambda_{QCD}\com m_{c}\approx 2\cdot 10^{-11}\mathrm{eV}\com m_{c}^{-1}\approx 10\mathrm{km} \nn\\
\bar{\mu}&=\varphi_{0}\com m_{c}\approx 1.27\cdot 10^{-5}\mathrm{eV}\com m_{c}^{-1}\approx 1.6\mathrm{cm}\\
\bar{\mu}&=2\cdot 10^{-3}\mathrm{eV}\com m_{c}\approx H_{0}\com m_{c}^{-1}\approx H_{0}^{-1}\qquad \text{\cite{CWQ}}\; ,\nn
\end{align}
with $H_{0}=1.5\cdot 10^{-33}$eV the present value of the Hubble parameter. The value of $\bar{\mu}$ is not a property of the standard model of particle physics. Being related to the flow of a relevant parameter at the UV-fixed point, its computation involves properties of this fixed point. Furthermore, the present value of $\chi$ is needed, which may involve properties of a cosmological solution.

The issue of the cosmon mass is directly related to the question under which circumstances an exactly scale invariant standard model can be realized. The latter corresponds to a vanishing cosmon mass -- spontaneously broken exact gravity scale symmetry implies the presence of an exactly massless Goldstone boson, the dilaton. We will find that the issue depends crucially on the question if the gauge and Yukawa couplings of the SM correspond to relevant or irrelevant couplings at the UV-fixed point. Only if these couplings are irrelevant, with nonzero values given by their fixed point values, exact gravity scale symmetry can hold for the SM. If gauge or Yukawa couplings correspond to relevant couplings with UV-fixed point values zero, the flow of these couplings from zero to their non-zero values at the Planck scale induces a violation of gravity scale symmetry that forbids a massless Goldstone boson.

In short, the issue concerns the question of what happens when the momentum scale $\mu$ exceeds $\chi$. If dimensionless couplings continue to flow in this range, this induces the presence of an intrinsic scale $\bar{\mu}$. Indeed, for $\mu\gg\chi$ the presence of $\chi$ is irrelevant for the fluctuation effects since $\mu$ (or $k$) is already an IR-cutoff, such that the fluctuations with momenta $q^{2}\gtrsim\mu^{2}$ are not affected by $\chi$. Thus $\chi$ cannot appear as a scale for a running of couplings at $\mu\gg \chi$. Any nonzero running of dimensionless couplings must therefore involve another scale, e.g. an intrinsic scale $\bar{\mu}$. The vanishing of $\hat{\beta}_{\lambda}$ in eq.~\eqref{eq:76} requires that the running of $\lambda$ with $\mu$ stops once $\mu$ exceeds $\chi$.

We demonstrate this issue in a simplified model of QCD coupled to quantum gravity in presence of a scalar singlet $\chi$ (``dilaton quantum gravity''). We assume the presence of an ultraviolet fixed point, for which the graviton-dilaton part of the effective action takes the form \eqref{eq:16}. Close to the UV-fixed point the gauge coupling may either be a relevant coupling with fixed point at $g_{*}=0$, or an irrelevant coupling with fixed point $g_{*}\neq 0$. Examples for both scenarios have been found in quantum gravity computations, depending on the matter content of the model \cite{DHR,FLP,HR1,CE1,CLRPA,EV,EHW}. The two scenarios lead to quite different implications for the intrinsic scale $\bar{\mu}$.

Let us start with the somewhat simpler toy example of an asymptotically free gauge coupling, with flow equation approximated by
\begin{equation}\label{eq:PS17} 
\mu\p_{\mu} g^{2}=\begin{cases}
-b_{UV}g^{4} & \textit{for }\mu>\chi\\
-b_{0}g^{4} & \textit{for }\mu<\chi\; .
\end{cases}
\end{equation}
We assume that for $\mu>\chi$ additional degrees of freedom influence the running of $g$ and $b_{UV}$. This is not precisely quantum gravity, since the gravitational contribution is an anomalous dimension, $\mu\p_{\mu}g^{2}=\eta_{g}g^{2}$, $\eta_{g}<0$. The example demonstrates, however, important issues about the dependence on $\chi$ and the presence of an intrinsic scale. For the toy model $b_{UV}$ determines the flow close to the fixed point. The flow in this range differs from $\mu<\chi$ where only the gluon and quark fluctuations contribute, with $b_{0}$ the value of the standard model given by eq.~\eqref{eq:31}. The solution of eq.~\eqref{eq:PS17} reads
\begin{align}\label{eq:PS18} 
g^{-2}(\mu=\chi)&=g^{-2}(\chi)=g^{-2}(\Lambda)-\dfrac{b_{UV}}{2}\ln\left (\dfrac{\Lambda^{2}}{\chi^{2}}\right )\, ,\\
g^{-2}(\mu<\chi)&=g^{-2}(\chi)-\dfrac{b_{0}}{2}\ln\left (\dfrac{\chi^{2}}{\mu^{2}}\right )\; .
\end{align}
The scale $\Lambda$ is arbitrary in the range $\Lambda\geq \chi$. Together with $g^{-2}(\Lambda)$ it defines the value of the relevant parameter at some scale $\Lambda$. In particular, we may choose $\Lambda$ to be the present Planck mass, $\Lambda=M$, such that
\begin{equation}\label{eq:PS19} 
g^{-2}(\mu<\chi)=g^{-2}(M)-\dfrac{b_{UV}}{2}\ln\left (\dfrac{M^{2}}{\chi^{2}}\right )-\dfrac{b_{0}}{2}\ln\left (\dfrac{\chi^{2}}{\mu^{2}}\right )\; .
\end{equation}

The confinement scale, defined by $g^{-2}(\Lambda_{QCD})=0$, depends on $\chi$ according to
\begin{equation}\label{eq:PS20} 
b_{0}\ln\left (\dfrac{\chi^{2}}{\Lambda_{QCD}^{2}}\right )+b_{UV}\ln\left (\dfrac{M^{2}}{\chi^{2}}\right )=\dfrac{2}{g^{2}(M)}\; .
\end{equation}
Two limits are of interest. For $b_{UV}=0$ one has $\Lambda_{QCD}\sim\chi$,
\begin{equation}\label{eq:PS21} 
\Lambda_{QCD}=\chi\exp\left (-\dfrac{1}{b_{0}g^{2}(M)}\right )\; .
\end{equation}
This limit corresponds to the scale invariant standard model, cf. eq.~\eqref{eq:37} with $\bar{g}^{2}=g^{2}(M)$. This is the limit considered in ref.~\cite{CWQ}. On the other hand, for $b_{UV}=b_{0}$ the confinement scale is independent of $\chi$,
\begin{equation}\label{eq:PS22} 
\Lambda_{QCD}=M\exp\left (-\dfrac{1}{b_{0}g^{2}(M)}\right )\; .
\end{equation}
This is the limit assumed in refs.~\cite{PSW,CTH1,FGHW,THAM,CNW,FKDV1,FKDV2}.

For the general $\chi$-dependence of $\Lambda_{QCD}(\chi)$ we may define
\begin{equation}\label{eq:PS23} 
\bar{\Lambda}_{QCD}=\Lambda_{QCD}(\chi=M)\, ,
\end{equation}
and use this in order to eliminate $g^{2}(M)$,
\begin{equation}\label{eq:PS24} 
b_{0}\left [\ln\left (\dfrac{\chi^{2}}{\Lambda_{QCD}^{2}}\right )-\ln\left (\dfrac{M^{2}}{\bar{\Lambda}_{QCD}^{2}}\right )\right ]+b_{UV}\ln\left (\dfrac{M^{2}}{\chi^{2}}\right )=0\; .
\end{equation}
This yields
\begin{equation}\label{eq:PS25} 
\Lambda_{QCD}(\chi)=\bar{\Lambda}_{QCD}\left (\dfrac{\chi}{M}\right )^{1-\frac{b_{UV}}{b_{0}}}\; .
\end{equation}
Only for $b_{UV}=b_{0}$ the confinement scale becomes independent of $\chi$ and only for $b_{UV}=0$ it is precisely proportional to $\chi$.

The two limits lead to rather different consequences for the intrinsic mass scale $\bar{\mu}$. In the limit $b_{UV}=0$ the running gauge coupling only depends on $\mu/\chi$, cf. eq.~\eqref{eq:38}. In this case no intrinsic scale is generated, $\bar{\mu}=0$. The spontaneous breaking of the exact scale invariance induces a massless Goldstone boson. The case $b_{UV}=0$ implies that in the range $\mu\gg \chi$ the gauge coupling runs towards a fixed point $g_{*}\neq 0$ since the running stops for $\mu>\chi$. (Threshold effects smoothen the transition at $\mu=\chi$, but the running has to stop sufficiently far above the threshold.) For the other limit $b_{UV}=b_{0}$ the confinement scale appears as an intrinsic mass scale, $\bar{\mu}=\Lambda_{QCD}$. The cosmon has a typical mass $\approx 2\cdot 10^{-11}\mathrm{eV}$. For intermediate $b_{UV}$ the cosmon mass interpolates between the two limits. These simple findings demonstrate that the cosmon mass cannot be computed without assumptions about the properties of the UV-fixed point.

In quantum gravity, the situation is conceptually similar. For $\mu>\chi$ the running of the gauge coupling obeys
\begin{equation}\label{eq:5.32A} 
\mu\p_{\mu}g^{2}=\eta_{g}g^{2}-b_{UV}g^{4}\com \eta_{g}<0\, .
\end{equation}
For $b_{UV}>0$ the gauge coupling is asymptotically free. Its UV-fixed point value is $g_{*}=0$, and it is a relevant coupling, with critical exponent $\theta_{g}=-\eta_{g}>0$. The value of $\alpha=g^{2}/4\pi$ has to reach at $\mu=\chi=M$ a value around $1/40$, if the present value of $\chi$ is associated to the Planck mass (or a scale in the vicinity of it). Neglecting for simplicity the term $\sim b_{UV}g^{4}$ one has
\begin{equation}\label{eq:5.32B} 
g^{-2}(\mu=\chi)=g^{-2}(\Lambda)\left (\dfrac{\chi}{\Lambda}\right )^{-\eta_{g}}\, .
\end{equation}
This yields for $\mu<\chi$, identifying $\Lambda$ with $M$
\begin{equation}\label{eq:5.32C} 
g^{-2}(\mu)=\dfrac{1}{4\pi\alpha(M)}\left (\dfrac{\chi}{M}\right )^{-\eta_{g}}-\dfrac{b_{0}}{2}\ln\left (\dfrac{\chi^{2}}{\mu^{2}}\right )\, .
\end{equation}
One finds a confinement scale depending both on $\chi$ and $M$
\begin{equation}\label{eq:5.32D} 
\Lambda_{QCD}=\chi\exp\left \{-\dfrac{1}{4\pi\alpha(M)}\left (\dfrac{\chi}{M}\right )^{-\eta_{g}}\right \}\, .
\end{equation}
The dependence of $\Lambda_{QCD}/\chi$ on $\chi$ introduces an explicit breaking of scale symmetry and an intrinsic mass scale. Asymptotically free gauge couplings in asymptotically safe quantum gravity are not compatible with a scale invariant standard model.

On the other hand, the scale invariant standard model is predicted if gauge and Yukawa couplings are irrelevant couplings at the UV-fixed point, taking nonzero fixed point values $g_{*}\neq 0$. For the standard model this is possible if one adds for the high-energy flow a sufficient number of scalars and gauge bosons as, for example, in grand unification \cite{EHW}. In this case $b_{UV}$ is negative and a fixed point occurs for
\begin{equation}\label{eq:5.32E} 
g_{*}^{2}=\dfrac{\eta_{g}}{b_{UV}}\, .
\end{equation}
The critical exponent at the fixed point is indeed negative
\begin{equation}\label{eq:5.32F} 
\theta=-\eta_{g}+2b_{UV}g_{*}^{2}=b_{UV}g_{*}^{2}\, .
\end{equation}
For $g$ near $g_{*}$ the flow equation for $\mu>\chi$ can be approximated by
\begin{equation}\label{eq:PS26} 
\mu\p_{\mu}g^{2}=-\theta\left (g^{2}-g^{2}_{*}\right )\, .
\end{equation}
Following the discussion in sect.~\ref{sec:RelevantParameters} the irrelevant coupling $g-g_{*}$ has to be zero for a renormalizable quantum field theory valid for all momenta. The solution for $\mu\geq \chi$ is simply $g(\mu\geq \chi)=g_{*}$. The ratio $\Lambda_{QCD}/\chi$ is predicted in terms of the fixed point value $g_{*}$,
\begin{equation}\label{eq:PS27} 
\dfrac{\Lambda_{QCD}}{\chi}=\exp\left (-\dfrac{1}{b_{0}g_{*}^{2}}\right )\; .
\end{equation}
Scale symmetry is exact without an intrinsic mass scale ($\bar{\mu}=0$), and a Goldstone boson is present for any $\chi\neq 0$. We conclude that scale invariant QCD is realized if the gauge coupling takes a non-zero fixed point value at the UV-fixed point.

Similar considerations hold for the electroweak sector of the standard model. Consider first the dimensionless Yukawa couplings, quartic scalar coupling and electroweak gauge couplings. If their UV-flow close to the fixed point is characterized by fixed nonzero values, with deviations from the fixed point values being irrelevant parameters, the values of these couplings do not depend on $\mu$ or $\chi$ as long as $\mu\geq\chi$. For $\mu<\chi$ they can only depend on the ratio $\mu/\chi$. The running as a function of $\mu/\chi$ is governed by the usual $\beta$-functions of the standard model. No scale symmetry breaking occurs in this sector. On the other hand, if some of these couplings are relevant parameters, with values for $\Lambda=M$ different from the UV-fixed point values, then the couplings depend for $\mu<\chi$ both on $\mu/\chi$ and $M/\chi$, similar to eq.~\eqref{eq:PS19} for $b_{UV}\neq 0$ or to eq.~\eqref{eq:5.32B}. Eliminating the values of the couplings at $M$ in terms of a suitable intrinsic scale $\bar{\mu}$ the couplings depend on $\mu/\chi$ and $\bar{\mu}/\chi$. The dependence on $\bar{\mu}$ reflects scale symmetry breaking by the flow away from the fixed point. In the language of sect.~\ref{sec:SponAndExpSymBr} the function $\hat{\beta}$ in eq.~\eqref{eq:77A} differs from zero.

The most important question concerns the deviation from the critical surface of the vacuum electroweak phase transition. There may be more than one relevant or irrelevant coupling playing a role. Consider a minimum of the Higgs potential at $\rho=h^{\dagger}h=0$, with terms quadratic in $h$ of the form
\begin{equation}\label{eq:PS28} 
U=\bar{m}^{2}\rho+\tilde{\varepsilon}\chi^{2}\rho+\ldots\, .
\end{equation}
At the UV fixed point both $\bar{m}^{2}/k^{2}$ and $\tilde{\varepsilon}$ take fixed values. The critical surface of the electroweak phase transition is invariant with respect to the flow. We assume that the UV-fixed point is on the critical surface ($\gamma_{*}=0$).

There are several possible scenarios, depending on the two parameters $\bar{m}^{2}$ and $\tilde{\varepsilon}$ being relevant or irrelevant. If both $\bar{m}^{2}$ and $\tilde{\varepsilon}$ are irrelevant, the theory is exactly on the critical surface of the vacuum electroweak phase transition \cite{CWPFP,CWMY}. This would be an example of ``self organized criticality'' \cite{BORW}. An explanation why the observed Fermi scale is not the minimal possible one will be needed. The SM would be scale invariant. If $\bar{m}^{2}/k^{2}$ is irrelevant it can be replaced by its fixed point value for $\mu\geq \chi$. The $\chi$-dependence of $\varphi_{0}$ depends then on the running of the dimensionless coupling $\tilde{\varepsilon}$, as defined by the quartic vertex with two singlets and two doublets. If this running does not involve an intrinsic scale $\bar{\mu}$, or if $\bar{\mu}\ll\varphi_{0}$, one will find $\varphi_{0}\sim\chi$. Scale symmetry remains exact (or a very good approximation for $\bar{\mu}\ll\varphi_{0}$). This realizes the standard model with quantum scale symmetry discussed in sect.~\ref{sec:QScaleSymAndQuqnatumEffAction}. The flow of $\bar{m}^{2}$ and $\tilde{\varepsilon}$ for $\mu<\chi$ depends only on $\mu/\chi$ and involves the anomalous mass dimension \eqref{eq:43}. If both $\bar{m}^{2}$ and $\tilde{\varepsilon}$ are relevant parameters at the UV-fixed point, the $\chi$-dependence of $\varphi_{0}$ depends on the relative size of the two contributions $\bar{m}^{2}\rho$ and $\tilde{\varepsilon}\chi^{2}\rho$. If the term $\tilde{\varepsilon}\chi^{2}\rho$ dominates, the intrinsic scale associated to the running of $\bar{m}^{2}(k)$ (more precisely $(\bar{m}^{2}-\bar{m}^{2}_{*})/k^{2}$) leads only to small corrections of the scaling relation $\varphi_{0}\sim\chi$. On the other hand, if $\bar{m}^{2}\rho$ dominates, the Fermi scale becomes an intrinsic scale, $\bar{\mu}=\varphi_{0}$. In this case the present mass of the cosmon is of the order $\bar{\mu}^{2}/M$. One may also consider the possibility that the intrinsic scale symmetry breaking is not directly related to the potential for the Higgs scalar. This leads to scale symmetry breaking in extensions of the standard model \cite{HAWI,CNW,FKDV1,FKDV2,IOO,HUKO,
ANP,IO2,EJKS,FPS,CJL,HRRST,HAST,CAR1,
KHO,HAIO,HKLL,QUI,CTH2,KATAM,
HHLS,BHELI,KLSY}.

In summary, the mass of the cosmon depends on the ratio $\bar{\mu}/\chi$ of the intrinsic scale $\bar{\mu}$, as compared to the scale of spontaneous scale symmetry breaking $\chi$. In turn, the issue of an intrinsic scale depends on the properties of the UV-fixed point and cannot be settled by investigating properties of the standard model as an effective low energy theory. If the couplings of the standard model at the scale $\mu=\chi$ are given by the UV-fixed point values, the standard model is quantum scale invariant. No intrinsic scale $\bar{\mu}$ arises at this level. There may still be an intrinsic scale $\bar{\mu}$ generated in a different sector, for example in the sector of the scalar singlet $\chi$. As we will discuss later, this scale may be many orders of magnitudes smaller than $\Lambda_{QCD}$. 

As a second important lesson of this investigation we argue that the scale invariance of the (extended) standard model cannot simply be ``postulated'' by imposing field-dependent renormalization conditions. The condition that a coupling $\lambda(\mu=\chi)$ is independent of $\chi$, i.e. $\hat{\beta}_{\lambda}=0$ in eq.~\eqref{eq:76}, makes implicitly an assumption. Either there is no momentum range $\mu>\chi$ for which a continuum quantum field theory holds, as for example for a lattice theory. Or a continuum quantum field theory remains a valid description for $\mu>\chi$, and $\lambda$ is given by a fixed point value $\lambda_{*}$, typically for the UV-fixed point defining the quantum field theory.

\section{Scale symmetry in quantum gravity}\label{sec:V}

The functional integral approach of formulating quantum gravity as a quantum field theory for the metric has a long history \cite{FEY2,DEWI2,GTHV,BAVI1,BAVI2}. If quantum gravity can be understood as a renormalizable continuum quantum field theory for the metric or vierbein, an UV-fixed point must exist. For gravity it is very likely that interactions do not vanish at this fixed point. One therefore encounters asymptotic safety \cite{WEIAS,MRQG} rather than asymptotic freedom \cite{GW,POL}. Quantum gravity is then not perturbatively renormalizable. The definition of quantum gravity at or in the vicinity of such an UV-fixed point implies quantum scale symmetry in the UV, e.g. for $\mu\raw\infty$. The issue of scale symmetry at finite momenta depends on the values of the relevant parameters for small deviations from the UV-fixed point.

The scale symmetry associated to the UV-fixed point of the full theory for gravity and particle physics will be named ``gravity scale symmetry''. As we have discussed, it is distinct from particle scale symmetry for the standard model as an effective low energy theory, associated to the second order vacuum electroweak phase transition. We will concentrate our discussion on two alternative scenarios. For the first the Planck mass $M$ corresponds to a relevant parameter and is associated to an intrinsic scale \cite{MRQG,DP,SOU,RS}. For the second $M$ is given by a scalar field $\chi$. In this case the Planck mass reflects a spontaneous breaking of gravity scale symmetry, rather than an intrinsic scale. The second setting has been named ``dilaton quantum gravity'' \cite{PHW1,PHW2}.

We first investigate possible forms for a scale invariant effective action for quantum gravity. A scale invariant effective action is a direct consequence of an UV-fixed point which is used to define the quantum field theory. The issue of a scale invariant effective action is not yet clarified for quantum Einstein gravity. In contrast, simple scale invariant effective actions can be constructed for dilaton quantum gravity, where a scalar field $\chi$ is added to the metric. We subsequently discuss the issue of the UV-fixed point both from the point of view of two different perturbative expansions, and from results of functional renormalization. We address the vicinity of the fixed point with relevant parameters responsible for explicit scale symmetry breaking by intrinsic mass scales.

The second half of this section concentrates on the gravitational renormalization effects for the effective potential of scalar fields. The flow of the effective potential $U(k;\chi)$, both with the renormalization scale $k$ and the field $\chi$, involves aspects not encountered in the analysis of the flow of a finite number of couplings. An important role is played by the scaling form of the potential where $U/k^{4}$ only depends on $y=\chi^{2}/k^{2}$. This issue is therefore discussed in considerable detail. We first derive the scaling form of the potential for a simple ansatz for the flow equation, which accounts for the most important qualitative aspects. Subsequently, we present a full functional renormalization group study motivating our simple ansatz. Applied to the effective potential for the Higgs scalar we demonstrate that the ratio of masses of the Higgs scalar and the top quark can be predicted. We also address the gauge hierarchy problem from the perspective of the quantum gravity computation. 

In this section we use the metric as basic gravitational degree of freedom. Investigations with the vierbein instead of the metric lead to similar results. One could also extend the geometric setting by adding additional fields for torsion. Our discussion does not require that the metric or vierbein are the most fundamental degrees of freedom. It is well conceivable that a fundamental theory is formulated in terms of other degrees of freedom, for example only using fermions as in spinor gravity \cite{CWSGRA,CWSGLAT}. In such theories the metric and vierbein arise as composite or collective degrees of freedom. Whenever there is a large range of length scales between the Planck length and an even more microscopic ``compositness scale'' for the metric, the discussion of the UV-fixed point and gravity scale symmetry of this section is relevant.

\subsection{Quantum Einstein Gravity}\label{sec:Quantum_Einstein_Gravity} 

A derivative expansion of a diffeomorphism invariant effective action for the metric involves in second order in the derivatives two parameters, namely the (reduced) Planck mass $M$ and the cosmological constant $V$. In this approximation the effective action takes the Einstein-Hilbert form with a cosmological constant
\begin{equation}\label{eq:92} 
\Gamma=\int\,\sqrt{g}\left (-\dfrac{M^{2}}{2}R+V\right ).
\end{equation}
Here $M$ is related to Newton's gravitational constant by $G_{N}=(8\pi M^{2})^{-1}$ and we denote the cosmological constant by $V$, anticipating a more extended setting where it can be identified with a value of the effective potential for scalar fields. Typically $M^{2}$ and $V$ are relevant couplings.

The UV-fixed point for quantum gravity has been found within functional renormalization for the effective average action \cite{CWFR,MRCW,MRQG,DP,SOU,RS}. For this purpose one introduces an effective infrared cutoff which suppresses the contributions of fluctuations with momenta smaller than $k$, or a suitable covariant generalization thereof. The effective average action $\Gamma_{k}$ depends on $k$. It interpolates between the microscopic or classical action $S$ for $k\raw\infty$ where no fluctuations are included, and the quantum effective action $\Gamma$ for $k\raw 0$, where all fluctuations are included. The dependence of $\Gamma_{k}$ on the IR-scale $k$ obeys an exact functional renormalization group equation which takes a simple one-loop form \cite{CWFR,MRCW,MRQG}, see sect.~\ref{sec:Asymptotic_Safety}.

Within functional renormalization the parameters $M^{2}$ and $V$ depend on $k$. Already in a simple Einstein-Hilbert truncation which only retains the two flowing dimensionless parameters $M^{2}/k^{2}$ and $V/k^{4}$ a fixed point for the flow of these dimensionless couplings has been found \cite{MRQG,DP,SOU,RS}. Both $M^{2}/k^{2}$ and $V/k^{4}$ correspond to relevant couplings near the fixed point. The basic mechanism producing the fixed point is very simple and can be understood by the simplified flow equations 
\begin{align}
k\p_{k}M^{2}&=\p_{t}M^{2}=4c_{M}k^{2}\, ,\nn\\
\label{eq:93} k\p_{k}V&=\p_{t}V=4c_{V}k^{4}.
\end{align}
Here $c_{M}$ and $c_{V}$ are dimensionless, and we assume $c_{M}>0$. Eq.~\eqref{eq:93} implies for the dimensionless couplings ($t=\ln(k/k_{0})$)
\begin{align}
\p_{t}\left (\dfrac{M^{2}}{k^{2}}\right )&=-2\dfrac{M^{2}}{k^{2}}+4c_{M}\, ,\nn\\
\label{eq:94} \p_{t}\left (\dfrac{V}{k^{4}}\right )&=-4\dfrac{V}{k^{4}}+4c_{V}\; .
\end{align}
The fixed point occurs for
\begin{equation}\label{eq:95} 
\dfrac{M^{2}_{*}(k)}{k^{2}}=2 c_{M}\com \dfrac{V_{*}(k)}{k^{4}}=c_{V}\, .
\end{equation}
The general solution of the flow equations has two integration constants that we choose as
\begin{equation}\label{eq:96} 
M^{2}=M^{2}(k=0)\com V=V(k=0).
\end{equation}

If we assume in a first approximation constant values for $c_{M}$ and $c_{V}$ the general solution takes the simple form
\begin{equation}\label{eq:97} 
M^{2}(k)=M^{2}+2 c_{M} k^{2}\com V(k)=V+c_{V}k^{4}.
\end{equation}
As it should be, the UV-fixed point \eqref{eq:95} is approached for $k\raw\infty$. In functional renormalization the squared Planck mass is running quadratically with $k$. We comment on the meaning of the quadratic running in sect.~\ref{sec:6.18}.

Using the dimensionless ratios
\begin{equation}\label{eq:FPA} 
u=\dfrac{V}{k^{4}}\com w=\dfrac{M^{2}}{2k^{2}}\com \tilde{\lambda}=\dfrac{V}{M^{4}}=\dfrac{u}{4w^{2}}\, ,
\end{equation}
the flow equations read
\begin{align}\label{eq:FPB} 
\p_{t}u&=-4(u-c_{V})\com \p_{t}w=-2(w-c_{M})\, ,\nn\\
\p_{t}\tilde{\lambda}&=\dfrac{c_{V}}{w^{2}}-\dfrac{4c_{M}\tilde{\lambda}}{w}\, .
\end{align}
The general solution for constant $c_{M}$, $c_{V}$,
\begin{align}\label{eq:FPC}
&u=c_{V}+\dfrac{U_{0}}{k^{4}}\com w=c_{M}+\dfrac{\overline{M}^{2}}{2k^{2}}\, ,\nn\\ 
&\tilde{\lambda}=\dfrac{\tilde{\lambda}_{0}\overline{M}^{4}+c_{V}k^{4}}{(\overline{M}^{2}+2c_{M}k^{2})^{2}}\, ,
\end{align}
has two free integration constants $\overline{M}^{2}$ and
\begin{equation}\label{eq:FPD} 
\tilde{\lambda}_{0}=\dfrac{U_{0}}{\overline{M}^{4}}\, .
\end{equation}
It interpolates between the UV-fixed point \eqref{eq:95} for $k\raw\infty$,
\begin{equation}\label{eq:FPE} 
\text{UV:}\quad w_{*}=c_{M}\com\tilde{\lambda}_{*}=\dfrac{c_{V}}{4c_{M}^{2}}\, ,
\end{equation}
and an intermediate IM-fixed point
\begin{equation}\label{eq:FPF} 
\text{IM:}\quad w_{*}^{-1}=0\com \tilde{\lambda}_{*}=\tilde{\lambda}_{0}\, .
\end{equation}
For constant $c_{M}$ and $c_{V}$ the flow trajectories in the $(w,\tilde{\lambda})$-plane are given by
\begin{equation}\label{eq:FPG} 
\tilde{\lambda}=\dfrac{c_{V}}{4w^{2}}+\tilde{\lambda}_{0}\left (1-\dfrac{c_{M}}{w}\right )^{2}\, .
\end{equation}
We will find in sect.~\ref{sec:Infrared_QG} that the approximation of constant $c_{V}$ breaks down as the ratio
\begin{equation}\label{eq:FPH} 
v=\dfrac{u}{w}=4\tilde{\lambda}w
\end{equation}
approaches one. A reasonable qualitative approximation uses eq.~\eqref{eq:FPG} only for $v<1$ or $w<1/(4\tilde{\lambda})$, while for larger $w$ one may approximate
\begin{equation}\label{eq:FPI} 
\tilde{\lambda}=\dfrac{1}{4w}\, .
\end{equation}
Thus for $k\raw 0$, where $w\raw\infty$, the IR-fixed point is given by
\begin{equation}\label{eq:FPJ} 
\text{IR:}\quad w_{*}^{-1}=0\com \tilde{\lambda}_{*}=0\, .
\end{equation}

\begin{figure}[t!]
\includegraphics[scale=0.7]{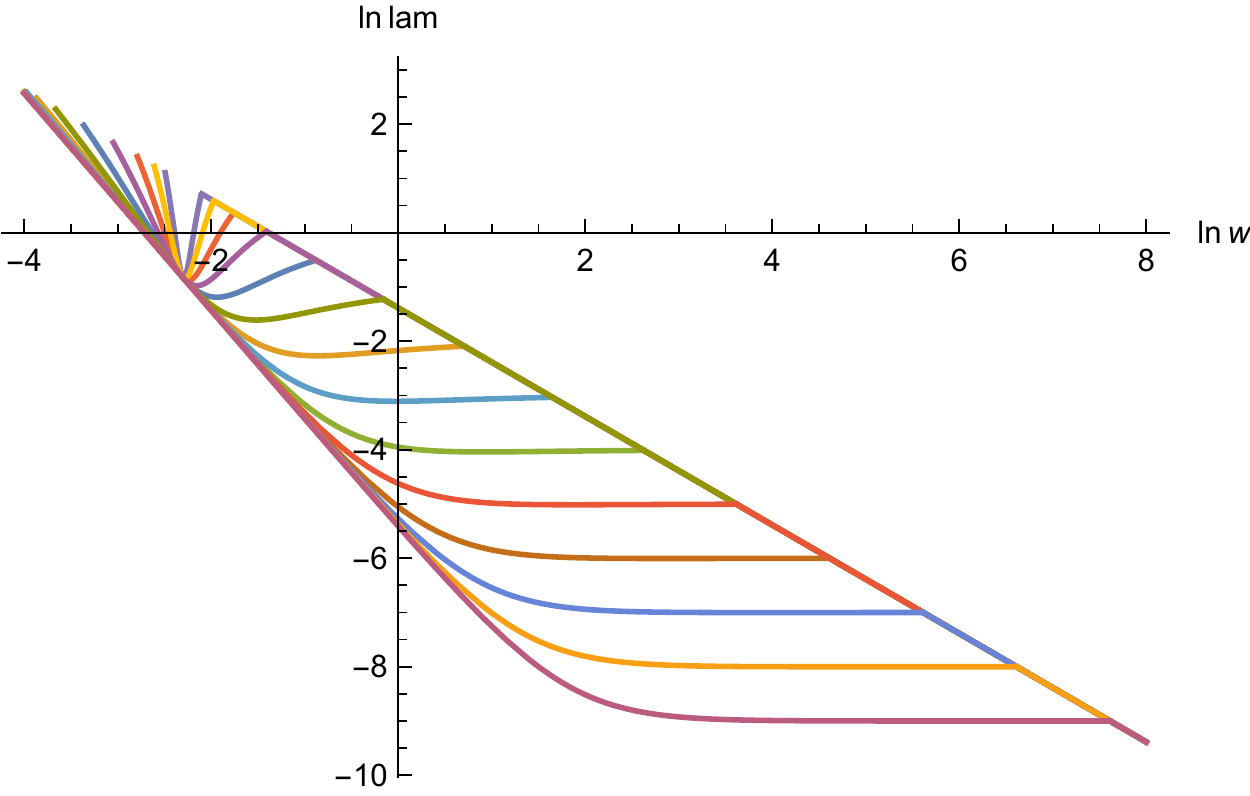}
\caption{Flow trajectories of $\ln(\tilde{\lambda})$ as function of $\ln(w)$. Parameters are $c_{M}=0.1$, $c_{V}=0.018$. All trajectories flow towards the IR-fixed point at $\tilde{\lambda}_{*}=0$, $w_{*}^{-1}=0$. The vicinity of the IR-fixed point is given by the upper diagonal straight line. The intermediate IM-fixed point corresponds to the horizontal lines. The lower diagonal reflects the flow of $\tilde{\lambda}\sim w^{-2}$ away from the UV-fixed point which is realized for small $\tilde{\lambda}_{0}$. The UV-fixed point is shared by all trajectories.}\label{fig:FL} 
\end{figure}

\begin{figure}[t!]
\includegraphics[scale=0.7]{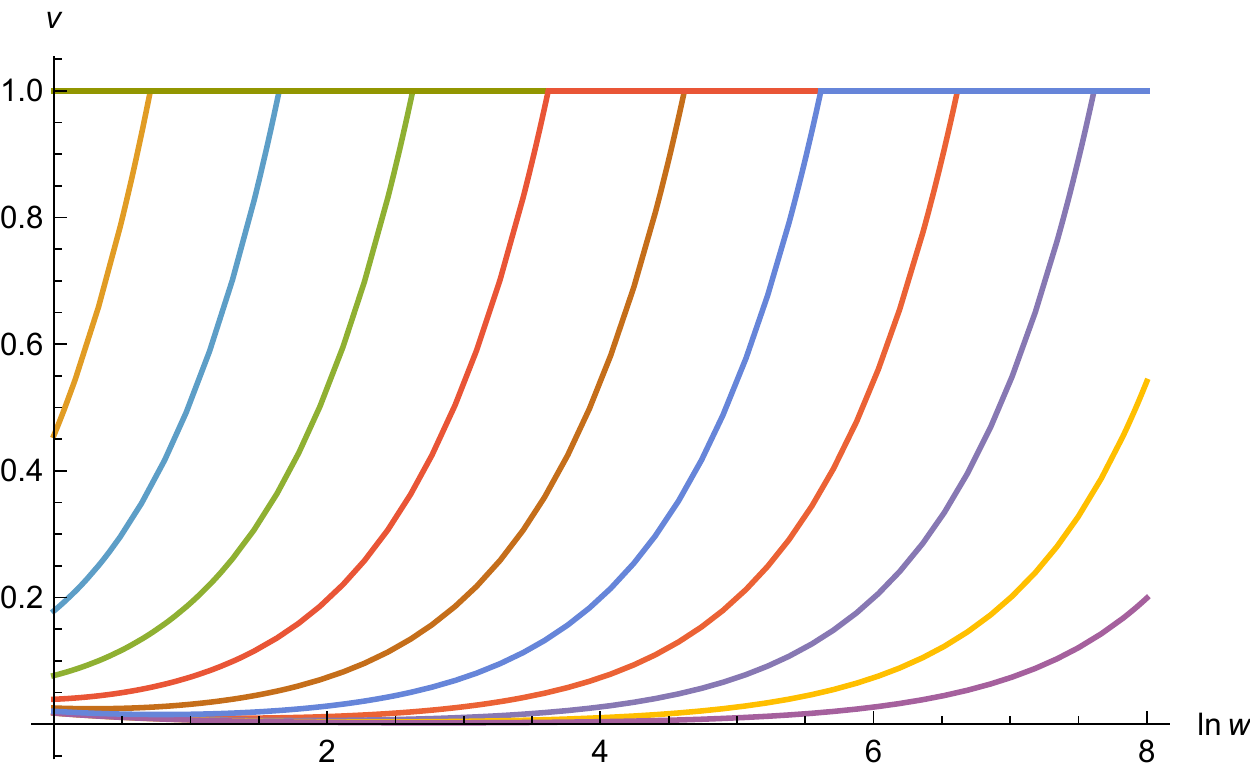}
\caption{Flow trajectories of $v$ as a function of $\ln(w)$, for $c_{M}=0.1$, $c_{V}=0.018$. For small $\tilde{\lambda}_{0}$ one observes first a decrease of $v$ to very small values, and a subsequent increase towards the infrared fixed point $v=1$.}\label{fig:FV} 
\end{figure}

We have depicted the flow trajectories in this approximation in Fig.~\ref{fig:FL}, taking $c_{M}=0.1$, $c_{V}=0.018$. All flow trajectories start for $k\raw\infty$ at the UV-fixed point. For small $\tilde{\lambda}_{0}$ they move away from the fixed point as $\tilde{\lambda}=u_{*}/(4w^{2})$, which corresponds to constant $u=u_{*}=c_{V}$. Subsequently, as $k$ is lowered further, they reach the IM-fixed point with constant $\tilde{\lambda}$. It corresponds to the horizontal lines, whose location depends on $\tilde{\lambda}_{0}$. Finally, for $k\raw 0$ all trajectories turn over to the IR-fixed point with $\tilde{\lambda}=1/(4w)$ or $v=1$. The corresponding trajectories for $v$ as a function of $w$ are shown in Fig.~\ref{fig:FV}. All trajectories end for $k\raw 0$ in the IR-fixed point $v=1$. For small $\tilde{\lambda}_{0}$ they decrease from the UV-fixed point towards a value close to zero, and increase subsequently to the IR-value. We discuss the flow of $\tilde{\lambda}$ and $v$ in detail in sects.~\ref{sec:Scal_Pot_Dil_Grav}-\ref{sec:Crossover_QuantumGravity}.

In dilaton quantum gravity $w$ becomes a function of a scalar field $\chi$, with a typical behavior
\begin{equation}\label{eq:FPK} 
w=\bar{w}+\dfrac{\xi}{2}\dfrac{\chi^{2}}{k^{2}}=\bar{w}+\dfrac{\xi y}{2}\, .
\end{equation}
For rather generic cosmological solutions $\chi$ increases with cosmic time $t$. The flow trajectory in fig.~\ref{fig:FL} is then followed towards large $w$ as $t$ increases. This is of direct relevance for dark energy since $\tilde{\lambda}\overline{M}^{4}$ corresponds (in the Einstein frame) to the potential energy of the cosmon, the pseudo-Goldstone boson of spontaneously broken scale symmetry. We will discuss in sect.~\ref{sec:Scale_Sym_Cosmo} a normalization of the scalar field where $\varphi$ is directly related to $\tilde{\lambda}$ by
\begin{equation}\label{eq:FPL} 
\dfrac{\varphi}{\overline{M}}=-\ln(\tilde{\lambda})\, .
\end{equation}
The vertical axis in Fig.~\ref{fig:FL} corresponds then to $-\varphi/\overline{M}$. In the context of dilaton quantum gravity the flow diagram in Figs.~\ref{fig:FL}, \ref{fig:FV} is of central importance for the discussion of dynamical dark energy.

Beyond the approximation of constant $c_{M}$ and $c_{V}$ we have to understand their dependence on the dimensionless ratios $M^{2}/k^{2}$ and $V/k^{4}$. As an example, we report the graviton contribution to the flow of $V$, computed from functional renormalization \cite{CWGFE,CWIRG,NAPER}. This ``graviton approximation'' accounts for the contribution of the traceless transversal tensor modes of the metric fluctuations. While remaining algebraically comparatively simple, this approximation contains already many important aspects of the UV-fixed point. The graviton approximation reads in the truncation \eqref{eq:92}
\begin{equation}\label{eq:105} 
\p_{t}V=\dfrac{5}{2}\int_{q}\, G_{g,k}(q^{2})\,\p_{t}R_{k}(q^{2})
\end{equation}
with IR-regularized propagator
\begin{equation}\label{eq:106} 
G_{g,k}(q^{2})=\left (\dfrac{M^{2}}{4}q^{2}+R_{k}(q^{2})-\dfrac{V}{2}\right )^{-1}
\end{equation}
involving the IR-cutoff function
\begin{equation}\label{eq:107} 
R_{k}(q^{2})=\dfrac{M^{2}k^{2}}{4}\, r_{k}(q^{2}/k^{2})\, .
\end{equation}
We employ
\begin{equation}\label{eq:108} 
\int_{q}=\int_{-\infty}^{\infty}\, \dfrac{\dif^{4}q}{(2\pi)^{4}}=\dfrac{k^{4}}{16\pi^{2}}\int_{0}^{\infty}\dif x\, x\com x=\dfrac{q^{2}}{k^{2}}\, ,
\end{equation}
and observe that the momentum integral is UV-finite if $r_{k}(x)$ falls off fast enough for large $x$. It is also IR-finite if $r_{k}(x)$ approaches for $x\raw 0$ a constant or increases, due to the form of the regulated propagator
\begin{equation}\label{eq:109} 
G_{g,k}=\dfrac{4}{M^{2}k^{2}}\left (x+r_{k}(x)-v\right )^{-1}\; .
\end{equation}
Here we define the dimensionless ratio $v$ by
\begin{equation}\label{eq:110} 
v=\dfrac{2V}{M^{2}k^{2}}\; .
\end{equation}
In eq.~\eqref{eq:105} the factor five arises from the five components of the traceless transverse mode. 

The r.h.s. of \eq{105} is a one-loop integral with propagator depending on the $k$-dependent variables $M^{2}(k)$ and $V(k)$. Expressing the flow as a dimensionless integral yields
\begin{equation}\label{eq:111} 
\p_{t}V=\dfrac{5 k^{4}}{16\pi^{2}}\ell_{0}(\tilde{w})\com \tilde{w}=-v\, ,
\end{equation}
with threshold function ($r(x)=r_{k}(x)$)
\begin{align}\label{eq:112} 
\ell_{0}(\tilde{w})&=\dfrac{8\pi^{2}}{k^{4}}\int_{q}\, G_{g,k}\,\p_{t}R_{k}\nn\\
&=\dfrac{1}{2}\int_{0}^{\infty}\dif x\, x\left (p(x)+\tilde{w}\right )^{-1}\, f(x)r(x)\, ,
\end{align}
where
\begin{equation}\label{eq:113} 
p(x)=x+r(x)
\end{equation}
and
\begin{equation}\label{eq:114} 
f(x)=\dfrac{\p_{t}R_{k}}{R_{k}}=2+\eta_{M}-2\dfrac{\p \ln(r)}{\p \ln(x)}\; ,
\end{equation}
with
\begin{equation}\label{eq:115} 
\eta_{M}=\dfrac{\p_{t}M^{2}}{M^{2}}=4c_{M}\dfrac{k^{2}}{M^{2}}\; .
\end{equation}
We identify the graviton contribution to $c_{V}$ as
\begin{equation}\label{eq:116} 
c_{V}^{(g)}=\dfrac{5}{64\pi^{2}}\,\ell_{0}(-v)\; .
\end{equation}

The threshold function $\ell_{0}(\tilde{w})$ has simple qualitative properties. For $\tilde{w}=0$ it is of the order one, with a part proportional to $\eta_{M}$. For large positive $\tilde{w}$ one finds a decay $\ell_{0}(\tilde{w})\sim\tilde{w}^{-1}$,
\begin{align}\label{eq:117} 
\ell_{0}(\tilde{w}\gg 1)&=\dfrac{m_{0}}{\tilde{w}}\nn\\
 m_{0}&=\int_{0}^{\infty}\dif x\, x\left [\left (1+\dfrac{\eta_{M}}{2}\right )r-\dfrac{\p r}{\p \ln(x)}\right ]\; ,
\end{align}
provided the function $r$ is chosen such that the integral $m_{0}$ is finite. Typically $r$ decays fast for $x\gg 1$, such that only the interval $[0,1]$ contributes effectively to the integral \eqref{eq:117}. For negative $\tilde{w}$ the threshold function is enhanced as compared to $\tilde{w}=0$. It may have a singularity at some critical $\tilde{w}_{cr}<0$. Details of the threshold function depend on the precise choice of $r_{k}(x)$. We will need the threshold function $\ell_{0}(-v_{*})$ at the fixed point, where
\begin{equation}\label{eq:118} 
v_{*}=\dfrac{c_{V}}{c_{M}}\com \eta_{M*}=2\, .
\end{equation}

There are further contributions to $c_{V}$ from the scalar metric fluctuations and the measure factor which combines contributions of gauge degrees of freedom and ghosts. Furthermore, the fluctuations of gauge bosons, fermions and scalars contribute to $c_{V}$. This holds similarly for $c_{M}$. It is not our aim here to report existing more complete results on $c_{V}$ and $c_{M}$. They depend on the assumed matter content, truncation and choice of infrared cutoff.

For a discussion of the vicinity of the UV-fixed point suitable variables are $v$ and
\begin{equation}\label{eq:119} 
w=\dfrac{M^{2}}{2k^{2}}\; .
\end{equation}
The flow for these dimensionless couplings reads
\begin{align}\label{eq:120} 
\p_{t}w&=\beta_{w}=-2 w+2c_{M}(w,v)\, ,\nn\\
\p_{t}v&=\beta_{v}=-2v-\dfrac{2c_{M}(w,v)v}{w}+\dfrac{4c_{V}(w,v)}{w}\; ,
\end{align}
where we use
\begin{equation}\label{eq:121} 
\eta_{M}=\dfrac{2c_{M}}{w}\; .
\end{equation}
(Note $c_{M}=c$ in eq.~\eqref{eq:55}.) The fixed point occurs for ($\eta_{M}(w_{*},v_{*})=2$)
\begin{equation}\label{eq:122} 
w_{*}=c_{M}(w_{*},v_{*})\, ,\, v_{*}=\dfrac{c_{V}(w_{*},v_{*})}{w_{*}}=\dfrac{c_{V}(w_{*},v_{*})}{c_{M}(w_{*},v_{*})}\, ,
\end{equation}
in accordance with eq.~\eqref{eq:95}. 

For the behavior close to the fixed point we first take the approximation where all couplings except $w$ and $v$ are taken as constants, e.g. we neglect the $\beta$-functions for the other couplings. The resulting reduced stability matrix $T$ in eq.~\eqref{eq:45} reads
\begin{align}\label{eq:123} 
T_{ww}&=-\dfrac{\p\beta_{w}}{\p w}=2-2\dfrac{\p c_{M}}{\p w}=2-2\dfrac{\p\ln(c_{M})}{\p \ln(w)}\nn\, ,\\
T_{wv}&=-\dfrac{\p\beta_{w}}{\p v}=-2\dfrac{\p c_{M}}{\p v}=-\dfrac{2w}{v}\dfrac{\p\ln(c_{M})}{\p \ln(v)}\, ,\nn\\
T_{vw}&=-\dfrac{\p\beta_{v}}{\p w}=-\dfrac{2c_{M}v}{w^{2}}+\dfrac{2v}{w}\dfrac{\p c_{M}}{\p w}+\dfrac{4c_{V}}{w^{2}}-\dfrac{4}{w}\dfrac{\p c_{V}}{\p w}\nn\\
&=\dfrac{2v}{w}\left (1+\dfrac{\p\ln(c_{M})}{\p\ln(w)}-2\dfrac{\p\ln(c_{V})}{\p\ln(w)}\right )\, ,\nn\\
T_{vv}&=-\dfrac{\p\beta_{v}}{\p v}=4+\dfrac{2v}{w}\dfrac{\p c_{M}}{\p v}-\dfrac{4}{w}\dfrac{\p c_{V}}{\p v}\nn\\
&=4+2\dfrac{\p\ln(c_{M})}{\p\ln(v)}-4\dfrac{\p\ln(c_{V})}{\p\ln(v)}\, ,
\end{align}
where all quantities have to be evaluated at the fixed point.

We need the eigenvalues $\theta_{i}$ of the stability matrix $T$. We employ
\begin{align}\label{eq:124} 
\tr(T)&=\theta_{1}+\theta_{2}=6+2y_{1}\, ,\nn\\
y_{1}&=\dfrac{\p\ln (c_{M})}{\p\ln(v)}-2\dfrac{\p\ln(c_{V})}{\p \ln(v)}-\dfrac{\p\ln(c_{M})}{\p\ln(w)}\, ,
\end{align}
and
\begin{align}\label{eq:125} 
\det(T)&=\theta_{1}\theta_{2}=8\left (1+y_{2}\right )\, ,\\
y_{2}&=\dfrac{\p\ln c_{M}}{\p\ln(v)}-\dfrac{\p\ln(c_{V})}{\p \ln(v)}-\dfrac{\p\ln(c_{M})}{\p\ln(w)}\nn\\
&\quad+\dfrac{\p\ln(c_{V})}{\p \ln(v)}\dfrac{\p\ln(c_{M})}{\p\ln(w)}-\dfrac{\p\ln(c_{V})}{\p \ln(w)}\dfrac{\p\ln(c_{M})}{\p \ln(v)}\nn\; .
\end{align}
The two eigenvalues are
\begin{equation}\label{eq:126} 
\theta=3+y_{1}\pm\sqrt{1-8y_{2}+6y_{1}+y_{1}^{2}}\; .
\end{equation}
For $y_{1}=y_{2}=0$ one recovers the solution \eqref{eq:97} with two relevant parameters, corresponding to $\theta_{1}=2$, $\theta_{2}=4$. Correspondingly, for $|y_{1}|\ll 1$, $|y_{2}|\ll 1$ both critical exponents remain positive
\begin{align}\label{eq:127} 
\theta_{1}&=2+4y_{2}-2y_{1}\, ,\nn\\
\theta_{2}&=4-4y_{2}+4y_{1}\; .
\end{align}

For $y_{1}>-3$, $y_{2}>-1$ both couplings remain relevant. As long as
\begin{equation}\label{eq:128}
\Delta=8y_{2}-6y_{1}-y_{1}^{2}<1
\end{equation}
both critical exponents are real. On the other hand, for $\Delta>1$ the two critical exponents are complex and conjugate to each other. The corresponding limit cycle behavior has been observed in the approximation of ref.~\cite{SOU,RS}. For $y_{1}<-3$ at least one of the couplings becomes irrelevant. For $y_{2}<-1$ one coupling is relevant, the other irrelevant, with both critical exponents real. Two irrelevant couplings would be realized for $y_{1}<-3$, $y_{2}>-1$. Realistic gravity is obtained if both $w$ and $v$ are relevant parameters, i.e. for $y_{1}>-3$, $y_{2}>-1$. One can then choose initial values close to the fixed point such that arbitrary $M^{2}$ and $V$ are obtained for $k\raw 0$. One of the intrinsic scales is arbitrary and sets the units - typically this is $M^{2}$. A tiny ratio $V/M^{4}$ needs tuning of the initial conditions to high accuracy.

The fluctuations of fermions, scalars and gauge bosons contribute to $c_{M}$ and $c_{V}$. These matter contributions are independent of $v$ and $w$, since these two parameters appear only in the propagator of the gravitational degrees of freedom. In lowest order these contributions are also independent of Yukawa couplings, gauge couplings and scalar self interactions, such that the restriction to a reduced stability matrix is justified in this respect. If the constant parts in $c_{M}$ and $c_{V}$ dominate, the quantities $y_{1}$, $y_{2}$ are small and eq.~\eqref{eq:127} becomes a valid approximation. Furthermore, for large $c_{M}$ the fixed point value $w_{*}$ becomes large, such that close to the fixed point $M^{2}(k)=2w_{*}k^{2}$ is large as compared to $k^{2}$. Gravity becomes weak in this limit, and even a perturbative expansion in $w^{-1}$ may become possible. 

The truncation of keeping only the two terms \eqref{eq:92} in the gravitational sector is, of course, only an approximation. A more general ansatz will involve additional couplings. The discussion of a fixed point now also comprises these couplings, and the stability matrix has to be extended accordingly. Investigations of rather extended truncations, including arbitrary functions $f(R)$, the squared Weyl tensor, or the Goroff-Sagnotti invariant have all found an UV-fixed point with similar properties as the Einstein-Hilbert truncation \cite{DP,LAURE,LIT2,NIRE,NIMA,MASAU,
COPE,COPERRA,BEMASA,BEMASAU2,
RSREV,BENCAR,DIEMO,FLNR1,CLPR,
Benedetti:2013jk,Demmel:2014sga,CKPR,FLNR2,DSZ,CKMPR,Ohta:2015efa,GKLS,Falls:2016msz,DPR,HAYA,Christiansen:2017bsy,FKLNR,deBrito:2018jxt,
GORS,AEREV,FLS}.

Within functional renormalization the UV-fixed point is seen in the limit $k\raw\infty$. The infrared-cutoff scale $k$ is, however, only an artificial scale introduced in order to perform the integration of fluctuations stepwise. At the end one is interested in the quantum effective action which obtains for $k\raw 0$. Since both $k^{2}$ and external momenta $p^{2}=\mu^{2}$ play the role of an effective IR-cutoff, the flow with $\ln(k)$ or $\ln(\mu)$ is rather similar. As a reasonable approximation only the largest IR-cutoff is effective. For a given $\mu$ the flow with $k$ occurs only for $k>\mu$, while for $k<\mu$ the flow with $k$ stops. Similarly, the flow with $\mu$ matters only for $\mu>k$, and stops for $\mu<k$. Qualitatively, we can interpret $\p_{t}$ as $k\p_{k}$ for $k>\mu$, and $\mu\p_{\mu}$ for $\mu>k$. This rough translation has been well tested for simple models, as pure scalar theories in various dimensions.

For Quantum Einstein Gravity the limit $k\raw 0$ is not yet understood in all aspects. First of all, the form of the quantum effective action for a theory defined exactly at the UV-fixed point is not known. It has to be scale invariant, which excludes the terms $\sqrt{g}M^{2}R$ or $\sqrt{g}V$. Natural terms could be $\sim C\sqrt{g}R^{2}$ or $\sim D\sqrt{g} C_{\mu\nu\rho\sigma}C^{\mu\nu\rho\sigma}$, with $C_{\mu\nu\rho\sigma}$ the Weyl-tensor. The couplings $C$ and $D$ are dimensionless, such that for constant $C$ and $D$ they are consistent with quantum scale symmetry. For our oversimplified ad hoc discussion in sect.~\ref{sec:RelevantParametersAndMassScales} a model defined precisely on the UV-fixed point would indeed lead to a constant $w^{*}=C$, with $M^{2}=0$ in eq.~\eqref{eq:58}. For this simplification one may replace $k^{2}$ by $R$, such that in eq.~\eqref{eq:55} the combination $c-\frac{dc^{2}}{4w}$ corresponds to $c_{M}$.

A gravity theory with constant non-zero $D$ has ghosts. The corresponding instabilities render such a quantum effective action unacceptable. The flow of $C$ and $D$ with $\mu$ has been investigated in perturbation theory \cite{STR1,STE,FRATSE,AVRABAR,STR2}. No fixed point has been found in the perturbative approximation, but hints may be seen in extrapolations \cite{NIED1,NIED2}. For the moment, the problem of the fixed point quantum effective action remains open. This is important because this fixed point effective action determines the graviton-graviton scattering in the high momentum limit \cite{DUNO,AADO}.

Away from the fixed point the limit $k\raw 0$ seems at first sight to be easier. For $k^{2}\ll M^{2}$ the contribution of the fluctuations to the flow of $M^{2}$ becomes small, $c_{M}k^{2}\ll M^{2}$, as long as $c_{M}$ does not become large. The running of $M^{2}$ with $k$ is then essentially stopped. The situation would be similar for $V$ if the cosmological constant would have its apparently natural magnitude $V\sim M^{4}$. This can indeed be realized for negative $V$. For $V>0$, however, a large enhancement of $c_{V}$ occurs as $v$ approaches a critical value $v_{cr}$. This is due to a singularity in the threshold function $\ell_{0}(-v)$ in eq.~\eqref{eq:116}. We will discuss this important aspect of ``infrared gravity'' later in sect.~\ref{sec:Infrared_QG}.

\subsection{Dilaton quantum gravity}

Dilaton quantum gravity adds to the metric a scalar field $\chi$ whose expectation value sets the Planck mass, and perhaps indirectly all other mass scales, in the present Universe. We will concentrate here on a real scalar field with discrete symmetry $\chi\raw-\chi$. Generalizations to a complex scalar or multicomponent scalar are possible. Dilaton quantum gravity offers the advantage that for $\chi\neq 0$ a scale invariant effective action for gravity is rather easily formulated. As we have discussed before, the effective action \eqref{eq:16} is scale invariant if $B$ and $\tilde{\lambda}$ are constants. For any solution with $\chi\neq 0$ scale symmetry is spontaneously broken. For example, for $\tilde{\lambda}=0$ the effective action \eqref{eq:16} induces gravitational field equations that have as solution Minkowski space with $\chi$ an arbitrary constant $\chi_{0}$. In the absence of any intrinsic mass scale (for a model defined precisely on the UV-fixed point) the scalar excitation $\ln(\chi/\chi_{0})$ behaves as a massless Goldstone boson -- the dilaton. We employ in this work the historical name ``dilaton quantum gravity'', even though a naming ``metron quantum gravity'' would be more precise.

The presence of the scalar metron field allows for dimensionless ratios as $D^{2}/\chi^{2}$ or $R/\chi^{2}$, with $D^{2}=D^{\mu}D_{\mu}$ formed from covariant derivatives. Using these ratios, a scale invariant effective action for dilaton quantum gravity can have a rich form, as
\begin{align}\label{eq:129} 
\cL&=\sqrt{g}\biggl \{-\dfrac{1}{2}\chi^{2}fR+\dfrac{1}{2}\p^{\mu}\chi f(B-6)\p_{\mu}\chi+\chi^{2}\tilde{\lambda}\chi^{2}\nn\\
&\quad-\dfrac{1}{2}RCR+\dfrac{1}{2}C^{\mu\nu\rho\sigma}DC_{\mu\nu\rho\sigma}+\tilde{\cL}_{GB}\biggl \}\, .
\end{align}
Here $C_{\mu\nu\rho\sigma}$ is the Weyl tensor
\begin{align}\label{eq:CON1} 
C_{\mu\nu\rho\sigma}&=R_{\mu\nu\rho\sigma}-\dfrac{1}{2}\left (g_{\mu\rho}R_{\nu\sigma}+g_{\nu\sigma}R_{\mu\rho}-g_{\mu\sigma}R_{\nu\rho}-g_{\nu\rho}R_{\mu\sigma}\right )\nn\\
&\quad +\dfrac{1}{6}R\left (g_{\mu\rho}g_{\nu\sigma}-g_{\mu\sigma}g_{\nu\rho}\right )\, .
\end{align}
Our conventions for the curvature tensor are
\begin{equation}\label{eq:CON2} 
[D_{\mu},D_{\nu}]V_{\rho}=R_{\mu\nu\rho}\,^{\lambda}V_{\lambda}\com R_{\mu\rho}=R_{\mu\nu\rho}\,^{\nu}\com R=g^{\mu\rho}R_{\mu\rho}\, 
\end{equation}
such that for euclidean signature the curvature scalar for a sphere is positive. The term
\begin{equation}
\tilde{\cL}_{GB}=R_{\mu\nu\rho\sigma}ER^{\mu\nu\rho\sigma}-4R_{\mn}ER^{\mn}+RER
\end{equation}
reduces to the Gauss-Bonnet term for constant $E$. It contributes to the field equations only if $E$ depends on fields or covariant Laplacians. In eq.~\eqref{eq:129} the symbols $f$, $B$, $\tilde{\lambda}$, $C$, $D$ and $E$ stand for dimensionless functions of ratios as $-\chi^{-2}D^{2}$, $\chi^{-2}R$ etc. The ansatz \eqref{eq:129} describes for a scale invariant model the most general propagator for the metric and $\chi$ in flat space. The most general vertices also involve possible terms with higher powers of the curvature tensor. We note that for $D=0$ and constant $B\geq 0$, $\tilde{\lambda}\geq 0$, $C\geq 0$, $E$ the theory has no ghosts, tachyons or other instabilities and constitutes a valid effective action for quantum gravity. One expects that this property is not lost for a suitable range of functions $f$, $B$, $\tilde{\lambda}$, $C$, $D$ and $E$. One may rescale the scalar field to obtain $f=1$. 

If we define the theory in the vicinity of the UV-fixed point the running of the functions $f$, $B$, $\tilde{\lambda}$, $C$ and $D$ will induce an intrinsic scale $\bar{\mu}$. In this case these functions depend, in addition, on the ratio $\bar{\mu}^{2}/\chi^{2}$. If the scales $\varphi_{0}$ or $\Lambda_{QCD}$ in the standard model are all proportional to $\chi$, reflecting spontaneous breaking of scale symmetry, it is possible that the largest intrinsic scale is induced by the running of couplings in the gravity-scalar sector. For example, a coupling function $\tilde{\lambda}=\bar{\mu}^{2}/\chi^{2}$ will result in a mass term for $\chi$ given by an effective potential $U(\chi)=\bar{\mu}^{2}\chi^{2}$. If $U(\chi)$ determines the present dark energy density one has for this case $\bar{\mu}$ of the order of the present Hubble parameter.

Quantum effective actions in the limit where dimensionless functions only depend on $\bar{\mu}^{2}/\chi^{2}$ have been investigated in ref.~\cite{CWIQ}. For $\bar{\mu}^{2}/\chi^{2}\raw 0$ this results in a scale invariant effective action with constant coefficients $f,B,\tilde{\lambda},C$ and $D$. Such scale invariant gravitational actions (sometimes considered as classical) have been discussed in refs.~\cite{STR1,STR2,NALI}. For a functional renormalization group study one has, in addition, the IR-cutoff scale $k$. All functions can now also depend on $k^{2}/\chi^{2}$. Typically, one may identify for the dimensionless functions $-D^{2}$ or $R$, or a linear combination thereof, with $\mu^{2}$. Then $f$, $B$, $\tilde{\lambda}$, $C$ and $D$ can be viewed as dimensionless couplings that depend on $\mu^{2}/\chi^{2}$, $\bar{\mu}^{2}/\chi^{2}$ and $k^{2}/\chi^{2}$. The discussion of the previous sections applies to these couplings. Typically, $\mu^{2}$ and $k^{2}$ both act as IR-cutoffs for the flow and only the largest one matters. The flow with $k$ at $\mu=0$ is similar to the flow with $\mu$ at $k=0$. Furthermore, $\bar{\mu}$ may induce an intrinsic IR-cutoff that stops the flow with $\mu$ or $k$ once they are smaller than $\bar{\mu}$.

We will discuss here dilaton quantum gravity in a simple form
\begin{equation}\label{eq:DG1} 
\cL=U-\dfrac{F}{2}R+\dfrac{K}{2}\p^{\mu}\chi\p_{\mu}\chi-\dfrac{C}{2}R^{2}\, ,
\end{equation}
with $U$, $F$, $K$ and $C$ functions of $y=\chi^{2}/k^{2}$. Here $k$ may later be replaced by an intrinsic scale $\bar{\mu}$, where the flow with $k$ effectively stops. The approximation \eqref{eq:DG1} corresponds to $\mu=0$, or to a situation where the dependence of $U$, $F$, $K$ and $C$ on derivatives and geometrical invariants can be neglected. For a derivative expansion the ansatz \eqref{eq:DG1} accounts for the most general terms involving up to two derivatives. For terms with four derivatives we have omitted $D$ and $E$ as well as terms with four derivatives of $\chi$. In terms of eq.~\eqref{eq:129} the effective action \eqref{eq:DG1} corresponds to
\begin{equation}\label{eq:DG2} 
F=f\chi^{2}\com U=\tilde{\lambda}\chi^{4}\com K=f(B-6)\, .
\end{equation}
Expressed by the dimensionless functions $u(\tilde{\chi})$, $w(\tilde{\chi})$, defined by
\begin{equation}\label{eq:DG3} 
u=\dfrac{U}{k^{4}}\com w=\dfrac{F}{2k^{2}}\, ,
\end{equation}
one has
\begin{equation}\label{eq:DG4} 
\cL=k^{4}\left \{u-w\tilde{R}+\dfrac{K}{2k^{2}}\p^{\mu}\tilde{\chi}\p_{\mu}\tilde{\chi}-\dfrac{C}{2}\tilde{R}^{2}\right \}\, ,
\end{equation}
where
\begin{equation}\label{eq:DG5} 
\tilde{R}=\dfrac{R}{k^{2}}\com \tilde{\chi}=\dfrac{\chi}{k}\, ,
\end{equation}
and
\begin{equation}\label{eq:DG6} 
u=\tilde{\lambda}\tilde{\chi}^{4}\com w=\dfrac{f}{2}\tilde{\chi}^{2}\, .
\end{equation}

First functional renormalization group studies for dilaton gravity \cite{PHW1,PHW2} have investigated a truncation of the form \eqref{eq:DG1}, with $C=0$. A fixed point corresponds to a scaling solution where the functions $f$, $B$ and $\tilde{\lambda}$ depend only on $k^{2}/\chi^{2}$, without the appearance of any intrinsic scale $\bar{\mu}$. For such a scaling solution the dimensionless couplings are fixed functions of 
\begin{equation}\label{eq:133} 
y=\dfrac{\chi^{2}}{k^{2}}\; ,
\end{equation}
e.g.
\begin{equation}\label{eq:131} 
k\p_{k}f\bigl |_{y}=0\com k\p_{k}B\bigl |_{y}=0\com k\p_{k}\tilde{\lambda}\bigl |_{y}=0\, .
\end{equation}
For a theory defined exactly at the fixed point the limit $k^{2}/\chi^{2}\raw 0$ or $y\raw \infty$ yields then the scale invariant effective action \eqref{eq:129} appropriate for the IR-fixed point at $\mu=0$. The value of the functions $f$, $B$, $\tilde{\lambda}$ at non-zero $k^{2}/\chi^{2}$ translates approximately to the scale invariant effective action \eqref{eq:129} with $k^{2}$ replaced by $\mu^{2}$.

For a search of scaling solutions one first derives the general flow equations for $f(y,k)$, $B(y,k)$, $\tilde{\lambda}(y,k)$, namely
\begin{equation}\label{eq:132} 
k\p_{k}f\bigl |_{y}=\beta_{f}\com k\p_{k}B\bigl |_{y}=\beta_{B}\com k\p_{k}\tilde{\lambda}\bigl |_{y}=\beta_{\lambda}\, .
\end{equation}
The flow generators or $\beta$-functions involve the functions $f$, $B$ and $\tilde{\lambda}$, as well as derivatives of these functions with respect to $y$. They also may depend directly on $y$. The three conditions for a scaling solution
\begin{equation}\label{eq:134} 
\beta_{f}=0\com \beta_{B}=0\com\beta_{\lambda}=0\, ,
\end{equation}
constitute three non-linear differential equations for the three functions $f(y)$, $B(y)$, $\tilde{\lambda}(y)$. The candidates for scaling solutions that have been found by numerical solutions of these three differential equations have the property
\begin{equation}\label{eq:135} 
\tilde{\lambda}(y)={u}(y)\, y^{-2}\, ,
\end{equation}
with ${u}(y)$ smoothly varying between $u_{0}={u}(y=0)$ and $u_{\infty}={u}(y\raw\infty)$. As a result, one has
\begin{equation}\label{eq:136} 
U(\chi)=\tilde{\lambda}\chi^{4}=\begin{cases}
u_{0}k^{4} & \textit{for }\chi\raw 0\\
u_{\infty}k^{4} & \textit{for }\chi\raw\infty\; .
\end{cases}
\end{equation}
For $k\raw 0$ the potential vanishes. Similarly, $B(y)$ shows a smooth interpolation between $B_{0}=B(y=0)$ and $B_{\infty}=B(y\raw\infty)$. Finally, $f$ behaves as
\begin{equation}\label{eq:137} 
f(y)=\begin{cases}
f_{0}y^{-1} & \textit{for }\chi\raw 0\\
\xi & \textit{for }\chi\raw\infty\; ,
\end{cases}
\end{equation}
resulting in
\begin{equation}\label{eq:138}
F^{2}(\chi)R=f\chi^{2}R=\begin{cases}
f_{0}k^{2}R & \textit{for }\chi\raw 0\\
\xi\chi^{2}R & \textit{for }\chi\raw\infty\; .
\end{cases}
\end{equation}
For $k^{2}/\chi^{2}\raw 0$ only the term $\sim \chi^{2}R$ survives. After a rescaling of $\chi$ the scale invariant effective action for a theory defined precisely on the fixed point corresponds to eq.~\eqref{eq:16}, with $\tilde{\lambda}=0$.

\subsection{Simple scale invariant effective actions}

The scale invariant effective action will always dominate the high momentum behavior if quantum gravity can be defined as a renormalizable continuum quantum field theory. We explore next a few properties of a simple ansatz for dilaton quantum gravity
\begin{equation}\label{eq:139} 
\cL=\sqrt{g}\left \{-\dfrac{1}{2}\chi^{2}R+\dfrac{1}{2}(B-6)\p^{\mu}\chi\p_{\mu}\chi-\dfrac{1}{2}C R^{2}\right \}\, ,
\end{equation}
with constant $B>0$, $C\geq 0$. In eq.~\eqref{eq:129} we have set $D=0$ such that no ghosts appear. We also set $\tilde{\lambda}=0$ such that neither the graviton propagator nor the scalar propagator shows a tachyonic instability, in accordance with the scaling solution of dilaton quantum gravity. We use the normalization $f=1$. The ansatz \eqref{eq:139} for a scale invariant effective action has no problematic features or inconsistencies.

For a solution of the field equations it is convenient to replace for the term $\sim R^{2}$ the curvature scalar by a scalar field $\phi$. We follow here the treatment of ref.~\cite{CWIQ}. The action \eqref{eq:139} is a particular case of a general class of modified gravity theories, namely $f(R)$ theories \cite{BUDA,HWA,CCT,CDTT,NOD,ADT,AGDT,FAR2,TSU,HUSA}, adapted to the presence of a scalar field $\chi$. (For a review and the relation to an equivalent formulation as coupled quintessence see ref.~\cite{CWMGCQ}.) Introducing in the functional integral an additional integration over a scalar field $\phi$ an equivalent description of the model \eqref{eq:139} becomes
\begin{align}\label{eq:140} 
\tilde{\cL}&=\sqrt{g}\biggl \{-\dfrac{\chi^{2}}{2}R+\dfrac{1}{2}(B-6)\p^{\mu}\chi\p_{\mu}\chi\nn\\
&\quad -\dfrac{1}{2}C R^{2}+\dfrac{C}{2}\left (\phi-R\right )^{2}\biggl \}\, .
\end{align}
Indeed, the Gaussian integration over $\phi$ only amounts to an irrelevant constant factor. The field equation derived from \eqref{eq:140} has the solution
\begin{equation}\label{eq:141} 
\phi=R\; ,
\end{equation}
which associates $\phi$ with the curvature scalar. The action \eqref{eq:140} can be written as
\begin{equation}\label{eq:142} 
\tilde{\cL}=\sqrt{g}\left \{-\left (\dfrac{\chi^{2}}{2}+C\phi\right )R+\dfrac{1}{2}(B-6)\p^{\mu}\chi\p_{\mu}\chi+U(\phi)\right \}\, ,
\end{equation}
with potential
\begin{equation}\label{eq:143} 
U(\phi)=\dfrac{C}{2}\phi^{2}\, .
\end{equation}
The field equations have a simple solution with Minkowski geometry and an arbitrary constant value $\chi_{0}$ of $\chi$,
\begin{equation}\label{eq:144} 
g_{\mu\nu}=\eta_{\mu\nu}\com \phi=0\com \chi=\chi_{0}\, .
\end{equation}

\subsection{Weyl scaling}\label{sec:Weyl_Scaling}

For a field dependent Planck mass $M^{2}=\chi^{2}+2C\phi$, the scalar fluctuations in the metric mix with the scalar fluctuations in $\chi$ and $\phi$ on the propagator level. ``Diagonalisation'' of the propagator is achieved by a Weyl scaling \cite{HW,RDI}. This conformal transformation of the metric transforms the coefficient of the curvature scalar to a constant $M^{2}$. More generally, for an effective action with a field dependent coefficient $F$ of the curvature scalar,
\begin{equation}\label{eq:144A} 
\tilde{\cL}_{R}=-\dfrac{1}{2}\sqrt{g}FR\, ,
\end{equation}
a Weyl scaling defines a new metric by a field dependent conformal transformation
\begin{equation}\label{eq:144B} 
g_{\mu\nu}=\omega^{2}\, g^{\prime}_{\mu\nu}\com \omega^{2}=\dfrac{M^{2}}{F}\, .
\end{equation}
In the new metric the curvature scalar $R^{\prime}$ formed with the metric $g^{\prime}_{\mu\nu}$ appears in the effective action with a constant Planck mass
\begin{equation}\label{eq:145A} 
\tilde{\cL}_{R}^{\prime}=-\dfrac{1}{2}\sqrt{g^{\prime}}M^{2}R^{\prime}\, .
\end{equation}
This constant Planck mass is not a parameter of the theory, but rather introduced by the Weyl transformation as a matter of convenience. The choice of the metric $g^{\prime}_{\mu\nu}$ is called the ``Einstein frame''.

While Weyl scaling has been used in cosmology since a long time - for an example of inflation cf. ref.~\cite{CWIHD1,CWIHD2} - it has been argued that conformal transformations of the classical action do not lead an equivalent quantum field theory. (For a review of the discussion see \cite{FAR1}.) Indeed, an equivalent quantum field theory needs in the functional integral not only the transformation of the classical action, but also a Jacobian from the transformation of the measure. If this Jacobian is not taken into account, which is usually not done, the quantum field theories related by field transformations of the classical action are inequivalent. The quantum effective action is a powerful tool for avoiding problem with Jacobians. Since the functional integral is already performed, no functional measure factor appears anymore. Expectation values of observables are expressed in terms of functional derivatives of the effective action. On the level of the effective action arbitrary field transformations can be performed, if observables are transformed correspondingly. On the level of the quantum effective action different choices of the metric or different ``frames'' are fully equivalent. Arbitrary Weyl transformations can be performed without changing the predictions for observations.

This point of view has been taken in ref.~\cite{CWVP}, where both the quantum effective action and the corresponding solutions of the quantum field equations are seen to be mapped consistently by conformal transformations. The map changes all quantities with dimension, as temperature or energy densities in cosmology. Detailed following work \cite{MASA,FHU,DEF,FLA,CPS,DSA,CHIYA,POVO2} has mapped many of the quantities relevant for cosmology, including correlation functions for primordial fluctuations \cite{CWPCFV,KAPATA} or different time variables \cite{CWEU}. The invariance of observables under field redefinitions on the level of the arguments of the quantum effective action has been named ``field relativity'' \cite{CWUWE}. For conformal transformations a particularly useful class of observables are those that remain invariant under the transformation \cite{CWVP,CPS,CWEU,CWPCFV,JKSV,JKM,KAPI}. Such observables have to be dimensionless quantities, typically ratios as galaxy distance over atom size or similar, but this condition is not always sufficient.

For arbitrary functions of a scalar field $\omega$ the curvature scalar $R^{\prime}$ in the Einstein frame is related to $R$ by
\begin{equation}\label{eq:146} 
R=\omega^{-2}\left \{R^{\prime}-6\ln(\omega)_{;}^{\mu}\ln(\omega)_{;\mu}-6(\ln(\omega))_{;}^{\mu}\,_{\mu}\right \}\, .
\end{equation}
The semicolons denote covariant derivatives in the Einstein frame. Using also for the determinant
\begin{equation}\label{eq:147} 
\sqrt{g}=\omega^{4}\sqrt{g^{\prime}}\; ,
\end{equation}
and omitting a total derivative, one obtains
\begin{equation}\label{eq:148} 
\sqrt{g}FR=\sqrt{g^{\prime}}\left \{M^{2}R^{\prime}-6M^{2}\p^{\mu}\ln(\omega)\p_{\mu}\ln(\omega)\right \}\, .
\end{equation}
The transformation of all other terms in the effective action is determined by the replacement of $g_{\mn}$ by $g^{\prime}_{\mn}$ according to eq.~\eqref{eq:144B}. It is often convenient to perform simultaneously field transformations on the other fields. For example, the transformation of a fermion (Grassmann) field $\psi$,
\begin{equation}\label{eq:149} 
\psi=\omega^{-\frac{3}{2}}\psi^{\prime}\, ,
\end{equation}
results for the standard fermion kinetic term in
\begin{equation}\label{eq:150} 
\tilde{\cL}_{F,kin}=i\sqrt{g}\bar{\psi}\gamma^{\mu}D_{\mu}\psi=i\sqrt{g^{\prime}} \bar{\psi}^{\prime}\gamma^{\mu}D_{\mu}^{\prime}\psi^{\prime}+\ldots\, ,
\end{equation}
where the dots indicate additional terms involving $\p_{\mu}\omega$, and $D_{\mu}^{\prime}$ is the covariant derivative in the Einstein frame. Gauge fields and the standard $F^{2}$-term in the effective action are invariant under conformal transformations.

As a first example, we consider the effective action of variable gravity \cite{CWVG}
\begin{equation}\label{eq:151} 
\tilde{\cL}=\sqrt{g}\left \{-\dfrac{1}{2}F(\chi)R+\dfrac{1}{2}K(\chi)\,\p^{\mu}\chi\p_{\mu}\chi+U(\chi)\right \}\, .
\end{equation}
This reads in the Einstein frame
\begin{equation}\label{eq:152} 
\tilde{\cL}=\sqrt{g^{\prime}}\left \{-\dfrac{M^{2}}{2}R^{\prime}+\dfrac{M^{2}K^{\prime}}{2\chi^{2}}\,\p^{\mu}\chi\p_{\mu}\chi+U^{\prime}(\chi)\right \}\, ,
\end{equation}
where
\begin{equation}\label{eq:153} 
K^{\prime}=\chi^{2}\left \{\dfrac{K}{F}+\dfrac{3}{2}\left (\dfrac{\p\ln(F)}{\p\chi}\right )^{2}\right \}\, ,
\end{equation}
and
\begin{equation}\label{eq:154} 
U^{\prime}=\dfrac{M^{4}U}{F^{2}}\, .
\end{equation}
In particular, the scale invariant effective action \eqref{eq:16} becomes, with $K=B-6$, $K^{\prime}=B$,
\begin{equation}\label{eq:155} 
\tilde{\cL}_{\chi R}=\sqrt{g^{\prime}}\left \{-\dfrac{M^{2}}{2}R^{\prime}+\dfrac{M^{2}B}{2}\p^{\mu}\ln(\chi)\p_{\mu}\ln(\chi)+\tilde{\lambda}M^{4}\right \}\, .
\end{equation}
This shows that such a model is stable provided $B\geq 0$, even for a negative kinetic term in the scale invariant frame, $K=B-6$. One also sees that the scale invariant quartic coupling $\tilde{\lambda}$ leads in the Einstein frame to a cosmological constant
\begin{equation}\label{eq:156} 
U^{\prime}=\tilde{\lambda}M^{4}\, .
\end{equation}

We next discuss the scale invariant effective action \eqref{eq:139} with constant $C$, $D=0$, which is equivalent to eq.~\eqref{eq:142}, \eqref{eq:143}. We take
\begin{equation}\label{eq:157} 
\omega^{2}=\dfrac{M^{2}}{\chi^{2}+2C\phi}
\end{equation}
and combine eq.~\eqref{eq:148} with the kinetic term for $\chi$,
\begin{equation}\label{eq:158} 
\sqrt{g}\p^{\mu}\chi\p_{\mu}\chi=\sqrt{g^{\prime}}\, \dfrac{M^{2}}{\chi^{2}+2C\phi}\p^{\mu}\chi\p_{\mu}\chi\, .
\end{equation}
The effective action \eqref{eq:142} reads in the Einstein frame
\begin{equation}\label{eq:159} 
\tilde{\cL}=\sqrt{g^{\prime}}\left \{-\dfrac{M^{2}}{2}R^{\prime}+V_{E}(\chi,\phi)+\cL_{kin}\right \}\, ,
\end{equation}
with $V_{E}=U^{\prime}$ and kinetic term
\begin{align}\label{eq:160} 
\cL_{kin}&=\dfrac{M^{2}}{2}\left (\chi^{2}+2C\phi\right )^{-2}\biggl \{ \left [B\chi^{2}+2(B-6)C\phi\right ]\, \p^{\mu}\chi\p_{\mu}\chi\nn\\
&\quad +12C\chi\p^{\mu}\chi\p_{\mu}\phi+6C^{2}\p^{\mu}\phi\p_{\mu}\phi\biggl \}\, .
\end{align}
The potential in the Einstein frame is given by
\begin{equation}\label{eq:161} 
V_{E}(\chi,\phi)=\dfrac{CM^{4}\phi^{2}}{2(\chi^{2}+2C\phi)^{2}}\, .
\end{equation}
For $C=0$ we recover eq.~\eqref{eq:155}, with $\tilde{\lambda}=0$. 

The presence of an explicit mass $M$ in the Einstein frame hides the presence of scale symmetry. The scale symmetry that was manifest by simple linear rescalings of fields in the original ``scaling frame'' appears in the Einstein frame in a different realization. Since $\omega$ is not a scale invariant function it transforms non-trivially, often in a rather complicated way. The metric in the Einstein frame inherits this complicated transformation. It is obvious that an understanding of the consequences of scale symmetry is done best in the original scaling frame. Properties that are easily understood in the scaling frame may appear ``unnatural'' in the Einstein frame. Nevertheless, the Einstein frame is convenient for a discussion of cosmological solutions. We will come back to the scale invariant action \eqref{eq:139} in our discussion of scale symmetry in cosmology in sect.~\ref{sec:Scale_Sym_Cosmo}.

For practical computations in cosmology the Einstein frame is often the most convenient one, since in the limit of scale symmetry both the Planck mass and particle masses are constant. (Note that the scaling frame is not the Jordan frame with constant particle masses.) Since the propagator for the scalar modes is diagonal, one can read the mass of the cosmon -- the pseudo Goldstone boson of spontaneously broken scale symmetry -- directly from the second derivative of the potential in the Einstein frame. For a single field $\sigma$ besides the metric, normalized such that the normalization of the scalar kinetic  term is canonical, it reads
\begin{equation}\label{eq:CM1} 
m_{c}^{2}=\dfrac{\p^{2}V_{E}}{\p\sigma^{2}}\, .
\end{equation}
For several scalar fields with canonical kinetic term one has to diagonalize the mass matrix. Eq.~\eqref{eq:CM1} has to be evaluated for the cosmological solution, which often corresponds to $\sigma\neq 0$.

\subsection{Einstein gravity with $R^{2}$-term}

Relevant parameters at the UV-fixed point induce intrinsic mass scales. The simplest example is quantum Einstein gravity where the Planck mass is a relevant parameter. We will discuss here a simplified form of the effective action where a scale invariant term $\sim R^{2}$ is supplemented by the scale symmetry breaking in terms of a term linear in $R$ with constant coefficient. No scalar field $\chi$ is introduced for this example.

An effective action for which the Einstein-Hilbert term is supplemented by a term $\sim R^{2}$,
\begin{equation}\label{eq:162} 
\tilde{\cL}=-\sqrt{g}\left \{\dfrac{\tilde{M}^{2}}{2}R+\dfrac{C}{2}R^{2}\right \}\, ,
\end{equation}
has been used by Starobinsky's seminal paper on inflation \cite{STAR}. Introducing the scalar field $\phi\sim R$ and performing a Weyl scaling one replaces in eq.~\eqref{eq:160} $\chi\raw \tilde{M}$ and omits all derivatives of $\chi$
\begin{align}\label{eq:163} 
\tilde{\cL}&=\sqrt{g^{\prime}}\biggl \{-\dfrac{M^{2}}{2}R^{\prime}+\dfrac{3C^{2}M^{2}}{(\tilde{M}^{2}+2C\phi)^{2}}\p^{\mu}\phi\p_{\mu}\phi\nn\\
&\quad +\dfrac{CM^{4}\phi^{2}}{2(\tilde{M}^{2}+2C\phi)^{2}}\biggl \}\, .
\end{align}
The standard normalization of the scalar kinetic term is obtained for
\begin{equation}\label{eq:164} 
\sigma=\sqrt{\dfrac{3}{2}}\, M\, \ln\left (1+\dfrac{2C\phi}{\tilde{M}^{2}}\right )\, ,
\end{equation}
which yields a standard scalar field coupled to gravity \cite{CWMGCQ}
\begin{equation}\label{eq:165} 
\tilde{\cL}=\sqrt{\bar{g}^{\prime}}\left \{-\dfrac{M^{2}}{2}R^{\prime}+\dfrac{1}{2}\p^{\mu}\sigma\p_{\mu}\sigma+V_{E}(\sigma)\right \}\, ,
\end{equation}
with potential
\begin{equation}\label{eq:166} 
V_{E}(\sigma)=\dfrac{M^{4}}{8C}\left [1-\exp\left (-\sqrt{\dfrac{2}{3}}\dfrac{\sigma}{M}\right )\right ]^{2}\, .
\end{equation}

Expanding the potential around its minimum at $\sigma=0$
\begin{equation}\label{eq:167} 
V_{E}=\dfrac{M^{2}}{12C}\sigma^{2}+\ldots
\end{equation}
reveals a scalar field with mass
\begin{equation}\label{eq:168} 
m_{\sigma}=\dfrac{M}{\sqrt{6C}}\, .
\end{equation}
This demonstrates that the model \eqref{eq:162} is stable for $C>0$ when expanded around the solution
\begin{equation}\label{eq:169} 
g_{\mu\nu}=\eta_{\mu\nu}\com \sigma=0\, .
\end{equation}
The scalar mass diverges for $C\raw 0$, such that the scalar field effectively decouples in this limit. One recovers Einstein gravity.

On the other hand, for $\sigma\raw\infty$ the potential becomes flat
\begin{equation}\label{eq:170} 
V_{E}(\sigma\raw\infty)=\dfrac{M^{4}}{8C}\left (1-2\exp\left (-\sqrt{\dfrac{2}{3}}\dfrac{\sigma}{M}\right )\right )\, .
\end{equation}
For the slope one finds for large $\sigma$ an exponential suppression
\begin{equation}\label{eq:171} 
\dfrac{M\p_{\sigma}V_{E}}{V_{E}}=\sqrt{\dfrac{8}{3}}\, \exp\left (-\sqrt{\dfrac{2}{3}}\dfrac{\sigma}{M}\right )\, .
\end{equation}
Starting initially with large enough $\sigma/M$ the evolution of $\sigma$ according to the cosmological solution corresponds to slow roll inflation. We will discuss the cosmology of this model in sect.~\ref{sec:Starobinski_Inflation}.

\subsection{Scale invariant $R^{2}$-gravity}

We next discuss in more detail scale invariant effective actions for gravity. They are important in two respects. First, they describe the high momentum behavior of gravity, for example the high momentum graviton scattering. In the high momentum limit the UV-fixed point is approached and the effective action becomes scale invariant. Second, a scale invariant effective action can often be associated to inflationary cosmology. It typically describes the very early stages of inflation. Realistic inflation requires an instability of a scale invariant solution which is responsible for an evolution away from scale invariance and finally for the end of inflation. Often this instability is associated to small scale violating effects. We therefore also discuss the scale invariant effective actions for gravity as limiting cases of effective actions that also contain intrinsic scales. Weyl scaling is an important tool for understanding the stability properties of solutions of scale invariant effective gravity actions.

Our first example is scale invariant $R^{2}$-gravity. It can be obtained as the limit of the model \eqref{eq:162} for $\tilde{M}=0$,
\begin{equation}\label{eq:171A} 
\tilde{\cL}=-\dfrac{C}{2}\sqrt{g}R^{2}\, .
\end{equation}
In this case the effective action is scale invariant. Eq.~\eqref{eq:165} remains valid, with
\begin{equation}\label{eq:172} 
\sigma=\sqrt{\dfrac{3}{2}}M\ln\left (\dfrac{\phi}{M^{2}}\right )\com V_{E}=\dfrac{M^{4}}{8C}\, .
\end{equation}
This is a theory of a massless free scalar field coupled to Einstein gravity with a cosmological constant $V_{E}$. We recall that the scale $M$ is introduced only by the redefinition of fields \eqref{eq:144B}, \eqref{eq:172}. The observable ratio
\begin{equation}\label{eq:173} 
\dfrac{V_{E}}{M^{4}}=\dfrac{1}{8C}
\end{equation}
does not involve $M$. For $\phi\raw 0$ the field transformations become singular, since $\omega=M/\sqrt{2\phi}$. In this limit flat space is approached in the original metric frame, $R\raw 0$, by virtue of eq.~\eqref{eq:141}. In the Einstein frame the cosmological solution remains a de Sitter space with $R^{\prime}=M^{2}/(4C)$. The value of $\phi$ or $\sigma$ is not fixed by the field equations. 

For any $\phi\neq 0$, corresponding to $R\neq 0$, scale symmetry is spontaneously broken. The massless scalar field is the corresponding Goldstone boson. We conclude that the field equations derived from the scale invariant theory \eqref{eq:171A} admit for $C>0$ a de Sitter solution. The value of $R$ is arbitrary. Any non-zero $R$ spontaneously breaks scale symmetry, and there exists an associated Goldstone boson. These features can be found equivalently by solving the field equation of the ``higher derivative model'' \eqref{eq:171A}.

We have obtained the scale invariant limit \eqref{eq:172} by taking in eqs.~\eqref{eq:162},\eqref{eq:163} the limit $\tilde{M}^{2}\raw 0$. The same limit is obtained if we take in \eq{163} the limit $\phi\raw\infty$ for fixed $\tilde{M}^{2}$. More precisely, this limit is reached if the dimensionless ratio $\tilde{M}^{2}/(2C\phi)$ goes to zero. The interpretation is simple. The effective action \eqref{eq:162} contains a term $\sim\tilde{M}^{2}$ that breaks scale symmetry explicitly, and a second term $\sim C$ that is scale invariant. A typical scale for spontaneous scale symmetry breaking is $CR$. The relative size of explicit scale symmetry breaking vs. spontaneous scale symmetry breaking is given by
\begin{equation}\label{eq:173A} 
r_{A}=\dfrac{\tilde{M}^{2}}{2CR}\quad\hat{=}\quad\dfrac{\tilde{M}^{2}}{2C\phi}\, .
\end{equation}
For $2C\phi\gg \tilde{M}^{2}$ the mass term for the scalar field is given by the second $\sigma$-derivative of \eq{170},
\begin{equation}\label{eq:174} 
m_{\sigma}^{2}=\dfrac{\p^{2}V_{E}}{\p\sigma^{2}}=-\dfrac{M^{2}}{6C}\exp\left (-\sqrt{\dfrac{2}{3}}\dfrac{\sigma}{M}\right )=-\dfrac{M^{2}\tilde{M}^{2}}{12 C^{2}\phi}\, .
\end{equation}
As expected, it vanishes $\sim r_{A}=\tilde{M}^{2}/(2C\phi)$. The mass term is negative, indicating an instability of the de-Sitter solution as required for inflationary cosmology.

In the other limit $r_{A}\raw\infty$ one approaches the flat space solution with $\phi=0$. The mass of the scalar field is given by \eq{168} with $M=\tilde{M}$. Taking subsequently $\tilde{M}\raw 0$ again results in a massless field. This suggests that the double limit $\tilde{M}^{2}\raw 0$, $\phi\raw 0$ can be taken without problems, such that for $\tilde{M}^{2}=0$ no discontinuity arises between the family of solutions with arbitrary $R\neq 0$, and the flat space solution with $R=0$.

Consider now directly $\tilde{M}^{2}=0$. For any given nonzero momenta, $p^{2}\neq 0$, the behavior is expected to be independent of $R$ if $R\ll p^{2}$. With respect to the scale set by $R$ the momenta are in the region of the ``high-momentum limit'', which should not depend on the detailed geometry for $R\ll p^{2}$ (and similar for other geometrical invariants). The limit $R\raw 0$ or $\phi\raw 0$ can therefore be taken without problems. Since for every $R\neq 0$ the Weyl transformation is well defined, one can employ the Einstein frame formulation \eqref{eq:165}, \eqref{eq:172} in order to see that the theory does not show any instabilities. Taking subsequently $R\raw 0$ will not change the situation, and we conclude that for any nonzero momenta the model is well behaved also for the flat space solution. In view of this, any discontinuity between the family of solutions with $R\neq 0$ and the flat space solution $R=0$ seems rather unlikely.

One may, nevertheless, retain some skepticism since the field transformation becomes singular in the double limit $\phi\raw 0$, $\tilde{M}^{2}\raw 0$. We also recall that mass terms or potentials are not dimensionless and therefore will depend on the metric frame. The ``observable physics'' should be formulated in terms of dimensionless ratios. It therefore seems reasonable to investigate the flat space solution directly in the original ``scaling frame'' where scale symmetry is manifest. Expanding the effective action \eqref{eq:171A} around flat space yields terms that involve four derivatives. For example, the effective action for the scalar ``conformal mode'' $\tilde{\sigma}$, defined by
\begin{equation}\label{eq:175} 
g_{\mn}=\mathrm{e}^{2\tilde{\sigma}}\eta_{\mn}\, ,
\end{equation}
reads \cite{STR2}
\begin{equation}\label{eq:176} 
\tilde{\cL}_{\sigma}=-18C\left (D^{2}\tilde{\sigma}-\p^{\mu}\tilde{\sigma}\p_{\mu}\tilde{\sigma}\right )^{2}\, ,
\end{equation}
with $D^{2}=\p^{\mu}\p_{\mu}$. This follows directly from \eq{146} and \eqref{eq:147}, with $\tilde{\sigma}=\ln(\omega)$ and $R^{\prime}=0$. It has been argued that this higher derivative scalar model is well behaved \cite{STR2}.

In ref.~\cite{STR2} the action \eqref{eq:171A} has been considered as part of a classical action that only contains dimensionless parameters, called ``agravity''. One then needs to add the effect of quantum fluctuations. In our approach the action \eqref{eq:162}, which contains the part \eqref{eq:171A}, is considered as an approximation to the quantum effective action. All fluctuation effects are included. Typically the value of $C$ corresponds to an ultraviolet fixed point value. The term $\sim C$ dominates the high momentum behavior. We conclude that for an effective action \eqref{eq:162} the high momentum behavior is well behaved. This is expected to include the high momentum scattering of all physical modes in the metric (e.g. the graviton) for this model.

For pure $R^{2}$-gravity the dimensionless coupling $C$ is only an overall multiplicative constant in the effective action. As long as there are no other terms in the effective action the role of the value of $C$ is reduced. In the scaling frame it does not appear in the field equations, and therefore not in the form of the cosmological solutions. Its appearance in the Einstein frame in the form of the cosmological constant \eqref{eq:172} arises from the $C$-dependence in the conformal scaling $\omega^{2}=M^{2}/(2C\phi)$. Using instead $\omega^{2}=M^{2}/(2\phi)$ and a different normalization for $\sigma$ would render all terms of the effective action proportional $C$ also in the Einstein frame. In the scaling frame the correlation function for metric fluctuations is $\sim (1/C)$, while irreducible vertices are $\sim C$. Without comparison this has not a clear physical meaning. The situation of the role of the coupling $C$ differs if the classical action is $\sim CR^{2}$. Then $C$ determines the importance of the fluctuations and enters in the $\beta$-functions for running couplings. On the level of the effective action the size of the coupling $C$ matters only if other terms are present to which the term $\sim CR^{2}$ can be compared.

\subsection{Scale invariance with variable Planck mass}

For our second example of a scale invariant effective action for gravity we return to the scale invariant effective action \eqref{eq:139}. Its form in the Einstein frame has already been computed, given by eqs.~\eqref{eq:159}-\eqref{eq:161}. The main difference to the effective action \eqref{eq:162} is the more complicated kinetic term and the dynamics of the $\chi$-field. For $B>6$ the model has a stable solution with Minkowski geometry and arbitrary constant $\chi_{0}$,
\begin{equation}\label{eq:177} 
g^{\prime}_{\mn}=\eta_{\mn}\com \chi=\chi_{0}\com\phi=0\, .
\end{equation}
For $C>0$, $B<6$ this solution becomes unstable and we restrict our discussion to $B>0$. The solution \eqref{eq:177} is a good approximation for late cosmology and may be reached in the infinite future.

Another simple solution is de Sitter space with arbitrary constant $\phi_{0}$ and $\chi=0$
\begin{equation}\label{eq:178} 
R^{\prime}=4V_{E}=\dfrac{M^{4}}{2C}\com\chi=0\com\phi=\phi_{0}\, .
\end{equation}
This solution is an interesting candidate for a ``beginning'' of the Universe in the infinite past. For particle masses $\sim\chi$, all particle masses vanish at the beginning. Also the metric of de Sitter space vanishes at the beginning. For $\chi^{2}/R\raw 0$ (in the original scaling frame \eqref{eq:139}) one approaches the UV-fixed point according to the discussion in sect.~\ref{sec:Flow_In_Field_Space}, identifying $\mu^{2}=R$.

The issue of stability is related to the properties of the kinetic term \eqref{eq:160}. Writing it in a matrix form, $i,j=(\chi,\phi)$, $\sigma_{1}=\chi$, $\sigma_{2}=\phi$
\begin{equation}\label{eq:179} 
\cL_{kin}=\dfrac{M^{2}}{2(\chi^{2}+2C\phi)^{2}}\tilde{K}_{ij}\p^{\mu}\sigma_{i}\p_{\mu}\sigma_{j}\, ,
\end{equation}
with
\begin{align}\label{eq:180} 
\tilde{K}_{\chi\chi}&=B\chi^{2}+2(B-6)C\phi\, ,\nn\\
\tilde{K}_{\chi\varphi}&=\tilde{K}_{\varphi\chi}=6C\chi\com \tilde{K}_{\varphi\varphi}=6C^{2}\, ,
\end{align}
we observe
\begin{equation}\label{eq:181} 
\det(\tilde{K})=6C^{2}(B-6)(\chi^{2}+2C\phi)\, .
\end{equation}
For $C>0$, $B<6$ the determinant $\det(\tilde{K})$ is negative, such that one eigenvalue is negative, the other positive. In the conformal limit $C=B=0$ no propagating scalar degree of freedom is left.

For understanding the vicinity of the solution \eqref{eq:178} a useful choice of scalar field variables for this model is
\begin{align}\label{eq:182} 
\tilde{\chi}&=\omega\chi=\dfrac{M\chi}{\sqrt{\chi^{2}+2C\phi}}\, ,\nn\\
\tilde{\varphi}&=-\sqrt{\dfrac{3}{2}}M\ln(\omega^{2})=\sqrt{\dfrac{3}{2}}M\ln\left (\dfrac{\chi^{2}+2C\phi}{M^{2}}\right )\, .
\end{align}
In terms of these fields one has
\begin{equation}\label{eq:183} 
V_{E}=\dfrac{M^{4}}{8C}\left (1-\dfrac{\tilde{\chi}^{2}}{M^{2}}\right )^{2}\, ,
\end{equation}
such that the potential becomes independent of $\tilde{\chi}$. The kinetic term mixes $\tilde{\chi}$ and $\tilde{\varphi}$
\begin{align}\label{eq:184} 
\tilde{\cL}_{kin}&=\dfrac{\sqrt{g^{\prime}}}{2}\biggl \{\left [1+\dfrac{(B-6)\tilde{\chi}^{2}}{6M^{2}}\right ]\p^{\mu}\tilde{\varphi}\p_{\mu}\tilde{\varphi}+(B-6)\p^{\mu}\tilde{\chi}\p_{\mu}\tilde{\chi}\nn\\
&\quad +\sqrt{\dfrac{2}{3}}\dfrac{(B-6)\tilde{\chi}}{M}\p^{\mu}\tilde{\chi}\p_{\mu}\tilde{\varphi}\biggl \}\, .
\end{align}

The solution \eqref{eq:178} corresponds to $\tilde{\chi}=0$ with arbitrary constant $\tilde{\varphi}=\tilde{\varphi}_{0}$. The field $\tilde{\varphi}$ corresponds to a free massless scalar field that can be associated with the Goldstone boson of spontaneously broken scale symmetry. The field equation for small $\tilde{\chi}$ reads
\begin{equation}\label{eq:185} 
(B-6)D^{2}\tilde{\chi}=-\dfrac{M^{2}\tilde{\chi}}{2C}+\dfrac{B-6}{6M^{2}}\left [\p^{\mu}\tilde{\varphi}\p_{\mu}\tilde{\varphi}-\sqrt{6}MD^{2}\tilde{\varphi}\right ]\tilde{\chi}\, .
\end{equation}
For constant $\tilde{\varphi}$, the solution $\tilde{\chi}=0$ is unstable for $B>6$ due to the negative mass term from the potential \eqref{eq:183}. For $B<6$ both mass term and kinetic coefficient are negative, resulting in stable behavior. We will discuss implications for inflation in the next section. There we will also diagonalize the kinetic term by a redefinition of the scalar fields.

\subsection{Flowing couplings and fixed points}

So far we have considered scale invariant effective actions that express important features of an UV-fixed point. We have used a constant $C$ that may be associated to the value of the function $C$ in eq.~\eqref{eq:129} at the fixed point. Away from a fixed point the functions $C$ or $D$ in eq.~\eqref{eq:129} will depend on the momentum scale $\mu$, or on the infrared cutoff scale $k$ in functional renormalization. In ranges where the flow of $C$ is slow enough and the effects of $D\neq0$ remain negligible, our previous discussion remains relevant. We next discuss the running with $\mu$ or $k$ and the evidence for the existence of an UV-fixed point.

We first discuss the flow of the dimensionless couplings $C$ and $D$ in the limit where they do not depend on scalar fields as $\chi$. The dependence of the corresponding function $C(q^{2})$ and $D(q^{2})$ on the momentum, evaluated in a flat space geometry, dominates the UV-behavior for $q^{2}\raw \infty$. The flow of $C$ and $D$ can be addressed within two different perturbative expansions, one for scales $\mu$ below $M$, the other for scales $\mu$ above $M$ and for large $C$ and $D$. Functional renormalization covers all scales $\mu$ and arbitrary values of $C$ and $D$. The functional renormalization flow of the Planck mass and the cosmological constant will be addressed in sect.~\ref{sec:Asymptotic_Safety}. It is not accessible to standard perturbation theory, but suitable resummations may help \cite{NIED1,NIED2}.

The fluctuations of massless scalars, fermions and gauge bosons induce a running of the couplings $C$ and $D$. For this effect only the coupling of these fields to the metric via their kinetic terms matter, whereas interactions among these fields are neglected. The perturbative computation amounts to a one-loop integral with propagators evaluated in a curved background \cite{BIDA,PATO,BOS}. In a similar fashion one may add the graviton contributions in a low energy effective theory of gravity \cite{DONO1,BUR,DONO2}.  This is possible in an expansion near flat space where the graviton propagator is assumed to be the one obtained from the Einstein-Hilbert action without a cosmological constant. In this case the squared Planck mass $M^{2}$ is only a multiplicative constant in the inverse graviton propagator and drops out in the one loop computation. One finds \cite{DONO3} for the flow of small $C$ and $D$
\begin{equation}\label{eq:6.91} 
\mu\p_{\mu}C=-\dfrac{1}{576\pi^{2}}\left \{18+\sum_{i=1}^{N_{s}}\, \left (6\xi_{i}-1\right )^{2}\right \}\, ,
\end{equation}
where the sum goes over $N_{s}$ real scalar fields with standard kinetic term and non-minimal coupling to gravity $\cL_{\xi}=-\xi_{i}R\chi_{i}^{2}/2$. Similarly, one obtains
\begin{equation}\label{eq:6.92} 
\mu\p_{\mu}D=\dfrac{1}{2880\pi^{2}}\left \{126+36N_{V}+9N_{F}+3N_{s}\right \}
\end{equation}
for $N_{V}$ gauge bosons, $N_{F}$ Weyl fermions and $N_{s}$ real scalars. (For the standard model one has $N_{V}=12$, $N_{F}=45$ and $N_{s}=4$.) For decreasing $\mu$ one finds an increase of $C$ and a decrease of $D$. For both $C$ and $D$ all individual contributions have the same sign, with opposite sign for $C$ and $D$.

Functional renormalization group investigations for the truncation \eqref{eq:162}, including in addition a cosmological constant $V$, have found a fixed point for the couplings $C$, $\tilde{M}^{2}/k^{2}$ and $V/k^{4}$ \cite{LAURE,FLNR2,FKLNR}. All three dimensionless couplings have been found to be relevant parameters. Extended truncations with high polynomials
\begin{equation}\label{eq:186} 
\tilde{\cL}=-\dfrac{\sqrt{g}}{2}\sum_{n=0}^{N}\, c_{n}R^{n}\, ,
\end{equation}
with $N$ as high as $34$, again find a fixed point with three relevant directions \cite{FLNR2,FKLNR}. As $N$ is varied from a few to $34$ the series shows rather satisfactory convergence properties. The picture of an UV-fixed point with $C$, $\tilde{M}^{2}/k^{2}$ and $V/k^{4}$ as relevant parameters is also consistent with several other truncations. For pure gravity, typical fixed point values $C_{*}$ are found in the range
\begin{equation}\label{eq:187} 
C_{*}=-6.3\cdot 10^{-4}\, .
\end{equation}

Since $C$ is dimensionless, its running with $k$ at low or zero curvature or covariant momentum should be the same as the running with a momentum or geometry scale $\mu$ for $k\ll\mu$. Thus one expects at $k=0$ an ultraviolet fixed point in the running of $C(\mu\raw\infty)$. This fixed point $C_{*}$ characterizes the ultraviolet behavior of the quantum effective action at high $\mu$. The flow of the relevant parameter associated to the Planck mass, $\tilde{M}^{2}(k)/k^{2}$, may induce an intrinsic scale ${M}=\tilde{M}(k=0)$, as in \eq{162}. We adopt here the working hypothesis that the ultraviolet behavior of the quantum effective action is characterized by a fixed point $C_{*}$. It is not clear at the moment if the negative value of $C_{*}$ in eq.~\eqref{eq:187} would pose a problem of instability in the scalar sector. For this issue a more complete study of the scalar propagator would be needed. Given the very small value \eqref{eq:187} it is also possible that in a more complete truncation the value $C_{*}$ turns out to be positive.

We will later assume that for the momentum range relevant for cosmology one can use the approximation of a constant $C$. A reasonable value of $C$ may be of the same order as $C_{*}$, but rather different values are possible as well. If $C$ is a relevant parameter, all values are, in principle, possible. We next ask if the perturbative expansion for large $C$ and $D$ is consistent with this picture.

Consider the action \eqref{eq:129} with $\tilde{\lambda}=0$ and constant $B$, $C$, $D$. We normalize the scalar field such that $f=1$. (Another often used parameterization is $f=1/(B-6)$. The disadvantage of the second parameterization is a negative value of the effective $M^{2}$ for $B<6$, which does not lead to stable gravity, in contrast to a negative coefficient of the scalar kinetic term which yields a stable theory for $C=0$, $B>0$. This matters close to the conformal value $B=0$, where $B-6$ is negative.) 

For large values of $C$ and $D$ a perturbative computation of the flow of couplings becomes possible \cite{STE,FRATSE,STR1,AVRABAR,STR2,BPSH}. The expansion in inverse powers $C^{-1}$ and $D^{-1}$ becomes accessible because the propagators of the gravitational degrees of freedom scale $\sim 1/Cq^{4}$, $\sim 1/Dq^{4}$. In this type of perturbation theory one obtains
\begin{equation}\label{eq:188} 
\dfrac{\p C}{\p\ln(\mu)}=-\dfrac{1}{48\pi^{2}}\biggl \{\dfrac{5}{6}+\dfrac{B^{2}}{12(B-6)^{2}}+\dfrac{15C}{D}+\dfrac{15C^{2}}{D^{2}}\biggl \}+\ldots
\end{equation}
where the dots stand for higher orders and contributions from other ``particle fluctuations'' beyond $g_{\mn}$ and $\chi$ that are the same as in eq.~\eqref{eq:6.92}. The negative sign of the r.h.s. of \eq{188} for $C/D>0$ implies that $C$ decreases towards the UV such that the perturbative expansion for large $C$ ceases to be valid. For large $C$, the increase towards the IR is consistent with a fixed point $C_{*}$ for which $C-C_{*}$ is a relevant parameter. The fixed point $C_{*}$ is, however, not in the domain of validity of the perturbative expansion in $C^{-1}$. If $B$ and $C/D$ can be neglected, and if we omit the contribution of other particles, the next order in the expansion takes the form \cite{STR2}
\begin{equation}\label{eq:189} 
\dfrac{\p C}{\p\ln(\mu)}=\beta_{C}=-\dfrac{5}{288\pi^{2}}+\dfrac{5}{27648\pi^{4}}C^{-1}\, .
\end{equation}
This $\beta$-function has formally a zero with fixed point value
\begin{equation}\label{eq:190} 
C_{*}=\dfrac{1}{96\pi^{2}}\, .
\end{equation}
Formally, the critical exponent would be
\begin{equation}\label{eq:191} 
\theta_{C}=-\dfrac{\p\beta_{C}}{\p C}\bigl |_{C_{*}}=\dfrac{5}{3}\, .
\end{equation}
The values \eqref{eq:190}, \eqref{eq:191} are outside the range of validity of perturbation theory. While opposite in sign, the small absolute value of $C_{*}$ is in qualitative agreement with the findings for functional renormalization. We note that the fixed point for $C$ may disappear once $\mu$ becomes smaller than $M$. There is then no contradiction with the low energy flow \eqref{eq:6.91}.

For large $D$ the perturbative running of $D$ obeys
\begin{equation}\label{eq:192} 
\dfrac{\p D}{\p\ln(\mu)}=2A_{D}\, ,
\end{equation}
with
\begin{equation}\label{eq:193} 
A_{D}=\dfrac{1}{160\pi^{2}}\left (c_{D}+N_{V}+\dfrac{1}{4}N_{F}+\dfrac{1}{12}N_{S}\right )\, .
\end{equation}
The contribution of the metric degrees of freedom depends only mildly on $C$
\begin{equation}\label{eq:194} 
c_{D}=\begin{cases}
\frac{133}{2} & \textit{for }C\raw\infty\\
\frac{199}{3} & \textit{for }C\raw 0\; .
\end{cases}
\end{equation}
We have included additional particles as $N_{V}$ vectors, $N_{F}$ fermions and $N_{S}$ scalars. We note the close similarity of the flow equation \eqref{eq:6.92} for small $\mu/M$ and eq.~\eqref{eq:193} for large $\mu/M$, the main difference being a multiplicative factor not far from one. In a flat geometry the function $D(q^{2})$ in \eq{129} increases for large $D$ with increasing $q^{2}$,
\begin{equation}\label{eq:195} 
D(q)=\bar{D}+A_{D}\,\ln\left (\dfrac{q^{2}}{M^{2}}\right )\, .
\end{equation}
The inverse coupling $D^{-1}$ is asymptotically free. We observe that a flow of $D$ with $q^{2}$ is not incompatible with the existence of a UV-fixed point and scale symmetry. For example, if $M^{2}=\chi^{2}$, one obtains a scaling solution for $D(q^{2}/\chi^{2})$. Within functional renormalization this extends to $\tilde{M}^{2}=M^{2}/k^{2}=\tilde{M}^{2}(\chi^{2}/k^{2})$.

If the region of large $D$ is relevant at high momenta, the propagator of the graviton (tensor mode) obeys asymptotically for $q^{2}\raw\infty$
\begin{equation}\label{eq:196} 
G_{\gamma}(q^{2})\sim \dfrac{1}{q^{4}(\bar{D}+A_{D}\ln(q^{2}/M^{2}))}\, .
\end{equation}
The question of stability of the graviton modes depends on the poles of $G_{\gamma}(q^{2})$ and therefore on the behavior of $G_{\gamma}$ for smaller $q^{2}$. Furthermore, the perturbative renormalization group equations are computed for $\chi=0$. For realistic gravity one has $\chi=0$ and an evaluation for nonzero $\chi$ is needed. The knowledge of $D(\mu,\chi)$ in the momentum range $q^{2}=\mu^{2}\approx \chi^{2}$ is rather poor. We discuss here qualitatively a few alternatives.

Assume that for $\chi=0$ the flow of $D$ reaches a fixed point with finite $D_{*}$. This fixed point would typically be IR-stable, e.g. $D-D_{*}$ is an irrelevant parameter. If the theory is defined at this fixed point, or if the running of $D$ has approached the fixed point, the graviton propagator for $\chi=0$ would read
\begin{equation}\label{eq:197} 
G_{\gamma}\sim \dfrac{1}{D_{*}q^{4}}\, .
\end{equation}
Such a propagator leads to a secular instability of solutions which is comparatively harmless. The issue of instability becomes more severe if a fixed point for $D$ is combined with a non-zero effective Planck mass, such that
\begin{equation}\label{eq:198} 
G_{\gamma}\sim \dfrac{1}{\chi^{2}q^{2}+\tilde{D}q^{4}}\, .
\end{equation}
This propagator has two poles at
\begin{equation}\label{eq:199} 
q^{2}=0\com q^{2}=-\dfrac{\chi^{2}}{\tilde{D}}=-m_{D}^{2}\, .
\end{equation}
We can write
\begin{equation}\label{eq:200} 
G_{\gamma}\sim \chi^{-2}\left (\dfrac{1}{q^{2}}-\dfrac{1}{q^{2}+m_{D}^{2}}\right )\, .
\end{equation}
The negative sign of the second term indicates a negative kinetic term for the mode with mass $m_{D}$. This is called a ghost. 

We recall that we are not using four-derivative gravity for the classical action in the definition of a quantum theory of gravity. In our approach the four derivative terms are an approximation to the form of the quantum effective action that contains already all effects of quantum fluctuations. In general, the momentum dependence of inverse propagators is not described by a polynomial in $q^{2}$ - in this sense it always contains ``non-local terms''. A simple example is the effective action for QED \eqref{eq:20}, \eqref{eq:27}. The logarithmic dependence on $q^{2}$ does not pose particular problems, while a Taylor expansion of the inverse propagator to order $q^{4}$ (or to any other finite order) would lead to ghosts. Nevertheless, a stability analysis for the time evolution of small deviations from a given cosmological solution or flat space is necessary for any proposed form of the effective action. See ref.~\cite{BHOL,SAL} for the case of constant $D$. Too strong instabilities are not acceptable for a realistic theory of gravity. The issue is similar in Einstein gravity for which $\chi^{2}$ is replaced by $M^{2}$. A flow of $D$ towards a fixed point has to be discussed in this respect. If $D$ becomes momentum independent one infers a ghost \eqref{eq:200} and one has to find out how dangerous is the associated instability. (This may also depend on a possible imaginary part of $D$.) 

It seems more likely that $D$ is indeed given at the UV-fixed point by a scaling function $D(q^{2}/\chi^{2})$ which does not lead to any ghost instability. For example, this scaling function could diverge for $q^{2}/\chi^{2}\raw\infty$. We next argue that a dangerous ghost instability is unlikely to occur. The vicinity of a pole has a strong influence on the running of couplings. Approaching a pole the propagator diverges and fluctuation effects can become very large. In functional renormalization group studies it is typically found that the functional flow avoids to enter region of strong instabilities \cite{TETW}. The problematic region of the propagator is in the range of negative $q^{2}<0$. For euclidean momenta $q^{2}\geq 0$ the problem is absent. The euclidean graviton propagator does not change sign in the region $q^{2}>0$, and for $q^{2}\raw 0$ the contribution $\sim\tilde{D}q^{4}$ in \eq{198} becomes unimportant. For a euclidean computation a fixed point $D_{*}$ for the momentum range $q^{2}\geq 0$ seems perfectly viable. In this case one should not extrapolate the polynomial form of $G_{\gamma}^{-1}$ to negative $q^{2}$, however. In this region a Minkowski space functional renormalization group investigation may become necessary since analytic continuation of an euclidean computation may be difficult. There are many possible functions $G_{\gamma}^{-1}(q^{2})$ that have a zero at $q^{2}=0$, no further zero for negative $q^{2}$, and approaching closely $G^{-1}\sim \chi^{2}q^{2}+\tilde{D}q^{4}$ for $q^{2}>0$. It seems well conceivable that propagators with reasonable stability properties fall into this class. There are various attempts to construct explicit models avoiding the ghost problem, see, for example, refs.~\cite{TOMB,BMS,MODE,BGKM,
HOREN,NAR1,NAR2}.

It is intriguing to see that the flow generators for $D$ are very similar for the two expansions for small $D$, cf. eq.~\eqref{eq:6.92} and for large $D$, cf. eqs.~\eqref{eq:192}, \eqref{eq:193}, with only a small shift in the numerical value for $A_{D}$. In particular, the contributions from massless vectors, fermions or scalars are the same. A fixed point in $D$ would have to be generated by fluctuation effects that are not seen in both perturbative approaches. It may therefore seem more plausible that the only fixed point is $D_{*}^{-1}=0$, and the flow away from the fixed point has the structure of eq.~\eqref{eq:192}. Eq.~\eqref{eq:196} would lead to a zero of $D$ at positive $q^{2}=m_{D}^{2}$, $m_{D}^{2}=M^{2}\exp(-\bar{D}/A_{D})$. Typically $m_{D}^{2}$ is tiny as compared to $M^{2}$. For small $q^{2}/M^{2}$ the graviton propagator is dominated by $(M^{2}q^{2})^{-1}$ for Einstein gravity. Indeed, for $M^{2}=\chi^{2}$ and $\chi\neq 0$ the inverse graviton propagator at small $q^{2}$ will be dominated by $\chi^{2}q^{2}$ and behave therefore very differently from $q^{4}$. Even if the UV-fixed point value $D_{*}$ would be finite, (which seems not very likely), the gravitational degrees of freedom decouple below the effective Planck mass, resulting in the flow \eqref{eq:6.92} for $q^{2}\ll M^{2}$. One should therefore understand the consequences of this flow for the graviton propagator.

With $A_{D}\ll 1$, $\overline{D}>0$ the inverse propagator
\begin{equation}\label{eq:CP1} 
G^{-1}=\chi^{2}q^{2}+\overline{D}q^{4}+A_{D}q^{4}\ln\left (\dfrac{q^{2}}{\chi^{2}}\right )
\end{equation}
has a zero at $q^{2}=0$ but no zero for real $q^{2}>0$. With
\begin{equation}\label{eq:CP2} 
q^{2}=r\ee^{-i\varphi}\com -\pi <\varphi\leq \pi\com x=\dfrac{r}{\chi^{2}}\, ,
\end{equation}
further zeros of $G^{-1}$ in the complex plane require the conditions
\begin{align}\label{eq:CP3} 
\cos(\varphi)&=-\dfrac{\sin(\varphi)}{\varphi}\left (\dfrac{\overline{D}}{A_{D}}+\ln\left (\dfrac{\sin(\varphi)}{A_{D}\varphi}\right )\right )\, ,\nn\\
x&=\dfrac{1}{A_{D}}\dfrac{\sin(\varphi)}{\varphi}\, .
\end{align}
For $\overline{D}/A_{D}\gg 1$ one finds solutions with
\begin{equation}\label{eq:CP4} 
\varphi=\pm (\pi-\varepsilon)\, ,
\end{equation}
where
\begin{align}\label{eq:CP5} 
\varepsilon&\approx \dfrac{\pi}{g_{D}-\ln(g_{D})+\ln(\pi)}\com g_{D}=\dfrac{\overline{D}}{A_{D}}+\ln\left (\dfrac{1}{\pi A_{D}}\right )\, ,\nn\\
x&\approx \dfrac{\varepsilon}{\pi A_{D}}\, .
\end{align}
The pole of the propagator at $q^{2}=-\chi^{2}/\overline{D}$ for $A_{D}=0$ splits for $A_{D}>0$ into two poles on both sides of the cut at the negative real axis
\begin{equation}\label{eq:CP6} 
q_{\pm}^{2}=-\dfrac{\chi^{2}}{\overline{D}}\left [1-\dfrac{A_{D}}{\overline{D}}\ln(\overline{D})\right ]^{-1}\left (1\pm i\dfrac{A_{D}\pi}{\overline{D}\left (1-\dfrac{A_{D}}{\overline{D}}\ln(\overline{D})\right )}\right )\, .
\end{equation}
The positive relative sign of the imaginary part occurs for $\varphi>0$, while the negative sign corresponds to $\varphi<0$. This simple example demonstrates that the issue of poles of the propagator depends on the precise form of functions as $D(q^{2})$. In the Minkowski domain of a negative real part of $q^{2}$ one expects a behavior of $G^{-1}(q)$ that is more complex than the simple logarithmic form of eq.~\eqref{eq:CP1}. For precision one presumably needs a direct computation with Minkowski signature since analytic continuation is subtle. Analytic continuation from euclidean $q^{2}$ to Minkowski signature is done by $q^{2}\raw q^{2}\ee^{-2i\alpha}$ with $\alpha$ increasing from zero to $\pi/2$. This gets problematic once one approaches the location of the pole of $G^{-1}$ at $q_{+}^{2}$.

For cosmology one typically needs the behavior at $\mu^{2}<\chi^{2}$, where $\mu^{2}$ may be set by a geometric quantity as $R$ instead of $q^{2}$ for flat space. Even for the inflationary epoch the observable effects of $D$ seem very small unless $D$ happens to be very large. Since the running of $D$ is rather slow, a constant $D$ may be a reasonable approximation. A constant $D$ does not contribute to the field equations for a homogeneous isotropic cosmology.

We finally extract the perturbative running of $B$ for large $C$ and $D$ from ref.~\cite{STR1},
\begin{equation}\label{eq:201} 
\dfrac{\p B}{\p\ln(\mu)}=\beta_{B}=-\dfrac{1}{16\pi^{2}}\left \{\dfrac{B(2B-9)}{9(B-6)C}-\dfrac{5(B-6)C}{D^{2}}\right \}\, .
\end{equation}
Fixed points and their properties depend on the ratio $C^{2}/D^{2}$. It is not clear what lessons can be drawn for the relevant ranges of couplings where the expansion in powers of $C^{-1}$, $D^{-1}$ is no longer valid, and for $\chi\neq 0$. A general feature that may be retained is a fixed point for which both $B$ and $C$ vanish simultaneously. For $B=C=0$ the model is invariant under \textit{local} scale transformations (local Weyl transformations). Such an enhanced symmetry always constitutes a partial fixed point if the regularization can be made consistent with the enhanced symmetry. In this case terms in $\beta_{B}$ that do not vanish for $B=0$ are expected to be proportional to $C$. Another general lesson is that fixed points for $B$ seem to arise rather naturally. For \eq{201} the zeros of $\beta_{B}$ occur for
\begin{equation}\label{eq:202} 
B_{*}=\dfrac{1}{2(2-y_{B})}\left \{9-12y_{B}\pm9\sqrt{1+\frac{8}{9}y_{B}}\right \}\, ,
\end{equation}
with
\begin{equation}
y_{B}=45\dfrac{C^{2}}{D^{2}}\, .
\end{equation}
For the relevant ranges of couplings the value of a fixed point for $B$ may, however, be rather different from \eq{202}.

At present, our knowledge of the dimensionless function $B$, $C$, $D$, $E$ in eq.~\eqref{eq:129} remains rather incomplete. This concerns both the scaling form at the fixed point, and the deviation from scale invariance due to relevant parameters, as reflected by the appearance of an intrinsic mass scale in the dimensionless functions. As a general lesson, the scale invariant effective action does not reduce to a few fixed point values of constant couplings $C$, $D$ etc. The scaling solution involves functions, as $C(q^{2}/\chi^{2})$, $D(q^{2}/\chi^{2})$. As far as we know, nothing excludes a behavior for $q^{2}/\chi^{2}\raw\infty$ where $C$ approaches a fixed point and $D$ diverges. A graviton propagator without additional poles could result in a stable theory. There is no need that the detailed behavior for $q^{2}\raw\infty$ and $q^{2}\raw -\infty$ is the same.

\subsection{Asymptotic safety}\label{sec:Asymptotic_Safety} 

Asymptotic safety for quantum gravity is based on the existence of an UV-fixed point that does not correspond to a free theory. For pure gravity this is the Reuter fixed point \cite{MRQG,DP,SOU,RS,LAUREU1}. The key ingredient is the scaling of the effective Planck mass $M^{2}(k)\sim k^{2}$ for high enough $k$, cf. eq.~\eqref{eq:97}. The strength of the gravitational interaction may be measured by the dimensionless Planck length $\ell_{p}(k)$, with
\begin{equation}\label{eq:AS1} 
\ell_{p}^{2}(k)=\dfrac{k^{2}}{M^{2}(k)}
\end{equation}
proportional to the running dimensionless ``Newton constant''. In perturbative gravity $M^{2}(k)=M^{2}$ is constant and $\ell_{p}^{2}(k)$ diverges for $k\raw\infty$. In contrast, for asymptotic safety the low energy increase of $\ell_{p}^{2}(k)$ is tamed for large $k$, tending to a constant limit
\begin{equation}\label{eq:AS2} 
\lim_{k\raw\infty}\, \ell_{p}^{2}(k)=\dfrac{1}{2c_{M}}=\dfrac{1}{2w_{*}}\, .
\end{equation}

The method for computing the scale dependence of the Planck mass is functional renormalization for the effective average action $\Gamma_{k}$ \cite{CWFR,MRCW,MRQG,TETWE,ELLW1,MOR,ELLW2}. For reviews related to particle physics and gravity see refs.~\cite{BTW,PAWREV,GIREV,NIRE}. All fluctuations for which the covariant Laplacian $-D^{2}$, or momenta $q^{2}$ in flat space, exceed $k^{2}$ are integrated out. The $k$-dependence of $\Gamma_{k}$ follows an exact flow equation \cite{CWFR}
\begin{equation}\label{eq:AS3} 
\p_{t}\Gamma_{k}=k\p_{k}\Gamma_{k}=\dfrac{1}{2}\mathrm{Str}\left \{[\Gamma_{k}^{(2)}+R_{k}]^{-1}k\p_{k}R_{k}\right \}\, .
\end{equation}
Here $R_{k}$ is a smooth IR-cutoff function which vanishes fast for $q^{2}/k^{2}\raw\infty$. (Here we employ the notation $q^{2}$ also for ``squared covariant momenta'' $-D^{2}$.) The second functional derivative $\Gamma_{k}^{(2)}$ denotes the field-dependent and $k$-dependent full inverse propagator. The trace Str includes a momentum integration (or summation of positions in position space) as well as a trace over all Lorentz and internal indices. For fermions it contains a minus sign due to the anticommuting nature of Grassmann variables. The r.h.s. of eq.~\eqref{eq:AS3} is called the flow generator. It is ultraviolet finite due to the fast decrease of $\p_{t}R_{k}$ and infrared finite due to the regulator term in the inverse propagator. The flow equation has a one-loop form, with momentum integral dominated by a small range around $q^{2}=k^{2}$. 

For quantum gravity and other local gauge theories one uses gauge fixing with an appropriate inclusion of a flow contribution from the Faddeev-Popov determinant, for example in the form of ghost degrees of freedom. It is a great advantage to use a physical gauge fixing which only acts on the gauge fluctuations without affecting the physical fluctuations \cite{CWGIF,RPWY}. In this case the contribution from gauge fluctuations and the Faddeev-Popov determinant can be combined into a simple measure contribution. The remaining part of the flow equation \eqref{eq:AS3} involves projectors on the physical modes. On the projected space of physical modes $\Gamma_{k}^{(2)}+R_{k}$ is invertible without gauge fixing. A gauge invariant flow equation \cite{CWGIF} evaluates the projected inverse propagator from the second functional derivative of a gauge invariant action $\bar{\Gamma}_{k}$ which only depends on a single macroscopic gauge field. Depending on the precise definition of $\Gamma_{k}$ and the macroscopic field this procedure may be exact or a rather accurate approximation.

The functional flow equation \eqref{eq:AS3} has to be approximated by suitable truncations of $\Gamma_{k}$. The UV-fixed point will depend on the truncation employed. Being non-perturbative, the convergence of a series of truncations is not guaranteed a priori. Nevertheless, the fixed point has been found in a rather wide class of truncations, as polynomial expansions in powers of the curvature scalar, inclusion of the Weyl tensor or the Sagnotti invariant, as well as systematic expansions in vertex functions. 

The basic reason for the existence of the fixed point is rooted in simple dimensional analysis. This implies a general structure of the flow equations similar to eq.~\eqref{eq:93}
\begin{equation}\label{eq:AS4} 
\p_{t}M^{2}=4c_{M}(v,g_{i})k^{2}\com \p_{t}V=4c_{V}(v,g_{i})k^{4}\, ,
\end{equation}
where the dimensionless coefficients $c_{M}(v,g_{i})$ and $c_{V}(v,g_{i})$ depend on $v=2V/(M^{2}k^{2})$ and the various additional dimensionless couplings $g_{i}$ of a given truncation. It is sufficient that the couplings $g_{i}$ admit fixed points $g_{i*}$, and that the coefficients
\begin{equation}\label{eq:AS5} 
c_{M*}(v)=c_{M}(v,g_{i*})\com c_{V*}(v)=c_{V}(v,g_{i*})
\end{equation}
are compatible with the existence of a solution of eq.~\eqref{eq:AS4},
\begin{equation}\label{eq:AS6} 
\dfrac{M^{2}}{k^{2}}=2c_{M*}(v)>0
\end{equation}
and
\begin{equation}\label{eq:AS7} 
v=\dfrac{c_{V*}(v)}{c_{M*}(v)}\, .
\end{equation}
Both the positivity of $c_{M*}(v)$ and the existence of a solution of eq.~\eqref{eq:AS7} places constraints on the particle content of a model. They seem to be obeyed for a large class of models, including the standard model coupled to gravity \cite{DEP,BPS,Meibohm:2015twa,Eichhorn:2018akn,Eichhorn:2018ydy}.

The issue of quantum scale symmetry appears for the effective average action $\Gamma_{k}$ in a modified form, since the IR-cutoff $k$ introduces a scale. We may consider the change of $\Gamma_{k}$ under an infinitesimal rescaling of $k$,
\begin{equation}\label{eq:ST1} 
\delta^{(k)}\Gamma_{k}=\delta k\p_{k}\Gamma_{k}\com \delta k=\delta\alpha\cdot k\, .
\end{equation}
A combination of this scaling with an infinitesimal scale transformation \eqref{eq:6} of $q^{2}=-g^{\mu\nu}D_{\mu}D_{\nu}$ leaves $q^{2}/k^{2}$ invariant,
\begin{equation}\label{eq:ST2} 
\delta q^{2}=2\delta\alpha\, q^{2}\com \delta\left (\dfrac{q^{2}}{k^{2}}\right )=0\, .
\end{equation}
As a result, the IR-cutoff function $R_{k}(q^{2})$ scales according to the scaling dimensions of the field bilinear that it multiplies, such that the infrared cutoff term in the effective average action is invariant under the combined scaling \eqref{eq:ST1}, \eqref{eq:ST2}. 

A ``scaling form'' of the effective average action is realized if the scale transformation of fields $\delta\Gamma_{k}$, combined with a scale transformation of $k$, $\delta^{(k)}\Gamma_{k}$, leaves $\Gamma_{k}$ invariant
\begin{equation}\label{eq:ST3} 
\delta\Gamma_{k}+\delta^{(k)}\Gamma_{k}=0\, .
\end{equation}
Here $\delta^{(k)}\Gamma_{k}$ has to be evaluated for fixed renormalized fields $\varphi_{R}$. We recall that the fields with the simple canonical scaling dimensions $d_{i}$ are the renormalized fields $\varphi_{R}^{(i)}$,
\begin{equation}\label{eq:ST4} 
\delta\Gamma_{k}=\delta\alpha\, \sum_{i}\, d_{i}\varphi_{R}^{(i)}\dfrac{\p\Gamma_{k}}{\p\varphi_{R}^{(i)}}\, .
\end{equation}
Correspondingly, the microscopic or bare fields may have non-canonical scaling dimensions, including an anomalous dimension. An example is a quartic scalar potential
\begin{equation}\label{eq:ST5} 
\Gamma_{k}=\int_{x}\,\sqrt{g}\, \left (\dfrac{m_{R}^{2}}{2}\varphi_{R}^{2}+\dfrac{\lambda_{R}}{8}\varphi_{R}^{4}\right )\, ,
\end{equation}
which transforms under the combined transformations as
\begin{align}\label{eq:ST6} 
&(\delta+\delta^{(k)})\Gamma_{k}\nn\\
&=\delta\alpha\int_{x}\, \sqrt{g}\left (-m_{R}^{2}\varphi^{2}_{R}+\dfrac{1}{2}\p_{t}m_{R}^{2}\varphi_{R}^{2}+\dfrac{1}{8}\p_{t}\lambda_{R}\varphi_{R}^{4}\right )\, .
\end{align}
It is invariant if
\begin{equation}\label{eq:ST7} 
\p_{t}m_{R}^{2}=2m_{R}^{2}\com \p_{t}\lambda_{R}=0\, ,
\end{equation}
as realized by constant values $\tilde{m}_{R*}^{2}$, $\lambda_{R*}$,
\begin{equation}\label{eq:ST8} 
m_{R}^{2}=\tilde{m}_{R*}^{2}k^{2}\com\lambda_{R}=\lambda_{R*}\, .
\end{equation}

Keeping fixed $k$ and performing a scale transformation only on the fields, $\Gamma_{k}$ is not invariant for any $k\neq 0$ even for the scaling solution
\begin{equation}\label{eq:ST9} 
\delta \Gamma_{k}=-\delta^{(k)}\Gamma_{k}\, .
\end{equation}
The r.h.s. of eq.~\eqref{eq:ST9} vanishes, however, for $k\raw 0$ if in this limit the couplings reach finite values. This guarantees that the scaling solution for $\Gamma_{k}$, which is associated to a fixed point, results in a scale invariant quantum effective action $\Gamma_{k=0}$ once all fluctuations are included and the IR-cutoff is removed. We note that $\delta^{(k)}\Gamma_{k}$ in eq.~\eqref{eq:ST1} can be expressed directly by the flow equation \eqref{eq:AS3} for $\Gamma_{k}$, once translated to fixed renormalized fields.

\subsection{Relevant parameters}

So far  we have encountered possible relevant parameters for the flow of dimensionless couplings away from the UV-fixed point, as $C-C_{*}$ or $B-B_{*}$. Other relevant couplings may concern parameters with dimension of mass. An example is a field-independent contribution to the effective Planck mass, which replaces $\chi^{2}R$ by $(\tilde{M}^{2}+\chi^{2})R$. The ratio $\tilde{M}^{2}/k^{2}$ is likely to be a relevant parameter. Other examples concern the cosmological constant $V$, or a mass term for $\chi$ $\sim\mu^{2}_{\chi}\chi^{2}$. In general, there is no reason why a quantum gravity theory should be defined exactly on the UV-fixed point. For any definition by the vicinity of a fixed point the flow of the relevant parameters will induce intrinsic mass scales. One then expects that scale symmetry is not an exact quantum symmetry. The appearance of intrinsic mass scales in the quantum effective action, often called dilatation anomaly, violates quantum scale symmetry.

We will denote the largest intrinsic mass scale $\bar{\mu}$. This value is a free parameter and not computable. It can be used to set the units for intrinsic parameters with units of mass. In case of a simultaneous spontaneous breaking of scale symmetry by a field $\chi$, the ratio $\bar{\mu}/\chi$ decides on the importance of the explicit scale symmetry breaking. For $\bar{\mu}/\chi\gg 1$ the dominant contribution to the observable particle masses is set by $\bar{\mu}$. In the opposite limit $\bar{\mu}/\chi\ll 1$ the dominant mass contribution arises from $\chi$. For small $\bar{\mu}/\chi$ one expects the presence of a pseudo Goldstone boson with mass $\sim\bar{\mu}^{2}/\chi$. As we discussed in sect.~\ref{sec:Mass_Pseudo_Goldstone} it is not clear a priori in which sector $\bar{\mu}$ appears. For the scale invariant standard model the intrinsic scale $\bar{\mu}$ may be in the sector of the scalar singlet coupled to gravity. For example, one may have $\tilde{M}^{2}\sim\bar{\mu}^{2}$, or $U\sim \bar{\mu}^{4}$, or both. Since $\bar{\mu}$ is arbitrary, we can set it to the vicinity of $10^{-3}$eV without invoking any small dimensionless coupling. In the next section we will discuss cosmological solutions for which the ratio $\bar{\mu}/\chi$ varies due to a time evolution of the scalar field $\chi$. Far in the past $\chi$ may have been of the order of $\bar{\mu}$ (say $10^{-3}$eV), while today it can be many orders of magnitude larger.

We next ask which parameters in a general effective potential $U(\chi)$ for the scalar field $\chi$ correspond to relevant parameters. For this purpose we extend the functional renormalization group investigation at the beginning of this section from a single cosmological constant $V$ to a $\chi$-dependent effective potential $U(\chi)$. Within functional renormalization, the contribution of the traceless transversal tensor mode to the flow of $U(\chi)$ is given by \eq{111}, replacing $V$ by an arbitrary $\chi$-dependent potential
\begin{equation}\label{eq:204} 
k\p_{k}U(\chi)=\dfrac{5f_{\eta}k^{4}}{16\pi^{2}}\ell_{0}\left (-\dfrac{2U(\chi)}{M^{2}(\chi)k^{2}}\right )\, ,
\end{equation}
with threshold function $\ell_{0}(\tilde{w})$ given by \eq{112}. The factor $f_{\eta}$ arises from a possible $k$-dependence of the factor $M^{2}(\chi)$ multiplying the IR-cutoff. It is of the order one and can be found explicitly in sect.~\ref{sec:Flow_Eff_Scalar_Pot_Quantum_Grav}. For $M^{2}\sim k^{2}$ one has $f_{\eta}=4/3$. With the definition of dimensionless functions
\begin{align}\label{eq:205} 
u(\chi)&=\dfrac{U(\chi)}{k^{4}}\com w(\chi)=\dfrac{M^{2}(\chi)}{2k^{2}}\nn\\
 v(\chi)&=\dfrac{2U(\chi)}{M^{2}(\chi)k^{2}}=\dfrac{u(\chi)}{w(\chi)}\, ,
\end{align}
and $\p_{t}=k\p_{k}$ \eq{204} reads
\begin{equation}\label{eq:206} 
\p_{t}u=-4u+\dfrac{5f_{\eta}}{16\pi^{2}}\ell_{0}\left (-\dfrac{u}{w}\right )\, .
\end{equation}

There will be further contributions to the flow of $u$ from other components of the metric fluctuations and particle fluctuations. We concentrate on those that depend only on $u$ and $w$, omitting for example a dependence on $\p u/\p\chi$ in the flow generator. In this approximation one has
\begin{equation}\label{eq:207} 
\p_{t}u=\beta_{u}=-4u+4c_{V}(u,w)\, ,
\end{equation}
with
\begin{equation}\label{eq:208} 
c_{V}(u,w)=\dfrac{5f_{\eta}}{64\pi^{2}}\ell_{0}\left (-\dfrac{u}{w}\right )+\ldots\; .
\end{equation}
In sect.~\ref{sec:Flow_Eff_Scalar_Pot_Quantum_Grav} we discuss the functional renormalization flow of the effective potential in more detail. Terms in the flow generator for $u$ involving field derivatives as $\p_{\chi}u$, $\p_{\chi}^{2}u$ appear only in the contribution from $\chi$-fluctuations. This contribution is small as compared to the others. Furthermore, for a small mass and self-interaction of $\chi$, as we will encounter in the following discussion, even the contribution of $\chi$-fluctuations is well approximated by a constant contribution to $c_{V}$. Eq.~\eqref{eq:207} is therefore a rather reliable approximation. It is our purpose to concentrate first on the qualitative features of the flow.

We may select a particular value $\bar{\chi}$ and follow the flow of $\bar{u}=u(\bar{\chi})$. We also define $\bar{w}=w(\bar{\chi})$. We first concentrate on Einstein gravity coupled to a scalar field where $M^{2}$ and therefore $w$ do not depend on $\chi$. We also consider the region close to the fixed point where $\bar{w}$ is approximately independent of $k$. For $k$-independent $\bar{w}$ the flow of $\bar{u}$ shows a partial fixed point for 
\begin{equation}\label{eq:209} 
\bar{u}_{*}=c_{V}(\bar{u}_{*},\bar{w})\; .
\end{equation}

Let us for a moment neglect the off-diagonal elements of the stability matrix (cf. sect.~\ref{sec:Quantum_Einstein_Gravity}). This holds in the approximation where $\bar{w}$ can be considered as a given constant. The critical exponent is given in this approximation by
\begin{equation}\label{eq:210} 
\theta=4-4\dfrac{\p c_{V}}{\p u}=4-A_{u}\; .
\end{equation}
We infer the existence of a relevant parameter unless the ``anomalous dimension'' $A_{u}=4\p c_{V}/\p u$ exceeds four. If we only include the graviton contribution to $c_{V}$ the anomalous dimension reads
\begin{equation}\label{eq:211} 
A_{u}=\dfrac{5 f_{\eta}}{16\pi^{2}\bar{w}}\ell_{1}\left (-\dfrac{u}{\bar{w}}\right )\, ,
\end{equation}
where we employ the threshold function
\begin{equation}\label{eq:212} 
\ell_{1}(\tilde{w})=-\dfrac{\p}{\p \tilde{w}}\ell_{0}(\tilde{w})\, .
\end{equation}
In this approximation $A_{u}$ depends on the ratio $\bar{u}/\bar w$ at the fixed point
\begin{equation}\label{eq:213} 
A_{u}=4\dfrac{\bar{u}}{\bar{w}}\dfrac{\ell_{1}\left (-\frac{\bar{u}}{\bar{w}}\right )}{\ell_{0}\left (-\frac{\bar{u}}{\bar{w}}\right )}\, .
\end{equation}
We observe that a constant contribution to $c_{V}$, as characteristic for the fluctuations of massless fermions, gauge bosons or scalars, does not contribute to $A_{u}$.

For $\bar{u}/\bar{w}$ approaching one the ratio $\ell_{1}/\ell_{0}$ is enhanced as compared to one. In this range $A_{u}>4$ seems not excluded. As a rough guide that reflects the qualitatives properties we may take
\begin{align}\label{eq:214} 
\ell_{0}(\tilde{w})&=\dfrac{c_{\ell}}{\bar{p}+\tilde{w}}\com \ell_{1}(\tilde{w})=\dfrac{c_{\ell}}{(\bar{p}+\tilde{w})^{2}}\nn\\
 \dfrac{\ell_{1}(\tilde{w})}{\ell_{0}(\tilde{w})}&=\dfrac{1}{\bar{p}+\tilde{w}}\, .
\end{align}
In this approximation the anomalous dimension $A_{u}$ exceeds four if
\begin{equation}\label{eq:215} 
\dfrac{\bar{u}}{\bar{w}}>\dfrac{\bar{p}}{2}\, .
\end{equation}
Details will depend on the precise particle content of the model, as well as on a more accurate treatment of the stability matrix. Our simple treatment reveals that there is no guarantee that the cosmological constant, which is related to $\bar{u}$, is always a relevant parameter. For example, it could be relevant for small $\bar{\chi}$ and irrelevant for large $\bar{\chi}$.

A similar discussion can be made for the $\chi$-dependence of the potential. Let us define the dimensionless mass parameter
\begin{equation}\label{eq:216} 
u_{1}(\chi)=k^{2}\dfrac{\p u}{\p\chi^{2}}=\dfrac{1}{k^{2}}\dfrac{\p U}{\p\chi^{2}}\com\bar{u}_{1}=u_{1}(\bar{\chi})\, .
\end{equation}
Its evolution obeys
\begin{align}\label{eq:217} 
\p_{t}\bar{u}_{1}&=2\bar{u}_{1}+k^{2}\dfrac{\p}{\p\chi^{2}}\p_{t}\bar{u}=-2\bar{u}_{1}+4k^{2}\dfrac{\p c_{V}}{\p \chi^{2}}\, .
\end{align}
Retaining only the graviton contribution one finds
\begin{align}\label{eq:218} 
\p_{t}\bar{u}_{1}&=-2\bar{u}_{1}+\dfrac{5f_{\eta}}{16\pi^{2}}\ell_{1}\left (-\dfrac{\bar{u}}{\bar{w}}\right )k^{2}\dfrac{\p}{\p\chi^{2}}\left (\dfrac{u}{\bar{w}}\right )\bigl |_{\bar{\chi}}\nn\\
&=(A_{u}-2)\bar{u}_{1}\, .
\end{align}
The fixed point is $\bar{u}_{1*}=0$, consistent with \eq{209} where $\bar{u}_{*}$ does not depend on $\bar{\chi}$ for a given constant $\bar{w}$. The critical exponent is now given by
\begin{equation}\label{eq:219} 
\theta_{1}=2-A_{u}\, ,
\end{equation}
and $\bar{u}_{1}$ becomes an irrelevant parameter for $A_{u}>2$.

Continuing in the same approximation to
\begin{equation}\label{eq:220} 
u_{2}=\dfrac{\p^{2} U}{(\p\chi^{2})^{2}}=k^{2}\dfrac{\p }{\p\chi^{2}}u_{1}\com\bar{u}_{2}=u_{2}(\chi)
\end{equation}
one has
\begin{align}\label{eq:221} 
\p_{t}\bar{u}_{2}&=2\bar{u}_{2}+k^{2}\dfrac{\p}{\p\chi^{2}}\p_{t}{u}_{1}\bigl |_{\bar{\chi}}=A_{u}\bar{u}_{2}\, .
\end{align}
The fixed point is at $\bar{u}_{2*}=0$, and $\bar{u}_{2}$ is an irrelevant parameter for all positive $A_{u}$. This simple approximation suggests that the scalar potential $U(\chi)$ is characterized by at most two relevant parameters. This number is reduced to one for $A_{u}>2$, and to zero for $A_{u}>4$.

\subsection{Scaling potential for dilaton gravity}\label{sec:Scal_Pot_Dil_Grav}

The simple discussion above gives a flavor which couplings may correspond to relevant parameters. It is insufficient if the Planck mass depends on $\chi$ and $k$. Before investigating the issue of relevant parameters we first have to generalize the fixed points $(\bar{u}_{*},\bar{w}_{*})$ to $\chi$-dependent functions. Indeed, in dilaton quantum gravity the fixed points for a few individual couplings are replaced by scaling solutions for dimensionless functions as $u_{*}(\chi^{2}/k^{2})$ or $w_{*}(\chi^{2}/k^{2})$. The $\chi$-dependence and $k$-dependence are related for models defined precisely at the UV-fixed point, such that no intrinsic mass scale $\bar{\mu}$ appears and quantum scale invariance is preserved. Indeed, in the absence of an intrinsic mass scale the dimensionless functions $u$ and $w$ can only depend on the dimensionless ratio
\begin{equation}\label{eq:222} 
y=\dfrac{\chi^{2}}{k^{2}}\, .
\end{equation}
(We have defined $U$ and $M^{2}$ at zero momentum in flat space. The scaling solution remains valid for $\mu,\bar{\mu}\ll k,\chi$, or for $\mu\sim k$.) Scaling solutions obey
\begin{equation}\label{eq:223} 
\p_{t}u\bigl |_{y}=0\com \p_{t}w\bigl |_{y}=0\, ,
\end{equation}
where the $k$-dependence is taken at fixed $y$ instead of fixed $\chi$.

We will discuss the scaling solution for $u$ in a simple approximation where we take $w_{*}(y)$ as a given function and neglect the influence of higher derivative gravitational invariants as $CR^{2}$. Of course, a proper treatment has to determine the scaling solutions for $u$, $w$, $B$ etc. simultaneously. Nevertheless, important lessons can be inferred from the investigation of the scaling solution for a single function. A restriction to a single scaling function makes the conceptual issues more easy to follow. 

Instead of $u$ one may equivalently investigate the ratio $v=u/w$. At fixed $\chi$ the graviton contribution to the flow of $v(\chi,k)$ can be extracted from \eq{204}
\begin{equation}\label{eq:224} 
\p_{t}v=\beta_{v}=-2v-\eta_{M}v+\dfrac{5f_{\eta}}{16\pi^{2}w}\ell_{0}(-v)\, ,
\end{equation}
with
\begin{equation}\label{eq:225} 
\eta_{M}(\chi)=\dfrac{\p_{t}M^{2}(\chi)}{M^{2}(\chi)}=2+\p_{t}\ln(w(\chi))\, .
\end{equation}
For the additional contributions to the flow of $v(\chi)$ we concentrate here on those parts that only involve $v$ and approximate
\begin{equation}\label{eq:226} 
\p_{t}v=-4v-\p_{t}\ln(w)\, v+\dfrac{4c_{V}(v)}{w}\, ,
\end{equation}
with
\begin{equation}\label{eq:227} 
c_{V}(v)=\dfrac{5f_{\eta}}{64\pi^{2}}\ell_{0}(-v)+\ldots\, .
\end{equation}
The dots denote the contributions from fluctuations beyond the graviton.

Our interest are scaling solutions for which $v$ and $w$ depend only on the dimensionless ratio $y=\chi^{2}/k^{2}$. Eq.~\eqref{eq:224} involves $k$-derivatives at fixed $\chi$. We have to translate to fixed $y$, using
\begin{equation}\label{eq:228} 
\p_{t}v\bigl |_{\chi}=\p_{t}v\bigl |_{y}+\dfrac{\p v}{\p y}\p_{t}\left (\dfrac{\chi^{2}}{k^{2}}\right )\bigl |_{\chi}=\p_{t}v\bigl |_{y}-2y\dfrac{\p v}{\p y}\, .
\end{equation}
One infers
\begin{equation}\label{eq:229} 
\p_{t}v\bigl |_{y}=-4v+2y\dfrac{\p v}{\p y}-\left (\p_{t}\ln(w)\bigl |_{y}-2\dfrac{\p\ln(w)}{\p\ln(y)}\right )v+\dfrac{4c_{V}(v)}{w}\, .
\end{equation}
Scaling solutions obey $\p_{t}v\bigl |_{y}=0$, and are therefore determined as solution of the differential equation
\begin{equation}\label{eq:230} 
\dfrac{\p v}{\p\ln(y)}=-\dfrac{1}{2}\beta_{v}=2v-\dfrac{\p\ln(w)}{\p\ln(y)}v-\dfrac{2}{w}c_{V}(v)\, ,
\end{equation}
or
\begin{equation}\label{eq:231} 
\dfrac{\p\ln(v)}{\p\ln(y)}=2-\dfrac{\p\ln(w)}{\p\ln(y)}-\dfrac{2}{wv}c_{V}(v)\, .
\end{equation}

We use as an ansatz for the fixed scaling function $w_{*}(y)$ a behavior found in functional renormalization gauge studies \cite{PHW1,PHW2},
\begin{equation}\label{eq:232} 
w_{*}(y)=\bar{w}+\dfrac{\xi y}{2}\, .
\end{equation}
This corresponds to $F(\chi)=2\bar{w}k^{2}$ for $\chi\ll k$, and $F(\chi)=\xi\chi^{2}$ for $\chi\gg k$. The transition between the two regimes occurs for $y=y_{F}$,
\begin{equation}\label{eq:233} 
y_{F}=\dfrac{2\bar{w}}{\xi}\com w_{*}(y)=\dfrac{\xi}{2}\left (y+y_{F}\right )\, .
\end{equation}
For this ansatz one has
\begin{equation}\label{eq:234} 
\dfrac{\p\ln(w)}{\p\ln(y)}=\dfrac{y}{y+y_{F}}\, ,
\end{equation}
and therefore
\begin{equation}\label{eq:235} 
\dfrac{\p\ln(v)}{\p \ln(y)}=\dfrac{y+2y_{F}}{y+y_{F}}-\dfrac{c_{V}(v)}{\xi(y+y_{F})v}\, .
\end{equation}
We further need the behavior of $c_{V}(v)$. We take
\begin{equation}\label{eq:236} 
c_{V}=\dfrac{a_{V}}{1-v}+b_{V}\, ,
\end{equation}
where the first term reflects the graviton contribution in the approximation \eqref{eq:214}, with $\bar{p}=1$ for simplicity, and the second reflects the contribution of fluctuations of particles to $\p_{t}U$ that are independent of $U$. The form \eqref{eq:236} accounts for the suppression of graviton contributions for large negative $v$, and for the enhancement as positive $v$ approach a critical value. Quantitative values for $a_{V}$ and $b_{V}$ can be found in sect.~\ref{sec:Flow_Eff_Scalar_Pot_Quantum_Grav}, eq.~\eqref{eq:V17}. There we will establish that eq.~\eqref{eq:236} is a good approximation to the functional renormalization flow.

In summary, the most important features of the differential equation defining the scaling solution are given by
\begin{equation}\label{eq:237} 
\dfrac{\p \ln(v)}{\p \ln(y)}=\dfrac{y+2y_{F}}{y+y_{F}}-\dfrac{4a_{V}+4b_{V}(1-v)}{\xi(y+y_{F})v(1-v)}\, .
\end{equation}
The corresponding equation for $u(y)$ reads
\begin{equation}\label{eq:DS1} 
\dfrac{\p u}{\p\ln(y)}=-\beta_{u}=2u-2c_{V}=2u-2\bar{c}_{V}-2a_{V}\left [\dfrac{\xi(y+y_{F})}{2u}-1\right ]^{-1}\, ,
\end{equation}
with
\begin{equation}\label{eq:237D} 
\bar{c}_{V}=a_{V}+b_{V}\, .
\end{equation}
Many characteristic features of the scaling solution can be understood from the solutions of eq.~\eqref{eq:237} or \eqref{eq:DS1}.

\begin{figure}[t!]
\includegraphics[scale=0.7]{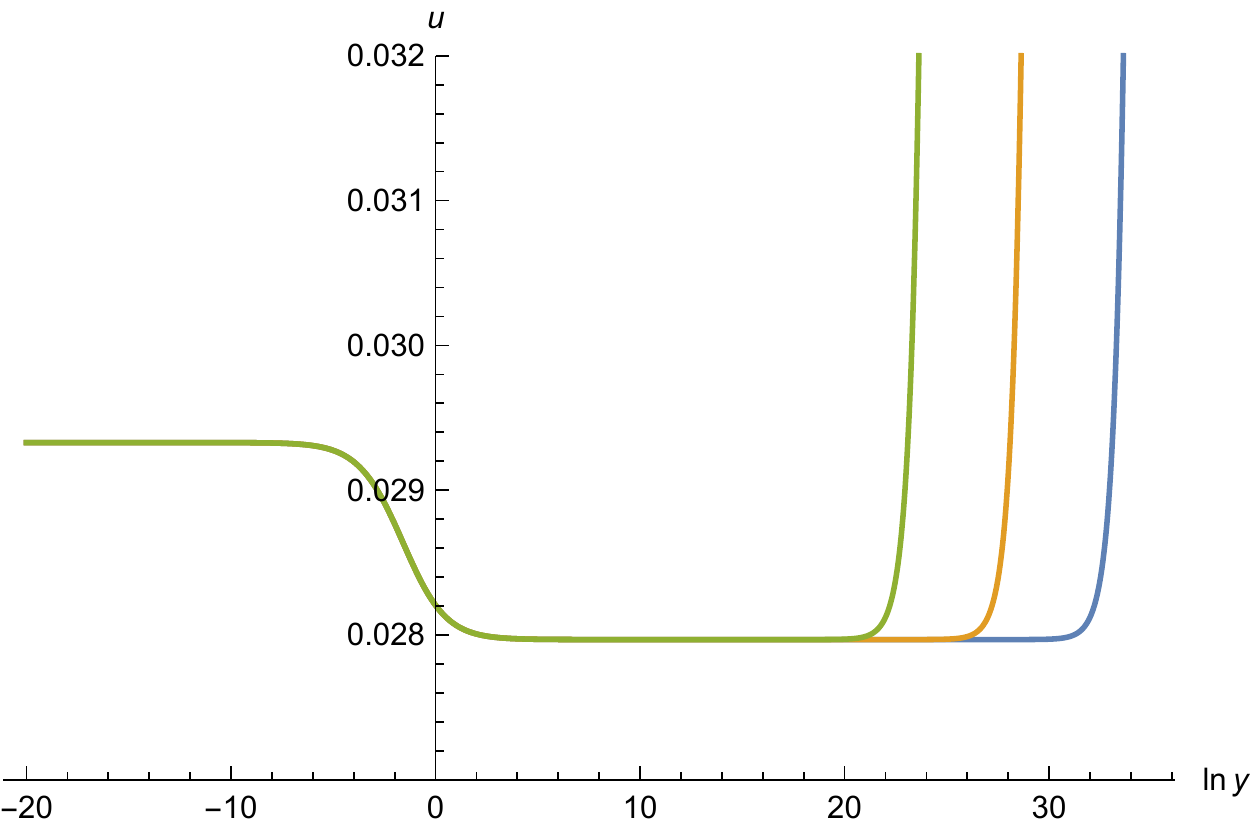}
\caption{Dimensionless potential $u$ as a function of $\ln(y)$. Parameters are $\xi=1$, $y_{F}=0.4$, $b_{V}=0.02$. We plot the solution of eq.~\eqref{eq:DS1} for three different values of the integration constant. One observes a transition near $y_{F}$, and a further transition at some large $y_{V}$, which depends on the integration constant. For the particular ``flat scaling solution'' the flat region extends to $y\raw\infty$, corresponding to $y_{V}\raw\infty$.}\label{fig:A} 
\end{figure}

We plot the numerical solution for $u$ as a function of $\ln(y)$ in Fig.~\ref{fig:A}. We show three solutions, corresponding to different values of the free integration constant which characterizes the general solution of eq.~\eqref{eq:237D}. One observes three different regions in $y$. For $y\raw 0$ the potential $u$ approaches a constant. For intermediate $y$ one finds a somewhat smaller constant, while for large $y$ the potential increases rapidly. We will next discuss these different regions in $y$ analytically.

\subsection{Crossover in dimensionless fields}\label{sec:Crossover_DimLess_Fields}

Scaling solutions typically show a crossover as a function of the variable $y=\chi^{2}/k^{2}$. At fixed $k$ this is directly related to a crossover in field space. If we replace $k$ by an intrinsic scale $\bar{\mu}$ we are close to the crossover situation discussed in sect.~\ref{sec:Crossover_Field_Space}. An important modification arises, however, from the possibility that the crossover may be a switch from a fixed point for $u$ to a fixed point for $v$ or for $\tilde{\lambda}=U/M^{4}$. 

We may demonstrate this for the range of large $y$. For $y+y_{F}\gg 2u/\xi$ we can approximate eq.~\eqref{eq:DS1} by
\begin{equation}\label{eq:DS2} 
\dfrac{\p u}{\p\ln(y)}=2u-2\bar{c}_{V}-\dfrac{4a_{V}u}{\xi(y+y_{F})}\, .
\end{equation}
A fixed $u_{\infty}=u(y\raw\infty)$ is reached for $y\raw\infty$,
\begin{equation}\label{eq:DS3} 
u_{\infty}=\bar{c}_{V}\, .
\end{equation}
There is a particular solution that connects to $u_{\infty}$ for $y\raw\infty$, which can be obtained by a Taylor expansion in inverse powers of $y$,
\begin{equation}\label{eq:DS4} 
u_{\infty}(y)=\bar{c}_{V}\left (1+\dfrac{4a_{V}}{3\xi y}+\ldots\right )\, .
\end{equation}
This corresponds to the scaling solution in ref.~\cite{PHW1,PHW2}.

The general solution of \eq{DS2} involves one free integration constant. In particular, the solutions in the neighborhood of $u_{\infty}(y)$ are given by
\begin{equation}\label{eq:DS5} 
u(y)=u_{\infty}(y)+\delta y^{2}\, ,
\end{equation}
with $\delta$ a free integration constant. For any $\delta\neq 0$ they move away from $u_{\infty}(y)$ as $y$ increases. The size of $\delta$ determines the value $y_{cr}$ for which $u(y)-u_{\infty}$ is of similar size as $u_{\infty}$, such that for $y>y_{cr}$ the solution $u_{\infty}(y)$ does not remain a valid approximation anymore. The behavior of the general solution can easily be seen if we also neglect for large $y$ the last term in \eq{DS2}. The differential equation
\begin{equation}\label{eq:DS6} 
\dfrac{\p u}{\p\ln(y)}=2u-2\bar{c}_{V}
\end{equation}
has the general solution
\begin{equation}\label{eq:DS7} 
u=\bar{c}_{V}+\delta y^{2}\, .
\end{equation}
It describes a crossover from a fixed point of $u$ for $y^{2}\ll \bar{c}_{V}/\delta$
\begin{equation}\label{eq:DS8} 
u_{*}=\bar{c}_{V}\, ,
\end{equation}
to a fixed point of $\tilde{\lambda}$, given by
\begin{equation}\label{eq:DS9} 
\tilde{\lambda}=\dfrac{U}{M^{4}}=\dfrac{u}{4w^{2}}\, ,
\end{equation}
for $y^{2}\gg \bar{c}_{V}/\delta$. The value of the fixed point for $\tilde{\lambda}$ depends on the integration constant $\delta$,
\begin{equation}\label{eq:DS10} 
\lambda_{*}=\dfrac{\delta}{\xi^{2}}\, .
\end{equation}
The crossover value for $y$ is given by
\begin{equation}\label{eq:DS11} 
y_{V}=\sqrt{\dfrac{\bar{c}_{V}}{\delta}}\, .
\end{equation}

The approximation \eqref{eq:DS2} remains valid as long as $c_{V}$ in \eq{236} can be approximated by $\bar{c}_{V}$, i.e. as long as $v\ll 1$. As $y$ increases the increase of $u\sim \delta y^{2}$ will violate this condition at some value $y_{IR}$,
\begin{equation}\label{eq:DS12} 
y_{IR}=\dfrac{\xi}{2\delta}\, .
\end{equation}
Indeed, for $u=\delta y^{2}$ and $y\gg y_{F}$ one finds an increase of $v$ according to
\begin{equation}\label{eq:DS13} 
v=\dfrac{u}{w}=\dfrac{2\delta y}{\xi}\, .
\end{equation}
For $v$ near one the flow enters the IR-regime with an approximate fixed point $v_{*}=1$. We will discuss this infrared regime in sect.~\ref{sec:Infrared_QG}. In the range of large $y$ one observes two crossovers. Around $y_{V}$ the flow crosses from the fixed point \eqref{eq:DS8} for $u$ to the fixed point \eqref{eq:DS10} for $\tilde{\lambda}$. Subsequently, it crosses around $y_{IR}$ to the IR-fixed point for $v$ at $v=1$.

The UV-fixed point for $u$ is reached in the limit $y\raw 0$. It occurs for ($w=\bar{w}=\xi y_{F}/2$)
\begin{equation}\label{eq:DS14} 
u_{*}=\dfrac{1}{2}\left (\bar{w}+b_{V}-\sqrt{\left (\bar{w}-b_{V}\right )^{2}-4a_{V}\bar{w}}\right )\, .
\end{equation}
This fixed point is attractive in the UV-direction. All solutions of \eq{DS1} that can be extended to $y\raw 0$ end for $y=0$ in this fixed point. Consider now the solutions \eqref{eq:DS5} for the region of large $y$. There is no obstacle to extending these solutions to $y=0$. In particular, the solution $u_{\infty}(y)$ for $\delta=0$, given by \eq{DS4}, can be continued to $y=0$. This particular scaling solution connects the UV-fixed point \eqref{eq:DS14} for $y\raw 0$ to the fixed point \eqref{eq:DS3} for $y\raw\infty$. In the neighborhood of this solution we may parametrize the one-parameter family of general scaling solutions by $\delta$, or, equivalently, by the value of $y_{V}$. More generally, all scaling solutions connect the UV-fixed point \eqref{eq:DS14} for $y\raw0$ to the IR-fixed point $v=1$ for $y\raw\infty$.

For this type of crossover between UV- and IR-fixed points the model has to be specified by the selection of a specific crossover trajectory. This is done by the specification of the free parameter $y_{V}$. Furthermore, the location on the crossover trajectory needs to be fixed. If we associate $k$ with an intrinsic scale $\bar{\mu}$, the location on the crossover trajectory will depend  on the value of the scalar field $\chi$, $y=\chi^{2}/\bar{\mu}^{2}$. This value may change in the course of the cosmological history, such that $y$ may effectively depend on cosmic time. 

\begin{figure}[t!]
\includegraphics[scale=0.7]{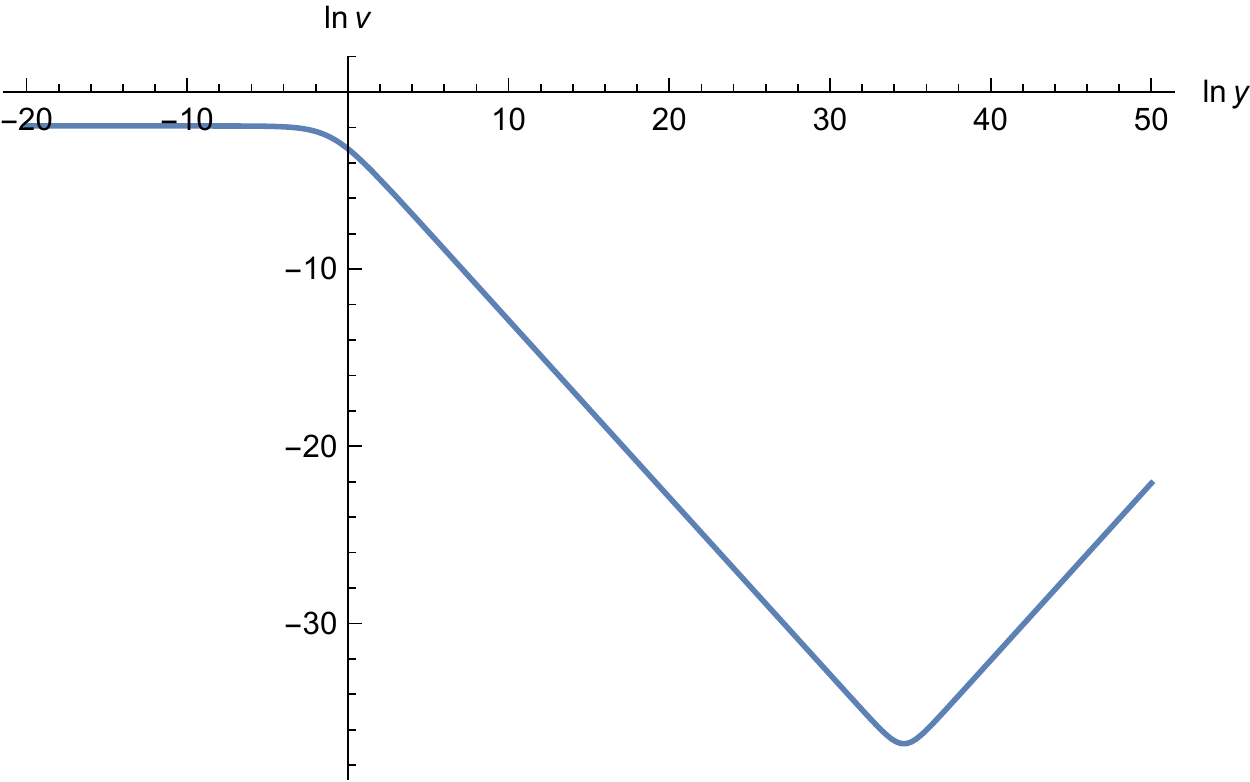}
\caption{Double logarithmic plot of $\ln(v)$ vs. $\ln(y)$. Parameters are as in Fig.~\ref{fig:A}. The two crossovers in $v(y)$ at $y_{F}$ and $y_{V}$ are clearly visible as changes in the slope. The crossover at $y_{IR}$ to the IR-behavior $\ln(v)=0$ (not shown in this figure) occurs once the curve reaches the axis at large $y$.}\label{fig:B} 
\end{figure}

The integration constant $y_{V}$ can be associated to a relevant parameter at the UV-fixed point for the flow with $\mu$ or $k$ at fixed $\chi$. It specifies the location of change in the properties of the scalar potential, while $y_{F}$ indicates a qualitative change in the gravitational interaction. In Fig.~\ref{fig:B} we plot $\ln(v)$ as a function of $\ln(y)$. The two crossovers at $y_{F}$ and $y_{V}$ are clearly visible. While $v$ decreases $\sim y^{-1}$ between $y_{V}$ and $y_{F}$, it increases $\sim y$ for $y>y_{F}$.

The parameter $y_{V}$ can be many orders of magnitude larger than $y_{F}$. Cosmology may depend strongly on the value of $y_{V}$. For $y_{V}\approx y_{F}$ the change of the potential occurs typically during or around the inflationary epoch. For $y_{V}\gg y_{F}$ this qualitative change happens at much later stages of cosmology. The curves in Fig.~\ref{fig:A} and Fig.~\ref{fig:B} all correspond to very large values of $y_{V}$. We will see that for $y_{V}\approx 10^{60}$ the crossover from a constant potential $U\sim \bar{\mu}^{4}$ to a potential $U\sim \chi^{4}$ occurs close to the present cosmological epoch.

\subsection{Quantitative scaling solutions}\label{sec:Quantitative_Scaling_solution} 

It is worthwhile to investigate the $y$-dependence of the scaling solutions, starting from $y=0$ and following the solutions to $y\raw\infty$. We will concentrate here on $v(y)$, as given by the one parameter family of solutions to the differential equation \eqref{eq:237}. There are several regions in $y$ and $v(y)$ where the differential equation \eqref{eq:237} has simple analytic solutions. We first consider $y\raw 0$ where \eq{237} becomes
\begin{equation}\label{eq:237A} 
\dfrac{\p v}{\p\ln(y)}=-\dfrac{1}{2}\beta_{v}=2v-\dfrac{2}{\bar{w}}\left (\dfrac{a_{V}}{1-v}+b_{V}\right )\, .
\end{equation}
This equation has two fixed points if
\begin{equation}\label{eq:237B} 
a_{V}<\dfrac{(\bar{w}-b_{V})^{2}}{4\bar{w}}\, ,
\end{equation}
located at
\begin{equation}\label{eq:237C} 
v_{*}=\dfrac{1}{2\bar{w}}\left (\bar{w}+b_{V}\pm\sqrt{(\bar{w}-b_{V})^{2}-4\bar{w}a_{V}}\right )\, .
\end{equation}
For positive $\bar{c}_{V}$ both values $v_{*}$ are positive, while for negative $\bar{c}_{V}$ one fixed point value is negative, the other positive. 

The smaller $v_{*}$ (with the minus sign in \eq{237C}) is attractive for $y\raw 0$, the larger one gets approached for increasing $y$. We concentrate on the UV-attractive fixed point. In the vicinity of the fixed point the scaling solution obeys
\begin{equation}\label{eq:237E} 
\dfrac{\p v}{\p\ln(y)}=\tilde{A}_{v}(v-v_{*})\, ,
\end{equation}
with
\begin{equation}\label{eq:237F} 
\tilde{A}_{v}=-\dfrac{1}{2}\dfrac{\p \beta_{v}}{\p v}\bigl |_{v_{*}}=2-\dfrac{2a_{V}}{\bar{w}(1-v_{*})^{2}}=2-\dfrac{A_{u}}{2}\, .
\end{equation}
For the UV-attractive fixed point one has $A_{u}\leq 4$ and therefore $\tilde{A}_{v}\geq 0$. For $y\raw 0$ the behavior is given (for any $v(y)$ different from the larger $v_{*}$) by
\begin{equation}\label{eq:237G} 
v(y)=v_{*}+\hat{v}_{0}y^{\tilde{A}_{v}}\, .
\end{equation}
The integration constant $\hat{v}_{0}$ may be positive or negative. The corresponding effective potential behaves as
\begin{equation}\label{eq:237H} 
U(\chi)=\bar{w}k^{4}v=\bar{w}k^{4}\left [v_{*}+\hat{v}_{0}\left (\dfrac{\chi^{2}}{k^{2}}\right )^{\tilde{A}_{v}}\right ]\, .
\end{equation}

We observe that for $\hat{v}_{0}\neq 0$ the behavior for $\chi\raw 0$ is not analytic unless $\tilde{A}_{v}$ happens by accident to be an integer. Depending on the value of $\tilde{A}_{v}$ the derivatives $\p^{n}U/(\p\chi^{2})^{n}$ at $\chi=0$ will diverge for $n$ larger than some critical value. This is a priori not in contradiction with the existence of an UV-fixed point, since a well defined scaling solution exists. Defining ($\rho=\chi^{2}/2$)
\begin{equation}\label{eq:NFAB} 
\lambda_{\chi}=\dfrac{\p^{2}U}{\p\rho^{2}}=2\bar{w}\hat{v}_{0}\left (2-\dfrac{A_{u}}{2}\right )\left (1-\dfrac{A_{u}}{2}\right )\left (\dfrac{\chi}{k}\right )^{-A_{u}}\, ,
\end{equation}
we observe large $\lambda_{\chi}$ in a range of small $\chi$. In the range of large $\lambda_{\chi}$ our approximation can no longer be trusted. Indeed, the scalar fluctuations contribute to the flow equation a term $\sim \lambda_{\chi}^{2}$,
\begin{equation}\label{eq:NFAC} 
\p_{t}\lambda_{\chi}=A_{u}\lambda_{\chi}-C_{\chi}+d_{\chi}\lambda_{\chi}^{2}\, .
\end{equation}
This assures that $\lambda_{\chi}$ cannot diverge in the UV. At the UV-fixed point $\lambda_{\chi}$ is an irrelevant parameter that takes its fixed point value $\lambda_{*}\approx C_{\chi}/A_{u}$. The behavior \eqref{eq:237H} may still be valid for not too small $\chi$. For very small $\chi$ it is modified, however, such that the non-analyticity is avoided and $\lambda_{\chi}$ reaches its fixed point value.

For large $y\gg y_{F}$ and small $v\ll 1$ eq.~\eqref{eq:237} is approximated by
\begin{equation}\label{eq:6.158A} 
\dfrac{\p v}{\p y}=\dfrac{v}{y}-4\dfrac{\bar{c}_{V}}{\xi y^{2}}\, .
\end{equation}
This corresponds to \eq{DS6}. The solution (cf. \eqs{DS7}{DS11})
\begin{equation}\label{eq:6.158B} 
v=\dfrac{2\bar{c}_{V}}{\xi y}+\dfrac{2\delta y}{\xi}=\dfrac{2\bar{c}_{V}}{\xi y}\left (1+\dfrac{y^{2}}{y_{V}^{2}}\right )
\end{equation}
exhibits a minimum of $v(y)$ for
\begin{equation}\label{eq:6.158C} 
y_{min}=y_{V}\com v_{min}=v(y_{min})=4\dfrac{\bar{c}_{V}}{\xi y_{V}}\, .
\end{equation}

For finite $y_{V}$ the scaling $v\sim y$ continues until $v$ enters the IR-region where $v$ comes close to one such that the second term in \eq{237} becomes again important. This happens at $y_{IR}$, when $v(y_{IR})$ reaches a value of the order one. According to eq.~\eqref{eq:6.158B} one infers
\begin{equation}\label{eq:246} 
y_{IR}=\dfrac{\xi y_{V}^{2}}{2\bar{c}_{V}}\com \dfrac{y_{IR}}{y_{V}}=\dfrac{2}{v_{min}}\, .
\end{equation}
The extension of the transition region between $y_{V}$ and $y_{IR}$ can be substantial for large $y_{V}$ or small $v_{min}$. 

In the transition region between $y_{V}$ and $y_{IR}$ the $\chi$-dependence of $U(\chi)$ is quartic
\begin{equation}\label{eq:247} 
U(\chi)=\delta\chi^{4}=\dfrac{\bar{c}_{V}}{y_{V}^{2}}\chi^{4}\approx\dfrac{\xi^{2}v_{min}^{2}}{\bar{c}_{V}}\chi^{4}\, ,
\end{equation}
with a tiny coefficient for large $y_{V}$. The IR-region will be discussed below. We note that for all $v_{min}>0$ the IR-region is necessarily reached at some finite $y_{IR}$. An interesting limiting case is $v_{min}=0$. In this case \eq{DS3} implies for all $y\gg y_{F}$ a $\chi$-independent scalar potential
\begin{equation}\label{eq:248} 
U=\bar{c}_{V}k^{4}\, .
\end{equation}
This is the type of scaling function found in refs.~\cite{PHW1,PHW2}. The infrared region is never reached in this case.

The one-parameter family of solutions for large $y$, small $v$, has to be matched to the one-parameter family of solutions for $y\raw 0$. The integration constants $y_{V}$ (or $v_{min}$) for large $y$ and the integration constant $\hat{v}_{0}$ for small $y$ are not independent. The general solution of eq.~\eqref{eq:237} has at most one free integration constant, such that there must exist a functional relation expressing $\hat{v}_{0}$ as a function of $y_{V}$ and vice versa. A reasonable interpolation between the different limits is given for $y\ll y_{IR}$ by
\begin{equation}\label{eq:NFAA} 
\bar{v}(y)=\dfrac{2\bar{c}_{V}}{\xi}\left (\dfrac{1}{y+y_{F}}+\dfrac{y}{y_{V}^{2}}-\dfrac{y_{F}}{(y+y_{F})^{2}}\right )+\dfrac{v_{*}y_{F}^{2}}{(y+y_{F})^{2}}\, .
\end{equation}

Formally, one also finds scaling solutions with negative $\delta$ in \eq{DS7}. Such an unbounded potential will actually not be reached by the flow. In particular, the fluctuations of the scalar $\chi$ will prevent an unstable potential. They induce in $k\p_{k}U$ terms that involve $\p U/\p \chi^{2}$ and $\p^{2}U/(\p\chi^{2})^{2}$. Such terms have been neglected in the approximation of $c_{V}(v)$ depending only on $v$ in \eq{226}. We concentrate in the following on solutions with $\delta\geq 0$.

In our simple approximation we have found a one-parameter family of scaling solutions, with $y_{V}$ or $v_{min}$ parameterizing a given solution in this set. Since we have investigated a first order differential equation \eqref{eq:237} it is not surprising that possible solutions depend on one free integration constant, given by the value of $v(y)$ at some particular $y$. In our case we may take $v_{min}$ which corresponds roughly to $v(y)$ in the region near $y_{V}$.

In a more general approach to the functional renormalization flow one will encounter several (in principle even infinitely many) functions as $v(y)$, $w(y)$, $B(y)$, $C(y)$ etc. The system of nonlinear differential equations for the scaling solution typically involves these functions as well as their first and second derivatives. Local solutions of these differential equations will therefore depend on many free parameters. One expects that only a family of solutions with a finite number of parameters remains consistent in the whole range of $y$ between zero and infinity. 

This issue is related to the number of relevant parameters at the UV-fixed point. For any fixed $\chi$ the UV-limit corresponds to $y\raw 0$. The number of ``free'' initial conditions at very small $y$ is therefore bounded by the number of relevant parameters at the UV-fixed point. This is the maximal number of parameters whose values are not predicted by the presence of the fixed point. It is also the maximum of the number of free parameters that characterize the family of scaling solutions.

\subsection{Infrared quantum gravity}\label{sec:Infrared_QG} 

One often assumes that for momenta much below the Planck mass the fluctuations of the graviton can be neglected. In this naive picture gravity becomes classical below the Planck scale, and quantum gravity matters only near and above the Planck scale. In the functional renormalization group flow we can restrict $c_{V}$ to the pure contributions of the traceless transverse tensor mode in the metric, omitting the contributions from other particles, e.g. omitting the dots in \eq{227}. According to the naive picture the flow of the potential should then stop for large $y$, where $M^{2}(\chi)$ is much larger than $k^{2}$. We have seen that this is not the case.

First of all, a decoupling of graviton fluctuations arises from the smallness of the graviton induced interactions, not by decoupling of a heavy particle. The graviton is a light particle on all scales. The leading contribution to the flow of the potential $U(\chi)$ does not involve interactions. All massless particles give contributions to $\p_{t}U$ of the order $k^{4}$ - only the degrees of freedom count, not the interactions. If mass terms play no role, the overall strength of the contribution from graviton fluctuations is of the same order as the one from photon fluctuations. Contributions from particles with mass $m_{k}$ are for small $k$ of the order $k^{6}/m^{2}_{k}$ and therefore suppressed by a factor $k^{2}/m^{2}_{k}$ as compared to massless particles. In flat space the graviton propagator involves a mass term
\begin{equation}\label{eq:249} 
m_{g}^{2}=-\dfrac{2V}{M^{2}}\, ,
\end{equation}
as visible from \eq{106} for $R_{k}=0$,
\begin{equation}\label{eq:250} 
G_{g}(q^{2})=\dfrac{4}{M^{2}}\left (q^{2}-\dfrac{2V}{M^{2}}\right )^{-1}\, .
\end{equation}
For a negative cosmological constant $V$ the mass term $m_{g}^{2}$ is positive, explaining the suppression factor $(-v)^{-1}$ in $\ell_{0}(-v)$, cf. \eq{117}.

The central ingredient for the possible importance of graviton fluctuations far below the Planck scale is the observation that a positive cosmological constant $V$ entails in flat space a tachyonic negative mass term $m_{g}^{2}<0$. This renders perturbation theory for gravity in flat space unstable for all $V>0$. Functional renormalization tames this instability by the presence of the cutoff term $R_{k}$ in \eq{106}. As long as $(-m_{g}^{2})$ does not exceed the minimum of $P_{k}(q)=q^{2}+R_{k}(q)$ the momentum integral remains well defined even for $m^{2}_{g}<0$. The result is an enhancement in $\ell_{0}(-v)$ for $v>0$. As $v$ increases, this enhancement can become very strong. Thus for positive $V$ the graviton fluctuation effect on the flow of $V$ is no longer of similar strength as the effect from photon fluctuations. Graviton fluctuations are enhanced as compared to photon fluctuations. As $2V/M^{2}$ approaches the minimum of $P_{k}(q^{2})$ the graviton fluctuation become the dominant contribution to the renormalization of $V$.

The same arguments extend from a constant $V$ to a $\chi$-dependent potential $U(\chi)$. For the overall size of renormalization effects of the potential the graviton fluctuations are at least as important as the other fluctuations, and even dominate for positive $U$ if $v(\chi)=2U(\chi)/(M^{2}(\chi)k^{2})$ comes close to the minimum of $P_{k}$. This effect does only depend on the ratio $U/M^{2}$ which determines the mass term in the graviton propagator. A different question concerns the field dependence of the potential. For this issue the interactions play a role and gravitational fluctuation effects are suppressed by powers of $k^{2}/M^{2}$. They are negligible below the Planck scale unless compensated by a very strong enhancement factor due to the closeness of the instability.

In our previous discussion we have taken the enhancement of the graviton fluctuations into account by the pole structure of the threshold function $\ell_{0}(\tilde{w})$ in \eq{214}. We will investigate here more closely the behavior of the threshold function $\ell_{0}(\tilde{w})$ in the region of strong enhancement for negative $\tilde{w}$ \cite{CWGFE,CWIRG}. According to the definition \eqref{eq:112} it will depend on the precise shape of the cutoff function $R_{k}(q^{2})$. Suitable $R_{k}$ decay fast for $q^{2}>k^{2}$ such that the high momentum behavior is not affected, lead to $P_{k}=q^{2}+R_{k}$ with a minimum near $q^{2}=k^{2}$, and an increase for $q^{2}\raw 0$ such that the low momentum modes are suppressed efficiently.

For this general shape of $R_{k}$ the function $p(x)$ in \eq{113} has a minimum at $\bar{x}$, with $p(\bar{x})=\bar{p}$. Expanding in this region
\begin{equation}\label{eq:251} 
p(x)=\bar{p}+a(x-\bar{x})^{2}\, ,
\end{equation}
and approximating
\begin{equation}\label{eq:252} 
f(x)r(x)\approx f(\bar{x})\, r(\bar{x})=2\bar{s}\, ,
\end{equation}
one finds for $v$ in the vicinity of $\bar{p}$
\begin{equation}\label{eq:253} 
\ell_{0}(-v)=\dfrac{\pi\bar{x}\bar{s}}{\sqrt{a}}\, \left (\bar{p}-v\right )^{-\frac{1}{2}}\, .
\end{equation}
For $v$ close to $\bar{p}$ this accounts for a strong enhancement of the graviton fluctuations $\sim (\bar{p}-v)^{-1/2}$. In this region the graviton fluctuations dominate the flow of $v$ according to
\begin{equation}\label{eq:254} 
k\p_{k}v=\beta_{v}=-(2+\eta_{M})v+\dfrac{\bar{e}k^{2}}{M^{2}}(\bar{p}-v)^{-\frac{1}{2}}\, ,
\end{equation}
with
\begin{equation}\label{eq:255} 
\bar{e}=\dfrac{5\bar{x}\bar{s}}{8\pi\sqrt{a}}\com\eta_{M}=\dfrac{\p\ln(M^{2})}{\p \ln(k)}\, .
\end{equation}

\begin{figure}[t!]
\includegraphics[scale=0.6]{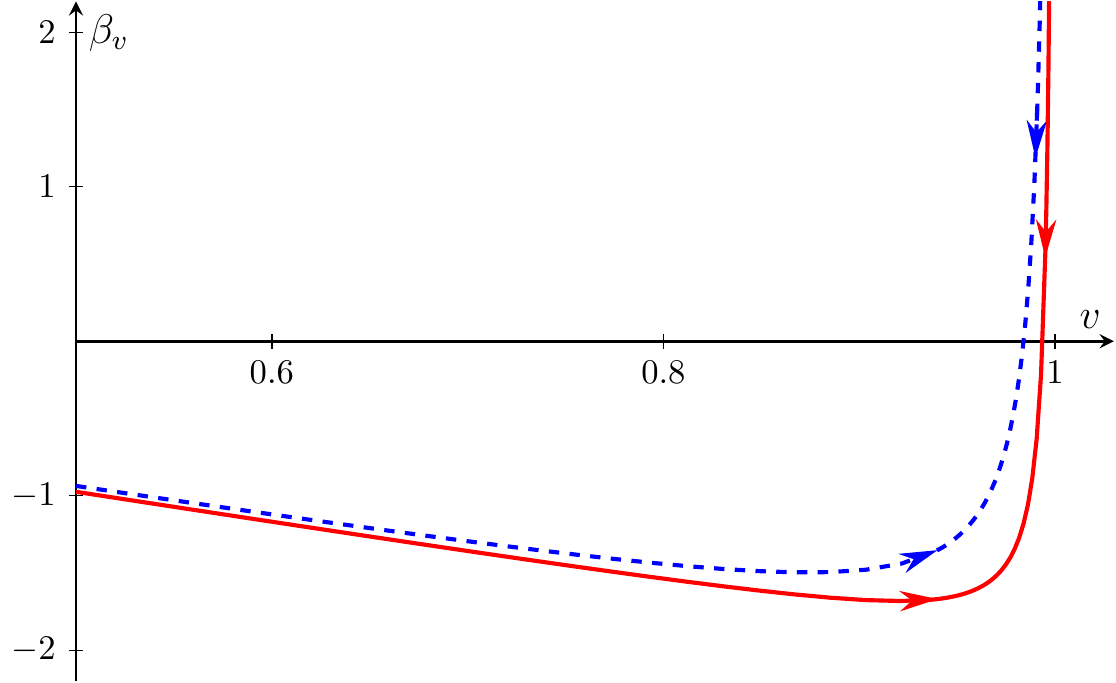}
\caption{$\beta$-function for $v$ for different values of $M^{2}$. From ref.~\cite{CWGFE}.}\label{fig:3} 
\end{figure}

Due to the strong enhancement factor the function $\beta_{v}$ is always positiv for $v$ close to $\bar{p}$. As a result, $v$ decreases with decreasing $k$ in this region. For $v<\bar{p}$ this means that $v$ moves away from the singularity at $\bar{p}$. In other words, if the flow initially starts in the region without instability, e.g. $v<\bar{p}$, the flow of $v$ can never reach the instability. The instability is ``repulsive'' - the flow equations account for the simple property that a stable microscopic theory will not lead to an unstable macroscopic theory. In the case of gravity this has been called \cite{CWGFE,CWIRG} the ``graviton barrier'': the dimensionless quantity $v$ can never exceed the value $\bar{p}$. Details of the IR-cutoff function will influence the precise shape of the barrier effect, e.g. the value of $\bar{p}$, and possibly also the power of the singularity, e.g. turning $(\bar{p}-v)^{-1/2}$ to $(\bar{p}-v)^{-1}$ \cite{CWGFE}. For a large class of suitable cutoff functions this does not influence the existence of the barrier.

\begin{figure}[t!]
\includegraphics[scale=0.6]{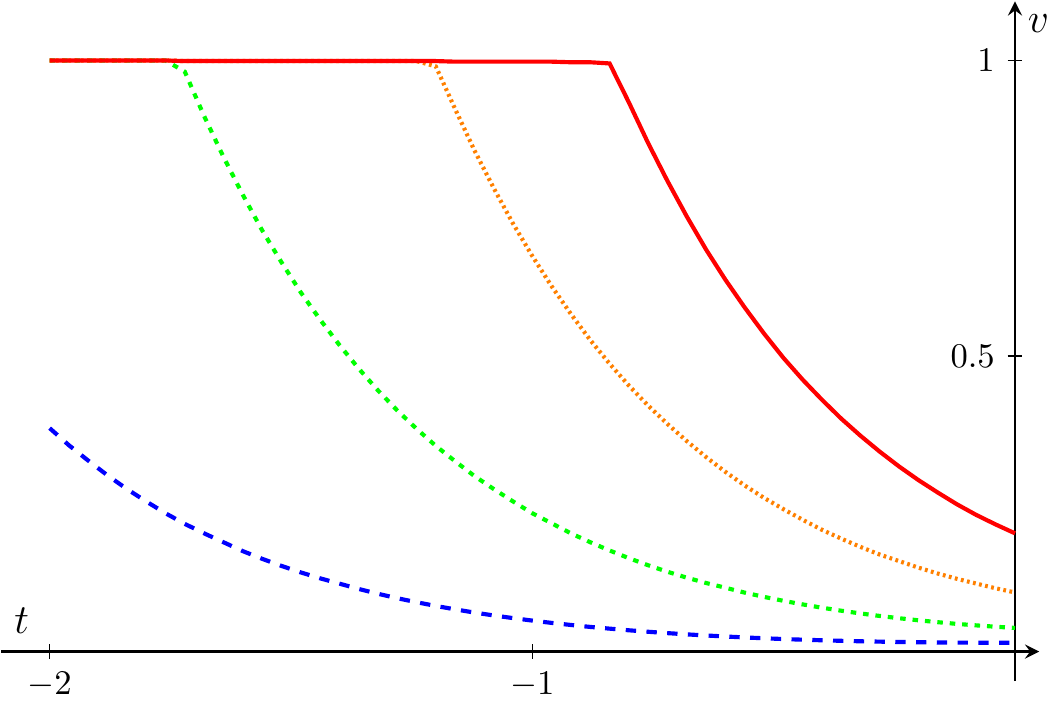}
\caption{Flow of $v$ for differential initial values. As $k$ is lowered, the infrared value $v=1$ is always approached. From ref.~\cite{CWGFE}.}\label{fig:4} 
\end{figure}

The behavior \eqref{eq:254} close to the singularity holds for arbitrary functions $v(\chi)$ and arbitrary $M^{2}(\chi)$ or $w(\chi)=M^{2}(\chi)/2k^{2}$. If the graviton enhancement occurs in the region where $M^{2}(\chi)$ is almost independent of $k$ we can neglect $\eta_{M}$. In this region the flow equation for $v(\chi)$ becomes
\begin{equation}\label{eq:256} 
k\p_{k}v=\beta_{v}=-2v+\dfrac{\bar{e}}{2w}(\bar{p}-v)^{-\frac{1}{2}}\, .
\end{equation}
We show $\beta_{v}$ for different values of $\bar{e}/w$ and $\bar{p}=1$ in Fig.~\ref{fig:3}. We observe that the zero of $\beta_{v}$ moves closer to $\bar{p}$ as $w$ increases. This shows that the stronger suppression by larger $w$, or smaller $k^{2}/M^{2}$, has to be compensated by a stronger enhancement, as provided by a value of $v$ closer to the instability. For a solution of the flow equation \eqref{eq:256}, with $w\sim k^{-2}$, one finds that the zero of $\beta_{v}$ is approached rapidly
\begin{equation}\label{eq:257} 
v=\bar{p}-\left (\dfrac{\bar{e}k^{2}}{2\bar{p}M^{2}}\right )^{2}\, ,
\end{equation}
or
\begin{equation}\label{eq:258} 
U=\dfrac{\bar{p}k^{2}M^{2}}{2}-\dfrac{\bar{e}^{2}k^{6}}{8\bar{p}^{2}M^{2}}\, .
\end{equation}
Even though the zero of $\beta_{v}$ is not an exact partial fixed point because of the $k$-dependence of $w$, it can be treated to a good approximation as  a $k$-dependent partial fixed point. We show the $k$-dependence of $v$ according to the solution of \eq{256} in Fig.~\ref{fig:4}.

For small values of $k^{2}/M^{2}$ the ratio $U/M^{2}$ comes very close to the universal IR-value
\begin{equation}\label{eq:259} 
\dfrac{U}{M^{2}}=\dfrac{\bar{p}k^{2}}{2}\, .
\end{equation}
The dimensionless ratio $U/M^{4}$ approaches zero for $k\raw 0$, 
\begin{equation}\label{eq:260} 
\dfrac{U}{M^{4}}=\dfrac{\bar{p}k^{2}}{2M^{2}}=\dfrac{\bar{p}}{4w}\, .
\end{equation}
In dilaton quantum gravity, with $M^{2}=\xi \chi^{2}$ for large $\chi^{2}/k^{2}$, the graviton barrier has an important consequence for the behavior of the potential at large $\chi$. The effective potential cannot increase faster than $\chi^{2}$. If the region of a strong graviton enhancement is reached, the asymptotic behavior is quadratic in $\chi$
\begin{equation}\label{eq:261} 
U_{as}=\dfrac{1}{2}\bar{p}\xi k^{2}\chi^{2}\, .
\end{equation}
Only potentials that are smaller than $U_{as}$ are allowed. For the example of a $\chi$-independent potential $U\sim k^{4}$ the region of strong graviton enhancement is never reached. We will see that the bound \eqref{eq:261} for the increase of $U$ with large $\chi$ has important consequences for the cosmological constant problem.

The graviton enhancement plays also an important role for the behavior of the scaling solution of dilaton quantum gravity as $y\raw \infty$. For $y>y_{IR}$ the function $c_{V}(v)$ is dominated by the enhanced graviton contribution
\begin{equation}\label{eq:262} 
c_{V}(v)\approx \dfrac{5}{64\pi^{2}}\ell_{0}(-v)\approx\dfrac{\bar{e}}{2}(\bar{p}-v)^{-\frac{1}{2}}\, ,
\end{equation}
such that the scaling equation for $v(y)$ becomes
\begin{equation}\label{eq:263} 
\dfrac{\p v}{\p\ln(y)}=v-\dfrac{2c_{V}(v)}{w}=v-\dfrac{\bar{e}}{2\xi y}(\bar{p}-v)^{-\frac{1}{2}}\, .
\end{equation}
Similar as for the $k$-dependence before, $v$ is rapidly attracted to a $y$-dependent partial fixed point at $v$ close to $\bar{p}$. Equating the two terms on the r.h.s. of \eq{263} yields
\begin{equation}\label{eq:264} 
v(y)=\bar{p}-\left (\dfrac{\bar{e}}{2\xi\bar{p}y}\right )^{2}=\bar{p}-\left (\dfrac{\bar{e}k^{2}}{2\bar{p}M^{2}}\right )^{2}\, ,
\end{equation}
in agreement with \eq{257}.

\subsection{Crossover in quantum gravity}\label{sec:Crossover_QuantumGravity} 

For the scaling solution the overall behavior of the scalar potential as a function of $\chi$ at fixed $k$ is a crossover from an UV-fixed point for $\chi\raw 0$ to an IR-fixed point for $\chi\raw\infty$. The behavior for $\chi\raw 0$ resembles Einstein gravity with constant $U\sim k^{4}$. On the other end the IR-behavior for $\chi\raw\infty$ is given by the fixed point behavior of infrared quantum gravity, with $U\sim k^{2}\chi^{2}$. An exception is the particular scaling solution for $y_{V}\raw \infty$, for which the potential for $\chi\raw\infty$ is flat, $U\sim k^{4}$. The crossover between the two fixed points may occur in several steps. The first crossover scale may be in the gravitational sector at $y_{F}$, when the Planck mass starts to be independent of $k$, proportional to $\chi$. Correspondingly, the potential turns from the behavior at small $\chi$, $U\sim k^{4}(1+c\chi^{4-A})$ again to a behavior $U\sim k^{4}$ but now with a different coefficient
\begin{equation}\label{eq:XAX} 
U=\bar{c}_{V} k^{4}\, .
\end{equation}
This behavior may extend to a large range in $y=\chi^{2}/k^{2}$, until for $\chi^{2}>y_{V}$ an increase $\sim \chi^{4}$ joins finally the asymptotic increase which is quadratic, $U\sim k^{2}\chi^{2}$.

For the observable dimensionless ratio $\tilde{\lambda}=U/M^{4}$ the crossover in the scaling solution is an example of the situation discussed in sect.~\ref{sec:Flow_Mom_Fixed_Field} if we replace $k$ by $\mu$. For all values of $\chi$ scale symmetry is an exact symmetry if the $\mu$-dependence of interactions obeys the scaling solution. Here $\mu$ is some physical IR-cutoff, arising from external momenta or a curved geometry. 

We have encountered a one-parameter family of scaling solutions, parameterized by the value $y_{V}$ at which the transition from $U\sim k^{4}$ to $U\sim \chi^{4}$ occurs. So far, we have implicitly assumed that every one of these solutions can be realized, in combination with an appropriate scaling function for $M^{2}(\chi)/k^{2}$ and other quantities. This is not guaranteed. The flow equation for $w=M^{2}(\chi)/(2k^{2})$ depends on $u=U(\chi)/k^{4}$. As a result, there will be a different scaling solution $w_{*}(y)$ for every $y_{V}$, if it exists at all. It is well conceivable that scaling solutions exist for a certain range of $y_{V}$, and not outside this range. It is possible that these restrictions are so strong, that out of the family of candidate scaling solutions a single one is selected, or even none exists.

In this respect we observe the existence of two classes of scaling solutions with a very different behavior of $u$ and $v$ for $\chi\raw\infty$. The first class consists of the single particular scaling solution for $y_{V}\raw\infty$. In this case the behavior $u=\bar{c}_{V}$, $U\sim k^{4}$ remains valid for all $\chi\raw\infty$. We call this solution the ``flat scaling solution''. For $M^{2}(\chi)\sim\chi^{2}$ the flat scaling solution implies $\lim_{\chi\raw\infty}v=0$. For all other candidate scaling solutions one has the asymptotic behavior $U\sim \chi^{2}$ and $v\raw 1$ for $\chi\raw\infty$. These solutions with a finite value of $y_{V}$ are called ``non-flat scaling solutions''. Imagine now that an accompanying scaling solution $w_{*}(y)$ exists for the non-flat scaling solutions only in a range of large $y$, but cannot be extended to $y\raw 0$. In this case the solutions with finite $y_{V}$ are not true scaling solutions associated to the UV-fixed point. The only remaining scaling solution is the flat scaling solution in this case. It would be singled out by the absence of scalar self interactions and mass terms for $\chi\raw\infty$. This is precisely the type of scaling solution found by the numerical solution of coupled quantum gravity flow equations in the investigation of dilaton gravity in refs.~\cite{PHW1,PHW2}.

In the limit $k\raw 0$ the potential simply vanishes for the flat scaling solution. For a realistic phenomenology one needs deviations from the scaling solution due to a relevant coupling. This relevant coupling can be the value of the flat potential for large $y$. For the flat scaling solution the value of $v$ is tiny for large $y$, decreasing for large $y$ as
\begin{equation}\label{eq:NSO1} 
v=\dfrac{2\bar{c}_{V}}{\xi y}\, .
\end{equation}
We can approximate $v=0$ in the threshold function for the gravitational coupling, such that flow equation for $U$ simply becomes
\begin{equation}\label{eq:NSO2} 
\p_{t}U=\dfrac{(N_{eff}+2)k^{4}}{32\pi^{2}}=4\bar{c}_{V}k^{4}\, ,
\end{equation}
with $N_{eff}$ accounting for the particle degrees of freedom that are effective at a given $k$ (for details see sect.~\ref{sec:Flow_Eff_Scalar_Pot_Quantum_Grav}). A particular solution is given by
\begin{equation}\label{eq:NSO3} 
U=\bar{c}_{V}\left (k^{4}+\bar{\mu}^{4}\right )\, ,
\end{equation}
where the first term corresponds to the flat scaling solution, and the second involves an intrinsic mass scale $\bar{\mu}$ reflecting the presence of a relevant parameter. For $k\raw 0$ the potential becomes for large $\chi$ simply constant,
\begin{equation}\label{eq:NSO4} 
\lim_{k\raw 0}U_{k}(\chi)=\bar{c}_{V}\bar{\mu}^{4}\, .
\end{equation}
In the large $\chi$ range the flow stops effectively at $k_{f}=\bar{\mu}$. The potential for $k\raw 0$ obtains from the scaling solution at $k=k_{f}$.

Other possibilities for the appearance of an intrinsic mass scale are conceivable as well. For example, the flow of the non-flat scaling solution may stop effectively at some scale $k=\bar{\mu}$. We can then use the scaling solution where $k$ is replaced by an intrinsic mass scale $\bar{\mu}$. In this case scale symmetry is explicitly broken by the scale $\bar{\mu}$. Nevertheless, for $\chi\raw 0$ and $\chi\raw\infty$ the coupling $\tilde{\lambda}$ becomes independent of the scale $\bar{\mu}$. At the corresponding UV- and IR-fixed points scale symmetry becomes again an exact symmetry, as depicted in fig.~\ref{fig:2}. Furthermore, for large $y_{V}$ we observe a large range in $y$ where $\tilde{\lambda}(y)$ is independent of $k$ and therefore of $\bar{\mu}$. We may associate this range with an approximate fixed point. Scale symmetry holds to a good approximation.

One may quantitatively measure the violation of scale symmetry in the gravitational sector by the scale dependence of the dimensionless ratio
\begin{equation}\label{eq:265} 
\tilde{\lambda}(\chi)=\dfrac{U(\chi)}{M^{4}(\chi)}=\dfrac{k^{2}v(\chi)}{2M^{2}(\chi)}\, .
\end{equation}
Replacing $k$ by $\bar{\mu}$ one finds for the range $y\lesssim y_{F}$ from \eq{237C}
\begin{equation}\label{eq:266} 
\tilde{\lambda}=\dfrac{v_{*}+\hat{v}_{0}\left (\dfrac{\chi^{2}}{\bar{\mu}^{2}}\right )^{2-\frac{A_{v}}{2}}}{4\bar{w}+2\xi\chi^{2}/\bar{\mu}^{2}}\, .
\end{equation}
Thus $\tilde{\lambda}$ becomes indeed independent of $\bar{\mu}$ in the limit $\chi\raw 0$, such that scale symmetry is realized,
\begin{equation}\label{eq:184A} 
\tilde{\lambda}=\dfrac{v_{*}}{4\bar{w}}=\dfrac{u_{*}}{4\bar{w}^{2}}\, .
\end{equation}
As $y=\chi^{2}/\bar{\mu}^{2}$ increases, the scale symmetry violation associated to a relevant parameter increases. In the range $y_{F}\lesssim y\lesssim y_{V}$, with
\begin{equation}\label{eq:267} 
\tilde{\lambda}=\dfrac{v_{min}\bar{\mu}^{2}}{2\xi\chi^{2}}+\dfrac{2\bar{c}_{V}\bar{\mu}^{4}}{\xi^{2}\chi^{4}}\, ,
\end{equation}
the relative change of $\tilde{\lambda}$ is substantial, but $\tilde{\lambda}$ itself decreases rapidly to tiny values as $\chi$ increases. For $y_{V}\lesssim y\lesssim y_{IR}$ the dimensionless ratio $\tilde{\lambda}$ stays at a constant value
\begin{equation}\label{eq:XBX} 
\tilde{\lambda}=\dfrac{\delta}{\xi^{2}}=\dfrac{\bar{c}_{V}}{\xi^{2}y_{V}^{2}}=\dfrac{v_{min}^{2}}{\bar{c}_{V}}\, .
\end{equation}
This can be a very small number for large $y_{V}$. For example, $y_{V}\approx 10^{60}$ implies $\tilde{\lambda}\approx 10^{-120}$. In the range $y>y_{V}$ scale symmetry is again realized to a good approximation. Finally, for $y\gtrsim y_{IR}$ one finds the flow towards the IR-fixed point at $\tilde{\lambda}=0$,
\begin{equation}\label{eq:269} 
\tilde{\lambda}=\dfrac{\bar{p}\bar{\mu}^{2}}{2\xi\chi^{2}}\, .
\end{equation}
We encounter three (approximate) fixed points for $\tilde{\lambda}$, given by \eqs{184A}{XBX} and $\tilde{\lambda}=0$. They are connected by two crossovers.

After Weyl scaling the cosmological constant in the Einstein frame is given by $\tilde{\lambda}M^{4}$. The flow of $\tilde{\lambda}$ towards zero is therefore of key importance for the understanding of the cosmological constant problem. So far, the theoretical understanding both of the possible scaling solutions and the deviations from the scaling solution by relevant parameters is still insufficient. Nevertheless, important qualitative insights are already available. The resulting consequences for cosmology will be discussed in sects.~\ref{sec:Cosm_Scaling_Sol}, \ref{sec:7.8}.

\subsection{Flow of effective scalar potential in quantum gravity}\label{sec:Flow_Eff_Scalar_Pot_Quantum_Grav} 

The effective potential for scalar fields plays a central role in particle physics and cosmology. For particle physics it determines the physics of electroweak symmetry breaking and the Higgs boson. Extensions of the standard model as grand unification need the effective potential for scalar fields as a central ingredient for the spontaneous breaking of the grand unified symmetry. In cosmology, both the physics of inflation and dynamical dark energy are determined by the scalar potential. We address here the functional renormalization group flow of the effective scalar potential $U(\varphi)$ including the effect of gravitational fluctuations.

The general flow generator $\tilde{\zeta}$ for the dependence of $U$ on the IR-cutoff scale $k$ can be split into contribution from the graviton ($\tilde{\pi}_{2}$), physical scalar fluctuations ($\tilde{\pi}_{s}$), a measure contribution (gauge fluctuations of the metric and ghosts) related to diffeomorphism symmetry ($\tilde{\eta}$), as well as further contributions from gauge bosons ($\tilde{\pi}_{g}$) and fermions ($\tilde{\pi}_{f}$),
\begin{equation}\label{eq:V1} 
\p_{t}U=k\p_{k}U=\tilde{\zeta}=\tilde{\pi}_{2}+\tilde{\pi}_{s}+\tilde{\eta}+\tilde{\pi}_{g}+\tilde{\pi}_{f}\, .
\end{equation}
We employ here a physical gauge fixing which acts only on the gauge fluctuations of the metric \cite{CWGIF} and a corresponding decomposition of the metric fluctuations into physical and gauge fluctuations \cite{CWFL3}. For simplicity, we also take the Litim infrared cutoff function \cite{LITC}.

Close to the UV-fixed point the dominant contribution to the field-dependence is generated by the graviton fluctuations. With the Litim cutoff the graviton contribution reads \cite{CWGFE}
\begin{equation}\label{eq:V2} 
\tilde{\pi}_{2}=\dfrac{5}{24\pi^{2}}\left (1-\dfrac{\eta_{g}}{8}\right )\dfrac{k^{4}}{1-v}\, ,
\end{equation}
with
\begin{equation}\label{eq:V3} 
v(\rho)=\dfrac{2U(\rho)}{F(\rho)k^{2}}\, .
\end{equation}
Here we employ the truncation
\begin{equation}\label{eq:V4} 
\Gamma_{k}=\int_{x}\,\sqrt{g}\left (U(\rho)-\dfrac{1}{2}F(\rho)R+\ldots\right )
\end{equation}
where $U$ and $F$ depend on appropriate invariants $\rho$ formed from scalar fields. Eq.~\eqref{eq:V2} corresponds to eq.~\eqref{eq:112} for
\begin{equation}\label{eq:V5} 
\ell_{0}(\tilde{w})=\dfrac{1}{2(1+\tilde{w})}\, .
\end{equation}
The field dependent Planck mass $M^{2}(\chi)$ is generalized to $F(\rho)$. We choose the IR-cutoff in the metric sector to be proportional to $F(\rho)$. For $k$-dependent $F(\rho)$ this entails a contribution to $\p_{t}R_{k}$, that we parametrize by
\begin{equation}\label{eq:V6} 
\eta_{g}(\rho)=-\p_{t}\ln\left (\dfrac{F(\rho)}{k^{2}}\right )=-\p_{t}\ln(w)\com w=\dfrac{F}{2k^{2}}\, .
\end{equation}
(For $F(\rho)$ independent of $k$ one has $\eta_{g}=2$, recovering eq.~\eqref{eq:204}.) Eq.~\eqref{eq:V2} also holds for more general truncations, provided we approximate the graviton propagator by eq.~\eqref{eq:250} with $V\raw U(\rho)$, $M^{2}\raw F(\rho)$.

The scalar contribution mixes the physical scalar fluctuation in the metric with fluctuations of scalars not associated to the metric. For the example of $N$ real scalars $\varphi_{a}$ with $O(N)$ symmetry one finds \cite{RPWY}
\begin{align}\label{eq:V7} 
\tilde{\pi}_{s}&=\dfrac{k^{4}}{24\pi^{2}}\biggl [\left (1-\dfrac{\eta_{g}}{8}\right )\left (1+u^{\prime}+2\tilde{\rho}u^{\prime\prime}\right )\nn\\
&\hphantom{=\dfrac{k^{4}}{24\pi^{2}}\biggl [}+\dfrac{3}{4}\left (1-\dfrac{\eta_{\phi}}{6}\right )\left (1-\dfrac{v}{4}\right )\biggl ]\nn\\
&\hphantom{=\dfrac{k^{4}}{24\pi^{2}}}\, \times \left [\left (1-\dfrac{v}{4}\right )\left (1+u^{\prime}+2\tilde{\rho}u^{\prime\prime}\right )+\dfrac{3\tilde{\rho}(u^{\prime})^{2}}{2w}\right ]^{-1}\nn\\
&\hphantom{=\dfrac{k^{4}}{24\pi^{2}}}\,+\dfrac{(N-1)\left (1-\frac{\eta_{\phi}}{6}\right )}{32\pi^{2}(1+u^{\prime})}\, .
\end{align}
Here we employ $\rho=\frac{1}{2}\sum_{a=1}^{N}\,\varphi_{a}^{2}$,
\begin{equation}\label{eq:V8} 
\tilde{\rho}=\dfrac{Z_{\varphi}\rho}{k^{2}}\com u(\tilde{\rho})=\dfrac{U(\rho)}{k^{4}}\, ,
\end{equation}
where $Z_{\varphi}$ is the scalar wave function renormalization (coefficient of kinetic term) and $\eta_{\phi}=-\p_{t}\ln(Z_{\phi})$. Primes denote derivatives with respect to $\tilde{\rho}$. The measure term takes the simple form \cite{RPWY}
\begin{equation}\label{eq:V9} 
\tilde{\eta}=-\dfrac{1}{8\pi^{2}}\, .
\end{equation}

The contribution of gauge boson fluctuations takes the form
\begin{equation}\label{eq:V10} 
\tilde{\pi}_{g}=\dfrac{k^{4}}{32\pi^{2}}\sum_{i=1}^{N_{B}}\, \left (\dfrac{3}{1+w_{i}^{(B)}}-1\right )\, .
\end{equation}
The sum extends over the different gauge bosons, with total number $N_{B}$. The first term arises from the physical fluctuations. The threshold function involves the masses of the gauge bosons $M_{i}(\phi)$ in presence of constant scalar fields $\phi$,
\begin{equation}\label{eq:V11} 
w_{i}^{(B)}=\dfrac{M_{i}^{2}(\phi)}{k^{2}}\, .
\end{equation}
These masses are proportional to the gauge couplings, $M_{i}^{2}=g_{i}^{2}a_{i}(\phi)$, with $a_{i}(\phi=0)=0$. The second contribution is the measure contribution associated to the local gauge symmetry, employing the physical Landau gauge. Finally, the fermion contribution is negative
\begin{equation}\label{eq:V12} 
\tilde{\pi}_{f}=-\dfrac{k^{4}}{16\pi^{2}}\sum_{j=1}^{N_{f}}\, \dfrac{1}{1+w_{j}^{(F)}}\, ,
\end{equation}
with
\begin{equation}\label{eq:V13} 
w_{j}^{(F)}=\dfrac{M_{j}^{2}(\phi)}{k^{2}}
\end{equation}
and fermion masses $M_{j}(\phi)$ involving Yukawa couplings. The sum extends over $N_{f}$ Weyl or Majorana fermions. In the limit of vanishing gauge and Yukawa couplings one has the simple relation
\begin{equation}\label{eq:V14} 
\tilde{\pi}_{g}+\tilde{\pi}_{f}=\dfrac{k^{4}}{16\pi^{2}}\left (N_{B}-N_{F}\right )\, .
\end{equation}

The flow near the UV-fixed point is dominated by the gravitational fluctuations if gauge, Yukawa and quartic scalar couplings are small close to the fixed point. This holds except for an overall additive constant that we denote by $V=U(\rho=0)$. In the limit of vanishing matter interactions and for $\tilde{\rho}(u^{\prime})^{2}\ll \tilde{F}$ one obtains
\begin{equation}\label{eq:V15} 
\p_{t}U=\dfrac{k^{4}}{24\pi^{2}}\left (1-\dfrac{\eta_{g}}{8}\right )\left (\dfrac{5}{1-v}+\dfrac{1}{1-\frac{v}{4}}\right )+\dfrac{(N-4)k^{4}}{32\pi^{2}}\, ,
\end{equation}
with
\begin{equation}\label{eq:V16} 
N=N_{S}+2N_{B}-2N_{F}\, ,
\end{equation}
and $N_{S}$ the total number of scalar fields, e.g. $N_{S}=N$ for $O(N)$-models. From eq.~\eqref{eq:V15} we can extract the quantitative values for the ansatz used in eq.~\eqref{eq:236}, namely
\begin{align}\label{eq:V17} 
a_{V}&=\dfrac{5}{96\pi^{2}}\left (1-\dfrac{\eta_{g}}{8}\right )=\dfrac{5f_{\eta}}{128\pi^{2}}\, ,\nn\\
b_{V}&=\dfrac{1}{128\pi^{2}}\left (N-4+\dfrac{4}{3}\left (1-\dfrac{\eta_{g}}{8}\right )\right )\, ,
\end{align}
where $\eta_{g}$ is taken as constant in the range of interest. We observe that the neglected mixing between the gravitational scalar and other scalars concerns only subleading degrees of freedom. Furthermore, it vanishes for
\begin{equation}\label{eq:V18} 
3\tilde{\rho}(u^{\prime})^{2}\ll 2w\left (1-\dfrac{v}{4}\right )\left (1+u^{\prime}+2\tilde{\rho}u^{\prime\prime}\right )\, .
\end{equation}
This is well obeyed for ranges of an almost flat potential, $u^{\prime}\raw 0$, as well as close to the origin, $\tilde{\rho}\raw 0$. For $u^{\prime}=\lambda_{c}\tilde{\rho}$, $w=\xi\tilde{\rho}/2$ the inequality \eqref{eq:V18} reads
\begin{equation}\label{eq:V19} 
3\lambda_{c}^{2}\tilde{\rho}^{3}\ll \xi\tilde{\rho}\left (1+3\lambda_{c}\tilde{\rho}\right )\, ,
\end{equation}
which is well obeyed in the range $\lambda_{c}\tilde{\rho}\ll\xi$ relevant for the discussion in sect.~\ref{sec:Crossover_DimLess_Fields}, \ref{sec:Quantitative_Scaling_solution}. Errors due to deviations from the approximation \eqref{eq:236} are small. 

The flow for the field -- dependence of the potential obtains by taking a derivative of the flow equation \eqref{eq:V15},
\begin{equation}\label{eq:V20} 
\p_{t}\left (\dfrac{\p U}{\p \rho}\right )=A\left [\dfrac{\p U}{\p \rho}-U\dfrac{\p \ln(F)}{\p \rho}\right ]\, ,
\end{equation}
with
\begin{equation}\label{eq:V21} 
A=\dfrac{\left (1-\dfrac{\eta_{g}}{8}\right )}{96\pi^{2}w}\left [\dfrac{20}{(1-v(\rho))^{2}}+\dfrac{1}{(1-\frac{v(\rho)}{4})^{2}}\right ]\, .
\end{equation}
(A similar relation holds replacing $\p/\p\rho$ by $\p/\p\varphi_{a}$.) For $v(\rho)>0$ the graviton contribution (first term in eq.~\eqref{eq:V18}) dominates the scalar contribution (second term in eq.~\eqref{eq:V18}) by more than a factor $20$. This justifies the graviton approximation advocated in ref.~\cite{CWGFE}. If one can further neglect the contribution from $\p\ln(F)/\p\rho$ the flow equation for the derivative is simply governed by the gravity induced anomalous dimension $A$
\begin{equation}\label{eq:V22} 
\p_{t}\left (\dfrac{\p U}{\p\rho}\right )=A\dfrac{\p U}{\p \rho}\, .
\end{equation}
In this limit any $\rho$-dependence of $U$ flows to zero for $A>0$. The fixed point potential becomes flat for $k\raw 0$.

The quantum gravity induced anomalous dimension $A$ is positive. Typically, it is found to be of the order one close to the UV-fixed point where $w(\rho)\gg1$ compensates the suppression factor $1/(2\pi^{2})$. This has an important consequence for quantum gravity predictions for the scalar sector. The second $\rho$-derivative of $U$ defines the quartic scalar coupling
\begin{equation}\label{eq:V23} 
\lambda=\dfrac{\p^{2}U}{\p\rho^{2}}\, .
\end{equation}
Taking the derivative of eq.~\eqref{eq:V20} yields the flow equation
\begin{equation}\label{eq:V24} 
\p_{t}\lambda=\beta_{\lambda}=A\lambda-C\, ,
\end{equation}
with
\begin{equation}\label{eq:V25} 
C=\left (A\dfrac{\p\ln(F)}{\p\rho}-\dfrac{\p A}{\p \rho}\right )\dfrac{\p U}{\p \rho}+AU\dfrac{\p^{2}\ln(F)}{\p\rho^{2}}\, .
\end{equation}
In the limit where the $\lambda$-dependence of $C$ is small eq.~\eqref{eq:V24} has a simple fixed point
\begin{equation}\label{eq:V26} 
\lambda_{*}=\dfrac{C}{A}\, .
\end{equation}
The critical exponent at the fixed point is negative
\begin{equation}\label{eq:V27} 
\theta_{\lambda}=-\dfrac{\p\beta_{\lambda}}{\p\lambda}\bigl |_{\lambda_{*}}=-A\, ,
\end{equation}
such that a quartic coupling is an irrelevant parameter.

This leads to a central quantum gravity prediction for particle physics \cite{SW,RPWY}. If quantum gravity is asymptotically safe, all quartic scalar couplings vanish at the UV-fixed point. Indeed, the computation of the gravitational contribution to the flow of a scalar coupling does not use any particular choice of scalar fields or particular quartic coupling. Models with many scalar fields as grand unified theories exhibit many different quartic scalar couplings. The gravity induced anomalous dimension $A$ is universal for the flow of all quartic couplings $\lambda_{i}$
\begin{equation}\label{eq:V28} 
\p_{t}\lambda_{i}=A\lambda_{i}-C_{i}\, .
\end{equation}

A more complete discussion does not change this simple finding. This is based on the observation that typically $C_{i}$ are small quantities, while $A$ is of the order one. The fixed point values $\lambda_{i*}=C_{i}/A$ therefore occur for very small values $\lambda_{i*}$. In principle, one needs the full stability matrix $T_{ij}$ in eq.~\eqref{eq:45}. For the sector of quartic couplings the stability matrix is to a good approximation diagonal, with diagonal elements given by $-A$,
\begin{equation}\label{eq:V29} 
T_{ij}=-A\delta_{ij}-\dfrac{\p A}{\p g_{j}}\lambda_{i*}+\dfrac{\p C_{i}}{\p g_{j}}\, ,
\end{equation}
with arbitrary other couplings $g_{j}$ (for example including $\xi=\p F/\p \rho$) and the derivative in eq.~\eqref{eq:V29} evaluated at the fixed point. The second term on the r.h.s. is small since $\lambda_{i*}$ is small, and third because $C_{i}$ and $\p C_{i}/\p g_{j}$ are small. Also the dependence of the flow of $g_{j}$ on $\lambda_{i}$ (reflected by $T_{ji}$) is typically small. Small off-diagonal terms contribute then only quadratically to the eigenvalues, such that the leading correction may arise from $\p C_{i}/\p\lambda_{i}$ (no sum over $i$)
\begin{equation}\label{eq:V30} 
\theta_{i}=-A+\dfrac{\p C_{i}}{\p\lambda_{i}}\, .
\end{equation}
This small correction does not affect the conclusion that all quartic scalar couplings are irrelevant. Their fixed point values are typically very small, and may be set to zero to a good approximation.

\subsection{Quantum gravity predictions for the Higgs \\
potential}\label{sec:6.17} 

As discussed in sect.~\ref{sec:Predictivity_Sev_Fixed_Point}, renormalizable couplings of the standard model that correspond to irrelevant couplings at the UV-fixed point can be predicted. The quartic coupling characterizing the self-interaction of the Higgs scalar is irrelevant. Its fixed point value sets the initial value for the flow in the effective low energy theory. Following the flow in the low energy theory down to the Fermi scale yields the quartic coupling at the Fermi scalar which determines the mass of the Higgs boson
\begin{equation}\label{eq:W1} 
M_{H}=\sqrt{2\lambda_{H}}\varphi_{0}\, ,
\end{equation}
with $\varphi=\sqrt{\rho_{0}}$ the Fermi scale. With the assumption that the low energy flow of $\lambda_{H}$ is well approximated by the standard model the mass of the Higgs boson mass has been predicted \cite{SW} to be $126$GeV, with a few GeV uncertainty. The range of this prediction agrees well with the experimentally observed value $M_{H}=125.15$GeV \cite{AAD1,CHAT1}.

\subsubsection{Quartic scalar coupling}\label{sec:Quartic_Scalar_Coupling}

In the flow equation~\eqref{eq:61} we can use $A_{H}=A$ as given by eq.~\eqref{eq:V21}. We approximate
\begin{equation}\label{eq:W1A} 
F=M^{2}(\chi)+\xi_{H}h^{\dagger}h+\tilde{M}^{2}k^{2}\, ,
\end{equation}
where the ``low energy Planck mass'' $M(\chi)$ may depend on the value of a scalar singlet field $\chi$. The last term in eq.~\eqref{eq:W1A} reflects the UV-fixed point, $F/k^{2}\raw \tilde{M}^{2}$ for $k\raw\infty$ (at fixed $\chi$). We neglect the dependence of $v$ and $\eta_{g}$ on $h^{\dagger}h$. The flow equation for $\lambda_{H}$ is then approximated by
\begin{equation}\label{eq:W2} 
\p_{t}\lambda_{H}=A\lambda_{H}-C_{H}=A\lambda_{H}-C_{H}^{(gr)}-C_{H}^{(p)}\, ,
\end{equation}
with
\begin{align}\label{eq:W3} 
A&=\dfrac{k^{2}}{48\pi^{2}(\tilde{M}_{*}^{2}k^{2}+M^{2})}\left (1-\dfrac{M^{2}}{4(M^{2}+\tilde{M}_{*}^{2}k^{2})}\right )\nn\\
&\quad \times \left (\dfrac{20}{(1-v)^{2}}+\dfrac{1}{(1-v/4)^{2}}\right )\, ,
\end{align}
and
\begin{equation}\label{eq:W4} 
C_{H}^{(gr)}=\dfrac{2A\xi_{H}}{F}m_{H}^{2}-\dfrac{\xi_{H}^{2}Av}{4w}\, ,
\end{equation}
where the mass term is given by
\begin{equation}\label{eq:W5} 
m_{H}^{2}=\dfrac{\p U}{\p (h^{\dagger}h)}\, .
\end{equation}

The contribution from particle fluctuations (fermions, scalars, gauge bosons) arises from $\tilde{\pi}_{s}^{\prime}+\tilde{\pi}_{f}+\tilde{\pi}_{g}$, where for $\tilde{\pi}_{s}^{\prime}$ the contribution of the scalar metric fluctuation is subtracted from $\tilde{\pi}_{s}$. In the one loop approximation $C_{H}^{(p)}$ is given by eq.~\eqref{eq:62}. For the standard model alone (without additional contributions from metric and other particle fluctuations) one has
\begin{align}\label{eq:W6} 
C_{H}^{(p)}&=-\beta_{\lambda}^{(SM)}\\
&\approx-\dfrac{1}{16\pi^{2}}\biggl \{12\lambda_{H}^{2}+12y_{t}^{2}\lambda_{H}-12y_{t}^{4}+\dfrac{9}{4}g_{2}^{2}\nn\\
&\quad +\dfrac{9}{10}g_{2}^{2}g_{1}^{2}+\dfrac{27}{100}g_{1}^{4}-\left (9g_{2}^{2}+\dfrac{9}{5}g_{1}^{2}\right )\lambda_{H}\biggl \}\nn\, ,
\end{align}
where $\beta_{\lambda}^{(SM)}$ is the $\beta$-function of the standard model and the approximate value corresponds to the one-loop approximation where only the top-quark Yukawa coupling $y_{t}$ is taken into account, with $g_{2}$ and $g_{1}$ the gauge couplings of $SU(2)$ and $U(1)$. (We have included in $C_{H}^{(p)}$ small contributions $\sim\lambda_{H}$ to the anomalous dimension from particle loops.)

The $k$-dependence of $F$ divides the flow into two regimes. For the high energy regime $k\gg k_{t}$ the term $\tilde{M}_{*}^{2}k^{2}$ dominates $F$, while for the low energy regime $k\ll k_{t}$ the coefficient $F$ becomes independent of $k$. The transition scale is determined by
\begin{equation}\label{eq:W7} 
k_{t}^{2}=\dfrac{M^{2}}{\tilde{M}_{*}^{2}}\com w=\dfrac{\tilde{M}_{*}^{2}}{2}\left (\dfrac{k_{t}^{2}}{k^{2}}+1\right )\, .
\end{equation}
If $M^{2}$ is dominantly a field independent constant the transition scale indicates a crossover where the UV-fixed point region is left and scale invariance is no longer valid. In contrast, for $M^{2}=\chi^{2}$ and has $k_{t}\sim \chi$. Scale invariance is not broken.

For the high energy regime one has $\eta_{g}=0$ and $A$ becomes independent of $k$. While the dependence of $v$ on $h^{\dagger}h$ can be neglected, $v$ still may depend on $\chi$. For the Higgs potential one is interested in the value of $\chi$ relevant for present cosmology. In this case $M(\chi)$ equals the observed Planck mass. For a positive effective particle number $N$ one finds $v>0$ in the high energy regime. A discussion of the value of $v$ at the fixed point and a quantitative estimate of $A$ in dependence on $\tilde{M}_{*}^{2}$ can be found in ref.~\cite{RPWY}. In the high energy regime the ratio $m_{H}^{2}/F$ is typically very small (see below). The second term in eq.~\eqref{eq:W4} dominates. It is negative for $v>0$. The particle contribution $C_{H}^{(p)}$ is negative if gauge bosons and scalars dominate, and positive if fermions dominate. Gauge couplings $g$ and Yukawa couplings $y$ may either be irrelevant and determined by a fixed point with nonzero $g_{*}^{2}$ and $y_{*}$ \cite{EHW}. As an alternative, the matter sector may be asymptotically free, with $g$ and $y_{t}$ being relevant couplings. In this case $g^{2}$ and $y_{t}$ increase with $k$. Given the small observed values, this increase is, however, rather slow. To a good approximation one finds a ``sliding partial fixed point'' for $\lambda_{H}$, where $C_{H}$ in eq.~\eqref{eq:V26} depends weakly on $k$ through the $k$-dependence of $g$ and $y$.

For $C_{H}<0$ the fixed point occurs for a negative value $\lambda_{H*}$. For small enough $|\lambda_{H*}|$ this does not indicate an instability of the potential. Indeed, a Taylor expansion in powers of $h^{\dagger}h$ is only valid for small enough $h^{\dagger}h$. For large $h^{\dagger}h$ the fixed point potential does no behave as $\lambda_{H}(h^{\dagger}h)^{2}/2$. It rather saturates at a field independent value, $U\sim k^{4}$. Indeed, for $h^{\dagger}h\raw \infty$ some of the gauge bosons and fermions simply decouple -- they no longer contribute to $\tilde{\pi}_{g}+\tilde{\pi}_{f}$ because of the suppression by their mass terms. In this limit the structure of the flow equation is again given by eq.~\eqref{eq:V15}, with $N$ replaced by an effective number of particles that remain light for $h^{\dagger}h\raw\infty$.

For the low energy regime the contribution of the metric fluctuations to the flow of $U$ can be neglected except for a field-independent constant. The flow of $\lambda_{H}$ is given by the $\beta$-function of the low energy effective theory. Since the non-gravitational couplings are small, the latter can be computed in perturbation theory. As a simple approximation we consider a scenario where the flow for $k<k_{t}$ is given by the flow in the standard model. The ``initial value'' $\lambda_{H}(k_{t})$ is set by the fixed point of the high energy flow, e.g. $\lambda_{H}(k_{t})=C_{H}/A$. Corrections to this scenario from a better resolution of the transition from the high- to the low-energy regime, including a possible grand unification, will be treated as ``threshold effects'' and discussed later.

The standard model flow of $\lambda_{H}$ is strongly influenced by the flow of the top quark Yukawa coupling, which obeys
\begin{equation}\label{eq:W8} 
\p_{t}y_{t}=\dfrac{1}{16\pi^{2}}\biggl \{\dfrac{9}{2}y_{t}^{3}-\left (8g_{3}^{2}+\dfrac{9}{4}g_{2}^{2}+\dfrac{17}{20}g_{1}^{2}\right )y_{t}\biggl \}\, ,
\end{equation}
with $g_{3}$, $g_{2}$, $g_{1}$ the gauge couplings of $SU(3)$, $SU(2)$ and $U(1)$, respectively. The ratio $\lambda_{H}/y_{t}^{2}$ is attracted towards an approximate partial IR-fixed point \cite{CWMHB,CWPFP,SCHW}, according to
\begin{align}\label{eq:W9} 
\p_{t}\left (\dfrac{\lambda_{H}}{y_{t}^{2}}\right )&=\dfrac{3y_{t}^{2}}{4\pi^{2}}\biggl \{\left (\dfrac{\lambda}{y_{t}^{2}}\right )^{2}+\dfrac{1}{4}\left (\dfrac{\lambda}{y_{t}^{2}}\right )-1\nn\\
&\quad \hphantom{\dfrac{3y_{t}^{2}}{4\pi^{2}}\biggl \{}+\delta\left (\dfrac{\lambda}{y_{t}^{2}}\right )+\eta\biggl \}\, ,
\end{align}
with
\begin{align}\label{eq:W10} 
\delta&=\left (\dfrac{4}{3}g_{3}^{2}-\dfrac{3}{8}g_{2}^{2}-\dfrac{1}{120}g_{1}^{2}\right )/y_{t}^{2}\, ,\nn\\
\eta&=\left (\dfrac{3}{16}g_{2}^{4}+\dfrac{3}{40}g_{2}^{2}g_{1}^{2}+\dfrac{9}{400}g_{1}^{4}\right )/y_{t}^{4}\, .
\end{align}
In practice, $\eta=0.067$ is small and $\delta\approx 1.9$ may be approximated by a constant, reflecting the slow evolution of $y_{t}^{2}/g_{3}^{2}$. For constant $\delta$ and $\eta$ the fixed point occurs for
\begin{equation}\label{eq:W11} 
\left (\dfrac{\lambda_{H}}{y_{t}^{2}}\right )_{*}=\dfrac{1}{8}\biggl \{\sqrt{65+8\delta+16\delta^{2}-64\eta}-1-4\delta\biggl \}=x_{0}
\end{equation}
recovering the result of ref.~\cite{CWMHB} for $\delta=\eta=0$. Inserting for $\delta$ the value at the Fermi scale, $\delta\approx 1.9$, one finds $x_{0}\approx 0.37$. In this approximation the fixed point ratio between Higgs scalar mass and top quark mass
\begin{equation}\label{eq:W12} 
\dfrac{M_{H}}{m_{t}}=\sqrt{2x_{0}}=0.86
\end{equation}
is not very for from the observed value, $M_{H}/m_{t}=0.722$.

For $\eta=0$ and constant $\delta$ the flow equation for
\begin{equation}\label{eq:W13} 
x=\dfrac{\lambda_{H}}{y_{t}^{2}}
\end{equation}
takes the form
\begin{equation}\label{eq:W14} 
\p_{s}x=(x-x_{0})(x-x_{1})\, ,
\end{equation}
with variable $s$ defined by
\begin{equation}\label{eq:W15} 
\dfrac{\dif s}{\dif t}=\dfrac{3y_{t}^{2}}{4\pi^{2}}\, ,
\end{equation}
where
\begin{equation}\label{eq:W16} 
x_{0}+x_{1}=-\dfrac{1}{4}\left (1+4\delta\right )\approx -2.15\, .
\end{equation}
The solution is similar to eqs.~\eqref{eq:64}, \eqref{eq:65}, with $t$ replaced by $s$. Any interval of high energy couplings, say $x(k_{t})$ between zero and infinity, is mapped at the Fermi scale to a corresponding infrared interval \cite{CWMHB}. In our case we start with $x(k_{t})\approx 0$. At the Fermi scale, one therefore predicts $x$ and the corresponding ratio $M_{H}/m_{t}$ close to the lower bound of the IR-interval. This has led to the prediction $M_{H}=126$GeV, with a few GeV uncertainty. The uncertainty is due to threshold effects, higher loops, and the precise relation between the observed masses $m_{t}$ and $M_{H}$ and the ``running masses'' at the Fermi scale.

\subsubsection{Prediction for the mass of the top quark}

The flow equation for the standard model have been computed to three loop order. We may define a benchmark solution $x_{B}(k)$ by the solution of the three-loop $\beta$-function with initial condition of vanishing $\lambda_{H}$ at the Planck scale
\begin{equation}\label{eq:W17} 
x_{B}(k=M)=0\, .
\end{equation}
The result for the ratio of pole masses is \cite{BKKS,DVEE,BDG}
\begin{equation}\label{eq:W18} 
\dfrac{M_{H}}{m_{t}}=0.730\, .
\end{equation}
For the measured value $M_{H}=125.15\,$GeV the benchmark solution predicts for the top quark a pole mass
\begin{equation}\label{eq:W19} 
m_{t}=171.5\, \mathrm{GeV}\, .
\end{equation}
The translation of this mass to the top quark mass as measured by experiment, $m_{t}\approx 173\,$GeV, may still involve a few GeV uncertainty \cite{ADM}.

By fixing the value of $\lambda_{H}(k_{t})$ quantum gravity predicts the ratio of Higgs boson mass over top quark mass. At present, the mass of the Higgs scalar, $M_{H}=125.15\,$GeV, is better known than the mass of the top quark. We may therefore turn the original prediction \cite{SW} of the Higgs scalar mass around and predict the value of the mass of the top quark, given the measured value of $M_{H}$. The prediction will to some extent depend on the assumptions about particles with mass smaller than $k_{t}$. In order to pin down uncertainties we will  employ the benchmark solution and consider small deviations from it.

We linearize in small deviations from the benchmark solution
\begin{equation}\label{eq:W20} 
y=x-x_{B}\com \p_{s}y=A_{y}y\, .
\end{equation}
The anomalous dimension
\begin{equation}\label{eq:W21} 
A_{y}=2x_{B}-x_{0}-x_{1}=2x_{B}+\dfrac{1}{4}+\delta
\end{equation}
is positive, $A_{y}>0$. As a result, neighboring solutions are attracted towards the benchmark solution as $k$ is lowered. For an estimate how a small deviation from the benchmark solution at some given scale $k_{in}$ is reflected in a change of $x$ at the Fermi scale, we need neither a high precision for $A_{y}$ nor non-linear effects in $\delta$. For this estimate no full three-loop running is necessary. It is sufficient to use in eq.~\eqref{eq:W21} for $x_{B}(k)$ the one-loop result. This provides for a simple method of computing changes in the prediction for $m_{t}$ due to threshold effects, e.g. values of $\lambda_{H}(k_{t})$ different from zero, values of $k_{t}$ different from $M$, and the precise crossing of the threshold near $k_{t}$. Also effects of possible intermediate scales between $k_{t}$ and the Fermi scale can be estimated in this way.

The benchmark solution obeys in one loop order eq.~\eqref{eq:W14}, with solution
\begin{equation}\label{eq:W22} 
x_{B}(k)=\dfrac{x_{0}+fx_{1}}{1+f}\, ,
\end{equation}
where ($|x_{1}|=-x_{1}$)
\begin{equation}\label{eq:W23} 
f(k)=\dfrac{x_{0}}{|x_{1}|}\exp\left \{(x-x_{0})\Delta s(k)\right \}
\end{equation}
and
\begin{equation}\label{eq:W24} 
\Delta s(k)=s(k)-s(k_{t})=-\dfrac{3}{4\pi^{2}}\int_{\ln(k/k_{t})}^{0}\dif t\, y_{t}^{2}(t)\, .
\end{equation}
This yields
\begin{equation}\label{eq:W25} 
A_{y}=(x_{0}-x_{1})\dfrac{1-f}{1+f}=(x_{0}-x_{1})g\, .
\end{equation}
The solution of eq.~\eqref{eq:W20} takes the form
\begin{equation}\label{eq:W26} 
y(k)=y(k_{in})\left (\dfrac{k}{k_{in}}\right )^{(x_{0}-x_{1})\bar{g}}\, ,
\end{equation}
with average $\bar{g}$
\begin{align}\label{eq:W27} 
\bar{g}&=\dfrac{1}{\ln\bigl (\frac{k_{in}}{k}\bigl )}\int_{\ln(k/k_{in})}^{0}\dif t\, g(t)\nn\\
&=\dfrac{1}{\ln\bigl (\frac{k_{in}}{k}\bigl )}\int_{\ln(k/k_{in})}^{0}\dif t\,\dfrac{1-f(t)}{1+f(t)}\, .
\end{align}
With positive $\bar{g}$, and positive $x_{0}-x_{1}$,
\begin{equation}\label{eq:W28} 
x_{0}-x_{1}=\dfrac{1}{4}\sqrt{65+8\delta+16\delta^{2}-64\eta}\approx 2.89\, ,
\end{equation}
the deviation from the benchmark solution $y(k)$ is smaller than the initial deviation $y(k_{in})$ if $k<k_{in}$.

We are interested in the change of masses at the Fermi scale and therefore in $y(\varphi_{0})$. We define
\begin{equation}\label{eq:W29} 
y(\varphi_{0})=\bar{R}(k_{in})y(k_{in})\, ,
\end{equation}
with
\begin{equation}\label{eq:W30} 
\bar{R}(k_{in})=\left (\dfrac{k_{in}}{\varphi_{0}}\right )^{-(x_{0}-x_{1})\bar{g}(k_{in})}\, ,
\end{equation}
and
\begin{equation}\label{eq:W31} 
\bar{g}(k_{in})=\dfrac{1}{\ln(k_{in}/\varphi_{0})}\int_{\ln(\varphi_{0}/k_{in})}^{0}\dif t\, \dfrac{1-f(t)}{1+f(t)}\, .
\end{equation}
For a numerical value one has $\overline{R}(M)=0.032$.

A change $\Delta\lambda_{H}$ at $k_{in}$ results for the observed top quark mass in a change
\begin{equation}\label{eq:W32} 
\Delta m_{t}=-R(k_{in})\dfrac{\varphi_{0}^{2}}{M_{H}^{2}}m_{t}\Delta \lambda(k_{in})
\end{equation}
with ``reduction factor''
\begin{equation}\label{eq:W33} 
R(k_{in})=\bar{R}(k_{in})\dfrac{y_{t}^{2}(\varphi_{0})}{y_{t}^{2}(k_{in})}\approx 0.22\, ,
\end{equation}
where the quantitative value applies to $k_{in}=M$. An increase of the top quark mass by $1$GeV needs lowering $\lambda_{H}(k_{in})$ as compared to the benchmark solution by
\begin{equation}\label{eq:W34} 
\Delta\lambda_{H}(k_{in})=-\left (\dfrac{1\mathrm{GeV}}{m_{t}}\right )\left (\dfrac{M_{H}^{2}}{\varphi_{0}^{2}}\right )\dfrac{1}{R(k_{in})}=-0.014\, .
\end{equation}
For the last quantitative value we have taken $k_{in}=M$.

At the UV-fixed point one has for the standard model
\begin{equation}\label{eq:QC1} 
\lambda_{H*}=\dfrac{C_{H}}{A}\approx\dfrac{3y_{t}^{2}}{4\pi^{2}A}-\dfrac{171\alpha^{2}}{50A}-\dfrac{\xi_{H}^{2}v}{4w}\, ,
\end{equation}
where we use in eq.~\eqref{eq:W6} $\lambda_{H}\approx0$ and $g_{2}^{2}\approx g_{1}^{2}=4\pi\alpha$ with $\alpha\approx 1/40$. The first two terms almost cancel. For the gravitational contribution \eqref{eq:W4} we neglect the first term since $m_{H}^{2}$ is very small close to the vacuum electroweak phase transition. The factor $w^{-1}=2/\tilde{M}^{2}_{*}(\chi)$ measures the strength of the gravitational interaction near the UV-fixed point. For not too small $\xi_{H}$ one predicts a negative value of $\lambda_{H*}$. Additional gauge bosons, say within a grand unified theory, go in the same direction since the coefficient of $\alpha^{2}$ in eq.~\eqref{eq:QC1} is enhanced. It is possible that a slightly negative value of $\lambda_{H*}$ around $-0.02$ is sufficient for shifting the prediction of the top quark mass by one or two GeV, according to eq.~\eqref{eq:W34}.

We recall in this context that a polynomial expansion of $U$ in powers of $h^{\dagger}h$ is only valid in a small range of $h^{\dagger}h/F$. A negative value of $\lambda_{H}$ at $h=0$ does not imply increasingly negative values of $U$ as $h^{\dagger}h\raw\infty$. Typically $U$ settles to a finite value for large $h^{\dagger}h$. It is conceivable that the effective UV-potential for $k\raw\infty$ has a minimum for $h^{\dagger}h\neq 0$. 

For lower $k$ the the last term in eq.~\eqref{eq:QC1} vanishes due to the decoupling of the metric fluctuations. The increasing value of the top quark Yukawa coupling will then drive $\lambda_{H}$ to positive values, according to the discussion in sect.~\ref{sec:Quartic_Scalar_Coupling}. There is typically a unique minimum of $U(h^{\dagger}h)$ in the limit $k\raw 0$. No issue of metastability of the standard model vacuum arises in this case. There could be interesting consequences for early cosmology, however. The potential in the Einstein frame, $V_{E}/M^{4}=U/F^{2}$ may reveal an interesting behavior in the Higgs direction.

Besides the fixed point in the quartic scalar coupling $\lambda_{H}$ the UV-fixed point may predict other relations between the couplings of the standard model. For example, if the ratio between the top-Yukawa coupling $y_{t}$ and the strong gauge coupling $g_{3}$ turns out to be an irrelevant parameter, asymptotic safety predicts at $k_{t}$ a value close to the fixed point value
\begin{equation}\label{eq:QC2} 
\dfrac{y_{t}^{2}(k_{t})}{g_{3}^{2}(k_{t})}\approx \dfrac{y_{t*}^{2}}{g_{3*}^{2}}\, .
\end{equation}
This ratio can then be extrapolated to the Fermi scale by the running in the low energy effective theory. In this case not only the ratio between the masses of Higgs boson and top quark can be predicted, but in addition also the ratio between top quark mass and W-boson mass. First positive indications in this direction have been found in ref.~\cite{AEHE1,AEHE2}.

\subsubsection{Prediction for the gauge hierarchy?}\label{sec:Pred_Gauge_Hier} 

A second important issue concerns the scalar mass term or, more precisely, the deviation $\delta$ from the critical surface of the second order vacuum electroweak phase transition discussed in sect.~\ref{sec:Scale_Sym_Vac_Electroweak_PT}. The gravitational contribution to the anomalous dimension \eqref{eq:PS10} is given by eq.~\eqref{eq:V21}, \eqref{eq:W3}. This dominates over the particle physics contribution $A_{(p)}$, as given by eq.~\eqref{eq:43} for the standard model or generalizations with an extended particle content. For simplicity, we focus on the gravitational contribution \eqref{eq:V21} for the flow in the high energy domain \cite{CWMY,ODYA}. (One may add $A_{(p)}$ without qualitative changes.) The flow of the dimensionless parameter $\gamma=\delta/k^{2}$ obeys
\begin{equation}\label{eq:CC1} 
\p_{t}\gamma=(A-2)\gamma\, .
\end{equation}
Comparison with eq.~\eqref{eq:V21} ($\tilde{F}=2w$) determines the parameter $b$ in eq.~\eqref{eq:FD2} as
\begin{equation}\label{eq:6.278A} 
b=Aw=\dfrac{1}{96\pi^{2}}\left (1-\dfrac{\eta_{g}}{8}\right )\left (\dfrac{20}{(1-v)^{2}}+\dfrac{1}{(1-v/4)^{2}}\right )\, .
\end{equation}

The coupling $\gamma$ is a relevant parameter for $A<2$, and irrelevant for $A>2$. For $A<2$ the Fermi scale $\varphi_{0}$ cannot be predicted. The ratio $\varphi_{0}/M$ is a free parameter whose value is needed for the specification of the model. In the limit where the vacuum electroweak phase transition can be taken as second order there is no constraint on $\varphi_{0}/M$. Deviations from this situation through the running strong gauge coupling place a lower bound on $\varphi_{0}/M$, as discussed in sect.~\ref{sec:Particle_Scale_Sym}. All values above this lowed bound are natural in the sense that small values are protected by scale symmetry.

On the other hand, for $A>2$ the deviation from the phase transition $\gamma$ is an irrelevant parameter. According to the general discussion in sect.~\ref{sec:Networks_FP_Crossover}, its value at the fixed point and at the transition scale $k_{t}$ can be predicted. It is given by the fixed point $\gamma_{*}=0$. An irrelevant coupling $\gamma$ has striking consequences. A model with $A>2$ predicts that nature is located exactly on the critical surface of the vacuum electroweak phase transition \cite{CWPFP,CWQR,CWMY,RPWY}. In the limit of a second order transition it predicts a vanishing Fermi scale $\varphi_{0}=0$. This is an example of self-organized criticality \cite{BORW}. This important property follows from the basic fact that the flow of couplings can never leave a critical surface. The structure of the flow generator \eqref{eq:CC1} has always a zero for $\gamma=0$. Only the value of $A$ depends on the flow of the other dimensionless couplings. If $\gamma$ is driven to zero by the short-distance fluctuations, it stays at zero for all $k$.

On the one hand, models with $A>2$ offer the intriguing possibility of an understanding of the gauge hierarchy by predicting $\varepsilon=\varphi_{0}^{2}/M^{2}=0$ in a first approximation. On the other hand, this prediction may be ``too good'' -- the predicted parameter $\varepsilon$ may turn out to be more than three orders of magnitude smaller than the observed value $\varepsilon\approx 5\cdot 10^{-33}$. Beyond the approximation of a second order phase transition the value of $\varepsilon$ is predicted to be at the lower bound for $\varphi_{0}$ permitted by the flowing couplings of the low energy theory. A simple estimate of non-perturbative QCD-effects situates this bound below $100$MeV. In this case an UV-fixed point with $A>2$ is not compatible with the standard model as an effective low energy theory. Possible solutions of this dilemma could be at both ends of the flow \cite{CWMY}. Either the range of $k$ for which $A>2$ holds is not valid to arbitrarily short distances. The form of quantum gravity discussed here may be valid far beyond $M$, but nevertheless be replaced by another short distance theory for $k$ many orders of magnitude above $M$. As a low energy alternative, the low energy effective theory may contain particles beyond the standard model with masses around the Fermi scale. Scale symmetry breaking by running couplings in this extended sector could lead to the observed Fermi scale.

The value of $A$ in eq.~\eqref{eq:V21} depends on the particle content of a microscopic model, as specified by $N$, and on the value of $\tilde{M}_{*}^{2}$ at the fixed point. Quantitative relations can be found in ref.~\cite{RPWY}. The dependence of $\tilde{M}_{*}^{2}$ on the particle content is needed in order to assess for which models $A>2$ holds, and for which ones not.

\subsubsection{Anomalous mass dimension}\label{sec:6.17.4} 

For a more precise understanding of the flow of $\gamma$ we follow the flow of the potential derivative (in the limit where $\tilde{\eta}_{g}=0$ and $\p F/\p\ln(\rho)=0$)
\begin{equation}\label{eq:AM1} 
\p_{t}\dfrac{\p U}{\p\rho}=A\dfrac{\p U}{\p\rho} +\dfrac{3k^{2}}{8\pi^{2}}\left (1+\dfrac{m_{t}^{2}}{k^{2}}\right )^{-2}y_{t}^{2}\, .
\end{equation}
Here we include besides the gravitational contribution the one from the top quark, with squared mass $m_{t}^{2}=y_{t}^{2}\rho$ and $\rho=h^{\dagger}h$. The top contribution arises from the derivative of eq.~\eqref{eq:V12}, with a factor $3$ from three colors and a factor $2$ since the top quark is a Dirac spinor. Gauge bosons and scalars can be added similarly. They do not change the qualitative picture. We define the renormalized scalar mass term
\begin{equation}\label{eq:AM2} 
m_{R}^{2}=Z^{-1}_{H}\dfrac{\p U}{\p \rho}\biggl |_{\rho=0}\, ,
\end{equation}
with $Z_{H}$ the wave function renormalization multiplying the kinetic term $\cL_{kin}=Z_{H}\p^{\mu}h^{\dagger}\p_{\mu}h$ and scalar anomalous dimension
\begin{equation}\label{eq:AM3} 
\eta_{H}=-\p_{t}\ln(Z_{H})\, .
\end{equation}

From eq.~\eqref{eq:AM2} one infers
\begin{equation}\label{eq:AM4} 
\p_{t}m_{R}^{2}=(A+\eta_{H})m_{R}^{2}+\dfrac{3k^{2}y_{t}^{2}}{8\pi^{2}}\, .
\end{equation}
Taking a further $\rho$-derivative and defining the renormalized quartic Higgs coupling
\begin{equation}\label{eq:AM5} 
\lambda_{H}=Z_{H}^{-2}\dfrac{\p^{2}U}{\p\rho^{2}}\biggl |_{\rho=0}
\end{equation}
yields the corresponding pieces in the flow equation \eqref{eq:W2}, \eqref{eq:W6},
\begin{equation}\label{eq:AM6} 
\p_{t}\lambda_{H}=(A+2\eta_{H})\lambda_{H}-\dfrac{3y_{t}^{4}}{4\pi^{2}}\, .
\end{equation}

In the limit of constant $y_{t}^{2}$ the general structure of the flow equation \eqref{eq:5.9} is directly visible, with $B=3y_{t}^{2}/(8\pi^{2})$. The critical trajectory obeys
\begin{equation}\label{eq:6.286A} 
m_{R*}^{2}(k)=\dfrac{3y_{t}^{2}k^{2}}{8\pi^{2}(2-A-\eta_{H})}\, .
\end{equation}
With $m_{R}^{2}=m_{R*}^{2}+\delta$, $\gamma=\delta/k^{2}$ one finds indeed
\begin{equation}\label{eq:6.286B} 
\p_{t}\delta = (A+\eta_{H})\delta\com \p_{t}\gamma=(A+\eta_{H}-2)\gamma\, .
\end{equation}
The total anomalous mass dimension $A+\eta_{H}$ includes besides the gravitational contribution a contribution $\eta_{H}$ from the scalar wave function renormalization. (In eq.~\eqref{eq:5.9} we have collected in $A$ all contributions to the anomalous mass dimension, in contrast to eq.~\eqref{eq:AM4} where $A$ only denotes the gravitational contribution.) For a $k$-dependent $y_{t}^{2}$ the critical trajectory is somewhat modified, while the anomalous mass dimension $A+\eta_{H}$ remains the same.

An additional contribution to the anomalous mass dimension arises from the fluctuations of the Higgs doublet. In the limit \eqref{eq:V18} the scalar contribution \eqref{eq:V7} is composed of a part from  the scalar fluctuations in the metric, that is incorporated in the metric contribution \eqref{eq:V21}, and a standard scalar contribution
\begin{align}\label{eq:AM7} 
\p_{t}U&=\dfrac{1}{32\pi^{2}}\left (1-\dfrac{\eta_{H}}{2}\right )\biggl [\dfrac{Z_{H}k^{6}}{Z_{H}k^{2}+\p U/\p\rho+2\rho\p^{2}U/\p\rho^{2}}\nn\\
&\quad +\dfrac{3Z_{H}k^{6}}{Z_{H}k^{2}+\p U/\p\rho}\biggl ]\, .
\end{align}
One infers an additional contribution to eq.~\eqref{eq:AM4},
\begin{equation}\label{eq:AM8} 
\Delta \p_{t}m_{R}^{2}=-\dfrac{3k^{2}}{16\pi^{2}}\left (1-\dfrac{\eta_{H}}{2}\right )\lambda_{H}\left (1+m_{R}^{2}/k^{2}\right )^{-2}\, .
\end{equation}
The scalar contribution to the anomalous mass dimension therefore reads
\begin{align}\label{eq:AM9}
A_{S}&=\dfrac{\p}{\p m_{R}^{2}}\Delta\p_{t}m_{R}^{2}\nn\\
&=\dfrac{3\lambda_{H}}{8\pi^{2}}\left (1-\dfrac{\eta_{H}}{2}\right )\left (1+m_{R}^{2}/k^{2}\right )^{-3}\, .
\end{align}
In lowest order in perturbation theory one can neglect $\eta_{H}$ and $m_{R}^{2}/k^{2}$ in eq.~\eqref{eq:AM9}. Thus eq.~\eqref{eq:AM9} accounts the first term in eq.~\eqref{eq:43}. (The other terms involving gauge and Yukawa couplings correspond to $\eta_{H}$.)

This computation shows in a direct way how the anomalous mass dimension obtains from the general flow equation for the scalar effective potential. The gravitational contribution is just one part of it. However, in view of realistic sizes of Yukawa couplings, gauge couplings and quartic scalar coupling it dominates by far. As it should be, the part from particle fluctuations agrees with the perturbative result in the limit where the particles are massless. The scalar anomalous mass dimension is actually only a particular diagonal element in a larger stability matrix. An off-diagonal element of the stability matrix arises from the term $\sim y_{t}^{2}$ in eq.~\eqref{eq:AM2}. As long as $A$ is large the effect of the small off-diagonal elements is negligible.

\subsection{Quadratic and logarithmic running}\label{sec:6.18} 

The functional renormalization flow \eqref{eq:AM4} of the scalar mass term is quadratic. The same holds for the running squared Planck mass \eqref{eq:93}. No quadratic running is seen in perturbative investigations using dimensionless regularization. One should therefore discuss the physical meaning and status of the quadratic running. We will do this first for the simple case of the flowing scalar mass term. The generalization to the flowing Planck mass will be straight forward. 

We first emphasize that the quadratic running concerns only the position of the critical hypersurface in coupling constant space. Once one has specified the model to be on or close to the critical surface, only the ``logarithmic'' running \eqref{eq:6.286B} of the deviation $\delta$ from the critical surface remains, given by the anomalous dimension $A+\eta_{H}$. The anomalous dimension is computable in perturbation theory with dimensional regularization and the two results agree. More precisely, renormalization group improvement by a solution of the flow equation translates the ``logarithms'' to a power like behavior with small power, given for constant $A$ by eq.~\eqref{eq:5.16A}. As we have argued in sect.~\ref{sec:Anomalous_Mass_Dim}, only $A$ appears in the observable quantities. From the point of view of the renormalizable theory valid for the universality class which corresponds to the critical (or nearly critical) theory, all observables can be computed with dimensional regularization. No trace of the quadratic running appears.

The quadratic running matters if one wants to translate a given specified microphysical model to observable macrophysics. In particular, it concerns the location of the critical surface in some microphysical or ``bare'' coupling constant space. For particle physics this is not a very relevant question, since we do not know the precise microphysics. All observables are computable in terms of the renormalized couplings present in the universality class corresponding to a given renormalizable theory. This differs for condensed matter physics. As an example, we may take a gas of ultracold atoms, for which the microphysical couplings are well known, given by atomic physics. A Bose-Einstein condensate or similar collective phenomena typically occur on length scales much larger than typical length scales in atomic physics. In particular, for many interesting systems a second order phase transition happens at some critical temperature $T_{c}$. For temperatures below $T_{c}$ a Bose-Einstein condensate forms and the atomic gas becomes a superfluid.

Two types of questions arise. The first concerns the computation of $T_{c}$ from the microphysical parameters. For this issue the ``quadratic running'' is crucial. Omitting it, no reasonable picture can be obtained. A second type of question concerns universal critical exponents and amplitudes at or near $T_{c}$. They are given by a renormalizable theory, for which the ``quadratic running'' no longer matters. Only the flow of small deviations from the critical surface of the phase transition is relevant. If the couplings would be small, one could employ perturbation theory with dimensional regularization for this second type of questions.

For quantum gravity, the situation is analogous. The quadratic running of the squared Planck mass $M^{2}$ does not translate directly to observable quantities. It drops out from the observable quantities that only depend on the universality class of a given model. The quadratic running would be needed, however, if some day we would have a well defined microscopic theory, perhaps formulated in different degrees of freedom. It would appear in the translation of the microscopic parameters to the relevant couplings at the UV-fixed point.

The quantum effective action obtains by the solution of the functional renormalization group equation for $k\raw 0$. All observable quantities can be computed from this quantum effective action. In the limit $k\raw 0$ the infrared cutoff scale $k$ drops below all momenta or field values relevant for a given observable. Typically, these momenta and field variables effectively replace $k$ as an IR-cutoff. While non-zero $k$ often gives a qualitatively correct picture of the dependence on physical IR-cutoffs, the ``threshold behavior'' for $k$ moving from above to below physical IR-cutoffs is needed for quantitative accuracy. Both the quadratic running and the threshold behavior depend on the chosen form of the IR-cutoff function. If truncations would be exact, this scheme dependence is cancelled in the physical observables. For a given truncation, the scheme dependence can be employed for some type of error estimate.

\section{Scale symmetry in cosmology}\label{sec:Scale_Sym_Cosmo} 

Observation of anisotropies in the cosmic microwave background (CMB) indicates an almost scale invariant spectrum of the primordial cosmic fluctuations. These fluctuations are the seed of all structures in the Universe. The approximate scale invariance of the spectrum suggests the presence of one form or another of approximate scale symmetry in the physics of the ``beginning of the Universe'', which is usually associated to inflationary cosmology \cite{STAR,MUCHI,GUTH,LIN1,LIN2,SHACWCI,ALST}. (The ``beginning" could be only an effective beginning of our observable world if the Universe underwent even earlier stages not leaving observable imprints.)

We may associate the beginning with the UV-fixed point of quantum gravity, or the close vicinity of it. This provides naturally for a scale invariant quantum effective action and for scale invariance of the cosmological equations. If there are relevant parameters at the UV-fixed point, the Universe may evolve away from the fixed point as time increases. This is realized if some relevant parameter is associated with an instability of the  cosmological solutions in the vicinity of the fixed point. 

The intrinsic scale $\bar{\mu}$ associated to some relevant parameter appears in dimensionless functions parametrizing the effective action through dimensionless ratios. We investigate here cosmologies with a scalar field $\chi$ for which the relevant ratio is $\chi^{2}/\bar{\mu}^{2}$. Due to the associated dependence on $\chi$ the $\bar{\mu}$-dependence manifests itself in the cosmological field equations. A cosmological change of $\chi$ can therefore explore different parts of flow trajectories. We will find that typical cosmological solutions approach in the far past the UV-fixed point with $\chi\raw 0$, and in the far future the IR-fixed point with $\chi\raw\infty$. Moving away from the fixed point scale symmetry is only approximate. One therefore expects approximate scale symmetry of the correlation function and therefore approximate scale symmetry of the primordial cosmic fluctuation spectrum if the beginning epoch corresponds indeed to the vicinity of the UV-fixed point.

In this picture inflation is associated with the vicinity of the UV-fixed point. Once the Universe evolves sufficiently far away from the fixed point, scale symmetry ends to be a valid approximation. This ends the inflationary epoch. Since the observable cosmological fluctuations have been frozen long (many e-foldings) before the end of inflation, they show approximate scale symmetry. The association of inflation with physics close to the UV-fixed point of quantum gravity provides a natural explanation of the approximate scale invariance of the primordial fluctuation spectrum \cite{CWIQ}. One rather appealing scenario for the beginning of our world assumes that the UV-fixed point is reached in the infinite past. The Universe can exist forever. In this scenario of an ``Eternal Universe" \cite{CWEU} the notion of time-evolution and an effective arrow of time originates from the unstable behavior of arbitrary small deviations from the fixed point \cite{CWVG,CWIQ}. 

Our investigation will be based on explicit models, specified by an effective action with diffeomorphism symmetry and (approximate) scale symmetry. We consider models with a scalar field or a scalar degree of freedom as in $R^{2}$ gravity. For given parameters of these models all fluctuation quantities can be unambiguously computed. More qualitative arguments for the relation between the UV-fixed point and the fluctuation spectrum can be found in ref.~\cite{BONREU,BOSA}. At the UV-fixed point not only scale symmetry, but full conformal symmetry may be realized. This may provide a bridge to discussions of scale invariant spectra arising in models with conformal symmetry \cite{RUB1,LRR,HJK,CJKS}.

Several classes of inflationary models can be related directly to approximate gravity scale symmetry. The first is Starobinski inflation \cite{STAR}. It is based on an effective action 
\begin{equation}
\mathcal{L}=\sqrt{g}\left\lbrace -\frac{C}{2}R^{2}-\frac{M^{2}}{2}R+V \right\rbrace \label{eq:C1}.
\end{equation} 
Scale invariance is realized if the dimensionfull constants $M^{2}$ and $V$ can be neglected. For Starobinski inflation $M^{2}$ and $V$ are constants that may be associated to relevant parameters at some UV-fixed point.

A second family is cosmon inflation \cite{CWCI,CWVG,CWIQ}. Here $M^{2}$ and $V$ are functions of a scalar field $\chi$ and one adds a scalar kinetic term 
\begin{equation}
\mathcal{L}=-\dfrac{F}{2}R+\frac{1}{2}\sqrt{g}K \partial^{\mu}\chi\partial_{\mu}\chi+U-\dfrac{C}{2}R^{2}\label{eq:C2}.
\end{equation}
For scale invariant functions $F=\chi^{2}$, $U=\lambda \chi^{4}$ and constant $K=B-6$ the effective action is scale invariant and could be associated to the UV-fixed point. In this limit the difference to Starobinski inflation would merely be the presence of an additional Goldstone boson from the spontaneous breaking of the scale symmetry for $\chi=\chi_{0}\neq 0$. This does not matter for late stages of cosmology and does not influence the primordial fluctuation spectrum. If $F(\chi)$ and $U(\chi)$ are associated to relevant parameters they typically involve intrinsic scales, e.g.
\begin{eqnarray}
U&=&b\bar{\mu}^{2}\chi^{2}+c\bar{\mu}^{4},\nonumber\\
F&=&\chi^{2}+d\bar{\mu}^{2}\label{eq:C3}.
\end{eqnarray}
This form of $U$ and $F$ reflects our discussion of the scaling potential in dilaton quantum gravity in sect.~\ref{sec:Flow_Eff_Scalar_Pot_Quantum_Grav}. It obtains if we replace $k=c_{\mu}\bar{\mu}$, with constant $c_{\mu}$ chosen for convenience. Furthermore, the kinetic coefficient $K$ may become a function of $\chi/\bar{\mu}$.

In the infrared limit $\bar{\mu}^{2}/ \chi^{2}\rightarrow 0$ scale symmetry becomes exact if $K$ becomes independent of $\bar{\mu}$. This scale symmetry is now associated to an IR-fixed point. If $\chi$ increases towards infinity in the infinite future one expects the presence of a pseudo-Goldstone boson with a tiny mass - the cosmon. This cosmon is responsible for dynamical dark energy or quintessence \cite{CWQ}. An effective action \eqref{eq:C2} solves the cosmological constant problem dynamically since the observable ratio $U/F^{2}\rightarrow \bar{\mu}^{2}/\chi^{2}$ or $\bar{\mu}^{4}/\chi^{4}$ vanishes for $\chi\rightarrow \infty $. The details of the beginning phase for $\chi \rightarrow 0$ depend on the dimensionless parameters $C$, $K$, $a$, $b$ and $d$.

Higgs inflation \cite{BS1} can be associated to approximate scale invariance as well. With $\chi^{2}$ replaced by $h^{\dagger}h$ for the Higgs-doublet $h$,
\begin{eqnarray}
F&=& {M}^{2}+\xi_{H}\left(h^{\dagger}h/{M}^{2}\right)h^{\dagger}h,\nonumber\\ 
U&=&\frac{1}{2}\lambda_{H}\left(h^{\dagger}h/{M}^{2}\right)\left(h^{\dagger}h\right)^{2},\label{eq:C4}
\end{eqnarray}
scale symmetry becomes exact for ${M}^{2}/h^{\dagger}h\raw 0$ and constant $\lambda_{H}$ and $\xi_{H}$. Approximate scale symmetry is therefore realized in the region $h^{\dagger}h\gg {M}^{2}/\xi_{H}$ if the running of $\lambda_{H}$ and $\xi_{H}$ with $h^{\dagger}h/ {M}^{2}$ is slow.

Finally, inflation can also be realized for exactly scale invariant models, e.g. defined precisely on the UV-fixed point \cite{SHAZE,GRSZ,FHNR,CPR}. For scale invariant cosmon inflation one replaces $\bar{\mu}$ in eq.~(\ref{eq:C3}) by a second scalar field $\sigma$ and adds a kinetic term for $\sigma$. Scale invariant Higgs inflation is realized if ${M}$ is replaced by $\chi$ in eq.~(\ref{eq:C4}). For scale invariant inflation the dynamics that ends inflation and gives rise to the small scale violation in the primordial fluctuation spectrum is no longer directly related to the breaking of gravity scale symmetry by an intrinsic mass scale. The flatness of the potential in the direction ``orthogonal'' to the Goldstone boson is now no longer an automatic consequence of gravity scale symmetry. It is typically postulated or assumed. A flat potential can often be associated with an additional scale symmetry. This additional scale symmetry makes this assumed potential technically natural.

Functional renormalization group investigations of quantum gravity are not yet on the level of precision where one or several of these inflationary scenarios can be excluded. We take here a heuristic attitude and present a phenomenological comparison of the different inflation candidates. This comparison is facilitated by a common language which uses for all models a standardized exponential potential. The differences show then up in different coefficients of kinetic terms or ``kinetials'' for the inflaton. The kinetial has a very direct connection to the properties of the primordial fluctuation spectrum.

Inflationary cosmology may not be the only epoch for which approximate scale symmetry plays a central role. In many models scalar fields evolve dynamically during the cosmological epochs after nucleosynthesis. This includes models in string theories, modified gravity theories as $f\left(R\right)$-models and quintessence or dynamical dark energy. Observable dimensionless couplings like the fine structure constant $\alpha_{em}$, or ratios between electron and nucleon masses, $m_{e}/m_{n}$, or nucleon and Planck masses, $m_{n}/M$, often depend on the value of such scalar fields. The cosmic time evolution of the scalar induces a time variation of the fundamental constants. This variation is severely restricted by observations \cite{CWVN,UZREP}. The question arises why the interaction of the scalar with atoms is several orders of magnitude smaller than the gravitational interaction.

Scale symmetry provides for a simple solution for this problem. If the scalar can be associated with the pseudo-Goldstone boson of a spontaneously broken approximate scale symmetry, its coupling to atoms must be very small. In the limit of exact scale symmetry the Goldstone boson can only have derivative couplings to atoms. The dimensionless parameters $\alpha_{em}$, $m_{e}/m_{n}$ or $m_{n}/M$ are scale invariant in this case. As a consequence, they will not depend on the value of the scalar field. Spontaneously broken exact scale symmetry predicts a vanishing time variation of fundamental constants. If the IR-fixed point responsible for exact quantum scale symmetry is reached only in the infinite future for $\chi\left(t\rightarrow\infty\right)\rightarrow\infty$, the time variation is not yet exactly zero. It is tiny, however, due to the vicinity of the fixed point. Correspondingly, any time variation of dimensionless couplings is expected to be tiny.

We have seen in sect.~\ref{sec:V} that in the infrared limit for $\chi\rightarrow\infty$ a scalar potential $\tilde{\lambda}\left(\chi/\bar{\mu}\right)\chi^{4}$ has a fixed point at $\tilde{\lambda}_{\ast}=0$. For a normalization where $F=\chi^{2}$ the approach to the fixed point is bounded by the graviton barrier, $\tilde{\lambda} < \bar{\mu}^{2}/\chi^{2}$ \cite{CWGFE}. For any finite $\chi$ the potential does not yet vanish. It constitutes a part of dynamical dark energy. Spontaneously broken approximate scale symmetry is therefore of relevance for the present epoch of cosmology. It can be associated to the dynamics of dark energy.

We discuss in detail the cosmology associated to the scaling solution of dilaton quantum gravity for the case where $B$ is small. We will find that this constitutes a viable cosmological model of ``crossover quintessence''. As a possible alternative the relevant parameter responsible for an intrinsic scale may be associated  to beyond standard model physics. This result is a growing ratio of neutrino mass over electron mass. This is the essential ingredient for the attractive cosmological model of ``growing neutrino quintessence''. Both forms of quintessence are compatible with the present observational situation.

\subsection{Primordial cosmic fluctuations}

One of the most striking consequences of approximate scale symmetry is the almost scale invariant spectrum of cosmic fluctuations \cite{HARR,ZELD,PEEBYU}. We deal here with models that can be represented as inflationary cosmologies in the Einstein frame for the metric. The link between approximate quantum scale invariance close to a fixed point and the approximately scale invariant primordial fluctuation spectrum is more general, however. Primordial fluctuations in inflationary cosmology have been discussed in ref.~\cite{MUCHI,HAWF,STARF}. Important insights have been gained by concentrating on gauge invariant fluctuations \cite{SASIF,MUGIF}. Deriving the correlation function for the fluctuations directly from the inversion of the second functional derivative of the effective action \cite{CWFL1,CWFL2,CWFL3}, it has become possible to discuss the fluctuation problem in a formulation that does not depend on the choice of a frame for the metric \cite{CWPCFV,KAPI}. This concerns both issues usually related to the ``choice of the vacuum'' and the time evolution.

De Sitter space is invariant under $SO(1,4)$-transformations. This subgroup of the diffeomorphisms (general coordinate transformations) includes a rescaling of spacelike (cartesian) coordinates and conformal time by a constant factor. This type of coordinate scale symmetry holds for de Sitter space independently of the question if the effective action contains parameters with mass or length or not. It is therefore not directly related to quantum scale symmetry. In a sense, our investigation of the cosmological consequences of quantum scale symmetry explains why approximate de Sitter space is a solution of the cosmological field equations in the Einstein frame.

For the purpose of this report we refer to the literature for a general discussion of the fluctuation problem. We rather concentrate on the computation of the fluctuation properties -- spectrum and amplitude -- for several models with approximate scale symmetry, and their connection with a possible UV-fixed point. Instead of a fully frame invariant discussion we prefer here to work in the Einstein frame with a particular normalization of the scalar field such that all properties of the shape of the fluctuation spectrum can be extracted directly from the coefficient function of the kinetic term, the ``kinetial''.

Instead of switching to a standard kinetic term we choose the normalization of the scalar field $\varphi$ such that the potential is normalized as
\begin{equation}
V_{\text{E}}\left(\varphi\right)=M^{4}\exp\left(-\frac{\varphi}{M}\right).\label{eq:C5}
\end{equation}
The details of the model are then encoded in the ``kinetial'' $k^{2}\left(\varphi\right)$ \cite{CWVG,HEBCW}. For single field inflation the effective action in the Einstein frame reads for this normalization
\begin{equation}
\mathcal{L}=-\frac{M^{2}}{2}R+V_{\text{E}}\left(\varphi\right)+\frac{1}{2}k^{2}\left(\varphi\right)\partial^{\mu}\varphi\partial_{\mu}\varphi.\label{eq:C6}
\end{equation}
The advantage is that the value of $\varphi$ is directly related to the potential energy density. In turn, this determines directly the amplitude of the tensor fluctuations by the value of $\varphi$ when a given fluctuation is frozen. On the other hand, the spectral index $n$ and the ratio $r$ of tensor to scalar fluctuations only depend on the kinetial \cite{CWVG}. Different inflationary models are easily compared in this formulation.  

For a constant kinetial ($\p_{\varphi}k^{2}=0$) exponential potentials have been discussed early \cite{LUMA} as interesting candidates for inflation. In the context of inflation they arise naturally from dimensional reduction of higher dimensional theories \cite{CWEXBE1,CWEXBE2} and can be associated to inflation from higher dimensions \cite{SHACWCI}. The formulation with a kinetial has been used in the discussion of quintessence models \cite{HEBCW} or for inflationary models described as ``$\alpha$-attractors'' \cite{KLR1,KLR2,DOW,DMAR}.

Inflationary epochs can be associated to field-ranges when $k^{2}\left(\varphi\right)$ is large. Indeed, a large kinetial slows down the evolution of $\varphi$ and gives rise to a ``slow roll behavior". Following ref. \cite{CWVG} we can derive the slow roll parameters $\epsilon$ and $\eta$ from the properties of $k^{2}\left(\varphi\right)$, 
\begin{equation}
\epsilon\left(\varphi\right)=\frac{1}{2k^{2}\left(\varphi\right)}\label{eq:C11}
\end{equation}
and 
\begin{equation}
\eta\left(\varphi\right)=2\epsilon\left(\varphi\right)-M\frac{\partial\epsilon\left(\varphi\right)}{\partial\varphi}.\label{eq:C12}
\end{equation}

This follows \cite{CWVG} from the relation of $\varphi$ to the field $\sigma$ with canonical kinetic term 
\begin{equation}
\frac{\partial\sigma}{\partial\varphi}=k\left(\varphi\right)\, ,\label{eq:C13}
\end{equation}
and the usual definitions 
\begin{equation}
\epsilon=\frac{M^{2}}{2}\left(\frac{\partial\ln V}{\partial \sigma}\right)^{2}\, , \hspace{10mm}\eta=\frac{M^{2}}{V}\frac{\partial^{2}V}{\partial \sigma^{2}}\, .\label{eq:C14}
\end{equation}
The spectral index is given by
\begin{equation}
 n=1-6\epsilon+2\eta=1-2\left(\epsilon+M\frac{\partial\epsilon}{\partial\varphi}\right),\label{eq:C15}
\end{equation}
while the tensor to scalar ratio obeys
\begin{equation}
r=16\epsilon=\frac{8}{k^{2}}\, .
\end{equation}
Both $r$ and $n$ have a simple direct expression in terms of the kinetial.

Inflation lasts as long as $\epsilon$ and $\eta$ remain small. This epoch ends when $\epsilon$ becomes of the order one.
The field value at the end of inflation is given by
\begin{equation}
\epsilon\left(\varphi_{\text{f}}\right)=1, \hspace{10mm} k^{2}\left(\varphi_{\text{f}}\right)=\frac{1}{2}.\label{eq:C17}
\end{equation} 
The number of e-foldings before the end of inflation obtains as
\begin{equation}
N\left(\varphi\right)=\frac{1}{M}\int_{\varphi}^{\varphi_{\text{f}}}d\varphi 'k^{2}\left(\varphi '\right).\label{eq:C18}
\end{equation}
It can be again be expressed purely in terms of the kinetial. Inverting eq.~(\ref{eq:C18}) on obtains $n\left(N\right)$ and $r\left(N\right)$. Depending on the details of the heating period after inflation one has $N$ in the range $50-70$.

With 
\begin{equation}
M\frac{\partial N}{\partial \varphi}=-k^{2}\left(\varphi\right)\label{eq:C18A}
\end{equation}
we can write eq.~(\ref{eq:C12}) as 
\begin{eqnarray}
\eta &=& 2\epsilon - M\frac{\partial N}{\partial \varphi}\frac{\partial \epsilon}{\partial N}=k^{2}\frac{\partial \epsilon}{\partial N}+2\epsilon\nonumber\\
&=&\frac{1}{2\epsilon}\frac{\partial\epsilon}{\partial N}+2\epsilon=\frac{1}{2}\frac{\ln \epsilon}{\partial N}+2\epsilon.\label{eq:C18B}
\end{eqnarray}
This relates the spectral index to the $N$-dependence of $\epsilon$
\begin{equation}
n=1+\frac{1}{N}\frac{\partial\ln\epsilon}{\partial\ln N}-2\epsilon.\label{eq:C18C}
\end{equation}
If $\epsilon\sim N^{-\alpha}$ decays faster than $\sim N^{-1}$ the term $-2\epsilon$ can be neglected, such that 
\begin{equation}
n=1-\frac{\alpha}{N}.\label{eq:C18D}
\end{equation}
For $\epsilon \sim c_{\epsilon}/N$ one has 
\begin{equation}
n=1-\frac{1+2c_{\epsilon}}{N}, \hspace{10mm} \epsilon =\frac{c_{\epsilon}}{N}.\label{eq:C18E}
\end{equation}

All these relations hold independently of the particular model since neither the explicit form of the kinetial $k^{2}\left(\varphi\right)$ nor the value of $\varphi$ are needed. The only condition is the domination of the inflationary epoch by a single scalar field, i.e. ``single field inflation". Once $N\left(\varphi\right)$ is computed, one may also invert this relation and compute the kinetial $k^{2}\left(N\right)$ as a function of $N$. The spectral parameters $n$ and $r$ can be then found directly from $k^{2}\left(N\right)$,
\begin{equation}
r=\frac{8}{k^{2}\left(N\right)},\hspace{10mm} n=1-\frac{\partial \ln k^{2}\left(N\right)}{\partial N}-\frac{r}{8}.\label{eq:C18F}
\end{equation}

Finally, the amplitude of the scalar fluctuations is given by
\begin{equation}
\Delta^{2}=\frac{V_{E}}{24\pi^{2}\epsilon M^{4}}=\dfrac{k^{2}V_{E}}{12\pi^{2}M^{4}}\; .\label{eq:C19}
\end{equation}
It has been observed as
\begin{equation}
\Delta^{2}=2.065\times 10^{-9},\label{eq:C20}
\end{equation}
such that realistic models have to obey
\begin{equation}
\frac{V_{E}}{M^{4}}\approx 5\epsilon\times 10^{-7} \label{eq:C21}
\end{equation}
at the values of $\varphi$ for which the observable fluctuations are frozen. We can therefore estimate the value of $\varphi$ at which freezing of the observable fluctuations takes place,
\begin{equation}
e^{-\frac{\varphi}{M}}=5\epsilon\times 10^{-7}=2.5\times 10^{-7} k^{-2}\left(\varphi\right).\label{eq:C22}
\end{equation}
or
\begin{equation}
\frac{\varphi}{M}=15.2 + \ln\left(k^{2}\left(\varphi\right)\right).\label{eq:C23}
\end{equation}

\subsection{Starobinski inflation}\label{sec:Starobinski_Inflation} 

Starobinski inflation \cite{STAR} is based on the cosmological solutions of the field equations derived from the effective action \eqref{eq:C1}. The term $V$ in eq.~(\ref{eq:C1}) accounts for the present dark energy density. It is completely negligible during the inflationary epoch and we will omit it. Since $M$ in eq.~(\ref{eq:C1}) is not necessarily the observed Planck mass, we denote this constant by $\tilde{M}$, similar to \eq{162}. This parameter sets the mass scale. The model therefore contains only one dimensionless parameter $C$. Inflation is then described by the equivalent effective action \eqref{eq:163} with an explicit scalar field $\phi$. 

For the particular case of Starobinski inflation the relation between $\varphi$ and $\phi$ follows from
\begin{equation}
\varphi=-M\ln\left(\frac{V_{\text{E}}}{M^{4}}\right)=M\ln\left(\frac{2\left(\tilde{M}^{2}+2C\phi\right)^{2}}{C\phi^{2}}\right)\label{eq:C7}
\end{equation} 
or
\begin{equation}
\frac{\phi}{\tilde{M}^{2}}=\sqrt{\frac{2}{C}}\left[\exp\left(\frac{\varphi}{2M}\right)-\sqrt{8C}\right]^{-1}\, .\label{eq:C8}
\end{equation}
With
\begin{equation}
\partial_{\mu}\varphi=-\frac{2\tilde{M}^{2}M}{\phi\left(\tilde{M}^{2}+2C\phi\right)}\partial_{\mu}\phi\, , \label{eq:C9}
\end{equation}
we can extract from \eq{163}
\begin{equation}
k^{2}\left(\varphi\right)=\frac{3C^{2}\phi^{2}}{2\tilde{M}^{4}}=\frac{3C}{\left(\exp\left(\frac{\varphi}{2M}\right)-\sqrt{8C}\right)^{2}}.\label{eq:C10}
\end{equation}

We observe that $k^{2}\left(\varphi\right)$ diverges for $\varphi/M=\ln\left(8C\right)$. This corresponds to the limit $C\phi/\tilde{M}^{2}\rightarrow\infty$. In this limit the $R^{2}$-term dominates in eq.~(\ref{eq:C1}) and scale invariance becomes exact. Inflation can be associated to this range of $R^{2}$-domination. Indeed, employing the original formulation \eqref{eq:C1} and using the association $\phi=R$ in eq.~\eqref{eq:141}, the kinetial takes the expression
\begin{equation}\label{eq:7.29A} 
k^{2}=\dfrac{3}{2}\left (\dfrac{CR}{M^{2}}\right )^{2}\, .
\end{equation}
The scale invariant effective action associated to the UV-fixed point corresponds to $M^{2}/(CR)\raw 0$ or $k^{2}\raw\infty$. Starobinski inflation is a simple example how a large kinetial and therefore inflation is associated to the vicinity of the UV-fixed point. Inflation ends when the intrinsic mass $M$ associated to a relevant parameter for the flow away from the fixed point becomes important. This is directly seen by the kinetial reaching $k^{2}=1/2$ for $CR\approx M^{2}$.

For a quantitative computation of $n$ and $r$ one needs the relation between $\varphi$ and $N$. Close to the critical value $\varphi_{0}=M\ln(8C)$ we expand
\begin{equation}
\varphi=\varphi_{0}+2M\delta, \hspace{10mm} e^{\frac{\varphi_{0}}{2M}}=\sqrt{8C},\label{eq:C24}
\end{equation}
such that 
\begin{equation}
k^{2}=\frac{3}{8\left(e^{\delta}-1\right)^{2}}.\label{eq:C25}
\end{equation}
From eq.~(\ref{eq:C18}) we extract the relation between the number of e-foldings before the end of inflation and $\delta$
\begin{eqnarray}
N &=&\frac{3}{4}\int_{\delta}^{\delta_{\text{f}}}\frac{d\delta '}{\left(e^{\delta '}-1\right)^{2}}=\frac{3}{4}\int dz \frac{1}{z^{2}\left(z+1\right)}\label{eq:C26}\\
&=&\frac{3}{4}\left\lbrace\frac{1}{e^{\delta}-1}+\ln\left(1-e^{-\delta}\right)-\frac{1}{e^{\delta_{\text{f}}}-1}-\ln\left(1-e^{-\delta_{\text{f}}}\right)\right\rbrace\nonumber,
\end{eqnarray}
where $z=e^{\delta}-1$. With $k^{2}\left(\delta_{\text{f}}\right)=\frac{1}{2}$ one obtains
\begin{eqnarray}
N\left(\delta\right)&=&\frac{3}{4}\left\lbrace\frac{1}{e^{\delta}-1}+\ln\left(1-e^{-\delta}\right)\right\rbrace -N_{\text{f}}\label{eq:C27}\\
&=&\frac{3}{4}\left\lbrace\frac{1}{z}+\ln\left(1+\frac{1}{z}\right)\right\rbrace-N_{\text{f}}\nonumber,
\end{eqnarray}
with
\begin{equation}
N_{\text{f}}=\frac{2}{\sqrt{3}}+\ln\left(1+\frac{2}{\sqrt{3}}\right).\label{eq:C28}
\end{equation}
We conclude that the relation between the kinetial and $N$ is independent of the parameter $C$.

For $N\gg1$ one needs $z\ll 1$ and we can compute $z\left(N\right)$ iteratively
\begin{equation}
z=\left[\frac{4\left(N+N_{f}\right)}{3}-\ln\left(1+\frac{4\left(N+N_{f}\right)}{3}\right)\right]^{-1}.\label{eq:C29}
\end{equation}
Hence $z\left(N\right)$ is small for $N\approx 60$. With $\epsilon=1/\left(2k^{2}\right)$ and $k^{2}=3/8z^{2}$ the slow roll parameter $\epsilon\left(N\right)$ becomes
\begin{equation}
\epsilon=\left(\frac{4z^{2}}{3}\right)=\frac{3}{4\left(N+N_{\text{f}}\right)^{2}}+\dotsc.\label{eq:C30}
\end{equation}
As well known, the tensor to scalar ratio $r$ is tiny for $N$ around $60$
\begin{equation}
r\approx\frac{12}{N^{2}}\approx 0.0034\, .\label{eq:C31}
\end{equation}
With eq.~(\ref{eq:C18B}) one finds
\begin{equation}
\eta=-\frac{1}{N+N_{\text{f}}}\, ,\label{eq:C32}
\end{equation}
such that the spectral index for the primordial scalar fluctuations reads
\begin{equation}\label{eq:C33} 
n=1-\frac{2}{N+N_{\text{f}}}\approx 0.966\, .
\end{equation}
The values of $n$ and $r$ are well compatible with the present observational bounds. 

The amplitude of the fluctuations fixes the parameter $C$. We can replace for the computation of the amplitude $\varphi=\varphi_{0}$, such that 
\begin{equation}
\frac{V_{E}}{M^{4}}=e^{-\frac{\varphi_{0}}{M}}=\frac{1}{8C}\, .\label{eq:C34}
\end{equation}
This expresses $C$ in terms of the observed value \eqref{eq:C20} for $\Delta$
\begin{equation}
C=\frac{1}{192\pi^{2}\epsilon\Delta^{2}}\approx\frac{N^{2}}{144 \pi^{2}\Delta^{2}}\approx 1.2\cdot 10^{9}\, .\label{eq:C35}
\end{equation}
Relatively large values of $C$ are needed for obtaining a realistic amplitude.

One may ask if such large values are compatible with asymptotic safety \cite{GORS}. As we have assumed before, $C$ may be a relevant parameter. It can therefore be considered as a free parameter. For the large value $C$ in eq.~(35) the dependence of $C$ on the renormalization scale $k$ (not to be confounded with the kinetial) is in the perturbative range
\begin{equation}
C=C\left(\bar{k}\right)+a_{\text{C}}\ln\left(\frac{k^{2}}{\bar{k}^{2}}\right )\, .\label{eq:C36}
\end{equation}
There we left the coefficient $a_{\text{C}}$ free in view of the unknown matter contributions. The running of $C$ is negligible for the large value of $C$ needed, such that the use of constant $C$ is well justified. For a discussion of Starobinski inflation in the context of asymptotic safety see refs.~\cite{BOPLA,BOPS}.

\subsection{Cosmon inflation}\label{sec:Cosmon_Infl} 

For Starobinski inflation the term $\sim CR^{2}$ plays no role for late cosmology. Except for the Higgs boson, there is no dynamical scalar field in the late stages of cosmology. This holds, in particular, after the QCD-``phase transition" when all scalars are frozen. Dark energy is not dynamical but rather given by a cosmological constant. This situation typically changes for variable gravity where the Planck mass is given by an additional scalar field $\chi$. The ``cosmon" $\chi$ may be responsible for both inflation and present dynamical dark energy. As we have discussed before, it can interpolate between the UV-fixed point for $\chi\rightarrow0$ and IR-fixed point for $\chi \rightarrow \infty$.

The UV-limit of cosmon inflation can be based on the effective action \eqref{eq:139}. In the presence of the term $\sim CR^{2}$ variable gravity involves two scalar fields, the cosmon $\chi$ and the field $\phi \sim R$, as visible in the equivalent UV-action \eqref{eq:142}, \eqref{eq:143}. In the limit when $B$ is constant or depends only on $\chi^{2}/R$ or $\chi^{2}/\phi$, the effective action \eqref{eq:142} is scale invariant. This is a particular case of scale invariant inflation. Due to the exact quantum scale symmetry one of the scalar fields is a Goldstone boson. It settles early in cosmology to a constant value $\chi_{0}$. In the limit where all intrinsic mass scales can be neglected as compared to $\phi$ and $\chi^{2}$ the model realizes scale invariant Starobinski inflation. It differs from Starobinski inflation by a mixing of the kinetic terms for $\phi$ and $\chi$, as we will discuss in sect.~\ref{sec:Scale_Inv_Inf}. We observe that in this case gravity scale symmetry is not related to the end of inflation and the slope of the primordial fluctuation spectrum. Its role is replaced by an effective scale symmetry which transforms the metric ( and fields for particles) but leaves $\chi$ invariant.

A new situation arises if we take the scale symmetry violation in the cosmon sector into account. The intrinsic scale $\bar{\mu}$ in eq.~(\ref{eq:C3}) changes the dynamics both for early and late cosmology. With eq.~(\ref{eq:C3}) we generalize the effective action \eqref{eq:142} to
\begin{align}
\tilde{\mathcal{L}}&=\sqrt{g}\biggl \{ -\left(\frac{\chi^{2}}{2}+\frac{d\bar{\mu}^{2}}{2}+C\phi\right)R+\frac{1}{2}\left(B-6\right)\partial^{{\mu}}\chi\partial_{{\mu}}\chi \nonumber\\
&\quad +\frac{C}{2}\phi^{2}+b\bar{\mu}^{2}\chi^{2}+c\bar{\mu}^{4}\biggl \}\label{eq:CI1}\, .
\end{align}
Inflation based on this effective action has been studied in refs.~\cite{CWIQ,PZHS}. For $\bar{\mu}^{2}=0$ the effective action in the Einstein frame is given by \eq{159}. The new terms modify the effective potential in the Einstein frame, cf. eqs.~\eqref{eq:159}-\eqref{eq:161},
\begin{equation}
V_{\text{E}}=\frac{M^{4}}{\left(\chi^{2}+d\bar{\mu}^{2}+2C\phi\right)^{2}}\left(\frac{C}{2}\phi^{2}+b\bar{\mu}^{2}\chi^{2}+c\bar{\mu}^{4}\right)\, .\label{eq:CI2}
\end{equation}
In the following we take $b=1$.

For $\chi=0$ and large $\phi$ the potential approaches the flat potential of Starobinski inflation. Furthermore, extrema or saddle points of $V_{\text{E}}$ give rise to cosmological solutions with constant $\phi$ and $\chi$, with a de Sitter geometry for $V_{\text{E}}>0$. With 
\begin{align}
\frac{\partial V_{\text{E}}}{\partial \phi}&=\frac{CM^{4}}{\left(\chi^{2}+d\bar{\mu}^{2}+2C\phi\right)^{3}}\biggl \{\left(\phi-4\bar{\mu}^{2}\right)\chi^{2}\\
&\quad -4c\bar{\mu}^{4}+d\bar{\mu}^{2}\phi\biggl \}\label{eq:CI3}
\end{align}
and
\begin{align}
\frac{\partial V_{\text{E}}}{\partial \chi}&=-\frac{2M^{4}\chi}{\left(\chi^{2}+d\bar{\mu}^{2}+2C\phi\right)^{3}}\biggl\{\bar{\mu}^{2}\left(\chi^{2}-2C\phi\right)\nn\\
&\quad +C\phi^{2}+\left(2c-d\right)\bar{\mu}^{4}\biggl \}\, ,\label{eq:CI4}
\end{align}
the two extremum conditions can be obeyed for
\begin{equation}
\chi=0, \hspace{10mm} \phi=\frac{4c}{d}\bar{\mu}^{2}\, .\label{eq:CI5}
\end{equation}
Another solution for $d>2c$ is given by 
\begin{equation}
\phi=2\bar{\mu}^{2}, \hspace{10mm} \chi^{2}=\left(d-2c\right)\bar{\mu}^{2}\, .\label{eq:CI6}
\end{equation}
The vicinity of the stationary fixed point solutions (\ref{eq:CI5}), (\ref{eq:CI6}) are candidates for inflationary cosmology. For positive $c$, $d$ and $C$ no other stationary solution with finite $\phi$ and $\chi$ exists. 

An interesting class of solutions reaches in the infinite past (for conformal time $\eta \rightarrow -\infty$) the limit $\chi\rightarrow0$, $\phi\rightarrow\infty$. In this limit one has 
\begin{align}\label{eq:CI7}
V_{\text{E}}&=M^{4}\Biggl[\frac{1}{8C}-\frac{\chi^{2}+d\bar{\mu}^{2}}{8C^{2}\phi}+\frac{1}{4C^{2}\phi^{2}}\biggl (\left(1+\frac{3d}{4C}\right)\bar{\mu}^{2}\chi^{2}+\nn \\
&+\quad \left(c+\frac{3d^{2}}{8C}\right)\bar{\mu}^{4}+\frac{3}{8C}\chi^{4}\biggl )\Biggr]\, .
\end{align}
For large finite $\phi$ the $\phi$-derivative of $V_{\text{E}}$ is positive
\begin{equation}
\frac{\partial V_{\text{E}}}{\partial \phi}=\frac{M^{4}\left(\chi^{2}+d\bar{\mu}^{2}\right)}{8C^{2}\phi^{2}}\, ,\label{eq:CI8}
\end{equation}
such that $\phi$ has a tendency to decrease slowly. As long as $\chi^{2}\ll d\bar{\mu}^{2}$ the field $\chi$ only plays a minor role. For any $\chi\neq0$ the field $\chi$ starts to increase, however, due to the negative derivative
\begin{equation}
\frac{\partial V_{\text{E}}}{\partial \chi}=-\frac{M^{4}\chi}{4C^{2}\phi}\, .\label{eq:CI9}
\end{equation}

A beginning epoch with large $\phi$ and small $\chi$ resembles in many aspects Starobinski inflation, with $\tilde{M}^{2}$ replaced by $d\bar{\mu}^{2}$. In fact, for $\chi=0$ the only difference between the effective action (\ref{eq:C1}) and (\ref{eq:CI1}) is the ratio $V_{E}/M^{4}$ receiving an additional factor involving $c\bar{\mu}^{4}$ 
\begin{equation}\label{eq:7.52AA} 
V_{E}=V_{E}^{(S)}\left (1+\dfrac{2c\bar{\mu}^{4}}{C\phi^{2}}\right )\, .
\end{equation}
We note, however, that this ratio only appears in terms $\sim\phi^{-2}$ and is therefore subleading. For $d>0$, $c>0$ the potential on the $\chi=0$ axis has a minimum given by eq.~\eqref{eq:CI5}, with
\begin{equation}\label{eq:7.53AA} 
\dfrac{V_{E}}{M^{4}}=\dfrac{1}{8C+d^{2}/c}\, .
\end{equation}
The stability of the axis $\chi=0$ with respect to small non-zero $\chi$ depends on the sign of $a$
\begin{equation}
\dfrac{\p^{2}V_{E}}{\p\chi^{2}}\biggl |_{\chi=0}=-\dfrac{2aM^{4}}{(d\bar{\mu}^{2}+2C\phi)^{3}}\, ,
\end{equation}
where
\begin{equation}
a=C\phi^{2}-2C\phi\bar{\mu}^{2}+(2c-d)\bar{\mu}^{4}\, .
\end{equation}
The $\chi=0$ axis is unstable for $a>0$, as always realized for large $\phi$. 

The main difference between cosmon inflation and Starobinski inflation appears at later stages when the dynamics of $\chi$ can no longer be neglected. This matters if the last sixty e-foldings of inflation are associated to an epoch when $\chi$ matters. The spectral parameters $n$ and $r$ may then deviate from the values of Starobinski inflation. If $\phi$ no longer dominates during this epoch the ratio $V_{\text{E}}/M^{4}$ is substantially smaller than $1/\left(8C\right)$. The value of $C$ will no longer be related directly to the amplitude of the primordial fluctuations. 

Another interesting class of approximate solutions corresponds to the limit $\mid 2C\phi\mid \ll \chi^{2}+d\bar{\mu}^{2}$, $C\phi^{2}/2\ll\bar{\mu}^{2}\chi^{2}+c\bar{\mu}^{4}$.  In this limit we can neglect $\phi$ and effectively set $C$ to zero. The corresponding version of cosmon inflation has been discussed extensively in refs.~\cite{CWCI,CWVG,CWIQ,RUBCW}. It is again a single field inflation with potential in the Einstein frame
\begin{equation}
V_{\text{E}}=\frac{M^{4}\left(\bar{\mu}^{2}\chi^{2}+c\bar{\mu}^{4}\right)}{\left(\chi^{2}+d\bar{\mu}^{2}\right)^{2}}\, ,\label{eq:CI10}
\end{equation}
and kinetic term 
\begin{equation}
\mathcal{L}_{\text{kin}}=\frac{M^{2}}{2}\left\lbrace \frac{B-6}{\chi^{2}+d\bar{\mu}^{2}}+\frac{6\chi^{2}}{\left(\chi^{2}+d\bar{\mu}^{2}\right)^{2}}\right\rbrace \partial^{\bar{\mu}}\chi\partial_{\bar{\mu}}\chi\, .\label{eq:CI11}
\end{equation}

Defining
\begin{equation}
\varphi=-M\ln\left[\frac{\bar{\mu}^{2}\chi^{2}+c\bar{\mu}^{4}}{\left(\chi^{2}+d\bar{\mu}^{2}\right)^{2}}\right]\label{eq:CI12}
\end{equation}
one arrives again at eq.~(\ref{eq:C6}), with kinetial
\begin{equation}
k^{2}\left(\varphi\right)=M^{2}\left\lbrace\frac{B-6}{\chi^{2}+d\bar{\mu}^{2}}+\frac{6\chi^{2}}{\left(\chi^{2}+d\bar{\mu}^{2}\right)^{2}}\right\rbrace\left(\frac{\partial \varphi}{\partial \chi}\right)^{-2}\, .\label{eq:CI13}
\end{equation}
Inserting
\begin{equation}
\frac{1}{M}\frac{\partial \varphi}{\partial \chi}=\frac{4\chi}{\chi^{2}+d\bar{\mu}^{2}}-\frac{2\bar{\mu}^{2}\chi}{\bar{\mu}^{2}\chi^{2}+c\bar{\mu}^{4}}\, ,\label{eq:CI14}
\end{equation}
one obtains
\begin{equation}
k^{2}\left(\varphi\right)=\frac{\left(B\chi^{2}+\left(B-6\right)d\bar{\mu}^{2}\right)\left(\chi^{2}+c\bar{\mu}^{2}\right)^{2}}{4\chi^{2}\left(\chi^{2}+\left(2c-d\right)\bar{\mu}^{2}\right)^{2}}\, ,\label{eq:CI15}
\end{equation}
where $\chi\left(\varphi\right)$ results from the inversion of eq.~(\ref{eq:CI12}). An inflationary epoch persists as long as $k^{2}\left(\varphi\right)\gg 1$. We recall that our description is valid only for a monotonic $V_{\text{E}}\left(\chi\right)$, or for a definite sign of $\partial\varphi/\partial \chi$. It breaks down for $\chi^{2}=\left(d-2c\right)\bar{\mu}^{2}$.

It is useful to define a dimensionless field variable \cite{CWCI} 
\begin{equation}
x=\frac{\chi^{2}}{d\bar{\mu}^{2}}=\frac{\chi^{2}}{m^{2}}, \hspace{10mm} \beta=\frac{c}{d}\, .\label{eq:CI16}
\end{equation}
In terms of this variable one has 
\begin{equation}
\frac{V_{\text{E}}}{M^{4}}=\frac{x+\beta}{d\left(x+1\right)^{2}}\, ,\label{eq:CI17}
\end{equation}
and 
\begin{equation}
\frac{\varphi}{M}=\ln\left(\frac{d\left(x+1\right)^{2}}{x+\beta}\right)\, .\label{eq:CI18}
\end{equation}
The kinetial reads
\begin{equation}
k^{2}=\frac{\left(Bx+B-6\right)\left(x+\beta\right)^{2}}{4x\left(x+2\beta-1\right)^{2}}\, .\label{eq:CI19}
\end{equation}
Inflation requires a range of $x$ for which $k^{2}\gg1$ in order to realize slow roll, and $V_{\text{E}}/M^{4}\ll 1$ in order to guarantee a small fluctuation amplitude. The slope of the fluctuation spectrum depends only on $B$ and $\beta$. One can then adjust $d$ in order to realize a realistic amplitude. We note that the constant term in the coefficient of the curvature scalar,
\begin{equation}
m^{2}=d\bar{\mu}^{2}\, ,\label{eq:CI20}
\end{equation}
and the scale $\bar{\mu}$ appearing in the scalar potential are presumably both relevant parameters. The values of these two intrinsic scales can differ substantially, such that large $d$ may find a natural explanation. A further natural suppression of $V_{\text{E}}/M^{4}$ arises for large $dx$ or $\chi^{2}\gg\bar{\mu}^{2}$. A small amplitude can be simply due to an end of inflation where $\chi^{2}/\bar{\mu}^{2}$ has already grown to large values.

For an investigation of the fluctuation spectrum we need the relation between $x$ and the number of e-foldings $N$ before the end of inflation (\ref{eq:C18})
\begin{eqnarray}
N\left(x\right)&=&\frac{1}{M}\int^{x_{\text{f}}}_{x} dx' k^{2}\left(x'\right)\frac{\partial \varphi}{\partial x'}\nonumber\\
&=&\int^{x_{\text{f}}}_{x}dx'k^{2}\left(x'\right)\left(\frac{2}{x'+1}-\frac{1}{x'+\beta}\right)\nonumber\\
&=&\int^{x_{\text{f}}}_{x}\frac{\left(Bx'+B-6\right)\left(x'+\beta\right)}{4x'\left(x'+1\right)\left(x'+2\beta-1\right)}\, .\label{eq:CI21}
\end{eqnarray}
There $x_{\text{f}}$ is determined by $k^{2}\left(x_{\text{f}}\right)=1/2$. Inverting the solution of eq.~(\ref{eq:CI21}), yields $k^{2}\left(N\right)$. The spectral parameters follow from eq.~(\ref{eq:C18F}).

Depending on $B$ and $\beta$ a large variety of inflationary models can be obtained. For our present purposes we only describe two limits.
For the first we take $\beta=1/2$. For $x_{\text{f}}$ this yields
\begin{equation}
\left(Bx_{\text{f}}+B-6\right)\left(x_{\text{f}}+\frac{1}{2}\right)^{2}=2x_{\text{f}}^{3}\, ,\label{eq:CI22}
\end{equation}
and eq.~(\ref{eq:CI21}) becomes
\begin{equation}
N\left(x\right)=\int^{x_{\text{f}}}_{x}dx'\frac{\left(Bx'+B-6\right)\left(x'+\frac{1}{2}\right)}{4x'^{2}\left(x'+1\right)}\, .\label{eq:CI23}
\end{equation}
The dominant contribution to the integral for $N\left(x\right)$ comes from the region of small $x$. If we neglect the variation of $B$ in this region we can approximate 
\begin{equation}
N\left(x\right)=\frac{B-6}{8}\int^{x_{\text{f}}}_{x}dx'\left(x'\right)^{-2}=\frac{B-6}{8x}-\bar{N}_{\text{f}}\, .\label{eq:CI24}
\end{equation}
In this approximation one finds for the kinetial
\begin{equation}
k^{2}=\frac{\left(Bx+B-6\right)\left(x+\frac{1}{2}\right)^{2}}{4x^{3}}\approx\frac{32\left(N+\bar{N}_{\text{f}}\right)^{3}}{\left(B-6\right)^{2}}\, .\label{eq:CI25}
\end{equation}
This yields 
\begin{equation}
r=\frac{\left(B-6\right)^{2}}{4\left(N+\bar{N}_{\text{f}}\right)^{3}}\, ,\hspace{10mm}
n=1-\frac{3}{N+\bar{N}_{\text{f}}}-\frac{r}{8}\, .\label{eq:CI26}
\end{equation}

As another limit we discuss the case where the relevant epoch during inflation occurs for large $\chi^{2}/\bar{\mu}^{2}$ such that $x\gg1$, $x\gg\beta$, $Bx\gg B-6$. In this limit one simply has 
\begin{equation}
k^{2}=\frac{B}{4}\, .\label{eq:CI27}
\end{equation}
The dynamics of inflation depends now explicitly on the flow of $B\left(\chi^{2}/\bar{\mu}^{2}\right)$. With $y=\chi^{2}/\bar{\mu}^{2}$ and
\begin{equation}
\frac{V_{\text{E}}}{M^{4}}=\frac{\bar{\mu}^{2}}{\chi^{2}}=\frac{1}{y}\com \frac{\varphi}{M}=\ln\frac{\chi^{2}}{\bar{\mu}^{2}}=\ln y\, , \label{eq:CI28}
\end{equation}
one has 
\begin{equation}
N=\frac{1}{4}\int^{y_{\text{f}}}_{y}dy'\frac{B\left(y'\right)}{y'}\, .\label{eq:CI29}
\end{equation}
Details depend on the shape of $B\left(y\right)$. If $B\left(y\right)$ has a plateau for $y>y_{\text{f}}$, which drops to small values $B<2$ rather steeply at $y_{\text{f}}$
\begin{equation}
B=\bar{B}\Theta\left(y_{\text{f}}-y\right)\, ,\label{eq:CI30}
\end{equation}
one obtains
\begin{equation}
N=\frac{\bar{B}}{4}\ln\frac{y_{\text{f}}}{y}\, .\label{eq:CI31}
\end{equation}
With $k^{2}=\bar{B}/4$ independent of $N$ eq.~(\ref{eq:C18F}) implies $n=1-\frac{r}{8}$. This is not compatible with observation.

Another possibility assumes that in the range relevant for the observed fluctuations $B$ behaves as \cite{CWIQ} 
\begin{equation}
B=\left(\frac{\bar{m}}{\chi}\right)^{\sigma}=b\left(\frac{\bar{\mu}}{\chi}\right)^{\sigma}=by^{-\frac{\sigma}{2}}\, ,\label{eq:CI32}
\end{equation}
with 
\begin{equation}
\bar{m}=b^{\frac{1}{\sigma}}\bar{\mu}\, ,\label{eq:CI33}
\end{equation}
a characteristic intrinsic scale where $B$ crosses from the UV-behavior for $\chi\rightarrow0$ to the IR-behavior for $\chi\rightarrow\infty$. According to eq.~(\ref{eq:CI32}) inflation would end for $B=2$. We may, however, restrict the validity of \eq{CI32} to a certain range of large $B$, say $B>\bar{B}$ and assume that this value is relevant for $\bar{N}$ e-foldings before the end of inflation. From eq.~(\ref{eq:CI29}) we can compute
\begin{equation}
N-\bar{N}=\frac{b}{4}\int_{y}^{\bar{y}}dy y^{-\left(1+\frac{\sigma}{2}\right)}=\frac{b}{2\sigma}\left(y^{-\frac{\sigma}{2}}-\frac{\bar{B}}{b}\right)\, ,\label{eq:CI34}
\end{equation}
with 
\begin{equation}
B\left(\bar{y}\right)=\bar{B}\com \bar{y}=\left(\frac{\bar{B}}{b}\right)^{-\frac{2}{\sigma}}\, .\label{eq:CI35}
\end{equation}
In turn, this yields
\begin{equation}
B=2\sigma\left(N-\bar{N}\right)+\bar{B}\, ,\label{eq:CI36}
\end{equation}
or 
\begin{equation}
\epsilon=\frac{2}{2\sigma\left(N-\bar{N}\right)+\bar{B}}\, .\label{eq:CI37}
\end{equation}

One finds, similar to ref.\cite{CWIQ},
\begin{equation}
r=\frac{32}{2\sigma\left(N-\bar{N}\right)+\bar{B}}\label{eq:CI38}
\end{equation}
and
\begin{equation}
n=1-\frac{2\sigma+4}{2\sigma\left(N-\bar{N}\right)+\bar{B}}\, .\label{eq:CI39}
\end{equation}
This implies a relation between $n$ and $r$ that is independent of $\bar{B}$ and only depends on $\sigma$
\begin{equation}
n=1-\frac{\sigma+2}{16}r\, .\label{eq:CI40}
\end{equation}
We may define $\sigma$ by the running of $B$, evaluated at the field that corresponds to $N$,
\begin{equation}
\sigma\left(N\right)=\frac{\partial\ln B}{\partial \bar{m}}\vert_{N} =\frac{\partial \ln B}{\partial \ln \bar{\mu}}\vert_{N}\, .\label{eq:CI41}
\end{equation}
For $1-n=0.03$ one has $r\approx 1/\left(2\sigma +4\right)$ such that $\sigma > 3$ is needed for $r<0.1$. 
Spectral properties compatible with present observational bounds can be obtained for a suitable flow of $B\left(y\right)$ between $y_{\text{f}}$ and $\bar{y}$. 

A small amplitude of the primordial fluctuations obtains rather naturally if $\chi^{2}\left(N\right)\gg \bar{\mu}^{2}$ or $y\left(N\right)\gg 1$. With eqs.~(\ref{eq:C19}),~(\ref{eq:CI28}) the amplitude is proportional to 
\begin{equation}
\frac{V_{\text{E}}}{M^{4}}=\frac{1}{y\left(N\right)}=\left(\frac{B}{b}\right)^{\frac{2}{\sigma}}=B\left(N\right)^{\frac{2}{\sigma}}\frac{\bar{\mu}^{2}}{\bar{m}^{2}}\, ,\label{eq:CI42}
\end{equation}
or
\begin{equation}
\Delta^{2}=\frac{1}{24\pi^{2}\epsilon\left(N\right)y\left(N\right)}=\frac{1}{12\pi^{2}}B\left(N\right)^{1+\frac{2}{\sigma}}\frac{\bar{\mu}^{2}}{\bar{m}^{2}}\, .\label{eq:CI43}
\end{equation}
It is sufficient that the crossover scale $\bar{m}$ in the flow of the kinetic term is sufficiently large as compared to the characteristic scale $\bar{\mu}$ for the potential.

For cosmon inflation the transition between a possible early stage of Starobinski-type inflation (\ref{eq:CI7}) to a late stage of cosmon-dominated inflation (\ref{eq:CI10}) depends on the particular cosmological solution. A family of solutions may be labeled by an ``initial value'' $\chi_{in}=\chi(\eta_{in})$, with ``initial conformal time'' $\eta_{in}$ taking some very large negative value. (The solution can be extended to arbitrary large negative $\eta_{in}$, with $\chi_{in}\raw 0$.) For $\chi_{in}=0$ one has $\chi(\eta)=0$ also for all later $\eta$, since the potential derivative \eqref{eq:CI9} vanishes for $\chi=0$. Any non-zero $\chi_{in}$ will lead to an increase of $\chi(\eta)$ for later $\eta$ until at some transition time $\eta_{tr}$ one has $\chi^{2}(\eta_{tr})\approx C\phi(\eta_{tr)}$ or $\bar{\mu}^{2}\chi^{2}(\eta_{tr})\approx C\phi^{2}(\eta_{tr})$. At this transition time cosmology changes from Starobinski inflation to one of the inflationary scenarios corresponding to the potential \eqref{eq:CI10}. The amplitude and spectrum of the primordial fluctuations depend on $\eta_{tr}$. If $\eta_{tr}$ happens to be later than $60$ e-foldings before the end of inflation, the predictions for the fluctuations are the same as for Starobinski inflation. On the other hand, if $\eta_{tr}$ is before $60$ e-foldings before the end of inflation, the spectrum is determined by one of the models corresponding to the potential \eqref{eq:CI10}. There is a continuous interpolation between the two limits.

The transition time $\eta_{tr}$ depends on the particular solution or $\chi(\eta_{in})$. For a given $\chi(\eta_{in})$ the transition time is earlier for smaller $C$. Indeed, the transition can happen in rather early stages of inflation if $C$ is not large \cite{CWIQ}. Depending on the choice of parameters and initial conditions a rather rich variety of inflation models can be accommodated within the general setting of cosmon inflation. It is not the purpose of this report to discuss inflationary models in detail. A quantum gravity computation may be able to determine some of the parameters of cosmon inflation since the number of relevant couplings at the UV-fixed point is typically smaller than the number of parameters used here for cosmon inflation. This demonstrates the potential predictive power of quantum gravity for inflation.

\subsection{Higgs inflation}\label{sec:Higgs_Infl} 

For Higgs inflation \cite{BS1,IRST,EGR,BKS,BS2,GBFR2,BGS2,SIHEWI,BEMAS,BAES,
BLT,POCA,BGS3,BMSS,BKKSS,ASA,HAKO,FGH,BS4,BRS2,JR3,RERU,FUPO,EEN,GMP,EJMN,FUMA,RAWA,BPR3,EERT}, for a recent review see \cite{JR1}, the relevant parts of the quantum effective action are given by
\begin{eqnarray}
\mathcal{L}&=&-\frac{1}{2}\left({M}^{2}+\xi_{H}\left(\frac{h^{\dagger}h}{{M}^{2}}\right)\right)R+\frac{1}{2}\lambda_{H}\left(\frac{h^{\dagger}h}{{M}^{2}}\right)\left(h^{\dagger}h\right)^{2}\nonumber\\
&+&D_{\mu}h^{\dagger}D^{\mu}h\, .\label{eq:HI1}
\end{eqnarray}
Here ${M}$ is the reduced Planck mass and one often neglects the flow of $\xi (h^{\dagger}h/{M}^{2})$. The quartic Higgs coupling $\lambda_{H}$ depends on $(h^{\dagger}h/{M}^{2})$ through the perturbative $\beta$-function of the standard model. A possible contribution of gravitational fluctuations to the flow of $\lambda_{H}$ is often neglected, but may be relevant \cite{IRST,BRMS}. The Planck mass ${M}$ is the only intrinsic mass scale for Higgs inflation. If one replaces ${M}$ by a scalar field one obtains a scale invariant version of Higgs inflation \cite{GRSZ}, which will be discussed in sec.~\ref{sec:Scale_Inv_Inf}. The effective action \eqref{eq:HI1} with ${M}$ replaced by $\chi$ has been introduced in ref. \cite{CWQ}.

Higgs inflation occurs for a range of the Higgs field $h$ where $\xi_{H}h^{\dagger}h\gg {M^{2}}$. The scale violation $\sim {M}^{2}R$ in the first term in eq.~\eqref{eq:HI1} becomes small in this region. In the limit of constant $\xi_{H}$ and $\lambda_{H}$ the effective action \eqref{eq:HI1} becomes scale invariant in this limit. This is the origin of the almost scale invariant primordial fluctuation spectrum. Inflation ends when the violation of scale symmetry becomes substantial for $\xi_{H}h^{\dagger}h\approx {M}^{2}$. Only the physical Higgs scalar (radial mode) in the Higgs-doublet matters. The Goldstone bosons are absorbed by the Higgs mechanism. We will replace in the following $h^{\dagger}h$ by $h^{2}$ and $D^{\mu}h^{\dagger}D_{\mu}h$ by $\partial^{\mu}h\partial_{\mu}h$, with $h$ a real scalar field.

In the Einstein frame the potential reads
\begin{equation}
\frac{V_{\text{E}}}{M^{4}}=\frac{\lambda_{H}\tilde{h}^{4}}{2\left(1+\xi_{H}\tilde{h}^{2}\right)^{2}}, \hspace{10mm} \tilde{h}=\frac{h}{M}\, , \label{eq:HI2}
\end{equation}
where $\lambda_{H}$, $\xi_{H}$ depend on the dimensionless field $\tilde{h}$. For $\tilde{h}\rightarrow \infty$ the potential develops a flat plateau 
\begin{equation}
\frac{V_{\text{E}}}{M^{4}}\left(\tilde{h}\rightarrow\infty\right)\raw\frac{\lambda_{H}\left(\tilde{h}\right)}{2\xi_{H}^{2}\left(\tilde{h}\right)}\left (1-\dfrac{2}{\xi_{H}\tilde{h}^{2}}\right )\raw \dfrac{R_{H}}{2}\, ,\label{eq:HI3}
\end{equation}
with
\begin{equation}
R_{H}=\dfrac{\lambda_{H}}{\xi_{H}^{2}}\, .
\end{equation}
For the kinetic term one obtains 
\begin{equation}
\mathcal{L}_{\text{kin}}=M^{2}\frac{1+\xi_{H}\tilde{h}^{2}+a_{H}^{2}\xi_{H}^{2}\tilde{h}^{2}}{\left(1+\xi_{H}\tilde{h}^{2}\right)^{2}}\partial^{\mu}\tilde{h}\partial_{\mu}\tilde{h}\, ,\label{eq:HI4}
\end{equation}
with
\begin{equation}
a_{H}^{2}=3\left(1+\frac{1}{2}\frac{\partial \ln\xi_{H}}{\partial \ln \tilde{h}}\right)^{2}\, .\label{eq:HI5}
\end{equation}

 We assume that $V_{E}/M^{4}$ is a monotonic function of $k$ and employ again the field $\varphi$ with normalization \eqref{eq:C5}. With
\begin{equation}
\dfrac{\p\varphi}{\p\tilde{h}}=-\dfrac{M\xi_{H}\tilde{h}}{1+\xi_{H}\tilde{h}^{2}}B_{H}\, ,
\end{equation}
the kinetial in eq.~\eqref{eq:C6} becomes
 \begin{equation}\label{eq:HI6}
k^{2}\left(\varphi\right)=\dfrac{2C_{H}}{B_{H}^{2}}\, ,
\end{equation}
with 
\begin{equation}\label{eq:HI7}
C_{H}=3\left (1+\dfrac{1}{2}\dfrac{\p\ln(\xi_{H})}{\p\ln(\tilde{h})}\right )^{2}+\dfrac{1+\xi_{H}\tilde{h}^{2}}{\xi_{H}^{2}\tilde{h}^{2}}
\end{equation}
and
\begin{equation}\label{eq:HI8}
B_{H}=\dfrac{\p\ln(R_{H})}{\p\ln(\tilde{h})}+\dfrac{4}{\xi_{H}\tilde{h}^{2}}\left (1+\dfrac{1}{4}\dfrac{\p\ln(\lambda_{H})}{\p\ln(\tilde{h})}\right )\, .
\end{equation}

In the limit of large $\xi_{H}\tilde{h}^{2}$ an important quantity is the running of the ratio $R_{H}=\lambda_{H}/\xi_{H}^{2}$. In previous approaches the contribution of gravitational fluctuations is neglected, $\xi_{H}$ is taken constant, and $\lambda_{H}$ follows the perturbative flow induced by the standard model of particles. In general, gravitational contributions matter, however, and we will discuss this issue below. For a first approach we take scale independent $\lambda_{H}$ and $\xi_{H}$, as assumed to be given by values at some UV-fixed point. In this limit the kinetial simplifies
\begin{equation}\label{eq:HI9} 
k^{2}=\dfrac{\xi_{H}(3\xi_{H}+1)\tilde{h}^{4}}{8}\, .
\end{equation}
The relation between $\varphi$ und $\tilde{h}$ becomes
\begin{equation}\label{eq:HI10} 
\dfrac{\p\varphi}{\p \tilde{h}}=-\dfrac{4M}{\xi_{H}\tilde{h}^{3}}\, .
\end{equation}
Inflation ends for $k^{2}=1/2$ or
\begin{equation}\label{eq:HI11} 
\tilde{h}^{4}_{f}=\dfrac{4}{\xi_{H}(3\xi_{H}+1)}\, .
\end{equation}
We can relate the value of $\tilde{h}$ with the number $N$ of $e$-foldings before the end of inflation by \eq{C18},
\begin{equation}\label{eq:HI12}
N=-\dfrac{1}{2}(3\xi_{H}+1)\int_{\tilde{h}}^{\tilde{h}_{f}}\dif\tilde{h}^{\prime}\, \tilde{h}^{\prime}=\dfrac{1}{4}(3\xi_{H}+1)\tilde{h}^{2}-\overline{N}_{f}\, ,
\end{equation}
with
\begin{equation}\label{eq:HI13}
\overline{N}_{f}=\dfrac{1}{4}(3\xi_{H}+1)\tilde{h}^{2}_{f}=\dfrac{1}{2}\sqrt{\dfrac{3\xi_{H}+1}{\xi_{H}}}\, .
\end{equation}
Inserting $\tilde{h}^{2}=4(N+\overline{N}_{f})/(3\xi_{H}+1)$ yields
\begin{equation}\label{eq:HI14} 
k^{2}(N)=\dfrac{2\xi_{H}}{3\xi_{H}+1}(N+\overline{N}_{f})^{2}\, .
\end{equation}

One infers for the spectral parameters
\begin{equation}\label{eq:HI15} 
r=\dfrac{4(3\xi_{H}+1)}{\xi_{H}(N+\overline{N}_{f})^{2}}
\end{equation}
and
\begin{equation}\label{eq:HI16} 
n=1-\dfrac{2}{N+\overline{N}_{f}}-\dfrac{r}{8}\, .
\end{equation}
The spectral parameters become very close to Starobinski inflation with approximate values \eqref{eq:C31}, \eqref{eq:C33}. Indeed, the expression for $\varepsilon(N)$ is the same up to a factor $1+1/(3\xi_{H})$, and up to the precise meaning of $\overline{N}_{f}$. (For the subleading corrections $\sim \overline{N}_{f}/N$ one would need an improvement of the rough estimate \eqref{eq:HI13}.) The fluctuation amplitude
\begin{equation}\label{eq:HI17} 
\Delta^{2}\approx \dfrac{\lambda_{H}(N+\overline{N}_{f})^{2}}{12\pi^{2}\xi_{H}(3\xi_{H}+1)}\approx 2\cdot 10^{-9}
\end{equation}
requires a very small value of $\lambda_{H}/\xi_{H}^{2}\approx 2\cdot 10^{-10}$.

Another interesting limit takes $\xi_{H}\raw 0$, with $\p\ln(\xi_{H})/\p\ln(\tilde{h})=0$. In this limit one finds
\begin{equation}\label{eq:HI9A}
B_{H}=\dfrac{4+\dfrac{\p\ln(\lambda_{H})}{\p\ln(\tilde{h})}}{\xi_{H}\tilde{h}^{2}}\com C_{H}=\dfrac{1}{\xi_{H}^{2}\tilde{h}^{2}}\, .
\end{equation}
and therefore a kinetial
\begin{equation}\label{eq:HI10A}
k^{2}(\varphi)=\dfrac{2\tilde{h}^{2}}{\left (4+\dfrac{\p\ln(\lambda_{H})}{\p\ln(\tilde{h})}\right )^{2}}\, .
\end{equation}
Since no Weyl scaling is involved for $\xi_{H}=0$, the kinetial is directly obtained from the relation between $\tilde{h}$ and $\varphi$
\begin{equation}\label{eq:HI11A} 
k^{2}(\varphi)=2M^{2}\left (\dfrac{\p\tilde{h}}{\p \varphi}\right )^{2}\com \dfrac{\p\varphi}{\p\tilde{h}}=-\dfrac{M}{\tilde{h}}\left (4+\dfrac{\p\ln(\lambda_{H})}{\p\ln(\tilde{h})}\right )\, .
\end{equation}
The dynamics of a possible inflationary phase depends on the function $\lambda_{H}(\tilde{h})$. In the limit of constant $\lambda_{H}$ one would have 
\begin{equation}\label{eq:HI12A} 
N+\overline{N}_{f}=\dfrac{\tilde{h}^{2}}{4}\, ,
\end{equation}
such that the corresponding spectral parameters,
\begin{equation}\label{eq:HI13A} 
r=\dfrac{16}{N+\overline{N}_{f}}\com n=1-\dfrac{3}{N+\overline{N}_{f}}\, ,
\end{equation}
are not compatible with observation.

An increase of the effective potential $U\sim h^{4}$ for $h\raw \infty$ contradicts, however, the graviton barrier discussed in sect.~\ref{sec:V}. In the range $\tilde{h}^{2}\gg 1$ the graviton fluctuations dominate the behavior of the Higgs-potential $U(\rho)$, $\rho=h^{2}/2$. As we have discussed in sect.~\ref{sec:6.17} they tend to make the potential flat, $\p^{2}U/\p\rho^{2}\raw 0$, $\p U/\p\rho\raw 0$. A reasonable approximation for the range of large $\tilde{h}$ may be
\begin{equation}\label{eq:HI14A}
U=\mu_{H}^{4}-\dfrac{\nu_{H}}{h^{2}}\, ,
\end{equation}
corresponding to
\begin{equation}\label{eq:HI15A}
\lambda_{H}=\dfrac{2\mu_{H}^{4}}{M^{4}}\tilde{h}^{-4}-\dfrac{2\nu_{H}}{M^{6}}\tilde{h}^{-6}\, .
\end{equation}
The parameters $\mu_{H}$ and $\nu_{H}$ violate scale symmetry. They should be such that the potential smoothly matches the perturbative potential of the standard model for $\tilde{h}^{2}\approx 1$. Typically, the dimensionless ratios
\begin{equation}\label{eq:HI16A} 
u_{H}=\dfrac{2\mu_{H}^{4}}{M^{4}}\com v_{H}=\dfrac{2\nu_{H}}{M^{6}}
\end{equation}
are very small.

With \eq{HI15A} one finds for $\xi_{H}\raw 0$ the kinetial
\begin{equation}\label{eq:HI17A} 
k^{2}(\varphi)=\dfrac{\tilde{h}^{2}(u_{H}\tilde{h}^{2}-v_{H})^{2}}{2v_{H}^{2}}=\dfrac{\tilde{h}^{2}(\tilde{h}^{2}-\gamma_{H})^{2}}{2\gamma_{H}^{2}}\, ,
\end{equation}
with
\begin{equation}\label{eq:HI18} 
\gamma_{H}=\dfrac{v_{H}}{u_{H}}=\dfrac{\nu_{H}}{\mu_{H}^{4}M^{4}}\, .
\end{equation}
This yields
\begin{equation}\label{eq:HI19} 
N+\overline{N}_{f}=\dfrac{1}{4\gamma_{H}}\left (\tilde{h}^{4}-2\gamma_{H}\tilde{h}^{2}\right )
\end{equation}
and therefore
\begin{equation}\label{eq:HI20} 
r=2\sqrt{\gamma_{H}}\overline{N}^{-\frac{3}{2}}\left (1+\dfrac{1}{2}\sqrt{\dfrac{\gamma_{H}}{\overline{N}}}\right )^{-1}\, ,
\end{equation}
with
\begin{equation}\label{eq:HI21} 
\overline{N}=N+\overline{N}_{f}+\dfrac{\gamma_{H}}{4}\, .
\end{equation}
For $\gamma_{H}\lesssim 1$ the tensor amplitude is small. The spectral index,
\begin{equation}\label{eq:HI22} 
n=1-\dfrac{3}{2\overline{N}}\, ,
\end{equation} 
seems compatible with observation. For a realistic amplitude,
\begin{equation}\label{eq:HI23} 
\Delta^{2}=\dfrac{2u_{H}}{\sqrt{\gamma_{H}}}\overline{N}^{\frac{3}{2}}=\dfrac{4\mu_{H}^{6}\overline{N}^{\frac{3}{2}}}{\sqrt{\nu_{H}}M^{3}}\approx 2\cdot 10^{-9}\, ,
\end{equation}
one needs $\mu_{H}$ sufficiently small as compared to $M$. 

This short discussion demonstrates that a realistic Higgs inflation does not necessarily require a value $\xi_{H}\gg 1$. It reveals, however, that an understanding of the role of gravitational fluctuations is crucial. They determine the behavior of the kinetial for large $\tilde{h}^{2}$. For large ranges of parameters they are important for the values of $\tilde{h}^{2}$ relevant for inflation. It seems possible that close to an UV-fixed point both $\lambda_{H}(\tilde{h}^{2})$ and $\xi_{H}(\tilde{h}^{2})$ correspond to irrelevant parameters and are therefore predictable.

\subsection{Scale invariant inflation}\label{sec:Scale_Inv_Inf}

For scale invariant inflation no intrinsic mass scale plays a role during the inflationary epoch. Quantum scale invariance can be considered as an exact symmetry. If a single scalar field has a non-vanishing cosmological value, it is a Goldstone boson. Its evolution settles early to a constant value. Inflation, if realized, does not end for a scale invariant model with a single scalar field. One therefore needs at least two physical scalar degrees of freedom. An example is ``scale invariant Starobinski inflation''. The constant Plank mass for Starobinski inflation is replaced by a scalar field $\chi$. An alternative is scale invariant Higgs inflation \cite{GRSZ,BKRS1,GRS,KR,CPR,JR1}, or the introduction of an independent inflaton field \cite{FHR1,FHR2,FHNR}.

\subsubsection{Scale invariant Starobinski inflation}

Scale invariant Starobinski inflation starts from the simple scale invariant effective action \eqref{eq:139}. Within the more general setting of cosmon inflation it is realized if for some epoch the values of $\phi$ and $\chi$ are such that all terms involving $\bar{\mu}^{2}$ can be neglected in \eqs{CI1}{CI2}. In the Einstein frame the effective action for scale invariant Starobinski inflation is given by eqs.~\eqref{eq:159}-\eqref{eq:161}. We can replace the field $\phi$ by $\varphi$
\begin{align}\label{eq:SI1} 
\varphi&=-M\ln\left (\dfrac{V_{E}}{M^{4}}\right )\nn\\
&=2M\left (\ln(\chi^{2}+2C\phi)-\ln(\phi)-\dfrac{1}{2}\ln\left (\dfrac{C}{2}\right )\right )\, ,
\end{align}
or
\begin{equation}\label{eq:SI2} 
\p_{\mu}\phi=\dfrac{2\phi}{\chi}\p_{\mu}\chi-\dfrac{\phi(\chi^{2}+2C\phi)}{2M\chi^{2}}\p_{\mu}\varphi\, .
\end{equation}
The kinetic term becomes
\begin{widetext}
\begin{align}\label{eq:SI3} 
\cL_{kin}&=\dfrac{M^{2}}{2}\biggl \{\left (B+\dfrac{12C\phi}{\chi^{2}}\right )\left (\chi^{2}+2C\phi\right )^{-1}\p^{\mu}\chi\p_{\mu}\chi -\dfrac{6C\phi}{M\chi^{3}}\p^{\mu}\chi\p_{\mu}\varphi+\dfrac{3C^{2}\phi^{2}}{2M^{2}\chi^{4}}\p^{\mu}\varphi\p_{\mu}\varphi\biggl \}\nn\\
&=\dfrac{1}{2}\biggl \{\dfrac{M^{2}}{\chi^{2}}\left [(B-6)\left (1-\sqrt{8C}\exp\left (-\dfrac{\varphi}{2M}\right )\right )+6\right ]\p^{\mu}\chi\p_{\mu}\chi +3C\exp\left (-\dfrac{\varphi}{M}\right )\left (1-\sqrt{8C}\exp\left (-\dfrac{\varphi}{2M}\right )\right )^{-2}\p^{\mu}\varphi\p_{\mu}\varphi\nn\\
&\quad-\dfrac{3\sqrt{8C}M}{\chi}\exp\left (-\dfrac{\varphi}{2M}\right )\left (1-\sqrt{8C}\exp\left (-\dfrac{\varphi}{2M}\right )\right )^{-1}\p^{\mu}\chi\p_{\mu}\varphi\biggl \} \, .
\end{align}
\end{widetext}
It can also be obtained directly from \eq{148}, using the relation between $\omega$ and $\chi$, $\varphi$,
\begin{align}\label{eq:SI4} 
\dfrac{V_{E}}{M^{4}}&=\exp\left (-\dfrac{\varphi}{M}\right )=\dfrac{1}{8C}\left (1-\dfrac{\chi^{2}\omega^{2}}{M^{2}}\right )^{2}\, ,\nn\\
\omega^{2}&=\dfrac{M^{2}}{\chi^{2}}\left (1-\sqrt{8C}\exp\left (-\dfrac{\varphi}{2M}\right )\right )\, .
\end{align}

The kinetic term can be brought to a diagonal form
\begin{equation}\label{eq:SI5} 
\cL=\dfrac{1}{2}\biggl \{k^{2}(\varphi)\p^{\mu}\varphi\p_{\mu}\varphi+A(\varphi)\p^{\mu}\tau\p_{\mu}\tau\biggl \}\, ,
\end{equation}
replacing $\chi$ by $\tau$,
\begin{equation}\label{eq:SI6} 
\ln\left (\dfrac{\chi}{M}\right )=f(\varphi,\tau)\, .
\end{equation}
Indeed, the mixed term $\sim \p^{\mu}\varphi\p_{\mu}\tau$ is eliminated if
\begin{equation}\label{eq:SI7} 
M\dfrac{\p f}{\p\varphi}=\dfrac{3(1-z)}{2z[(B-6)z+6]}\, ,
\end{equation}
where we employ the shorthand
\begin{equation}\label{eq:SI8} 
z(\varphi)=1-\sqrt{8C}\exp\left (-\dfrac{\varphi}{2M}\right )=\dfrac{\chi^{2}}{\chi^{2}+2C\phi}\, ,
\end{equation}
and the relation
\begin{equation}\label{eq:SI9} 
\dfrac{1}{\chi}\p_{\mu}\chi=\dfrac{\p f}{\p \varphi}\p_{\mu}\varphi+\dfrac{\p f}{\p\tau}\p_{\mu}\tau\, .
\end{equation}
Inserting \eqs{SI5}{SI7} one obtains for the kinetial for $\varphi$
\begin{equation}\label{eq:SI10} 
k^{2}(\varphi)=\dfrac{3(B-6)(1-z)^{2}}{8z[(B-6)z+6]}\, .
\end{equation}
For $(B-6)z\gg 6$ one recovers the limit of Starobinski inflation \eqref{eq:C10},
\begin{equation}\label{eq:SI11} 
k^{2}(\varphi)=\dfrac{3(1-z)^{2}}{8z^{2}}\, .
\end{equation}

The field $\tau$ has only derivative couplings and we identify it with the Goldstone boson. The coefficient $A(\varphi)$ for the kinetic term for $\tau$ reads
\begin{equation}\label{eq:SI12} 
A^{2}(\varphi)=M^{2}[(B-6)z+6]\left (\dfrac{\p f}{\p\tau}\right )^{2}\, .
\end{equation}
One may choose
\begin{equation}\label{eq:SI13} 
f(\varphi,\tau)=f_{0}(\varphi)+\dfrac{1}{\sqrt{6}M}\tau\, ,
\end{equation}
such that
\begin{equation}\label{eq:SI14} 
A^{2}(\varphi)=1+\dfrac{B-6}{6}z\, ,
\end{equation}
which is canonical for $(B-6)z\ll 6$. We observe that the field equations for $\tau$ always have a solution for constant $\tau=\tau_{0}$. Once $\tau$ has settled at $\tau_{0}$ it can be neglected. Only in possible early stages, when $\p_{\mu}\tau\neq 0$, the evolution of $\tau$ can have an influence on the evolution of $\varphi$ due to a term $\sim (\p A/\p\varphi)\p^{\mu}\tau\p_{\mu}\tau$ in the field equation for $\varphi$. In the following we assume that $\tau$ has reached a constant value for the inflationary stages relevant to observation, and neglect its evolution. 

The presence of the cosmon field modifies the inflationary dynamics by the mixing of the kinetic terms. The kinetial \eqref{eq:SI10} differs from the one of pure Starobinski inflation. Large values of $k^{2}(\varphi)$ correspond to inflationary epochs. We concentrate first on $(B-6\geq 0,z\geq 0)$
\begin{align}\label{eq:SI15} 
\chi^{2}&\ll |2C\phi|\com z\approx 0\com k^{2}\approx\dfrac{B-6}{16z}\, ,\nn\\
\dfrac{V_{E}}{M^{4}}&=\exp\left (-\dfrac{\varphi}{M}\right )\approx\dfrac{1}{8C}\, .
\end{align}
This may correspond to the immediate vicinity of the UV-fixed point where scale symmetry is associated to massless particles, $\chi\raw 0$. As $\chi$ and $z$ increase, there may be later a transition to Starobinski inflation once $z>6/(B-6)$, or to cosmon inflation as discussed in sect.~\ref{sec:Starobinski_Inflation}, if an intrinsic mass scale $\bar{\mu}$ starts to play a role. In the second case much smaller values of $C$ are possible, since $z$ may be close to one at the time when the observable fluctuations freeze. Without the effect of intrinsic mass scales one has for $\chi^{2}\gg |2C\phi|$, $z\approx 1$,
\begin{equation}\label{eq:SI16} 
\dfrac{V_{E}}{M^{4}}=\dfrac{(1-z)^{2}}{8C}\, ,
\end{equation}
and
\begin{equation}\label{eq:SI17} 
k^{2}(z\raw 1)=\dfrac{3(B-6)(1-z)^{2}}{8B}\, .
\end{equation}
For $B>6$ the kinetial \eqref{eq:SI17} is not large. We conclude that for $C$ substantially smaller than the value \eqref{eq:C35} the intrinsic mass scale $\bar{\mu}$ has to play a role for realistic cosmon inflation. 

\subsubsection{Scale invariant Higgs inflation}\label{sec:Scale_Inv_Higgs_Infl} 

For scale invariant Higgs inflation \cite{GRSZ,BKRS1,GRS,KR,CPR,JR1,RUBS1,SAL} the effective action is given by
\begin{align}\label{eq:TI1} 
\cL&=-\dfrac{1}{2}\left (\chi^{2}+\xi_{H}(\tilde{h})\, h^{\dagger}h\right )R+\dfrac{1}{2}\lambda_{H}(\tilde{h})(h^{\dagger}h-\varepsilon(\tilde{h})\chi^{2})^{2}\nn\\
&\quad +D^{\mu}h^{\dagger}D_{\mu}h+\dfrac{1}{2}(B(\tilde{h})-6)\p^{\mu}\chi\p_{\mu}\chi\, ,
\end{align}
with
\begin{equation}\label{eq:TI2} 
\tilde{h}^{2}=\dfrac{h^{\dagger}h}{\chi^{2}}\, .
\end{equation}
For the present stage of the Universe $\varepsilon(\tilde{h})$ is tiny and the term $\sim \xi_{H}h^{\dagger}h$ is negligible. This may be different for the inflationary stage in early cosmology. Omitting gravitational fluctuations the running of $\varepsilon$ with $\tilde{h}$ is small. It is given by \eq{43} with $\mu$ replaced by $\tilde{h}$. We will assume here that even in presence of gravitational fluctuations $\varepsilon(\tilde{h})$ remains small enough in the range of $\tilde{h}$ relevant for inflation, such that terms involving $\varepsilon(\tilde{h})$ can be neglected. (See ref.~\cite{RUBCW} for a discussion of a scenario where $\varepsilon(\tilde{h})$ becomes important in early cosmology.)

In the Einstein frame the potential is given by eq.~\eqref{eq:HI2}. With
\begin{equation}\label{eq:TI3} 
\omega^{2}=\dfrac{M^{2}}{\chi^{2}(1+\xi_{H}\tilde{h}^{2})}\, ,
\end{equation}
the kinetic term \eqref{eq:HI4} is supplemented by additional derivatives of $\chi$,
\begin{align}\label{eq:TI4} 
&\cL_{kin}=M^{2}\biggl \{\dfrac{1+\xi_{H}\tilde{h}^{2}+a_{H}^{2}\xi_{H}^{2}\tilde{h}^{2}}{(1+\xi_{H}\tilde{h}^{2})^{2}}\p^{\mu}\tilde{h}\p_{\mu}\tilde{h}\\
&+\dfrac{1}{2\chi^{2}}\left (\dfrac{B-6+2\tilde{h}^{2}}{1+\xi_{H}\tilde{h}^{2}}+6\right )\p^{\mu}\chi\p_{\mu}\chi\nn\\
&+\dfrac{2\tilde{h}}{\chi(1+\xi_{H}\tilde{h}^{2})}\left [1+3\xi_{H}\left (1+\dfrac{1}{2}\dfrac{\p\ln(\xi_{H})}{\p\ln(\tilde{h})}\right )\right ]\p^{\mu}\chi\p_{\mu}\tilde{h}\biggl \}\nn\, .
\end{align}
The relation between $\varphi$ and $\tilde{h}$ is the same as for Higgs inflation as discussed in sect.~\ref{sec:Cosmon_Infl}, now with $\tilde{h}=h/\chi$ instead of $\tilde{h}=h/M$. The only effect of the field $\chi$ arises from the diagonalization of the kinetic term. For a constant $\chi=M$ one recovers Higgs inflation as discussed in sect.~\ref{sec:Cosmon_Infl}.

The diagonalization of the kinetic term is achieved by defining similar to \eqs{SI6}{SI13}
\begin{equation}\label{eq:TI5} 
\ln\left (\dfrac{\chi}{M}\right )=f_{0}(\tilde{h})+g(\tau)\, ,
\end{equation}
provided $f_{0}(\tilde{h})$ obeys
\begin{equation}\label{eq:TI6} 
\dfrac{\p f_{0}}{\p\tilde{h}}=-\dfrac{2\tilde{h}\left [1+3\xi_{H}\left (1+\dfrac{1}{2}\dfrac{\p\ln(\xi_{H})}{\p\ln(\tilde{h})}\right )\right ]}{B+2\tilde{h}^{2}+6\xi_{H}\tilde{h}^{2}}\, .
\end{equation}
This yields 
\begin{align}\label{eq:TI7} 
k^{2}(\varphi)&=\dfrac{2M^{2}}{1+\xi_{H}\tilde{h}^{2}}\left (\dfrac{\p\tilde{h}}{\p\varphi}\right )^{2}\biggl \{1+\dfrac{a_{H}^{2}\xi_{H}^{2}\tilde{h}^{2}}{1+\xi_{H}\tilde{h}^{2}}\nn\\
&\quad -\dfrac{2(1+\xi_{H}a_{H}^{2})\tilde{h}^{2}}{B+2\tilde{h}^{2}+6\xi_{H}\tilde{h}^{2}}\biggl \}\nn\\
&=\dfrac{2\tilde{C}_{H}}{B_{H}^{2}}
\end{align}
with
\begin{equation}\label{eq:TI8}
\tilde{C}_{H}=\dfrac{(B+6\xi_{H}\tilde{h}^{2})(1+\xi_{H}\tilde{h}^{2}+a_{H}^{2}\xi_{H}^{2}\tilde{h}^{2})-2\xi_{H}a_{H}^{2}\tilde{h}^{2}}{\xi_{H}^{2}(B+6\xi_{H}\tilde{h}^{2}+2\tilde{h}^{2})\tilde{h}^{2}}
\end{equation}
replacing $C_{H}$ in \eq{HI6}, and $B_{H}$ given by \eq{HI8}.

Several limits are of interest. For the first limit we consider $B\ll 6\xi_{H}\tilde{h}^{2}$, such that
\begin{equation}\label{eq:TI9}
\tilde{C}_{H}=\dfrac{3(1+\xi_{H}\tilde{h}^{2}+a_{H}^{2}\xi_{H}^{2}\tilde{h}^{2})-a_{H}^{2}}{\xi_{H}(3\xi_{H}+1)\tilde{h}^{2}}\, .
\end{equation}
If we further consider $\xi_{H}\gg 1$ and the range $\xi_{H}\tilde{h}^{2}\gg 1$, this simplifies further to
\begin{equation}\label{eq:TI10}
\tilde{C}_{H}=a_{H}^{2}\, ,
\end{equation}
which equals the corresponding limit of $C_{H}$ in \eq{HI7}. One recovers Higgs inflation as discussed in sect.~\ref{sec:Cosmon_Infl}. The second limit considers $B\gg 6\xi_{H}\tilde{h}^{2}$, $B\gg 2\tilde{h}^{2}$, where
\begin{equation}\label{eq:TI11}
\tilde{C}_{H}=\dfrac{1+\xi_{H}\tilde{h}^{2}+a^{2}_{H}\xi_{H}^{2}\tilde{h}^{2}}{\xi_{H}^{2}\tilde{h}^{2}}=C_{H}
\end{equation}
again equals $C_{H}$ such that the mixing in the kinetic terms \eqref{eq:TI4} becomes negligible. Away from the two limits there are quantitative but no qualitative differences between $\tilde{C}_{H}$ and $C_{H}$.

Another interesting limit is given by $\xi_{H}\raw 0$, $\p\ln(\xi_{H})/\p\ln(\tilde{h})\raw 0$. While $B_{H}$ is again given by \eq{HI9A}, one finds for $\tilde{C}_{H}$
\begin{equation}\label{eq:TI12}
\tilde{C}_{H}=\dfrac{B}{\xi_{H}^{2}\tilde{h}^{2}(B+2\tilde{h}^{2})}\, .
\end{equation}
Therefore the kinetial
\begin{equation}\label{eq:TI13}
k^{2}=\dfrac{2B\tilde{h}^{2}}{B+2\tilde{h}^{2}}\left (4+\dfrac{\p\ln(\lambda_{H})}{\p\ln(\tilde{h})}\right )^{-2}
\end{equation}
is reduced by a factor $B/(B+2\tilde{h}^{2})$ as compared to \eq{HI10A}. For large enough $B$ the relevant range of $\tilde{h}^{2}$ obeys $2\tilde{h}^{2}\ll B$, such that the presence of the additional scalar $\chi$ induces only a small modification of Higgs inflation in the limit of small $\xi_{H}$ as well. For small $B$ the differences between scale invariant Higgs inflation and Higgs inflation with fixed $M^{2}$ are more substantial. 

One concludes that scale invariant Higgs inflation is rather similar to Higgs inflation with an intrinsic mass scale $M$. For scale invariant Higgs inflation the almost scale invariant primordial fluctuation spectrum is not directly related to gravity scale symmetry. Since there is no intrinsic mass scale, the end of inflation cannot be associated with the evolution away from the UV-fixed point due to a relevant parameter. For scale invariant Higgs inflation one encounters another type of scale symmetry where in \eq{TI1} the Higgs doublet and the metric scale according to $h\raw \tilde{\alpha} h$, $g_{\mn}\raw \tilde{\alpha}^{-2}g_{\mn}$, while $\chi$ remains invariant. This is the same transformation as particle scale symmetry, but now realized for a region where $\chi^{2}\ll\xi_{H}h^{\dagger}h$. This version of scale symmetry is broken substantially once $\chi^{2}\approx \xi_{H}h^{2}$ or $\xi_{H}\tilde{h}^{2}\approx 1$, as well as by the dependence of $\lambda_{H}$ and $\xi_{H}$ on $\tilde{h}^{2}$. The end of inflation is associated to a substantial violation of this scale symmetry. We observe that this scale symmetry becomes exact in the limit $\tilde{h}^{2}\raw\infty$, provided $\lambda_{H}$ and $\xi_{H}$ reach fixed values in this limit. In this limit $B_{H}$ approaches zero, the kinetial $k^{2}(\varphi)$ diverges, and the fluctuation spectrum becomes scale invariant, $\varepsilon\raw 0$, $r\raw 0$, $n\raw 1$.

\subsubsection{General scale invariant two-field inflation}

Other versions of scale invariant inflation employ besides $\chi$ an additional scalar field as a type of inflaton. For the formal treatment this is very similar to scale invariant Higgs inflation. Only the physical meaning of $h^{2}$ is changed. It is no longer the squared Higgs doublet, but the square of an appropriate new scalar field. Correspondingly, the restrictions on the potential related to the observed properties of the Higgs boson no longer apply.

By virtue of scale symmetry the effective potential can always be written as
\begin{equation}\label{eq:GI1} 
U=\dfrac{1}{2}\lambda_{S}(\tilde{h}^{2})h^{4}\com \tilde{h}^{2}=\dfrac{h^{2}}{\chi^{2}}\, ,
\end{equation}
such that $\lambda_{S}$ replaces $\lambda_{H}$ as used for scale invariant Higgs inflation. For example, a potential quadratic in $h$ reads
\begin{equation}\label{eq:GI2} 
U=\dfrac{\gamma}{2}\chi^{2}h^{2}\com\lambda_{S}=\dfrac{\gamma}{\tilde{h}^{2}}\, .
\end{equation}
We also employ a general coupling to the curvature scalar, replacing formally $\xi_{H}(\tilde{h}^{2})$ by $\xi_{S}(\tilde{h}^{2})$. All formal steps are the same as for scale invariant Higgs inflation. In particular, the potential in the Einstein frame only depends on the dimensionless $\tilde{h}^{2}$ according \eq{HI2}
\begin{equation}\label{eq:GI3} 
\dfrac{V_{E}}{M^{4}}=\dfrac{\lambda_{S}(\tilde{h}^{2})\, \tilde{h}^{4}}{2(1+\xi_{S}(\tilde{h}^{2})\,\tilde{h}^{2})}\, .
\end{equation}
The diagonalization of the kinetic term proceeds as for scale invariant Higgs inflation. One ends with a kinetial $k^{2}(\varphi)$ given by \eq{TI7}, with $\tilde{C}_{H}$ specified by \eq{TI8} and $B_{H}$ by \eq{HI8}.

This yields general formulae for scale invariant models with two scalar fields. The kinetial reads
\begin{equation}\label{eq:GI4} 
k^{2}(\varphi)=\dfrac{(a_{1}+a_{2}B)\tilde{h}^{2}}{16b^{2}(B+6\xi_{S}\tilde{h}^{2}+2\tilde{h}^{2})}\, ,
\end{equation}
where
\begin{align}\label{eq:GI5} 
a_{1}&=12\xi_{S}\tilde{h}^{2}\biggl [(3\xi_{S}^{2}+\xi_{S})\tilde{h}^{2}\nn\\
&\quad +(3\xi_{S}^{2}\tilde{h}^{2}-1)\left (\dfrac{\p\ln(\xi_{S})}{\p\ln(\tilde{h})}+\dfrac{1}{4}\left (\dfrac{\p\ln(\xi_{S})}{\p\ln(\tilde{h})}\right )^{2}\right )\biggl ]\nn\\
a_{2}&=2\left [1+\xi_{S}\tilde{h}^{2}+3\xi_{S}^{2}\tilde{h}^{2}\left (1+\dfrac{1}{2}\dfrac{\p\ln(\xi_{S})}{\p\ln(\tilde{h})}\right )^{2}\right ]\, ,
\end{align}
and
\begin{align}\label{eq:GI6} 
b&=1+\dfrac{1}{4}\dfrac{\p\ln(\lambda_{S})}{\p\ln(\tilde{h})}\nn\\
&\quad +\dfrac{\xi_{S}^{2}\tilde{h}^{2}}{4}\left (\dfrac{\p\ln(\lambda_{S})}{\p\ln(\tilde{h})}-2\dfrac{\p\ln(\xi_{S})}{\p\ln(\tilde{h})}\right )\, .
\end{align}

With
\begin{equation}\label{eq:GI7} 
\dfrac{\p\varphi}{\p\tilde{h}}=-\dfrac{4Mb}{\tilde{h}(1+\xi_{S}\tilde{h}^{2})}
\end{equation}
one relates the number of $e$-foldings $N$ to $\tilde{h}$ via
\begin{equation}\label{eq:GI8} 
N=-\dfrac{1}{8}\int_{\tilde{h}}^{\tilde{h}_{f}}\dif h^{2}\dfrac{a_{1}+a_{2}B}{(1+\xi_{S}h^{2})(B+6\xi_{S}h^{2}+2h^{2})b}\, .
\end{equation}
Solving $h^{2}(N)$ by use of \eq{GI8} one obtains $r(N)$ and $n(N)$ via \eq{C18F}. The amplitude is given by
\begin{equation}\label{eq:GI9} 
\Delta^{2}=\dfrac{\lambda_{S}\tilde{h}^{4}k^{2}}{24\pi^{2}\tilde{h}^{2}(1+\xi_{S}\tilde{h}^{2})}\, .
\end{equation}

With three functions $\lambda_{S}(\tilde{h})$, $\xi_{S}(\tilde{h})$ and $B(\tilde{h})$ a very large class of scale invariant inflationary models can be constructed. It is obvious that scale invariance alone is not sufficient to constrain the inflationary parameters effectively. Additional assumptions are needed for this purpose. For example, ref.~\cite{FHR1,FHR2,FHNR} assumes that $\xi_{S}$ and $B$ are independent of $\tilde{h}$ and
\begin{equation}\label{eq:GI10} 
\lambda_{S}=c_{1}+c_{2}\tilde{h}^{2}+c_{3}\tilde{h}^{4}\, .
\end{equation}
(In ref.~\cite{FHR1,FHR2,FHNR} the notation differs from the present one by a constant rescaling of $\chi$.) For suitable choices of the five parameters $B$, $\xi_{S}$, $c_{1}$, $c_{2}$, $c_{3}$ realistic inflationary scenarios can be found. We recall, however, that the approximate scale invariance of the primordial fluctuation spectrum is not related to gravity scale symmetry. It rather results from an effective scale symmetry realized for particular choices of the parameters. Also the smallness of $\Delta^{2}$ needs typically to be adjusted by a suitable condition on the parameters.

We have omitted in most of the discussion of scale invariant inflationary models the role of the Goldstone boson $\tau$. It settles early to an arbitrary constant value. Its fluctuations do not induce curvature fluctuations or non-Gaussian fluctuations \cite{GBRSZ,FHNR}. Effectively, the Goldstone boson plays no role for the inflationary dynamics. 

\subsection{Cosmological scaling solutions}\label{sec:Cosm_Scaling_Sol}

Cosmology with scalar fields is often described by scaling solutions \cite{CWQ,RAPE,CWQ2,VILI,COLIWA,CDS,LAMSS}, sometimes also called tracker solutions or cosmic attractor solutions. Scale symmetry is a central ingredient for the understanding of such scaling solutions since it permits to establish simple relations only based on the dimensions of relevant quantities. Famous examples for such scaling relations in other fields of physics are the Widom scaling form \cite{WID} of the equation of state for universal critical phenomena in statistical physics, or the Kolmogorov scaling \cite{KOL} for turbulence. Such scaling relations exist for exact scale symmetry, as well as for the presence of an intrinsic scale $\bar{\mu}$ that reflects the explicit breaking of scale symmetry. (For critical phenomena $\bar{\mu}$ corresponds to the deviation of the temperature from the critical temperature.) The discussion of scaling relations is best done in a scale invariant frame for the metric. It is somewhat obscured in the Einstein frame since the presence of a fixed mass scale $M$, introduced only by the Weyl transformation, makes arguments based on dimensions more complicated. Once a scaling solution is established in a given frame it can be transposed, or course, to arbitrary frames.

For scaling solutions different terms in the effective action are proportional to each other for a long time. These terms include the effective scalar potential $U(\chi)$, the curvature term $\sim\chi^{2}R$, the scalar kinetic term $\cL_{kin}$ and the energy density of matter or radiation $\rho$. (Other possible scaling solutions involving terms as $R^{2}$ are not discussed here.) If all four terms contribute, a scaling solution is characterized by
\begin{equation}\label{eq:S1} 
U\sim\chi^{2}R\sim\cL_{kin}\sim\rho\, .
\end{equation}
We concentrate on homogeneous and isotropic cosmologies where the curvature scalar is given by the Hubble parameter $H$,
\begin{equation}\label{eq:S2} 
R\sim H^{2}+\p_{t}H\com\cL_{kin}\sim (\p_{t}\chi)^{2}+3H\p_{t}\chi\, .
\end{equation}
For other possible scaling solutions a subset of the four terms in eq.~\eqref{eq:S1} is small and can be neglected. The overall scale is set by the cosmic time $t$. Then dimensionless combinations as $Ht$ have to be expressed in terms of varying scales as $\chi$ or intrinsic scales as $\bar{\mu}$.

\subsubsection{Scaling solutions for exact scale symmetry}

We first discuss situations of exact scale symmetry ($\bar{\mu}=0$). If $U(\chi)$ can be neglected ($U(\chi)=0$), one finds simple scaling solutions with
\begin{equation}\label{eq:S3} 
H=\eta t^{-1}\com\rho\sim\chi^{2}t^{-2}\, .
\end{equation}
The scalar kinetic term vanishes for large $t$ according to
\begin{equation}\label{eq:S4} 
\p_{t}\chi\sim t^{-3\eta}\, ,
\end{equation}
such that $\chi$ approaches a constant $\chi_{\infty}$ that we may identify with the Planck mass $M$. The solution \eqref{eq:S3} describes standard Friedman cosmology, supplemented by a Goldstone boson whose dynamical role can be neglected (after a possible short initial epoch). In contrast, for $U=\lambda\chi^{4}$ the scaling solutions are of the type
\begin{equation}\label{eq:S5} 
H^{2}\sim \chi^{2}\, ,
\end{equation}
with negligible $\rho$ for large time. With $\cL_{kin}$ vanishing asymptotically for large $t$ this amounts to a standard de Sitter solution, with $\lambda M^{4}$ playing the role of a cosmological constant. Again, the Goldstone boson plays no role.

\subsubsection{Scaling solutions with intrinsic mass scale}

Interesting cosmologies with variable $\chi$ (variable gravity \cite{CWVG}) arise in the presence of an intrinsic scale $\bar{\mu}$ in the effective potential. We first consider
\begin{equation}\label{eq:S6} 
U=\bar{\mu}^{2}\chi^{2}\, .
\end{equation}
The scaling solution implies then
\begin{equation}\label{eq:S7} 
H^{2}=b^{2}\bar{\mu}^{2}\com \rho\sim\bar{\mu}^{2}\chi^{2}
\end{equation}
and
\begin{equation}\label{eq:S8} 
\chi=\chi_{0}\exp\left (c\bar{\mu}t\right )\com \cL_{kin}\sim\bar{\mu}^{2}\chi^{2}\, ,
\end{equation}
with constant $b$, $c$, $\chi_{0}$. All four terms in \eq{S1} are therefore of the same order $\sim\bar{\mu}^{2}\chi^{2}$. The intrinsic scale sets a characteristic time scale $\bar{\mu}^{-1}$ for the whole cosmological evolution. 

The explicit scaling solutions of this type can be found in ref.~\cite{CWVG}. In the Einstein frame they describe a modification of Friedman cosmology by the presence of a constant fraction of early dark energy (EDE). Dark energy scales in the same way as the dominant radiation or matter component in $\rho$. By virtue of this scaling the tiny amount of dark energy in units of the Planck mass finds a simple explanation. Both matter or radiation and dark energy decrease simultaneously with increasing cosmic time. With
\begin{equation}\label{eq:S9} 
\dfrac{U}{\chi^{4}}=\dfrac{\bar{\mu}^{2}\chi^{2}}{\chi^{4}}=\dfrac{\bar{\mu}^{2}}{\chi^{2}}=\dfrac{\bar{\mu}^{2}}{\chi_{0}^{2}}\exp\left (-2c\bar{\mu}t\right )\, ,
\end{equation}
the effective cosmological constant vanishes asymptotically for $t\raw\infty$ \cite{CWQ}. Indeed, the dimensionless ratio between potential and fourth power of the (dynamical) Planck mass is the same in all frames. In the Einstein frame $U/\chi^{4}$ becomes $U/M^{4}$. The scaling solution predicts dark energy of the same order of magnitude as dark matter. The small present ratio $U/M^{4}\approx 7\cdot 10^{-121}$, associated to a tiny ``cosmological constant'', has the same simple explanation as the small present ratio for the dark matter density $\rho_{M}/M^{4}\approx 3\cdot 10^{-121}$. Both quantities decrease with time and are very small because the Universe is very old.

If the scaling solution would be valid until the present time, the dynamical dark energy would have at present the same time evolution as dark matter and baryons. The evolution of cosmology depends on the equation of state $w$ of dark energy, as given by
\begin{equation}\label{eq:EOSA} 
w=\dfrac{p}{\rho}\, .
\end{equation}
Here the pressure $p$ and energy density $\rho$ of dark energy correspond to the components of its energy-momentum tensor and should not be associated with a substance in thermodynamic equilibrium. For a spatially homogeneous scalar field in the Einstein frame, with a standard kinetic term for the scalar field, one has
\begin{equation}\label{eq:EOSB}
\rho=\dfrac{1}{2}(\p_{t}\varphi)^{2}+U(\varphi)\com p=\dfrac{1}{2}(\p_{t}\varphi)^{2}-U(\varphi)\, .
\end{equation}
If the scaling solution holds at present, one infers $w=0$, in clear contrast to the observed value $w\approx -1$. 

Observation indicates that at present the time variation of $\varphi$ is small, $(\p_{t}\varphi)^{2}\ll U(\varphi)$. Realistic models of dark energy therefore need an ingredient that ``kicks out'' the cosmological solution from the scaling solution in a comparatively recent past (typically at redshift $z\approx 5$.) A natural candidate are non-relativistic neutrinos \cite{ABW,CWGN} whose masses may reflect a relevant parameter and therefore an intrinsic scale in the beyond standard model physics \cite{CWIQ}. Another possibility is the crossover in the form of the scalar potential found in the discussion of scaling solutions for dilaton quantum gravity in sect.~\ref{sec:Scal_Pot_Dil_Grav}. We will discuss the stop of the scaling solution in sect.~\ref{sec:7.8}. 

Cosmological scaling solutions of the type \eqref{eq:S9} can be good approximations for the inflationary epoch, as well as for radiation and matter domination. The transition between such epochs are not described by scaling solutions. The present epoch corresponds to a transition period rather than a scaling solution. The details of the scaling solution \cite{CWVG,CWIQ} depend on the equation of state for the dominant energy density $\rho$ and the coefficient of the scalar kinetic term. This distinguishes between the inflationary, radiation- and matter-dominated epochs. For the matter dominated period it is crucial that no intrinsic mass scale is introduced by the particle masses, i.e. all particle masses have to scale $\sim\chi$. Deviations from this scaling, for example in the neutrino sector, typically end the scaling solution.

As a second example we take
\begin{equation}\label{eq:S10} 
U=\bar{\mu}^{4}\, .
\end{equation}
The scaling solution obeys
\begin{equation}\label{eq:S11} 
H^{2}\sim\dfrac{\bar{\mu}^{4}}{\chi^{2}}\com \rho\sim\bar{\mu}^{4}\com \chi\sim\bar{\mu}^{2}t\, .
\end{equation}
Again, the effective cosmological constant vanishes asymptotically for $t\raw\infty$
\begin{equation}\label{eq:S12} 
\dfrac{U}{\chi^{4}}=\dfrac{\bar{\mu}^{4}}{\chi^{4}}\sim\bar{\mu}^{4}t^{-4}\, .
\end{equation}
Models of this type have been discussed for pure scalar theories by Bertolami \cite{BER} and Ford \cite{FOR}, neglecting $\rho$. They can be extended to a radiation dominated epoch since massless particles do not introduce an intrinsic mass scale. Actually, the Hubble parameter vanishes for the scaling solution in the radiation dominated epoch, which is therefore characterized by a flat space geometry \cite{CWVG}. For the assumed constant particle masses the models of Bertolami and Ford do not admit, however, a realistic matter dominated epoch. Furthermore, a substantial variation of the ratio between nucleon mass and Planck mass contradicts observational limits. As noted in ref.~\cite{BER}, no realistic cosmology can be obtained on this basis. The situation changes profoundly \cite{CWQ2,CWQ} if particle masses scale $\sim\chi$, as dictated by scale symmetry. Then particle masses do not introduce an intrinsic mass scale and the scaling solution \eqref{eq:S11} can be extended to cover a realistic matter dominated epoch \cite{CWVG}. The model is again characterized by a constant fraction of EDE.

\subsubsection{Runaway cosmologies}

The scaling solutions are typical examples of ``runaway cosmologies'' for which $\chi(t)$ increases without settling at a given value. This circumvents Weinberg's no go theorem \cite{WEICC} for a natural explanation of a small cosmological constant, the latter being based on a static value of $\chi_{0}$ associated to a minimum of the scalar potential in the Einstein frame at $\chi_{0}$. In the infinite future $\chi$ grows to infinitely large values. 

While in the scale invariant frame the scaling solutions \eqref{eq:S7}, \eqref{eq:S8} and \eqref{eq:S11}, \eqref{eq:S12} look rather different, the physical behavior of cosmology is actually rather similar. Both models describe inflation, as well as radiation- and matter-dominated epochs with a constant fraction of EDE. In the Einstein frame both potentials \eqref{eq:S6} and \eqref{eq:S10} appear in the form of an exponential potential \cite{CWQ}. This extends to general potentials of the form
\begin{equation}\label{eq:S13} 
U=\bar{\mu}^{A}\chi^{4-A}\, ,
\end{equation}
where $A>0$ is the anomalous dimension for the dimensionless coupling $\tilde{\lambda}(\chi)=U/\chi^{4}$ near a fixed point at $\tilde{\lambda}_{*}=0$,
\begin{equation}\label{eq:S14} 
\mu\p_{\mu}\tilde{\lambda}=A\tilde{\lambda}\, .
\end{equation}
The existence of such a fixed point with positive $A$ solves the cosmological constant problem dynamically, as advocated in sect.~\ref{sec:V}.

The scaling solutions are a powerful mechanism for reducing the dark energy density to tiny values, similar in order of magnitude to the energy density of radiation or matter. In order to produce a dark energy density of the order observed today, $\rho_{h}\approx (2\cdot 10^{-3}\mathrm{eV})^{4}$, they have to last at least until rather recent cosmological times (on logarithmic scales), say redshift $z\approx 5$. If cosmology would move substantially away from the scaling solution much earlier, say at $t_{f}$, one typically would obtain a present dark energy density of the same order as at $t_{f}$, e.g. $\rho_{h}(t_{0})\approx \rho_{h}(t_{f})\approx\Omega_{h}\rho_{c}(t_{f})$, with $\rho_{c}(t_{f})$ the critical total energy density corresponding to a flat space-geometry. For an early $t_{f}$ one has $\rho_{c}(t_{f})$ much larger than $(2\cdot 10^{-3}\mathrm{eV})^{4}$. Unless $\Omega_{h}$ is tiny this restricts $t_{f}$ to be in the relatively recent past. For $\Omega_{h}=0.01$ the effective stop of the scalar field has to occur for a redshift $z_{f}\approx 5$.

In turn, the scaling solutions are possible only if quantum scale symmetry remains a good approximation in the relevant range of scales. This implies that an intrinsic scale $\bar{\mu}$ must be rather small on particle physics scales. In other words, $\Lambda_{QCD}$ and $\varphi_{0}$ must be proportional to $\chi$ with high accuracy. Coming back to the discussion in sect.~\ref{sec:Mass_Pseudo_Goldstone}, the dimensionless couplings of the standard model, normalized at a momentum scale $\mu=\chi$, have to be very close to $\chi$-independent fixed point values. In particular, this requires for dilaton quantum gravity that the gauge couplings are determined by non-zero fixed point values $g_{*}$ and correspond to irrelevant parameters \cite{EHW}. If, in contrast, the gauge couplings would be asymptotically free, the confinement scale would not be proportional to $\chi$, as given in \eq{PS25}. Cosmology would then differ strongly from the cosmological scaling solution at a time much earlier than needed for observation.

\begin{widetext}
\onecolumngrid
\begin{figure}[h]
\includegraphics[scale=0.45]{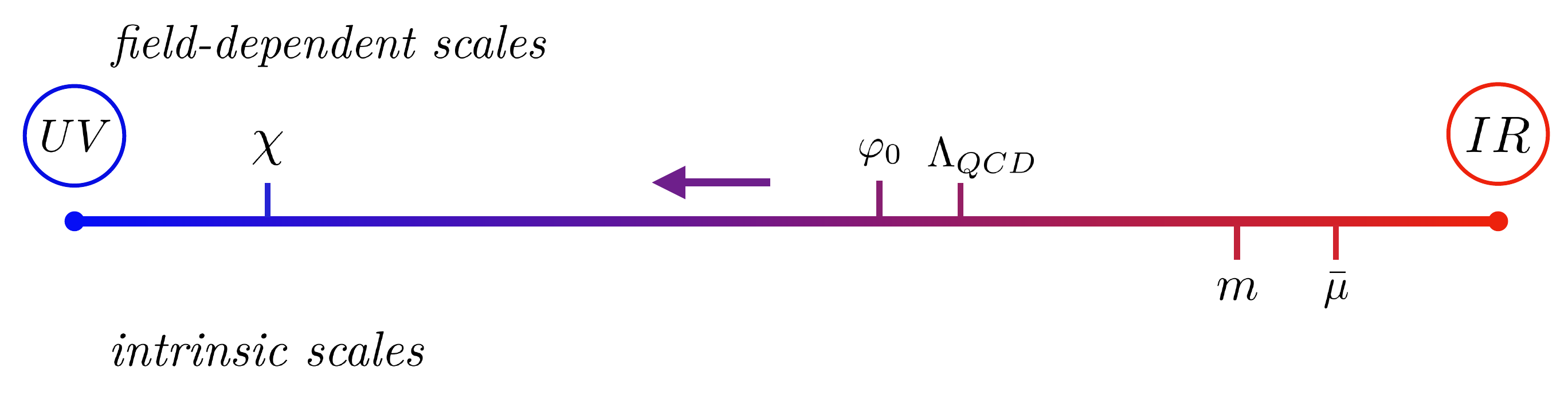}
\caption{Relative sliding of spontaneous and intrinsic scales in runaway cosmologies. Units are adapted to permit clear visualization. The arrow indicates the increase of $\chi$ with increasing cosmic time. The time-dependent Planck mass is given by $\chi(t)$, and $\varphi_{0}$, $\Lambda_{QCD}$ are proportional to $\chi$.}\label{fig:Y} 
\end{figure}
\end{widetext}
\twocolumngrid

\subsubsection{Sliding systems of scales}

During the cosmological evolution the system of scales induced by spontaneous scale symmetry breaking $\sim\chi$ slides relative to the intrinsic scales $\sim\bar{\mu}$ that arise from relevant parameters at the UV-fixed point. We have depicted this evolution in fig.~\ref{fig:Y}, where the arrow indicates the time-evolution of the cosmological increase of $\chi$. On the side of the intrinsic scales, $\bar{\mu}$ typically indicates the explicit scale symmetry breaking in the scalar potential. We have added a possible scale $m$ characterizing a crossover in the scalar kinetic term. While $m$ is proportional to $\bar{\mu}$, the ratio $m/\bar{\mu}$ may be large or small if the two scales arise from dimensional transmutation of two different relevant couplings, as discussed ins sect.~\ref{sec:Networks_FP_Crossover}. (We omit here for simplicity a possible $\chi$-dependence of $m$ and $\bar{\mu}$.) On the side of the ``sliding'' spontaneous scales, we show the Planck mass $\chi$, the Fermi scale $\varphi_{0}\sim\chi$ and the QCD-confinement scale $\Lambda_{QCD}\sim\chi$. In the early cosmology of the inflationary epoch $\chi$ typically started much smaller than $\bar{\mu}$, e.g. $\chi(t\raw-\infty)\raw 0$. During inflation it crosses $\bar{\mu}$ and inflation may end once $\chi$ crosses $m$. During the scaling solution in later cosmology also $\varphi_{0}$ and $\Lambda_{QCD}$ cross $m$ and $\bar{\mu}$, such that today the intrinsic scales are much smaller than the confinement scale. The standard model of particle physics is therefore almost scale invariant. (This concerns gravity scale symmetry - particle scale symmetry comes in addition.)

The actual value of $\bar{\mu}$ is arbitrary - it only sets the units of mass. Only dimensionless ratios as $\chi/\bar{\mu}$ are observable. The ratio $\chi/\bar{\mu}$ depends on time. It has increased to a huge value due to the continuous increase of $\chi$ during a very long cosmological period. The value of $\chi/\bar{\mu}$ today can be used to define the present cosmological epoch. Its value depends on the particular form of the cosmon effective potential. From
\begin{equation}\label{eq:S15} 
\dfrac{U(\chi(t_{0}))}{\chi^{4}(t_{0})}=8.3\cdot 10^{-121}\, ,
\end{equation}
and choosing units for which $\chi(t_{0})$ equals the Planck mass $M$, one infers for the potential \eqref{eq:S6} a value close to the present Hubble parameter
\begin{equation}\label{eq:S16}
\bar{\mu}=2\cdot 10^{-33}\mathrm{eV}\, .
\end{equation}
The characteristic time for the cosmological evolution is given by
\begin{equation}\label{eq:S17}
\bar{\mu}^{-1}\approx 10^{10}\mathrm{yr}\, .
\end{equation}
For the potential \eqref{eq:S10} one obtains
\begin{equation}\label{eq:S18}
\bar{\mu}=2.325\cdot 10^{-3}\mathrm{eV}\, ,
\end{equation}
with $\bar{\mu}^{4}$ the present dark energy density.

\subsubsection{Quantitative scaling solutions}

It is instructive to follow the time evolution of the geometry for the scaling solutions. For the potential \eqref{eq:S6} the scaling solutions
\begin{equation}\label{eq:S18A} 
H=b\bar{\mu}\com\chi=\chi_{0}\exp(c\bar{\mu} t)\com \rho=\bar{\rho}\bar{\mu}^{2}\chi^{2}\, ,
\end{equation}
are solutions of the field equations \cite{CWVG} for
\begin{equation}\label{eq:S18B} 
c^{2}=\dfrac{3}{3B-2}\left \{\sqrt{(1+\bar{\rho})^{2}-\dfrac{8\bar{\rho}}{3B}}-(1+\bar{\rho})+\dfrac{4}{3B}\right \}
\end{equation}
and
\begin{equation}\label{eq:S18C} 
b=-c+\dfrac{1}{\sqrt{3}}\left (1+\bar{\rho}+\dfrac{B}{2}c^{2}\right )\, .
\end{equation}
Scalar dominated scaling solutions with $\bar{\rho}=0$ exist for
\begin{equation}\label{eq:S18D} 
c^{2}=\dfrac{4}{B(3B-2)}\com b=\dfrac{B-2}{2}c
\end{equation}
and therefore for $B>2/3$. The universe is expanding ($b>0$) for $B>2$, and contracting ($b<0$) for $B<2$. We will see in the Einstein frame that $B=2$ coincides with the end of the inflationary epoch.

For the radiation dominated epoch one has
\begin{equation}\label{eq:S18E} 
\rho\sim a^{-4}
\end{equation}
and therefore
\begin{equation}\label{eq:S18F} 
\chi\sim a^{-2}\, .
\end{equation}
This requires
\begin{equation}\label{eq:S18G} 
b=-\dfrac{c}{2}\, .
\end{equation}
The proportionality factor $\bar{\rho}$ adapts to the condition \eqref{eq:S18G} and one obtains
\begin{equation}\label{eq:S18H} 
b=-\dfrac{1}{\sqrt{B}}\com c=\dfrac{2}{\sqrt{B}}\com \bar{\rho}=\dfrac{3(1-B)}{B}\, .
\end{equation}
In turn, this shows that this type of scaling solution exists only for
\begin{equation}\label{eq:S18I} 
B<1\, .
\end{equation}
The potential and kinetic energy of the scalar field constitute a homogeneous early dark energy component
\begin{equation}\label{eq:S18J} 
\rho_{h}=\bar{\mu}^{2}\chi^{2}+\dfrac{B}{2}\dot{\chi}^{2}\, ,
\end{equation}
with fraction
\begin{equation}\label{eq:S18K} 
\Omega_{h}=\dfrac{\rho_{h}}{\rho+\rho_{h}}=B\, .
\end{equation}
According to \eqs{S18G}{S18H} the universe is shrinking during the radiation dominated epoch. Similar considerations apply to the matter dominated epoch, for which the change of particle masses $\sim\chi$ introduces an additional term \cite{CWVN} in the energy momentum conservation law. The quantitative solution \cite{CWVG} shows that the universe is also shrinking during the matter dominated epoch. Despite the unusual geometry of a shrinking universe, the scaling solutions for the radiation and matter dominated epochs are compatible with observation. This is seen most easily in the Einstein frame, and discussed in more detail in sect.~\ref{sec:Dif_Pic_Universe}.

The evolution of geometry is different for a constant potential $U=\bar{\mu}^{4}$ \eqref{eq:S10}. For the radiation dominated epoch one now finds
\begin{equation}\label{eq:S18L} 
\rho=\bar{\rho}\bar{\mu}^{4}\sim a^{-4}\, ,
\end{equation}
which implies a flat Minkowski geometry
\begin{equation}\label{eq:S18M} 
b=0\, .
\end{equation}
One finds \cite{CWVG}
\begin{equation}\label{eq:S18N} 
\bar{\rho}=\dfrac{3(4-B)}{B}\, ,
\end{equation}
requiring
\begin{equation}\label{eq:S18O} 
B< 4\, .
\end{equation}
For the matter dominated epoch one finds \cite{CWVG} an expansion of the Universe according to
\begin{equation}\label{eq:S18P}
H=\dfrac{1}{3t}\, .
\end{equation}
While the behavior of geometry in the scaling frame is rather unusual, we will see that in the Einstein frame cosmology is given by a Friedman universe modified by early dark energy.

\subsubsection{Scaling solutions in the Einstein frame}

We finally briefly discuss the scaling solutions in the Einstein frame. Similar to our discussion of inflation we choose a normalization of the scalar field where the potential takes an exponential form \eqref{eq:C5}. For equivalent discussions with a standard normalization of the scalar kinetic term we refer to refs.~\cite{HOMSS,AMSS,AMS,GLSSS,RUBCW}. With
\begin{equation}\label{eq:S19}
\dfrac{\varphi}{M}=-\ln\left (\dfrac{U(\chi)}{\chi^{4}}\right )\, ,
\end{equation}
we only will keep the leading power of $\chi$ in $U(\chi)$. The potential $U=\bar{\mu}^{2}\chi^{2}$ corresponds to 
\begin{equation}\label{eq:S20}
\dfrac{\varphi}{M}=\ln\left (\dfrac{\chi^{2}}{\bar{\mu}^{2}}\right )\, .
\end{equation}
With a kinetic term \eqref{eq:16} the kinetial in the Einstein frame reads
\begin{equation}\label{eq:S21}
k^{2}(\varphi)=\dfrac{B}{4}\, .
\end{equation}
For the potential $U=\bar{\mu}^{4}$ one has
\begin{equation}\label{eq:S22}
\dfrac{\varphi}{M}=\ln\left (\dfrac{\chi^{4}}{\bar{\mu}^{4}}\right )\, ,
\end{equation}
while
\begin{equation}\label{eq:S23}
k^{2}(\varphi)=\dfrac{B}{16}\, .
\end{equation}
For a given $B$ the two models therefore only differ by a factor four in the kinetial and are qualitatively similar. 

The scaling solutions correspond to a constant kinetial $k^{2}$ or $B$, while a small $\varphi$-dependence of $k^{2}$ or a $\chi$-dependence of $B$ induces approximate scaling solutions. Scalar dominated scaling solutions (with radiation and matter neglected) exist for
\begin{equation}\label{eq:S24} 
k^{2}>\dfrac{1}{6}\, .
\end{equation}
For scaling solutions with radiation or matter the fraction in the potential and kinetic energy of the cosmon (early dark energy) amounts to
\begin{equation}\label{eq:S25}
\Omega_{h}=nk^{2}\, ,
\end{equation}
with $n=4$ for radiation domination and $n=3$ for matter domination. 

From $\Omega_{h}\leq 1$ one infers that such solutions exist only for
\begin{equation}\label{eq:S26}
k^{2}<\dfrac{1}{n}\, .
\end{equation}
If the condition \eqref{eq:S26} is obeyed, the scaling solution with radiation or matter turns out to be the attractor solution. Observational bounds restrict early dark energy to be less than around $1-2\%$ at the time of CMB-emission. This results in a bound
\begin{equation}\label{eq:BKA} 
k^{2}\lesssim 0.005\, ,
\end{equation}
with a corresponding bound on $B$. Perhaps the most characteristic signal of EDE is a reduction of the observed power of structures ($\sigma_{8}$) as compared to the CMB-prediction for $\Lambda$CDM models \cite{FEJOY,JSCW,PAW}. One percent EDE reduces $\sigma_{8}$ by around $5\%$ \cite{JSCW}.

During a scaling solution the transformation to a canonical kinetic term is very simple. With canonically normalized field
\begin{equation}\label{eq:S27}
\sigma=k\varphi
\end{equation}
the potential becomes
\begin{equation}\label{eq:S28}
U=M^{4}\exp\left (-\alpha\dfrac{\sigma}{M}\right )\com \alpha=\dfrac{1}{k}\, ,
\end{equation}
such that the bound \eqref{eq:S24} reads $\alpha^{2} < 6$, while \eq{S26} becomes $\alpha^{2}>n$. (The bounds on $\alpha$ or $k^{2}$ coincide with the bounds on $B$ discussed previously.)

\subsubsection{Mass of the cosmon}\label{sec:Mass_Cosmon}

The mass of the cosmon -- the pseudo Goldstone boson of spontaneously broken scale symmetry -- is given in the Einstein frame by eq.~\eqref{eq:CM1},
\begin{align}\label{eq:CM2} 
m_{c}^{2}&=\dfrac{\p^{2}V_{E}}{\p\sigma^{2}}=\left (k^{-1}\p_{\varphi}\right )^{2}V_{E}\nn\\
&=k^{-2}\left (\dfrac{\p^{2}V_{E}}{\p\varphi^{2}}-\dfrac{\p\ln(k)}{\p\varphi}\dfrac{\p V_{E}}{\p\varphi}\right )\nn\\
&=\dfrac{1}{M^{2}k^{2}}\left (1+M\p_{\varphi}\ln(k)\right )V_{E}\, .
\end{align}
The cosmon mass decreases with time as $V_{E}$ decreases. For the scaling solutions it is roughly given by the Hubble parameter,
\begin{equation}\label{eq:CM3} 
m_{c}\approx \dfrac{1}{M}\sqrt{V_{E}}\approx H\, .
\end{equation}
In eq.~\eqref{eq:CM3} we treat the kinetial as order one and omit it together with other factors of the order one.

Let us establish the relation to the general discussion of the mass of the pseudo-Goldstone boson in sect.~\ref{sec:Mass_Pseudo_Goldstone}. In the absence of an intrinsic scale $\bar{\mu}$ the potential in the scaling frame would have to be of the form $U=\tilde{\lambda}\chi^{4}$, with constant $\tilde{\lambda}$. The potential in the Einstein frame, $V_{E}=\tilde{\lambda}M^{4}$, (for a normalization where $F=\chi^{2}$) would be flat, and the cosmon mass term would vanish. In this limit the cosmon becomes the massless Goldstone boson of spontaneously broken scale symmetry -- the dilaton. (Actually, the graviton barrier and the stability of the scalar potential allow only for $\tilde{\lambda}=0$.) We conclude that the cosmon mass vanishes indeed in the limit $\bar{\mu}/\chi\raw 0$.

The relation between the scales of explicit symmetry breaking, spontaneous symmetry breaking, and the mass of the pseudo-Goldstone boson differs for runaway potentials from the case where the ground state of cosmological solution corresponds to the minimum of the effective potential. For a runaway potential in the Einstein frame of the form
\begin{equation}\label{eq:CM4} 
\dfrac{V_{E}}{M^{4}}=c\left (\dfrac{\bar{\mu}}{\chi}\right )^{\alpha}
\end{equation}
one has
\begin{equation}\label{eq:CM5} 
\dfrac{\p^{2}V_{E}}{\p\chi^{2}}\sim \dfrac{V_{E}}{\chi^{2}}\, .
\end{equation}
The mass term,
\begin{equation}\label{eq:CM6} 
m_{c}^{2}\sim \left (\dfrac{\p\chi}{\p\sigma}\right )^{2}\dfrac{V_{E}}{\chi^{2}}\, ,
\end{equation}
becomes for
\begin{equation}\label{eq:CM7} 
\sigma\sim M\ln\left (\dfrac{\chi}{M}\right )\com \dfrac{\p\sigma}{\p\chi}\sim \dfrac{M}{\chi}\, ,
\end{equation}
and a present value $\chi=M$,
\begin{equation}\label{eq:CM8} 
m_{c}^{2}\sim \dfrac{V_{E}}{\chi^{2}}\, .
\end{equation}
Comparing with the general relation $m_{c}^{2}\sim\tilde{\mu}^{4}/\chi^{2}$ with intrinsic scale $\tilde{\mu}$ and scale of spontaneous symmetry breaking $\chi$, one identifies
\begin{equation}\label{eq:CM9} 
\tilde{\mu}=V_{E}^{\frac{1}{4}}\left (\dfrac{\p\chi}{\p\sigma}\right )^{\frac{1}{2}}\, .
\end{equation}
For the particular case $U=c\bar{\mu}^{4}$, $V_{E}=c\bar{\mu}^{4}M^{4}/\chi^{4}$ one has $\tilde{\mu}\sim\bar{\mu}$ for $\chi=M$. On the other hand, $U=b\bar{\mu}^{2}\chi^{2}$, $V_{E}=b\bar{\mu}^{2}M^{4}/\chi^{2}$ yields $\tilde{\mu}\sim\sqrt{\bar{\mu}\chi}$. In both cases, $\tilde{\mu}$ is of the order $10^{-3}$eV and one recovers the last example in eq.~\eqref{eq:PS16}.

As an alternative to a potential $U=c\bar{\mu}^{4}$ it has been proposed \cite{SHAZEN} to consider a model with restricted diffeomorphism symmetry, namely volume preserving diffeomorphisms \cite{BUDRA2,BUDRA,SHAZE2}, and to take $U=0$. There seems to be no intrinsic mass scale and no explicit breaking of scale symmetry in this model. Still, the field equations and all observable quantities of this model are the same as for a model with full diffeomorphism invariance and explicit scale symmetry breaking by a potential $U=c\bar{\mu}^{4}$. The effects of the constant potential appear now in the form of a free integration constant. Instead of specifying the model by a relevant parameter, reflected in $U=c\bar{\mu}^{4}$, the latter is now a property of a particular solution of the field equations. It is not clear to us at present what are the precise differences between a model with restricted diffeomorphism symmetry and a partial gauge fixing in a model with full diffeomorphism symmetry, where the partial gauge fixing enforces $\sqrt{g}$ to be a given constant. In any case, field equations and therefore the mass of the cosmon are the same in both models. There is no exactly massless Goldstone boson in the model with restricted diffeomorphism symmetry. Since scale symmetry is spontaneously broken by the cosmological solution, this absence of an exact Goldstone boson needs additional understanding if no intrinsic mass scale is present.

\subsubsection{Beginning and end of scaling solution}

The transition from an early epoch of inflation with $k^{2}\gg 1$ to a scaling solution with radiation for $k^{2}<1/4$ can be realized by a crossover of the kinetial $k^{2}(\varphi)$ from large to small values. This type of cosmon inflation \cite{CWCI} realizes the general concept of quintessential inflation \cite{PEVI,HOMSS,SPOK,BRAMA}. For realizing this scenario $B(\chi)$ has to decrease from large values for $\chi\raw 0$, e.g. $k^{2}=8/r\gtrsim 80$, to small values for $\chi\gg\bar{\mu}$. With \eq{S21} or \eqref{eq:S23} the bound for $B(\chi\gg\bar{\mu})$ amounts to $B<1$ or $B<4$. For both cases the sign of the kinetic term \eqref{eq:16} in the scaling frame, $K=B-6$, is negative.

It is also possible that a crossover of $k^{2}(\varphi)$ from small values to large values for $\varphi\gtrsim\varphi_{c}$ ends the scaling solution. This type of ``leaping kinetic term'' can be employed to obtain \cite{HEBCW,CWCFP} realistic models of dynamical dark energy for which the present equation of state is close to $w=-1$. Once $k^{2}(\varphi)$ increases beyond the critical value $k^{2}_{c}=1/n$, the ratio between dark matter and quintessence starts to decrease and the universe turns to a scalar field dominated epoch. One may construct realistic dynamical dark energy models where $B(\chi)$ is large for $\chi \lesssim \bar{\mu}$, small in some intermediate range of $\chi$, and turns again large for very large $\chi$. This entails the sequence inflation, scaling solution, quintessence dominated universe. It is not clear if such a behavior of $B(\chi)$ can be obtained from quantum gravity, but one should keep in mind that $\chi$ may effectively correspond to some ``valley'' in a multi-scalar landscape. Phenomenological models of this type can be made viable \cite{DIMA}. Of course, the range of $\chi$ where the kinetial becomes large has to match the timing for the recent crossover to dark energy domination. 

As an alternative to an increase of $B(\chi)$ for large $\chi$ the scaling solution may be ended by a cosmic trigger event as for growing neutrino quintessence \cite{ABW,CWGN}. We will briefly discuss this issue in sect.~\ref{sec:7.8}.

Scaling solutions or approximate scaling solutions correspond to ranges of $\chi$ for which $B(\chi)$ varies slowly. Such ranges may be connected by crossover ranges with a faster change of $B(\chi)$. Correspondingly, cosmological scaling solutions may be separated by transition regions. For example, after the end of inflation cosmology typically enters an ``kination epoch'' during which the scalar kinetic energy dominates. A transition to the scaling solution with radiation domination requires that first enough entropy is produced to heat the Universe. The duration of the kination epoch depends on the details of the heating \cite{RUBCW}.

For a discussion of the evolution of the Universe one needs to understand the stability or instability of the scaling solutions with respect to small perturbations. Often this can be established \cite{CWQ} by simply investigating if a subdominant contribution is increasing or decreasing. A more general stability analysis \cite{CWQ2} for neighboring homogeneous and isotropic cosmologies can be performed in the general framework of non-linear differential equations \cite{VILI,COLIWA,LAC}. We also note that modified solutions exist if particle masses are not exactly proportional to $\chi$, but scale with a different power. In the Einstein frame they describe ``coupled quintessence'' \cite{CWQ2,LAC}.

\subsection{Time variation of fundamental ``constants''}

For cosmologies with varying scalar fields the ``fundamental constants'' may depend on the value of the scalar field and therefore on time \cite{JOR1,JOR2}. In early days this feature has been explored for a possible explanation of small dimensionless numbers, as the ratio of electron mass to Planck mass \cite{DIR}. Improving observational bounds it has become clear that the time variation of such mass ratios has to be tiny, much too small for an explanation of their small present values \cite{CWVP} -- see ref.~\cite{UZREP} for a review. At the same time, these strong bounds have become an important issue for models of dynamical dark energy or quintessence \cite{CWVN,CHI1,DAZA} or for string theories with light moduli fields \cite{DAPO,UZAN,DPV1,DPV2}. In string theories possible problems may be circumvented by mechanisms that give a sufficiently large mass to all scalar particles, such that they do not play a dynamical role for nucleosynthesis or later cosmology. This is not possible for dynamical dark energy since an almost massless scalar is the key ingredient for the time evolution of the dark energy density. The central problem to a dynamical solution of the cosmological constant problem by the time variation of a scalar field is not so much to find models where the ratio between the effective potential and the fourth power of the Planck mass vanishes asymptotically, but rather combine this with an almost constant ratio of nucleon mass over Planck mass \cite{CWVN}.

\subsubsection{Scale symmetry and varying couplings}

Quantum scale symmetry provides for a simple explanation for the absence of a time variation of fundamental constants, or why such a variation is very small \cite{CWTVC,CWCFP}. Indeed, exact quantum scale symmetry corresponds to a fixed point with a corresponding scaling solution for dimensionless couplings. As discussed in sect.~\ref{sec:Flow_In_Field_Space}, a renormalized dimensionless coupling $g$ can only depend on $\mu/\chi$, with $\mu$ some momentum or geometry scale, and $\chi$ the value of a scalar field that breaks scale symmetry spontaneously. In late cosmology (typically after inflation) $\mu/\chi$ vanishes to a very good approximation. We can set $\mu=0$ and infer that $g$ does not depend on $\chi$. Thus dimensionless couplings do not vary even though $\chi$ may depend on time.

This extends to dimensionless mass ratios as $m_{e}/M$ or $m_{n}/M$, with $m_{e}$ the electron mass and $m_{n}$ the nucleon mass. For $M=\chi$ exact quantum scale symmetry implies that both $m_{e}$ and $m_{n}$ are $\sim \chi$. Mass ratios remain constant despite a time variation of $\chi$. Mass ratios are invariant under Weyl scaling. In the Einstein frame with $\chi$-independent $M$ both $m_{e}$ and $m_{n}$ are therefore independent of $\chi$. With all dimensionless couplings and mass ratios independent of $\chi$, one finds that also all other dimensionless combinations as $b_{i}/m_{e}$ or $\sigma_{i}m_{e}^{2}$, with $b_{i}$ some binding energy and $\sigma_{i}$ some cross section, do not depend on $\chi$. All observable dimensionless quantities in particle physics are independent of $\chi$. In the Einstein frame $\chi$ decouples completely from particle physics, up to negligible effects of derivative couplings. In short, exact quantum scale symmetry implies the absence of a time variation of fundamental constants.

Cosmological scaling solutions that remain good approximations until a rather recent cosmological epoch require that the dimensionless couplings of the standard model correspond to irrelevant couplings at the UV-fixed point. They can therefore be predicted for a given model. For this situation of irrelevant standard model couplings the $\chi$-independence of dimensionless couplings and mass ratios is expected to be highly accurate. In models with exact scale symmetry the absence of time varying couplings is exact, unless there is an additional very light scalar field beyond the Goldstone boson. Any time variation of fundamental constants is related to the presence of an intrinsic scale $\bar{\mu}$. This scale is required in order to permit scaling solutions in the first place. If the presence of this intrinsic scale has some indirect effect on the standard model couplings it could lead to a tiny time variation of fundamental couplings. We discuss this possibility below.

\subsubsection{Observable ``fundamental constants''}

Discussing a possible time variation of ``fundamental constants'' needs a specification what is meant by this concept. Indeed, there are different types of what is usually called ``fundamental constants''. The first type are simple conversion factors, as the light velocity $c$ or the Planck constant $\hbar$. Both have been set to one in this report. Indeed, the light velocity only specifies the relation between the human length scale meter and time scale second. One could measure all distances in light-seconds (as actually done in practice) and never introduce the meter. It is very unlikely that a different civilization would use the meter - after all this is a purely historical unit. Also the second is a purely human historical unit, while frequencies of certain atomic transitions can be regarded as more universal. Being only a conversion factor $c$ can be taken as constant. It coincides with the velocity of light in vacuum in any theory with Lorentz invariance. In theories with spontaneously or explicitly broken Lorentz symmetry the physical velocity of light may be different in different directions. Still, any theory will need a conversion unit between distances in space and time, and this can be taken constant. Similar arguments apply to $\hbar$ which converts momentum units into inverse length units, according to the uncertainty relation of quantum mechanics. We can again fix $\hbar$ as a constant. Other conversion units are the Boltzmann constant converting temperature to energy units. By definition, conversion units do not depend on time.

A second type of ``fundamental constants'' are dimensionless parameters as the fine structure constant. They are related to particular couplings or parameters in a model. The observable dimensionless parameters correspond to renormalized couplings. They typically depend on the momentum or length scale $\mu$ of the observation. Those parameters can also depend on scalar fields as $\chi$, and they can vary with time if $\chi$ varies with time. If such dimensionless couplings are fundamental or not often depends on the level at which the theory is considered. The dimensionless parameters of nuclear physics may be reduced to the dimensionless parameters of the standard model.

The standard model also contains mass parameters as the proton mass $m_{p}$ or the electron mass $m_{e}$. In turn, they involve the confinement scale $\Lambda_{QCD}$ and the Fermi scale $\varphi_{0}$, plus dimensionless couplings as Yukawa couplings. Masses are not directly observable, however. One always needs another mass or length scale for comparison. A measurable quantity is the proton mass in units of the electron mass, or the dimensionless ration $m_{p}/m_{e}$. Here $m_{e}$ enters the measurement apparatus by fixing the size of the atoms and therefore the length units used in practice. In short, only dimensionless quantities as mass ratios are measurable. They can be treated as any other dimensionless renormalized coupling. In particular, mass ratios may depend on scales and scalar fields values and therefore on time.

\subsubsection{Different pictures of the Universe}\label{sec:Dif_Pic_Universe} 

The electron mass $m_{e}$ or related energy levels of atomic transitions set the units of time measurements. We have seen in sect.~\ref{sec:V} that masses do not remain invariant under conformal field transformations or Weyl scalings. One therefore expects different geometries for different frames for the metric.

Physical observables as expectation values, correlations or couplings can all be expressed in terms of the quantum effective action. Typically, they involve functional derivatives. The quantum effective action $\Gamma$ is a functional of fields. We can change the arguments of $\Gamma$ by non-linear field transformations, if the expressions for observables are transformed accordingly. Observables cannot depend on the choice of fields used to describe them. This well known feature of quantum field theory becomes simple only on the level of the quantum effective action. On the level of the functional integral a field redefinition has to be accompanied by a Jacobian for the measure, which is often inaccessible in practice.

Geometry is defined by the expectation value of the metric field. A non-linear field transformation of the metric changes the geometry. The geometry of the Universe is therefore, in general, not an observable quantity. Different pictures of the Universe, related by a non-linear field transformation of the metric, can all describe the same physical reality. Not only the coordinates used to describe a given geometry are a matter if convention. Also the geometry itself is a matter of the convention for the choice of the metric or the ``frame''. This ``field relativity'' \cite{CWUWE} is a much wider concept than invariance under general coordinate transformations. General coordinate transformations can be seen as a particular class of field transformations of the metric, embedded in more general field transformations that can change the geometry. Particular useful field transformations of the metric are field-dependent conformal transformations or Weyl scalings since they leave the observable mass ratios invariant.

Weyl transformations induce a change of the metric field which depends on the value of a scalar field $\chi$. For cosmologies with a constant expectation value of $\chi$ this only amounts to an overall constant rescaling of the metric, without further change of the geometry. The change of units for rods for length and time measurements is global. This changes profoundly for cosmologies with a time evolution of $\chi(t)$, as for inflation. Now different metric frames correspond to different geometries, while actually describing the same observable physics. In particular, for the scaling solutions discussed previously the geometry becomes frame-dependent even for the radiation or matter dominated universe. For the potential $U=\bar{\mu}^{2}\chi^{2}$ the Universe shrinks during the radiation and matter dominated epochs if geometry is described in the scaling frame. In the equivalent Einstein frame the Universe expands. The geometry in the two frames is qualitatively different. For example, the curvature scalar $R$ is not the same in the two frames.

Have we observed the expansion of the Universe by the redshift of light emitted in far distant galaxies? The answer is negative. What is observed is an increase of the dimensionless ratio of the distance between galaxies over the size of atoms. Instead of an expanding Universe with an increase of the intergalactic distances we can also attribute the increasing ratio to a shrinking of the size of atoms \cite{BRADI,NAAR,NAR,HONAR}. This is precisely what happens in the scaling frame for the potential $U=\bar{\mu}^{2}\chi^{2}$. The electron mass increases with time, and therefore the size of atoms shrinks. If atoms have been larger in the past, the emitted frequencies in the past are smaller, shifted towards the red. Since light for more and more distant galaxies has been emitted further and further in the past, we can understand the observed systematic trend in the redshift of distant galaxies.

One may perform similar discussions of other observable effects in the scaling frame. For example, in a shrinking Universe the temperature $T$ is increasing, instead of the decrease in an expanding Universe. Nevertheless, the particle masses $m_{p}$ are increasing even faster than the temperature, such that ratio $T/m_{p}$ is decreasing. This ratio does not depend on the frame, and we observe the same decrease as in the Einstein frame. The usual history of the Universe with nucleosynthesis, formation of atoms and liberation of the cosmic microwave radiation follows. 

There is actually no need to compute all processes in the scaling frame. The mathematical map to the equivalent Einstein frame is sufficient for taking any computation in the Einstein frame over to the scaling frame. For most practical computations of observable effects the Einstein frame is more convenient since constant particle masses and $M$ avoid the necessity to take their variation into account. For an understanding of scale symmetry and its consequences the scaling frame is more appropriate. In the scaling frame the approach to a fixed point for $\chi/\bar{\mu}\raw \infty$ is apparent, while hidden in the Einstein frame. In the Einstein frame many naive estimates of the role of quantum fluctuations, related to questions which is the ``natural size'' of quantities, lead to very inaccurate guesses and wrong statements \cite{CWEXP}.

\subsubsection{Time varying fundamental constants as test of explicit scale symmetry breaking}

An observable time variation of fundamental constants requires the presence of a very light dynamical scalar field. Its mass must be small enough such that its expectation value, on which the couplings may depend, can have changed enough since nucleosynthesis. The Hubble parameter at nucleosynthesis is around $10^{-15}$eV, such that the mass of the scalar field should not exceed this value very much. Spontaneously broken exact quantum scale symmetry implies a massless Goldstone boson - the dilaton. However, it also implies that the dimensionless couplings and mass ratios in the standard model are independent of the value of the scalar field $\chi$. There is therefore no time variation of fundamental constants. Furthermore, the Goldstone boson settles to a constant value very early in cosmology such that scaling solutions with a variation of $\chi$ in recent cosmology are not relevant.

Explicit breaking of quantum scale symmetry by an intrinsic mass scale is necessary for a time variation of fundamental constants if the light scalar field is associated to the pseudo Goldstone boson of spontaneously broken quantum scale symmetry. For small deviations from scale symmetry one expects a small time variation of fundamental constants. Observational bounds on the time variation of couplings may therefore translate to bounds on the possible explicit violations of scale symmetry.

On the other hand, explicit scale symmetry breaking does not necessarily lead to time varying couplings. A simple example is the scale symmetry breaking in the effective potential for the cosmon, say $U=\bar{\mu}^{2}\chi^{2}$ or $U=\bar{\mu}^{4}$. As we have seen, the intrinsic scale $\bar{\mu}$ leads to scaling solutions with varying $\chi$. The standard model couplings could, however, be independent of the ratio $\chi/\bar{\mu}$. This happens for dimensionless couplings that correspond to irrelevant parameters at the UV-fixed point. At a renormalization scale $\mu=\chi$ they are $\chi$-independent. As $\mu$ is lowered, they may flow according to the $\beta$-functions of the standard model. This flow stops, however, for $\mu$ smaller than the Fermi scale $\varphi_{0}$ or the confinement scale $\Lambda_{QCD}$. Since $\varphi_{0}$ and $\Lambda_{QCD}$ are proportional to $\chi$, the value of such couplings is then independent of $\chi$ for $\mu$ smaller $\Lambda_{QCD}$ or $m_{e}$. In consequence, it is not affected if the flow of beyond standard model parameters, as the cosmon potential, is stopped for $\mu$ smaller $\bar{\mu}$. In this case the dimensionless couplings and mass ratios of the standard model neither depend on $\chi$ nor on $\bar{\mu}$. There will be no time variation of fundamental constants despite the time variation of $\chi$. We conclude that the time variation of fundamental constants tests if the presence of an intrinsic scale $\bar{\mu}$ has an influence on the dimensionless couplings or mass ratios of the standard model.

As we have discussed briefly before, some type of crossover is necessary to end the scaling solution, switching cosmology to a behavior close to a cosmological constant (see sect.~\ref{sec:7.8}). This crossover occurs in the sector of beyond standard model (BSM) physics. If the intrinsic scale in BSM physics has some influence on the couplings of the standard model, there will be an indirect violation of gravity scale symmetry in the standard model sector as well. In turn, this can give rise to an observable time variation of fundamental constants. Since the effect is only indirect, one expects that observable effects are very small.

\subsubsection{Approach to the standard model fixed point}\label{sec:Approach_SM_FP} 

In principle, there are two epochs for which scale symmetry violation in the standard model sector can lead to time varying couplings. There is first a possible early epoch that ends the transition to the effective standard model. The time variation of couplings is expected to decrease in this epoch, since the partial fixed point characterizing the standard model is approached closer and closer. A second possible epoch is the onset of the crossover ending the scaling solution. In this epoch the time variation of couplings is increasing with time as long as the universe is only in the stage of the onset of the crossover. We discuss first the period of approach to the standard model fixed point. This also provides for a quantitative statement why the couplings of the standard model should be irrelevant at the UV-fixed point if the scaling solutions are employed for an explanation of dynamical dark energy and a solution of the cosmological constant problem.

Let us discuss the flow of some dimensionless standard model coupling $g$ with $\chi$ according to the discussion in sect.~\ref{sec:Flow_In_Field_Space}. We take $\mu=0$ and assume the vicinity of a fixed point $g_{*}$, such that
\begin{equation}\label{eq:VC1} 
\chi\p_{\chi}g=-A_{g}(g-g_{*})=\theta_{g}(g-g_{*})\, ,
\end{equation}
with $A_{g}$ the anomalous dimension and $\theta_{g}=-A_{g}$ the critical exponent. (For generalizations to several couplings see \eq{45}.) An irrelevant coupling ($\theta_{g}<0$, $A_{g}>0$) approaches the fixed point for increasing $\chi$
\begin{equation}\label{eq:VC3} 
g(\chi)=g_{*}+\left (\dfrac{\chi}{\bar{\mu}_{g}}\right )^{-A_{g}}v_{g}\, .
\end{equation}
Here the initial value $v_{g}=g(\chi=\bar{\mu}_{g})-g_{*}$ parameterizes the different flow trajectories. A nonvanishing flow with $\chi$ at $\mu=0$ indicates explicit scale symmetry breaking according to eq.~\eqref{eq:76}, and therefore introduces an intrinsic scale $\bar{\mu}_{g}$. The choice of $\bar{\mu}_{g}$ is arbitrary -- it is the pair $(v_{g},\bar{\mu}_{g})$ that specifies the trajectory defining the model.

If the standard model coupling $g$ is an irrelevant parameter at the UV-fixed point, the only possible trajectory for $A_{g}>0$ is given by $v_{g}=0$. Otherwise $g$ would diverge for $\chi\raw 0$. As generic for irrelevant couplings, $g$ is predicted to be equal $g_{*}$ at the value of $\chi_{tr}$ where gravity decouples effectively. It does not depend on $\chi$. (There may be some dependence on $\chi/\mu$ from the flow due to particles with mass much smaller than the Planck mass in the effective low energy theory. This flow stops for $\mu=\Lambda_{QCD}\sim \chi$, resulting only in an $\chi$-independent factor.) As stated before, $g$ is independent of $\chi$ if its an irrelevant coupling at the UV-fixed point, and a cosmic time variation of $g$ is absent. Positive $A_{g}$ and non-zero $v_{g}$ are only possible if the standard model fixed point $g_{*}$ is not the UV-fixed point, but only reached due to some high-scale crossover from the UV-fixed point to the standard model fixed point. On the other hand, for $A_{g}<0$ the coupling $g$ is a relevant parameter at the UV-fixed point. This happens, for example, if the standard model couplings are asymptotically safe at the UV-fixed point for quantum gravity. Marginal couplings do not obey eq.~\eqref{eq:VC3} since $A_{g}=0$. Their behavior is qualitatively similar to the case of very small $|A_{g}|$, with $A_{g}<0$ for marginally relevant couplings.

One is often interested in the relative change of $g$
\begin{equation}\label{eq:VC4} 
B_{g}=\dfrac{1}{g}\left (\chi\p_{\chi}g\right )=\dfrac{\p\ln(g)}{\p\ln(\chi)}=-A_{g}\left [1+\dfrac{g_{*}}{v_{g}}\left (\dfrac{\chi}{\bar{\mu}_{g}}\right )^{A_{g}}\right ]^{-1}\, .
\end{equation}
This is suppressed for small $A_{g}$, as well as for large positive $A_{g}$ by the huge term $\sim (\chi/\bar{\mu}_{g})^{A_{g}}$ in the denominator. This suppression provides for a general explanation why a time variation of fundamental constant is tiny or unobservably small \cite{UZREP} if $A_{g}>0$. In contrast, for negative $A_{g}$ the time variation of couplings may be substantial, beyond the observational bounds. For negative $A_{g}$ we may choose for $\bar{\mu}_{g}$ the value of $\chi$ corresponding to its cosmological value at nucleosynthesis. The value $v_{g}$ cannot be arbitrarily small. For example, the gauge decouplings at the time of nucleosynthesis have values substantially different from the vanishing fixed point values for asymptotic safety.

For an order of magnitude estimate we consider the relative change between nucleosynthesis and today
\begin{equation}\label{eq:VC5} 
\Delta g_{NS}=\dfrac{g_{NS}-g_{0}}{g_{0}}=\dfrac{v_{g}}{g_{0}}\left [\left (\dfrac{\chi_{NS}}{\bar{\mu}_{g}}\right )^{-A_{g}}-\left (\dfrac{\chi_{0}}{\bar{\mu}_{g}}\right )^{-A_{g}}\right ]\, ,
\end{equation}
where $\chi_{NS}$ and $\chi_{0}$ are the values of $\chi$ at the time of nucleosynthesis and today, respectively. One may take $v_{g}=g_{0}$ and $\bar{\mu}_{g}=\chi_{0}$,
\begin{equation}\label{eq:VC6} 
\Delta g_{NS}=\left (\dfrac{\chi_{0}}{\chi_{NS}}\right )^{A_{g}}-1\, ,
\end{equation}
and multiply the result by $v_{g}/g_{0}$ if needed. Let us consider the potential $U=\bar{\mu}^{2}\chi^{2}$ for which the value of $\chi_{NS}$ can be estimated by $\bar{\mu}^{2}M^{4}/\chi_{NS}^{2}=\Omega_{h}\rho_{NS}$ with $\rho_{NS}\approx (2\,\mathrm{MeV})^{4}$ a typical energy density during nucleosynthesis. Taking $\Omega_{h}=0.01$ this yields $\chi_{NS}=20$GeV, and therefore
\begin{equation}\label{eq:VC7}
\dfrac{\chi_{0}}{\chi_{NS}}=10^{17}\com x=\left (\dfrac{\chi_{0}}{\chi_{NS}}\right )^{A_{g}}=10^{17A_{g}}\, .
\end{equation}
For small $|A_{g}|\lesssim 0.005$ one expands
\begin{equation}\label{eq:VC11}
x=1+17\ln(10)A_{g}\com \Delta g\approx 17\ln(10)A_{g}\, .
\end{equation}
Taking for $g$ a gauge coupling with $g\approx 0.6$ one finds a relative change of $\alpha=g^{2}/4\pi$
\begin{equation}
\dfrac{\Delta \alpha}{\alpha}\approx 130 A_{g}\, .
\end{equation}

The observational bounds on the variation of the strong gauge coupling are around $|\Delta\alpha/\alpha|\lesssim 10^{-3}$, since a variation in $\alpha_{s}$ leads to a variation of $\Lambda_{QCD}$ and therefore to a variation of the nucleon mass in units of the Planck mass, $\Delta (m_{n}/M)$. This yields a rather severe bound on the anomalous dimension for the strong gauge coupling
\begin{equation}\label{eq:XB1} 
|A_{s}|\lesssim 10^{-5}\, .
\end{equation}
This bound applies for $g_{s}$ being a relevant coupling at the UV-fixed point. Typical values of the gravity induced anomalous dimension for the gauge coupling are of the order one, in clear discrepancy with the bound \eqref{eq:XB1}. This demonstrates quantitatively our statement that cosmological scaling solutions valid until the recent past require the standard model couplings to be irrelevant. For an irrelevant coupling the bound \eqref{eq:XB1} does not apply since $v_{g}=0$ and the couplings are strictly independent of $\chi$ unless influenced by additional intrinsic scales in the BSM sector.

\subsubsection{Time varying couplings from beyond standard model physics}

The scaling solution may end due to a crossover in the beyond standard model sector, cf. sect.~\ref{sec:7.8.2}. This sector is largely decoupled from the standard model sector, such that consequences of the onset of a crossover for the couplings of the standard model may be very small. One expects then at most a tiny time variation of couplings.

As an example, we may take the ``cascade'' or ``seesaw II'' mechanism for the generation of the light neutrino masses \cite{MACW,LSW,MOSE,SCHEVA}. We consider the effective scalar potential for the Higgs doublet $h$, a massive triplet $t$ and a singlet $\chi$
\begin{align}\label{eq:NVG1} 
U&=\dfrac{\lambda_{1}}{2}\left (h^{2}-\varepsilon_{H}\chi^{2}-\varepsilon_{t}t^{2}\right )^{2}+\dfrac{\lambda_{3}}{2}\left (h^{2}-\alpha_{t}t\chi\right )^{2}\nn\\
&+\lambda_{t}t^{4}+U_{0}(\chi)\, ,
\end{align}
where $h$ and $t$ refer to the electrically neutral components of the Higgs doublet and the triplet, respectively. The first term involves in the bracket an $SU(2)$-singlet with vanishing hypercharge, while the bracket in the second term corresponds to a triplet with the same hypercharge as two left handed neutrinos. This reflects the possibility to form from the two doublets either a singlet or a triplet. We take the expectation value of the doublet to be real, such that $h$ and $t$ denote the real parts of the complex doublet and triplet fields. The effective potential \eqref{eq:NVG1} allows for all quartic invariants formed from the neutral components $\chi$, $h$ and $t$. It does not involve an intrinsic mass scale provided $U_{0}=\lambda_{\chi}\chi^{4}$.

The leading terms for the triplet involve a linear term proportional to $h^{2}$, and a mass term $M^{2}_{t}$,
\begin{equation}\label{eq:NVG2} 
U_{t}=-\gamma_{t}h^{2}t+\dfrac{M^{2}_{t}}{2}t^{2}+\ldots\, ,
\end{equation}
with
\begin{equation}\label{eq:NVG3} 
\gamma_{t}=\lambda_{3}\alpha_{t}\chi\com M_{t}^{2}=\lambda_{3}\alpha_{t}^{2}\chi^{2}+\lambda_{1}\varepsilon_{t}\left (\varepsilon_{H}\chi^{2}-h^{2}\right )\, .
\end{equation}
The second contribution to $M_{t}^{2}$ is much smaller than the first one and can be neglected to a good approximation. Due to a large triplet mass term $M_{t}^{2}\gg \varphi_{0}^{2}$ the expectation value of the triplet is very small
\begin{equation}\label{eq:NVG4} 
t_{0}=\dfrac{\gamma_{t}\varphi_{0}^{2}}{M_{t}^{2}}=\dfrac{\varphi_{0}^{2}}{\alpha_{t}\chi}\, .
\end{equation}
With $\varphi_{0}^{2}=\varepsilon_{H}\chi^{2}$ the tiny gauge hierarchy parameter $\varphi_{0}/\chi=\sqrt{\varepsilon_{H}}$ appears with a square in the triplet expectation value
\begin{equation}\label{eq:NVG5} 
t_{0}=\dfrac{\varepsilon_{H}\chi}{\alpha_{t}}\, .
\end{equation}
This ``cascade'' of the gauge hierarchy parameter explains the tiny masses of the neutrinos, which receive a Majorana-mass contribution from a direct Yukawa coupling to the triplet \cite{MACW}.

More precisely, for $\lambda_{t}=0$ the potential minimum with respect to $h$ and $t$ obeys the conditions
\begin{equation}\label{eq:NVG6} 
h^{2}-\varepsilon_{H}\chi^{2}-\varepsilon_{t}t^{2}=0\com h^{2}-\alpha_{t}t\chi=0\, .
\end{equation}
We recover eq.~\eqref{eq:NVG4}. Insertion of eq.~\eqref{eq:NVG4} into the first equation \eqref{eq:NVG6} determines $\varphi_{0}$ by
\begin{equation}\label{eq:NEQ1} 
\varphi_{0}^{2}=\varepsilon_{H}\chi^{2}+\dfrac{\varepsilon_{t}\varphi_{0}^{4}}{\alpha_{t}^{2}\chi^{2}}\, .
\end{equation}
An expansion in powers of the small parameter $\varepsilon_{H}$ yields
\begin{equation}\label{eq:NEQ2} 
\varphi_{0}^{2}=\varepsilon_{H}\chi^{2}\left (1+\dfrac{\varepsilon_{t}\varepsilon_{H}}{\alpha_{t}^{2}}+\ldots\right )\, .
\end{equation}
The triplet expectation value results in a tiny relative shift of the effective parameter $\varepsilon_{H}$ that can be neglected. For $\lambda_{t}>0$ the triplet expectation value is slightly shifted, with $\p U/\p t=0$ implying
\begin{equation}\label{eq:NEQ3} 
t_{0}=\dfrac{\varphi_{0}^{2}}{\alpha_{t}\chi}-\dfrac{t_{0}}{\lambda_{3}\alpha_{t}^{2}\chi^{2}}\left \{4\lambda_{t}t_{0}^{2}-2\lambda_{1}\varepsilon_{t}\left (\varphi_{0}^{2}-\varepsilon_{H}\chi^{2}-\varepsilon_{t}t_{0}^{2}\right )\right \}\, .
\end{equation}
With
\begin{equation}\label{eq:NEQ4} 
\varphi_{0}^{2}=\varepsilon_{H}\chi^{2}+\varepsilon_{t}t_{0}^{2}+\delta\, ,
\end{equation}
the condition $\p U/\p h^{2}=0$ results in
\begin{equation}\label{eq:NEQ5} 
\delta=-\dfrac{t_{0}}{\lambda_{1}\alpha_{t}\chi}\left \{4\lambda_{t}t_{0}^{2}-2\lambda_{1}\varepsilon_{t}\delta\right \}\approx -\dfrac{4\lambda_{t}t_{0}^{3}}{\lambda_{1}\alpha_{t}\chi}\, .
\end{equation}
The shift in $t_{0}$ therefore amounts to
\begin{equation}\label{eq:NEQ6} 
t_{0}\approx \dfrac{\varphi_{0}^{2}}{\alpha_{t}\chi}-\dfrac{4\lambda_{t}t_{0}^{3}}{\lambda_{3}\alpha_{t}^{2}\chi^{2}}\, ,
\end{equation}
where we neglect tiny relative corrections $\sim t_{0}/\chi$.

At the relative minimum with respect to $h$ and $t$ the potential is given by
\begin{equation}\label{eq:NEQ7} 
U(\chi)=U_{0}(\chi)+\lambda_{t}t_{0}^{4}\approx U_{0}(\chi)+\dfrac{\lambda_{t}\varepsilon_{H}^{4}}{\alpha_{t}^{4}}\chi^{4}\, .
\end{equation}
We therefore identify $\tilde{\lambda}=U/\chi^{4}$ (for $F=\chi^{2}$) as
\begin{equation}\label{eq:NEQ8} 
\tilde{\lambda}=\dfrac{U_{0}(\chi)}{\chi^{4}}+\dfrac{\lambda_{t}\varepsilon_{H}^{4}}{\alpha_{t}^{4}}\approx \dfrac{U_{0}(\chi)}{\chi^{4}}+7\cdot 10^{-130}\, \dfrac{\lambda_{t}}{\alpha_{t}^{4}}\, .
\end{equation}
It is the potential $U(\chi)$ in eq.~\eqref{eq:NEQ7} that is discussed in sects.~\ref{sec:Scal_Pot_Dil_Grav} -- \ref{sec:Flow_Eff_Scalar_Pot_Quantum_Grav}. It can be many orders of magnitude smaller than the contribution of individual invariants.

If we only take the partial minimum with respect to the triplet and insert eq.~\eqref{eq:NVG5} into $U$ one obtains (up to tiny corrections) the effective potential for the Higgs doublet and $\chi$,
\begin{equation}\label{eq:NVG7} 
U=U_{0}(\chi)+\dfrac{1}{2}\left (\lambda_{1}+\lambda_{3}\right )\left (h^{2}-\varepsilon_{H}\chi^{2}\right )^{2}\, .
\end{equation}
We identify in eq.~\eqref{eq:11}
\begin{equation}\label{eq:NVG8} 
\lambda_{H}=\lambda_{1}+\lambda_{3}\, .
\end{equation}
The term $\sim\lambda_{3}$ in eq.~\eqref{eq:NVG1} contributes a quartic term $\sim\lambda_{3}h^{4}$. This term was not included in the discussion of ref.~\cite{CWGN}. Its presence matters for the discussion of time varying couplings.

The dimensionless coupling $\alpha_{t}$ concerns only the triplet sector. Suppose that in the range of $\chi$ of interest it flows with $\chi$ as
\begin{equation}\label{eq:NVG9} 
\dfrac{\p}{\p\ln(\chi)}\alpha_{t}=-2G\, .
\end{equation}
Since this flow occurs for $\mu=0$ it reflects a scale symmetry violation due to a relevant coupling in the beyond standard model sector, with flow trajectory 
\begin{equation}\label{eq:7.229A} 
\alpha_{t}\left (\dfrac{\chi}{\bar{\mu}_{t}}\right )=F-G\ln\left (\dfrac{\chi^{2}}{\bar{\mu}_{t}^{2}}\right )
\end{equation}
introducing some intrinsic scale $\bar{\mu}_{t}$. In consequence, the triplet expectation value $t_{0}$ will not scale precisely proportional to $\chi$ even for constant $\varepsilon_{H}$,
\begin{equation}\label{eq:NVG10} 
\dfrac{t_{0}}{\chi}=\dfrac{\varepsilon_{H}}{\alpha_{t}(\chi/\bar{\mu}_{t})}\, .
\end{equation}
This results in a $\chi$-dependence of the ratio between neutrino mass and electron mass. For a Yukawa coupling $H_{\nu}$ of the neutrinos to the triplet one has
\begin{equation}\label{eq:NVG11} 
m_{\nu}=H_{\nu}t_{0}=\dfrac{H_{\nu}\varphi_{0}^{2}}{\alpha_{t}\chi}=\dfrac{H_{\nu}\sqrt{\varepsilon_{H}}}{\alpha_{t}}\,\varphi_{0}\, ,
\end{equation}
such that we can identify in eq.~\eqref{eq:59} $M_{B-L}=\alpha_{t}\chi$. With $m_{e}=y_{e}h$ and $G>0$, the neutrino masses increase with $\chi$ faster than the electron mass,
\begin{equation}\label{eq:NEQ9} 
\dfrac{m_{\nu}}{m_{e}}=\dfrac{H_{\nu}\sqrt{\varepsilon_{H}}}{y_{e}\alpha_{t}(\chi/\bar{\mu}_{t})}=\dfrac{H_{\nu}\sqrt{\varepsilon_{H}}}{y_{e}\left (F-G\ln(\chi^{2}/\bar{\mu}_{t}^{2}\right )}\, .
\end{equation}

Despite the change of the ratio $t_{0}/\chi$ the ratio $\varphi_{0}/\chi=\sqrt{\varepsilon_{H}}$ remains independent of $\chi$ if $\varepsilon_{H}$ does not depend on $\chi$. Naively, one may think that in presence of the cubic term $\gamma_{t}h^{2}t$ a change $\Delta t$ of the triplet expectation value results in a change of the mass term $\sim\gamma_{t}\Delta t$ for $h$, and therefore in a change of the Fermi scale $\Delta\varphi_{0}^{2}\sim (\gamma_{t}/\lambda_{H})\Delta t$. The particular form of the potential \eqref{eq:NVG1}, which makes it compatible with a tiny $U(\chi)$ even in the presence of nonzero $\varphi_{0}$ and $t_{0}$, shields this leading effect. It remains to be seen at what level subleading effects could induce a small time variation of the ratio between Fermi scale and Planck mass. In the present approximation the dominant subleading effect arises from the $\alpha_{t}$-dependence of $\varphi_{0}$ in eq.~\eqref{eq:NEQ2}. Being suppressed by $\varepsilon_{H}\approx 5\cdot 10^{-33}$ it is tiny. It remains open if an extended setting can induce larger variations of $\varphi_{0}/\chi$ with $\chi$.

\subsection{Dynamical dark energy}\label{sec:7.8} 

Dynamical dark energy or quintessence involves the cosmological expectation value of a scalar field $\varphi$ changing in the present epoch. We will discuss it in the Einstein frame, where the homogeneous dark contribution to the energy density arises from the scalar effective potential $V=V_{E}$ and kinetic term with prefactor $k^{2}$,
\begin{equation}\label{eq:Q1} 
\rho_{h}=V(\varphi)+\dfrac{k^{2}(\varphi)}{2}\dot{\varphi}^{2}\, .
\end{equation}
Here $V$ and $k^{2}$ characterize the quantum effective action, with all fluctuation effects already included, and $\dot{\varphi}=\p_{t}\varphi$ is the time derivative in a Robertson-Walker metric. By definition the expectation value $\varphi$ is an average over space and therefore homogeneous in space, depending only on $t$. This is the origin of the label $h$ (for homogeneous) -- other often encountered labels are $Q$ (for quintessence) or $\varphi$ (for scalar field).

The time evolution of the scalar field determines its equation of state,
\begin{equation}\label{eq:Q2} 
w_{h}=\dfrac{p_{h}}{\rho_{h}}\com p_{h}=-V(\varphi)+\dfrac{k^{2}(\varphi)}{2}\dot{\varphi}^{2}\, .
\end{equation}
An accelerated expansion of the universe occurs if the overall equation of state for the sum of all its constituents obeys $w<-1/3$. With a matter equation of state $w_{m}=0$, an accelerated expansion requires an even more negative $w_{h}$. In the limit $k\dot{\varphi}\raw 0$ one finds $w_{h}=-1$. Indeed, for negligible $\dot{\varphi}$ the constant potential $V(\varphi)$ plays the role of a cosmological constant. The general setting of dynamical dark energy has a smooth limit to a cosmological constant, given by $k^{2}\dot{\varphi}^{2}/V\raw 0$. 

By an appropriate rescaling of $\varphi$ we may bring any monotonic potential to a standard form \eqref{eq:C5}, $V=M^{4}\exp(-\varphi/M)$. The particular dynamics is then encoded in the kinetial $k^{2}(\varphi)$ \cite{CWVG,HEBCW}. We have seen in sect.~\ref{sec:Cosm_Scaling_Sol} that for $k^{2}<1/3$ one can obtain a matter dominated universe as a scaling solution with a fraction of early dark energy, $\Omega_{h}=3k^{2}$. The equation of state for this scaling solution is $w_{h}=0$. While the scaling solution offers a very attractive explanation why today $\rho_{h}/M^{4}$ is tiny and why dark energy and dark matter are of similar size, it is not compatible with the observed value of $w_{h}$ close to $-1$. Something has to stop effectively the time evolution of $\varphi$ at a redshift $z_{c}$ when $\Omega_{h}\rho_{cr}(z_{c})$ was of the order of the present observed $\rho_{h}$. Here $\rho_{cr}(z)=3M^{2}H^{2}(z)$ is the critical energy density. This stopping ends the scaling solution. If efficient enough, it turns cosmology for $z<z_{c}$ effectively close to the standard cosmology with a cosmological constant $V(\varphi(z_{cr}))$.

There are two possibilities for the end of the scaling solution. For the first the kinetial $k^{2}(\varphi)$ increases and exceeds the value $1/3$. Cosmology turns then to a scalar field dominated universe \cite{CWQ}. We discuss this possibility in sect.~\ref{sec:7.8.1}. The increase of $k^{2}$ corresponds to a flattening of the scalar potential if we use a kinetic term with standard normalization. As an alternative, the effective stop or slowing down of the evolution can be triggered by the coupling of $\varphi$ to a matter component, as for growing neutrino quintessence. This alternative is discussed in sect.~\ref{sec:7.8.2}. Both versions of an effective stop of the scaling solution correspond to ``freezing quintessence'' \cite{ELIN}. There exist alternative models of ``thawing quintessence'' \cite{HEBCW,ELIN} that we will not discuss here. We also recall that a large number of modified gravity theories can be represented as quintessence models. The first model of quintessence was obtained from a simple modification of gravity by replacing the Planck mass by a scalar field \cite{CWQ}. Indeed, a large class of modified gravity theories whose physical particle content involves a scalar in addition to the graviton as, for example, $f(R)$-theories, is equivalent \cite{CWMGCQ} to coupled quintessence, with suitable scalar couplings to matter.

In the context of our discussion of quantum scale symmetry the transition from a cosmic scaling solution to dark energy domination is associated to a crossover between two (approximate) fixed points. It concerns some stage of the crossover in the scalar singlet potential from the UV-fixed point in Fig.~\ref{fig:FL} to the IR-fixed point or, alternatively, a crossover in the beyond the standard model sector. This crossover is the reason for the end of the cosmic scaling solution, similar to the end of inflation triggered by the crossover from the UV- to the SM-fixed point. For crossover quintessence discussed in sect.~\ref{sec:7.8.1} the crossover corresponds to the change from the flow away from the UV-fixed point (lower diagonal in Fig.~\ref{fig:FL}) to the intermediate fixed point (horizontal lines in Fig.~\ref{fig:FL}). The growing neutrino quintessence model discussed in sect.~\ref{sec:7.8.2} can be realized either in the vicinity of the UV-fixed point (lower diagonal in Fig.~\ref{fig:FL}) or in the vicinity of the IR-fixed point (upper diagonal in Fig.~\ref{fig:FL}). The crossover does not concern the scalar potential in this case. It happens in the beyond standard model sector which manifests itself for cosmology by a growing ratio of neutrino over electron mass.

\subsubsection{Crossover quintessence}\label{sec:7.8.1}

Crossover quintessence \cite{CWCRQU} can be described by a crossover from a small kinetial $k^{2}(\varphi)$ to a large value \cite{HEBCW,CWCFP,CWVSC}. It relies on the simple observation that for large enough $k^{2}$ the cosmology evolves towards a scalar dominated epoch. The crossover may either be imprinted directly in the behavior of the kinetic term, for example as a result of the solution of a renormalization group equation for this quantity. It can also arise from a crossover in the potential, which is translated to $k^{2}(\varphi)$ by the non-linear field redefinition of $\varphi$ which brings the potential to canonical exponential form.

To illustrate the second case we start from a normalization of the scalar field $\sigma$ with a canonical kinetic term and some potential $V(\sigma)$,
\begin{equation}\label{eq:QQ1} 
\cL_{s}=\dfrac{1}{2}\p^{\mu}\sigma\p_{\mu}\sigma+V(\sigma)\, .
\end{equation}
The scalar field $\varphi$ with a normalized exponential potential is related to $\sigma$ by
\begin{equation}\label{eq:QQ2} 
\varphi=-M\ln(V(\sigma)/M^{4})\, .
\end{equation}
With
\begin{equation}\label{eq:QQ3} 
\p_{\mu}\varphi=-M\p_{\sigma}\ln(V)\p_{\mu}\sigma
\end{equation}
one finds the kinetial
\begin{equation}\label{eq:QQ4} 
k^{2}(\varphi)=\dfrac{1}{M^{2}}(\p_{\sigma}\ln(V))^{-2}\, .
\end{equation}
Obviously, $k^{2}$ grows very large if an extremum or flat region of $V(\sigma)$ is approached. If $V(\sigma)$ has a minimum at $\sigma_{min}$, the scaling solution may end by $\sigma$ approaching $\sigma_{min}$ \cite{ALSKO,FHSW}. (For a description of oscillations around $\sigma_{min}$ the normalization $\sigma$ is more convenient than $\varphi$.) A non-trivial minimum with the needed tiny value of $V(\sigma_{min})$ and the necessary small second order derivative $\p_{\sigma}^{2}V(\sigma_{min})$ may, however, be difficult to understand in a natural way.

An alternative is the approach to a flat region in $V(\sigma)$. In the discussion of the scaling solution for dilaton quantum gravity in sect.~\ref{sec:Crossover_DimLess_Fields}, \ref{sec:Crossover_QuantumGravity} we have found a crossover for $y=\chi^{2}/k^{2}$ near $y_{V}$, cf. eq.~\eqref{eq:DS7}, that translates in the Einstein frame to a crossover to a flat region. We investigate here the cosmological consequences of the scaling solution for dilaton quantum gravity. We consider the case of large enough $y_{V}$ such that the infrared region is not yet reached for the present cosmological epoch, $y\ll y_{IR}$. Replacing $k$ by $\bar{\mu}$ the effective action corresponding to the scaling solution reads for large $\chi$ in the scaling frame
\begin{equation}\label{eq:7.236A} 
\cL=-\dfrac{\xi\chi^{2}}{2}R+\dfrac{K}{2}\p^{\mu}\chi\p_{\mu}\chi+U(\chi)\, ,
\end{equation}
with
\begin{equation}\label{eq:7.236B} 
U(\chi)=\bar{c}_{V}\bar{\mu}^{4}\left (1+\dfrac{\chi^{4}}{\chi_{V}^{4}}\right )\, .
\end{equation}
Here $\chi_{V}$ connects to the parameter $y_{V}$ that characterizes the particular scaling solution by $\chi_{V}^{2}=y_{V}\bar{\mu}^{2}$.

For a phenomenological discussion we transform to the Einstein frame, where the potential $V$ obtains as
\begin{equation}\label{eq:QQ5} 
\tilde{\lambda}=\dfrac{V}{M^{4}}=\dfrac{U}{\xi^{2}\chi^{4}}=\dfrac{\bar{c}_{V}}{\xi^{2}}\left (\dfrac{1}{y_{V}^{2}}+\dfrac{\bar{\mu}^{4}}{\chi^{4}}\right )=\tilde{\lambda}_{0}+\dfrac{\tilde{\mu}^{4}}{\chi^{4}}\, .
\end{equation}
(Recall that the dimensionless ratio $\tilde{\lambda}$ does not depend on the choice of frame.) For large enough $\chi$ the potential becomes flat and the kinetial $k^{2}$ therefore grows to large values. The dimensionless parameter $y_{V}$ corresponds to a relevant parameter. It can be very large if the flow of the potential stays for many orders of magnitude in the ``UV-region''. In consequence,
\begin{equation}\label{eq:QQ6} 
\tilde{\lambda}_{0}=\dfrac{\bar{c}_{V}}{\xi^{2}y_{V}^{2}}=\dfrac{\bar{c}_{V}\bar{\mu}^{4}}{\xi^{2}\chi_{V}^{4}}
\end{equation}
can be a tiny quantity near $10^{-120}$. For $\chi\ll \sqrt{y_{V}}\bar{\mu}$ cosmology can follow the scaling solution of sect.~\ref{sec:Cosm_Scaling_Sol} if the scalar kinetic term is in the appropriate range. During the scaling solution the dark energy density decreases $\sim\chi^{-4}\sim t^{-2}$. The scaling solution ends once $\tilde{\lambda}$ makes a crossover to a constant $\tilde{\lambda}_{0}$, which corresponds to a flat region in $V$.

For a computation of the kinetial $k^{2}(\varphi)$ we need the scalar kinetic term in the Einstein frame. Starting from variable gravity according to eq.~\eqref{eq:151} and neglecting a possible $\chi$-dependence of $K$ for the relevant region of the crossover, one arrives at eq.~\eqref{eq:155},
\begin{equation}\label{eq:QQ7} 
\cL_{kin}=\dfrac{M^{2}B}{2}\p^{\mu}\ln(\chi)\p_{\mu}\ln(\chi)\, ,
\end{equation}
with
\begin{equation}
B=\dfrac{K}{\xi}+6\, .
\end{equation}
Here we have assumed that the relevant range corresponds to $y\gg y_{F}$, such that $F=\xi\chi^{2}$ in eq.~\eqref{eq:151}. 

With constant $B$ one identifies
\begin{equation}\label{eq:QQ8} 
\sigma=\sqrt{B}M\ln\left (\dfrac{\chi}{M}\right )\com \chi=M\exp\left (\dfrac{\sigma}{\sqrt{B}M}\right )\, ,
\end{equation}
such that
\begin{equation}\label{eq:QQ9} 
V=M^{4}\left [\tilde{\lambda}_{0}+\dfrac{\tilde{\mu}^{4}}{M^{4}}\exp\left (-\dfrac{4\sigma}{\sqrt{B}M}\right )\right ]\, ,
\end{equation}
where
\begin{equation}\label{eq:QQ10} 
\tilde{\mu}^{4}=\dfrac{\bar{c}_{V}\bar{\mu}^{4}}{\xi^{2}}\, .
\end{equation}
From eq.~\eqref{eq:QQ4} one infers the kinetial
\begin{align}\label{eq:QQ11} 
k^{2}&=\dfrac{B}{16}\left (1+\dfrac{\tilde{\lambda}_{0}M^{4}}{\tilde{\mu}^{4}}\exp\left (\dfrac{4\sigma}{\sqrt{B}M}\right )\right )^{2}\nn\\
&=\dfrac{B}{16}\left (1+\dfrac{\tilde{\lambda}_{0}\chi^{4}}{\tilde{\mu}^{4}}\right )^{2}=\dfrac{B}{16}\left (1+\dfrac{\chi^{4}}{\chi_{V}^{4}}\right )^{2}\, .
\end{align}
The crossover happens for $\chi$ reaching $\chi_{V}$,
\begin{equation}\label{eq:QQ12} 
\chi_{V}=\tilde{\mu}\left (\tilde{\lambda}_{0}\right )^{-\frac{1}{4}}=\sqrt{y_{V}}\bar{\mu}\, .
\end{equation}
For $\chi\ll\chi_{V}$ one recovers eq.~\eqref{eq:S21} for the scaling solution. This scaling solution ends for $\chi\approx\chi_{V}$. Subsequently, $k^{2}(\varphi)$ increases and one enters a scalar dominated universe. With the standard exponential normalization of the potential for the scalar field $\varphi$ the details of the model are incorporated in the kinetial. The latter depends on two dimensionless parameters, namely $B$ and
\begin{equation}\label{eq:QQ13} 
\gamma_{V}=\dfrac{\chi_{V}}{M}=\sqrt{y_{V}}\dfrac{\bar{\mu}}{M}\, .
\end{equation}

The intrinsic scale $\bar{\mu}$ sets the units of mass. We want to choose units such that the present value of $\chi$ equals the Planck mass. This fixes $\bar{\mu}$ in these units. For its computation we use the present potential energy of the scalar field, as given by
\begin{equation}\label{eq:QQ14} 
\dfrac{V}{M^{4}}=\tilde{\lambda}_{0}\left (1+\dfrac{\chi_{V}^{4}}{\chi^{4}}\right )=\tilde{\lambda}_{0}\left (1+\dfrac{\gamma_{V}^{4}M^{4}}{\chi^{4}}\right )\, .
\end{equation}
The present cosmological time $t_{0}$ is specified by $\chi(t_{0})=M$, such that the present potential contribution to the dark energy density becomes
\begin{equation}\label{eq:QQ15} 
V(t_{0})=\tilde{\lambda}_{0}M^{4}\left (1+\gamma_{V}^{4}\right )=(2.325\cdot 10^{-3}\mathrm{eV})^{4}\dfrac{\Omega_{h}^{(0)}(1-w_{h}^{(0)})}{1.4}\, .
\end{equation}
Here $\Omega_{h}^{(0)}=\Omega_{h}(t_{0})$ and $w_{h}^{(0)}=w_{h}(t_{0})$ are the present dark energy fraction and equation of state, reflecting the part of $V$ in the critical energy density. The numerical factor arises from the present critical energy density, as inferred from the present Hubble parameter. This relation fixes $\tilde{\lambda}_{0}$ in dependence on $\gamma_{V}$. In turn, the relations \eqref{eq:QQ5} and \eqref{eq:QQ15} fix the intrinsic scale $\tilde{\mu}$ according to
\begin{equation}\label{eq:SFA} 
\dfrac{V(t_{0})}{M^{4}}=\tilde{\lambda}_{0}+\dfrac{\tilde{\mu}^{4}}{M^{4}}=\tilde{\lambda}_{0}\left (1+\gamma_{V}^{4}\right )
\end{equation}
or
\begin{equation}\label{eq:SFB} 
\tilde{\mu}=(\tilde{\lambda}_{0})^{1/4}\gamma_{V}M\, .
\end{equation}
We recall that $\tilde{\mu}$ is not independently observable. The observable ratio $\chi(t_{0})/\tilde{\mu}$ determines which value of the varying $\chi$ corresponds to the present cosmological time.

Out of the three free parameters of the model, $\tilde{\lambda}_{0}$, $\gamma_{V}$ and $B$, the relation \eqref{eq:QQ15} determines one relation, such that only $\gamma_{V}$ and $B$ remain free. The parameter $B$ specifies the amount of early dark energy during the scaling solution. For a fixed $V(t_{0})/M^{4}$ and fixed $B$ the parameter $\gamma_{V}$ will then determine the precise timing in the crossover. For small $\gamma_{V}$ the crossover happens at small $\chi/M$ such that the present equation of state is expected close to $w_{h}=-1$. As $\gamma_{V}$ gets larger the difference from a cosmological constant will be more pronounced. 

The family of scaling solutions for dilaton gravity involves one relevant parameter that we may identify with $y_{V}$ or, equivalently, $\tilde{\lambda}_{0}$. Specification of $y_{V}$ selects a particular scaling solution. In contrast, the parameter $\gamma_{V}$ specifies for a given scaling solution which value of $y$ corresponds to the present cosmological time $t_{0}$,
\begin{equation}\label{eq:QQ16} 
y(t_{0})=\dfrac{y_{V}}{\gamma_{V}^{2}}\, .
\end{equation}
The status of the third parameter $B$ or, equivalently, $K/\xi$ is not yet clarified since we did not discuss the flow of the kinetic term. If $B$ corresponds to a relevant parameter it is a free parameter of the model. In contrast, for $B$ determined by irrelevant couplings it becomes, in principle, computable. We observe that $B=0$ or $K=-6\xi$ corresponds to an enhanced symmetry at the IR-fixed point for $\chi\raw\infty$, namely conformal symmetry. A crossover trajectory for $B$ from the UV-fixed point to the IR-fixed point could lead rather naturally to the small values of $B$ required for a realistic scaling solution with not too much early dark energy.

The use of the normalization \eqref{eq:QQ2} for $\varphi$ with standard potential $V=M^{4}\exp(-\varphi/M)$ has the advantage that the dynamical equations for the dimensionless energy fraction $\Omega_{h}=\rho_{h}/\rho_{c}$ and equation of state $w_{h}$ can be formulated in a closed form \cite{CWVSC} as
\begin{equation}\label{eq:QQ17} 
\p_{x}\Omega_{h}=-\Omega_{h}(1-\Omega_{h}) \left \{3w_{h}-\left (1+\dfrac{\ee^{x}}{a_{eq}}\right )^{-1}\right \}\, ,
\end{equation}
and
\begin{equation}\label{eq:QQ18} 
\p_{x}w_{h}=(1-w_{h})\left \{k^{-1}(\varphi)\sqrt{3\Omega_{h}(1+w_{h})}-3(1+w_{h})\right \}\, .
\end{equation}
The variable
\begin{equation}\label{eq:QQ19} 
x=\ln(a)
\end{equation}
is the logarithmic scale factor in the Robertson-Walker metric and $a_{eq}$ the scale factor at matter - radiation equality. For $a\ll a_{eq}$ the bracket on the r.h.s. of eq.~\eqref{eq:QQ17} amounts to $4w_{h}$, as appropriate for radiation domination, while for $a\gg a_{eq}$ it equals the value $3w_{h}$ for matter domination. The value of $\varphi$ needed in eq.~\eqref{eq:QQ18} is given by the Hubble parameter $H$,
\begin{equation}\label{eq:QQ20} 
\dfrac{H^{2}}{M^{2}}=\dfrac{2}{3}\exp\left (-\dfrac{\varphi}{M}\right )\left [\Omega_{h}(1-w_{h})\right ]^{-1}\, .
\end{equation}
The latter can be connected to the present value of Hubble parameter $H_{0}=H(t_{0})$, such that 
\begin{align}\label{eq:QQ21} 
\dfrac{\varphi}{M}&=4x-\ln(\ee^{x}+a_{eq})-\ln\left (\dfrac{3\Omega_{m}^{(0)}H_{0}^{2}}{M^{2}}\right )\nn\\
&\quad -\ln\left (\dfrac{(1-w_{h})\Omega_{h}}{2(1-\Omega_{h})}\right )\, .
\end{align}
For a given present matter density
\begin{equation}\label{eq:QQ22} 
\dfrac{\rho_{m}(t_{0})}{M^{4}}=\dfrac{3\Omega_{m}^{(0)}H_{0}^{2}}{M^{2}}
\end{equation}
and given radiation content expressed by $a_{eq}$, the system of the two equations \eqref{eq:QQ17}, \eqref{eq:QQ18} is closed by eq.~\eqref{eq:QQ21}.

In eq.~\eqref{eq:QQ18} we need the kinetial $k^{2}(\varphi)$. It can be found by combining eqs.~\eqref{eq:QQ11} and \eqref{eq:QQ14},
\begin{equation}\label{eq:KIN1} 
k^{2}(\varphi)=\dfrac{B}{16}\left (1-\tilde{\lambda}_{0}\mathrm{e}^{\frac{\varphi}{M}}\right )^{-2}\, .
\end{equation}
The kinetial diverges for $\varphi\raw\varphi_{c}$
\begin{equation}\label{eq:KIN2} 
\dfrac{\varphi_{c}}{M}=-\ln(\tilde{\lambda}_{0})\, ,
\end{equation}
reflecting the fact that $V/M^{4}$ is bounded from below by $\tilde{\lambda}_{0}$. With $\tilde{\lambda}_{0}(1+\gamma_{V}^{4})$ given by eq.~\eqref{eq:QQ15},
\begin{equation}\label{eq:KIN3} 
\tilde{\lambda}_{0}(1+\gamma_{V}^{4})=\dfrac{3\Omega_{h}^{(0)}(1-w_{h}^{(0)})H_{0}^{2}}{2M^{2}}=8.3\cdot 10^{-121}\, \dfrac{\Omega_{h}^{(0)}(1-w_{h}^{(0)})}{1.4}\, ,
\end{equation}
we can write
\begin{align}\label{eq:KIN4} 
k^{2}(\varphi)&=\dfrac{B}{16}\left (1-\exp\left (\dfrac{\varphi-\varphi_{c}}{M}\right )\right )^{-2}\nn\\
&=\dfrac{B}{16}\left (1-\dfrac{\exp\left (\dfrac{\varphi-\varphi_{0}}{M}\right )}{1+\gamma_{V}^{4}}\right )^{-2}\, .
\end{align}
Here we employ
\begin{equation}\label{eq:KIN5} 
\dfrac{\varphi_{c}}{M}=\dfrac{\varphi_{0}}{M}+\ln(1+\gamma_{V}^{4})\, ,
\end{equation}
with present value of $\varphi(t_{0})=\varphi_{0}$ given by
\begin{equation}\label{eq:KIN6} 
\dfrac{\varphi_{0}}{M}=276.5-\ln\left (\dfrac{\Omega_{h}^{(0)}(1-w_{h}^{(0)})}{1.4}\right )\, ,
\end{equation}
where the central value $276.5$ corresponds to $\Omega_{h}^{(0)}=0.7$, $w_{h}^{(0)}=-1$. 

\begin{figure}[t!]
\includegraphics[scale=0.7]{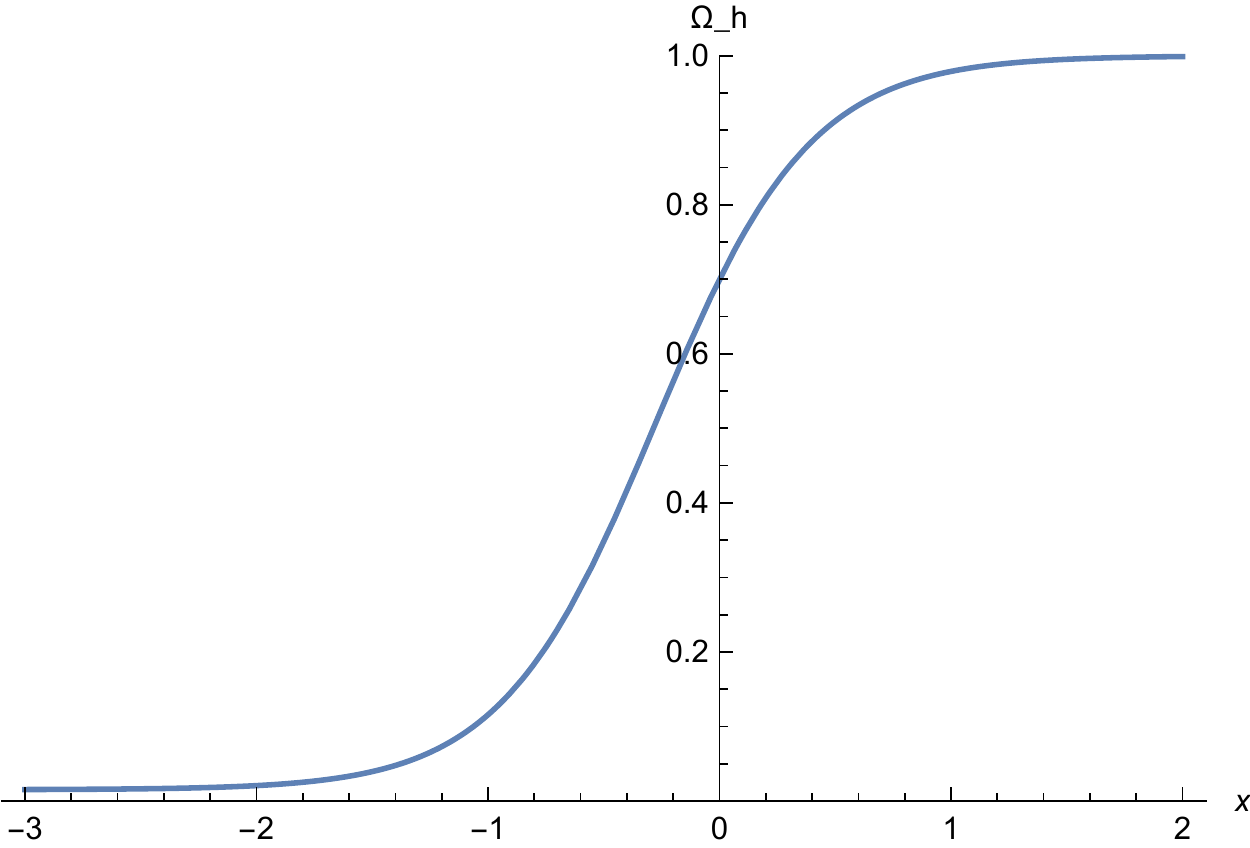}
\caption{Evolution of the dark energy fraction $\Omega_{h}$ as function of $x=\ln(a)$. The present dark energy fraction at $x=0$ amounts to $\Omega_{h}^{(0)}$, while in the future $\Omega_{h}$ will settle to one. The curve is rather similar to the one for a cosmological constant, except for a small fraction of early dark energy for negative $x$. Parameters are $B=0.08$, $\gamma_{V}=0.1$.}\label{fig:C} 
\end{figure}

\begin{figure}[t!]
\includegraphics[scale=0.7]{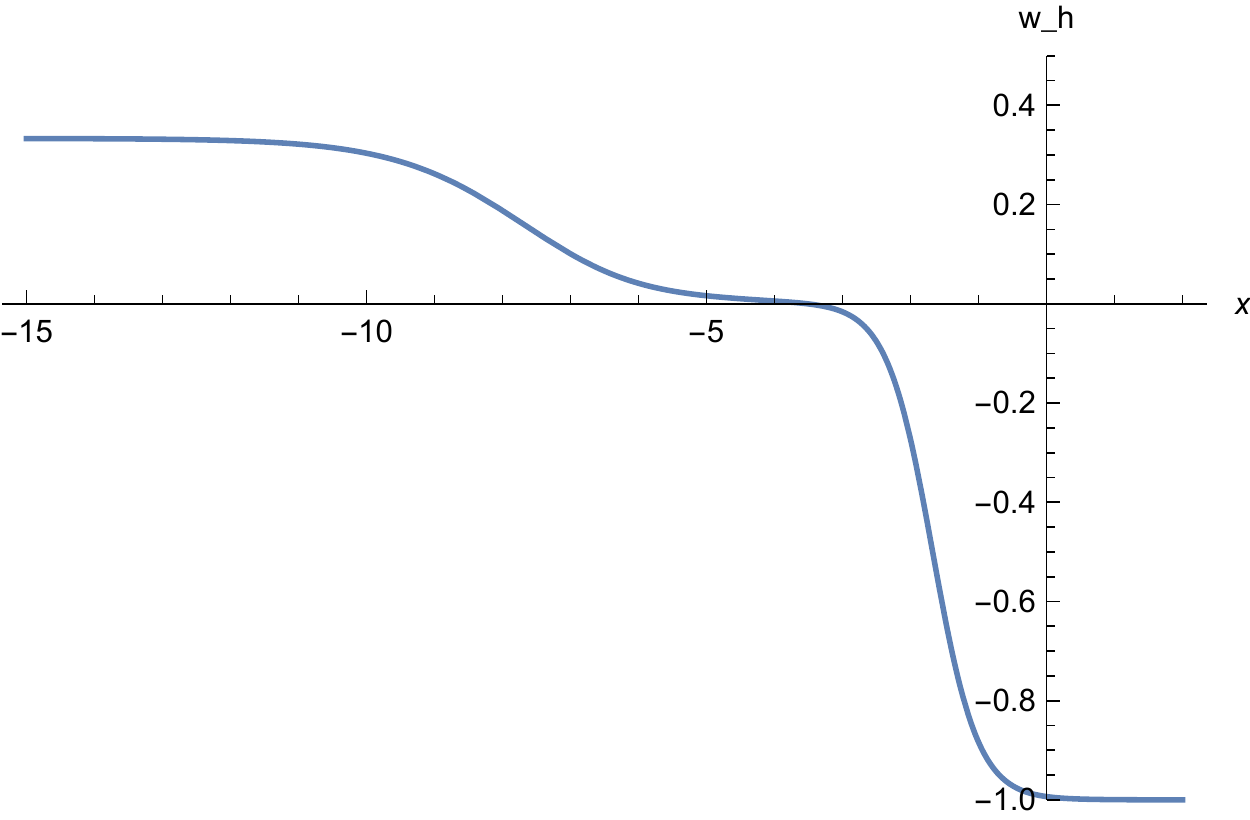}
\caption{Equation of state of dark energy $w_{h}$ as function of $x=\ln(a)$. One observes a time dependence of $w_{h}$, with a first crossover from $w_{h}=1/3$ for radiation domination to $w_{h}=0$ for matter domination, and a second crossover to $w_{h}=-1$ for dark energy domination. The present value at $x=0$ is $w_{h}^{(0)}=-0.993$. Parameters are $B=0.08$, $\gamma_{V}=0.1$.}\label{fig:D} 
\end{figure}

In Fig.~\ref{fig:C} we show the evolution of the dark energy fraction $\Omega_{h}$ as a function of $x=\ln(a)$. We have chosen $B=0.08$, corresponding to a fraction of early dark energy during matter domination of $\Omega_{e}=0.015$. The parameter $\gamma_{V}=0.1$ fixes the crossover at $\chi_{V}=0.1M$. One observes a smooth increase of $\Omega_{h}$ from a value close to $\Omega_{e}$ to a present value $\Omega_{h}^{(0)}=0.7$. In the future (for positive $x$) the dark energy fraction approaches one, as appropriate for a universe dominated by the potential energy of $\chi$. Fig.~\ref{fig:D} displays, for a wider range of $x$, the equation of state $w_{h}$. The parameters are the same as for Fig.~\ref{fig:C}. One sees the transition from a scaling solution in the radiation dominated universe ($w_{h}=1/3$) to the scaling solution of the matter dominated universe ($w_{h}=0$). This is followed in recent time by the crossover to a scalar potential dominated universe with $w_{h}=-1$. The present value of the equation of state is $w_{h}^{(0)}=-0.9932$, very close to a cosmological constant.

\begin{table}[t!]
\centering
\begin{tabular}{|c|c|c||c|c|c|}
\hline 
\multicolumn{3}{|c||}{\hspace{0.2cm}$\Omega_{e}=0.01$, $B=0.0533$\hspace{0.2cm}} & \multicolumn{3}{c|}{\hspace{0.2cm}$\Omega_{e}=0.015$, $B=0.08$\hspace{0.2cm}}  \\ 
\hline 
$\gamma_{V}$ & $\Omega_{h}^{(0)}$ & $w_{h}^{(0)}$ & $\gamma_{V}$ & $\Omega_{h}^{(0)}$ & $w_{h}^{(0)}$ \\ 
\hline 
\hspace{0.1cm}$0.0$ \hspace{0.1cm}& \hspace{0.1cm}$0.563$\hspace{0.1cm} & $-0.992$ & \hspace{0.1cm}$0.0$\hspace{0.1cm} & \hspace{0.1cm}$0.712$\hspace{0.1cm} & $-0.9935$ \\ 
\hline 
$0.1$ & $0.558$ & $-0.9918$ & $0.1$ & $0.700$ & $-0.9932$\\ 
\hline 
$0.2$ & $0.485$ & $-0.989$ & $0.2$ & $0.635$ & $-0.991$ \\ 
\hline 
$0.3$ & $0.30$ & $-0.976$ & $0.3$ & $0.420$ & $-0.978$ \\ 
\hline 
$0.5$ & $0.08$ & $-0.8836$ & $0.5$ & $0.118$ & $-0.886$ \\ 
\hline 
\end{tabular}
\caption{Present dark energy fraction $\Omega_{h}^{(0)}$ and equation of state $w_{h}^{(0)}$ for different values of the crossover parameter $\gamma_{V}$ and two values of $B$.}\label{tab:A} 
\end{table}

The dependence of the present dark energy fraction $\Omega_{h}^{(0)}$ and equation of state $w_{h}^{(0)}$ on the parameter $\gamma_{V}$ can be followed in table~\ref{tab:A}. We have chosen two values for the parameter $B$, corresponding to early dark energy fractions $\Omega_{e}=0.01$ and $0.015$ during matter domination. As expected, $\Omega_{h}^{(0)}$ decreases for increasing $\gamma_{V}$, since the onset of a potential dominated by a constant value $\tilde{\lambda_{0}}M^{4}$ occurs later. A maximum value of $\Omega_{h}^{(0)}$ is reached for the limit $\gamma_{V}\raw 0$. We observe that for $\Omega_{e}=0.01$ this maximum limit is below the observed value $\Omega_{h}^{(0)}\approx 0.7$. The corresponding value of $B$ is therefore not consistent with observation. The requirement $\Omega_{h}^{(0)}\approx 0.7$ places a lower limit on the fraction of early dark energy during matter domination. This can be seen from table~\ref{tab:B} where we show for $\gamma_{V}=0$ the dependence of $\Omega_{h}^{(0)}$ and $w_{h}^{(0)}$ on $B$.

\begin{table}[t!]
\centering
\begin{tabular}{|c|c|c|c|c|c|c|c|}
\hline 
$\Omega_{e}$ & 0.005 & 0.01 & 0.012 & 0.014 & 0.0145 & 0.015 & 0.02 \\ 
\hline 
$B$ & 0.0267 & 0.0533 & 0.064 & 0.0747 & 0.0773 & 0.08 & 0.107 \\ 
\hline 
\hspace{0.02cm}$\Omega_{h}^{(0)}$\hspace{0.02cm} & 0.32 & 0.563 & 0.63 & 0.68 & 0.7 & 0.712 & 0.805 \\ 
\hline 
\hspace{0.02cm}$w_{h}^{(0)}$\hspace{0.02cm} & \hspace{0.02cm}$-0.989$\hspace{0.02cm} & \hspace{0.02cm}$-0.992$\hspace{0.02cm} & \hspace{0.02cm}$-0.993$\hspace{0.02cm} & \hspace{0.02cm}$-0.993$\hspace{0.02cm} & \hspace{0.02cm}$-0.993$\hspace{0.02cm} & \hspace{0.02cm}$-0.993$\hspace{0.02cm} & \hspace{0.02cm}$-0.995$\hspace{0.02cm} \\ 
\hline 
\end{tabular} 
\caption{Upper bound on $\Omega_{h}^{(0)}$ as a function of $B$ or $\Omega_{e}$. The upper bound obtains for $\gamma_{V}=0$. We also display the corresponding lower bound for $w_{h}^{(0)}$.}\label{tab:B} 
\end{table}

Crossover quintessence makes two interesting predictions: First, the amount of early dark energy during matter domination should exceed a lower bound
\begin{equation}\label{eq:KIN7} 
\Omega_{e}\geq 0.0145\, .
\end{equation}
This is in the range that can be tested by cosmological observation. The dominant effect is a small reduction of the growth rate of cosmic inhomogeneities during the matter dominated universe \cite{FEJOY}. This reduces the power of observed structures, typically parametrized by $\sigma_{8}$, in comparison to the prediction of $\Lambda$CDM-cosmology based on the observed amplitude of the cosmic microwave anisotropies. As a rough rule \cite{JSCW}, one percent EDE reduces $\sigma_{8}$ by $5\%$. The bound \eqref{eq:KIN7} implies a reduction by $7$-$8\%$, compatible with the present status of the observations. Substantially higher values of $\Omega_{e}$ will be in conflict with the PLANCK-CMB data, and we take $\Omega_{e}=0.015$ as an approximate upper bound. The second prediction concerns the present equation of state of dark energy, $w_{h}^{(0)}$. It is predicted to be very close to the value of minus one for a cosmological constant. It is not arbitrarily close, however. For a given $B$ the value closest to $-1$ obtains for $\gamma_{V}\raw 0$. Despite the presence of two parameters $B$ and $\gamma_{V}$ one finds a close correlation between $\Omega_{h}^{(0)}$ and $w_{h}^{(0)}$. For $\Omega_{h}^{(0)}\approx 0.7$ and $\Omega_{e}\leq 0.015$ the crossover model predicts
\begin{equation}\label{eq:KIN8} 
w_{h}^{(0)}\approx -0.993\, .
\end{equation}
The predictions \eqref{eq:KIN7}, \eqref{eq:KIN8} may be somewhat modified if $B$ is not independent of $\chi$. We expect qualitatively similar results also for a mild $\chi$-dependence of $B$. 

A central ingredient for the emergence of the bound \eqref{eq:KIN7} is the fact that for $\gamma_{V}\raw 0$ the present dark energy fraction does not approach $\Omega_{h}^{(0)}=1$, but rather a value smaller than one which depends on the fraction in EDE. First of all, the evolution of $\chi$ is largely decoupled from the evolution of $\varphi$. The evolution of $\varphi$ maps the evolution of the potential $V_{E}$. It essentially stops after the crossover when $\chi>\chi_{V}$, since $V_{E}$ approaches the constant $V_{E}=\tilde{\lambda}_{0}M^{4}$. On the other hand, $\chi(t)$ can increase after $\chi_{V}$ is crossed without affecting much the almost constant value of $V_{E}$ and $\varphi$. This follows from the relations for $\chi\gg\chi_{V}$
\begin{equation}\label{eq:KIN9} 
\dfrac{V_{E}}{M^{4}}=\dfrac{U}{F^{2}}\com U\approx\lambda_{0}\chi^{4}\com F=\xi\chi^{2}\, .
\end{equation}
A change of $\chi$ does not affect $V_{E}$. For this reason it does not matter much if $\chi$ increases from $\chi_{t}\approx10^{-2}M$ to the present value $\chi=M$, or from $\chi_{V}\approx 10^{-8}M$ to $M$. Observational consequences depend only on the evolution of $\varphi$ which is very similar for both cases.

Second, for a cosmology close to the one for a cosmological constant the transition from matter domination to dark energy domination takes place in a finite interval $\Delta x$. As visible from Fig.~\ref{fig:C}, a ``complete transition'' takes about $\Delta x=2$, while the increase of $\Omega_{h}$ from almost zero to $\Omega_{h}^{(0)}=0.7$ takes about $\Delta x\approx 1.3$. The interval $\Delta x$ does not depend strongly on $\Omega_{e}$. On the other hand, $\Omega_{h}/\Omega_{m}$ increases after the stop of the evolution of $\varphi$ proportional to $a^{3}$. For the increase in the interval $\Delta x$ one therefore has
\begin{equation}\label{eq:GWA} 
\dfrac{\Omega_{h}^{(0)}}{\Omega_{e}}\approx\exp\left (3\Delta x\right )\approx 50\, .
\end{equation}
For given $\Omega_{h}^{(0)}$ this provides qualitatively for the lower bound for $\Omega_{e}$.

It is striking how the scaling solution of quantum gravity, or the deviation from it by a relevant parameter, leads to a rather predictive model of dynamical dark energy, with only a narrow range for early dark energy $\Omega_{e}\approx 0.015$ and present equation of state $w_{h}^{(0)}\approx -0.99$. The condition is a parameter $B\approx 0.08$. A value in this range guarantees the existence of the scaling solutions preceding the dark energy domination. It remains to be seen for a quantum gravity computation if $B$ is a relevant parameter that can take for large $\chi$ the value $\approx 0.1$, or if it is irrelevant and turns out to be in the allowed range or somewhere else. Particularly interesting would be a crossover of the function $B(\chi/\bar{\mu})$, with small values for large $\chi$ and large values for small $\chi$. In this case both inflation and the present dark energy can be described by the same scalar field $\chi$ \cite{CWIQ,CWCI}.

\subsubsection{Growing neutrino quintessence}\label{sec:7.8.2} 

The crossover responsible for the exit from the scaling solution may occur in physics beyond the standard model (BSM). One possibility is the sector of $B-L$-violation, which has important consequences for the dynamics of neutrinos. A decrease of the scale $M_{B-L}$ of spontaneous $B-L$-symmetry breaking results in an increase of the mass of the light (left handed) neutrinos according to eq.~\eqref{eq:59}. If the ratio $M_{B-L}/\chi$ decreases for increasing $\chi$, the ratio between neutrino mass $m_{\nu}$ and electron mass $m_{e}$ increases for increasing $\chi$. As a result, neutrinos have a nonvanishing coupling $\beta$ to the cosmon in the Einstein frame. By virtue of this coupling the cosmic neutrino background influences the dynamics of the cosmon field. Neutrinos can act as a cosmic trigger: as soon as they become non-relativistic the scaling solution ends and the time evolution of the cosmon field is substantially slowed down, making a transition to a cosmology rather close to the one for a cosmological constant.

Let us consider the dimensionless ratio
\begin{equation}\label{eq:GN1} 
\tau(k)=\dfrac{M_{B-L}(k)}{\chi}
\end{equation}
as a running coupling. Typically, $\tau(k)$ flows for $M_{B-L}<k<\chi$, while the flow stops for $k\ll M_{B-L}$, resulting for small $k$ in a $k$-independent value $\tau(0)=M_{B-L}/\chi$, with $M_{B-L}=M_{B-L}(k=0)$ the value related to the observed light neutrino masses by eq.~\eqref{eq:59}. In the absence of relevant couplings the scale symmetry of the UV-fixed point implies that $\tau(k=\chi)$ is independent of $\chi$. This implies $M_{B-L}\sim \chi$ and therefore a constant ratio $m_{\nu}/m_{e}$. A non-vanishing neutrino-cosmon coupling $\beta$ and the growing neutrino quintessence scenario require the presence of a relevant coupling that affects the ratio $\tau$.

Let us assume \cite{CWVG,CWGN} that the flow of this relevant coupling induces a $\chi$-dependence of $\tau(0)$ according to
\begin{equation}\label{eq:GN2} 
\dfrac{M_{B-L}}{\chi}=F_{B-L}-G_{B-L}\ln\left (\dfrac{\chi^{2}}{\bar{\mu}^{2}}\right )\, ,
\end{equation}
with positive $F_{B-L}$, $G_{B-L}$. Here $G_{B-L}$ corresponds to $G$ in eq.~\eqref{eq:NVG9}, with $M_{B-L}/\chi=\alpha_{t}$ and $F_{B-L}$ the integration constant specifying the particular flow trajectory. Correspondingly, the masses of the light neutrinos increase with $\chi$ as
\begin{equation}\label{eq:GN4} 
m_{\nu}=\dfrac{H_{\nu}\varphi_{0}^{2}(\chi)}{M_{B-L}(\chi)}=\dfrac{H_{\nu}\varepsilon_{H}\chi^{2}}{\tau(\chi)\chi}\, ,
\end{equation}
while the electron mass increases linearly $m_{e}=y_{e}\varphi_{0}=y_{e}\sqrt{\varepsilon_{H}}\chi$. Here we take for simplicity $H_{\nu}$, $y_{e}$ and $\varepsilon_{H}$ independent of $\chi$ and use equal $m_{\nu}$ for all neutrino flavors. This results in an increase of $m_{\nu}/m_{e}$ for decreasing $\tau(\chi)$,
\begin{equation}\label{eq:GN5} 
\dfrac{m_{\nu}}{m_{e}}=\dfrac{H_{\nu}\sqrt{\varepsilon_{H}}}{y_{e}\tau(\chi)}\, .
\end{equation}
We can indeed identify the ratio $\tau$ with the coupling $\alpha_{t}$ in eqs.~\eqref{eq:NVG1}, \eqref{eq:NEQ9}. With eq.~\eqref{eq:GN2} we can write
\begin{equation}\label{eq:GN6} 
\tau(\chi)=G_{B-L}\ln\left (\dfrac{\chi_{0}^{2}}{\chi^{2}}\right )\, ,
\end{equation}
with
\begin{equation}\label{eq:GN7} 
\chi_{0}=\bar{\mu}\exp\left \{\dfrac{F_{B-L}}{2G_{B-L}}\right \}
\end{equation}
indicating the scalar where $\tau$ becomes small.

As a typical measure for the strength of the $\chi$-dependence of $m_{\nu}/m_{e}$ we introduce the dimensionless parameter
\begin{equation}\label{eq:GN8} 
\tilde{\gamma}(\chi)=\dfrac{1}{2}\chi\dfrac{\p}{\p\chi}\left (\dfrac{m_{\nu}}{m_{e}}\right )=-\dfrac{1}{2}\dfrac{\p\ln(\tau)}{\p \ln(\chi)}=\dfrac{1}{\ln\bigl (\frac{\chi_{0}^{2}}{\chi^{2}}\bigl )}\, .
\end{equation}
We will not need the detailed form of the ansatz \eqref{eq:GN2}. Also $\chi$ will never be close enough to $\chi_{0}$ such that $\tilde{\gamma}(\chi)$ becomes huge. We only employ that $\tilde{\gamma}(\chi)$ increases with $\chi$ and reaches values of a few for a range of $\chi$ corresponding to present cosmology.

The dimensionless ratio $m_{\nu}/m_{e}$ does not depend on the frame and we can use eq.~\eqref{eq:GN8} in the Einstein frame as well. In the Einstein frame, the electron mass is $\chi$-independent, while the neutrino mass increases with $\chi$.

The numerical solution after the transition can be approximated by a constant ratio between the energy densities of the scalar field and neutrinos
\begin{equation}\label{eq:GQA} 
\dfrac{\Omega_{h}}{\Omega_{\nu}}=\tilde{\gamma}\, .
\end{equation}
There are oscillations between $\Omega_{h}$ and $\Omega_{\nu}$, while the sum $\Omega_{h}+\Omega_{\nu}$ increases without oscillations. The present neutrino energy density for a degenerate neutrino mass amounts to $\Omega_{\nu}(t_{0})=m_{\nu}(t_{0})/16$eV, with $m_{\nu}(t_{0})$ the present cosmological value of the neutrino mass. This results in a quantitative relation between the present dark energy density and the neutrino mass \cite{ABW}
\begin{equation}\label{eq:GQB} 
\rho_{h}^{\frac{1}{4}}(t_{0})=1.27\cdot\left (\dfrac{\tilde{\gamma}m_{\nu}(t_{0})}{\mathrm{eV}}\right )^{\frac{1}{4}}\cdot10^{-3}\mathrm{eV}\, .
\end{equation}
A present dark energy fraction $\Omega_{h}(t_{0})=0.7$ is reached for
\begin{equation}\label{eq:GQC} 
\tilde{\gamma}m_{\nu}(t_{0})=6.15\mathrm{eV}\, .
\end{equation}
For realistic neutrino masses the required present growth rate $\tilde{\gamma}$ is not very large. The correlation \eqref{eq:GQB} between the dark energy density and the neutrino masses is rather successful.

\begin{figure}[t!]
\includegraphics[scale=1.5]{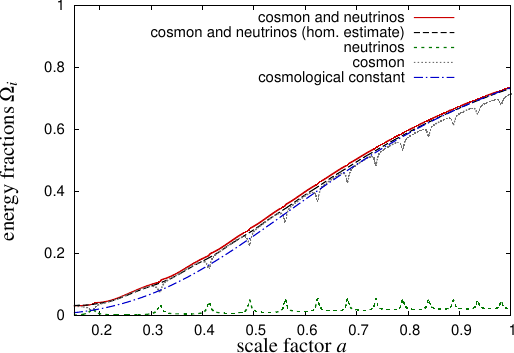}
\caption{Fractions of dark energy density and neutrino energy density as function of the scale factor. While the small neutrino energy density oscillates due to the oscillation of the neutrino mass, the combined cosmon and neutrino energy grows smoothly. This growth is well described by the neglection of backreaction effects (homogeneous estimate). It differs from the $\Lambda$CDM model only by a small fraction of early dark energy. From ref.~\cite{ABFPW}.}\label{fig:GN} 
\end{figure}

Since neutrino masses vary with $\varphi$ in the Einstein frame, the corresponding neutrino-cosmon coupling induces an attractive ``fifth force'' between neutrinos. It is typically more than a factor thousand stronger than the gravitational attraction. As a result, perturbations in the cosmic neutrino background grow very fast once neutrinos have become non-relativistic. They become non-linear at a redshift around one \cite{MPRW}. Subsequently large neutrino lumps form. For small present neutrino masses the lumps form and dissolve in an oscillatory way. The overall effect of the non-linear structures on the cosmic evolution remains small \cite{ABFPW}, as can be seen from Fig.~\ref{fig:GN}. Except for a small fraction of early dark energy according to the scaling solution of sect.~\ref{sec:Cosm_Scaling_Sol}, the overall cosmic evolution of this model is very similar to a cosmological constant. It remains to be seen if the lumps render the cosmic neutrino background observable. For present neutrino masses larger than around $0.5$eV the non-linear neutrino lumps are stable and do not dissolve anymore. Substantial backreaction effects of these structures on the cosmic evolution make it difficult to obtain an overall cosmic evolution compatible with observation \cite{CPW}.

The two models of dynamical dark energy discussed here may not be the only possible realisations of a crossover being responsible for the transition from a cosmic scaling solution to dark energy domination. Nevertheless, it is encouraging that both models give rise to realistic models of quintessence with possible interesting observable effects. In any case, a crossover in the flow of couplings is a rather interesting and natural possible explanation for a rapid qualitative change in the behavior of cosmology.

\section{Conclusions}\label{sec:Conclusion} 

We have discussed scale symmetry in particle physics, gravity and cosmology within a common framework. We focus on quantum scale symmetry, for which the effects of quantum fluctuations are taken into account in the form of the quantum effective action. Quantum fluctuations induce a flow of the renormalized dimensionless couplings as a function of a momentum scale or a field value. Exact quantum scale symmetry is realized at fixed points of this flow. A quantum field theory is defined at an UV-fixed point. This includes quantum gravity. Relevant parameters in the vicinity of the UV-fixed point generate a flow away from the fixed point as the momentum scale is lowered. All observable quantities are, in principle, computable as functions of the relevant parameters. By dimensionless transmutation the flow of these couplings sets the intrinsic mass or length scales of a given model.

The flow trajectory away from the UV-fixed point may pass very close to intermediate fixed points and end in an IR-fixed point. There is a good chance that this is realized by Nature. The intermediate fixed point is the SM-fixed point for the standard model as an ``effective low energy theory''. It is relevant if possible extensions of the SM-model become important only at energies much above the Fermi scale. On the other hand, for extensions near the Fermi scale the SM-fixed point may be replaced by a similar fixed point for the extended model. To each fixed point an effective quantum scale symmetry is associated: gravity scale symmetry for the UV-fixed point, particle scale symmetry for the SM-fixed point, and cosmic scale symmetry for the IR-fixed point. It is an interesting question if besides quantum scale symmetry the fixed points also induce extended symmetries, as conformal symmetry or local Weyl symmetry.

The approximate scale symmetry close to a fixed point has important consequences for particle physics.
\begin{enumerate}
\item A tiny ratio between the Fermi scale and the Planck scale (gauge hierarchy) is natural. It is protected by the particle scale symmetry associated to the (almost) second order vacuum electroweak phase transition.
\item Gravitational fluctuations turn the quartic Higgs coupling to an irrelevant parameter. This renders the ratio between the masses of the Higgs particle and the top quark predictable. The predicted very small value of the quartic Higgs coupling near the Planck scale has to be extrapolated to the Fermi scale by the perturbative flow of the ``low energy effective theory'' below the Planck mass. If this flow is well approximated by the flow in the standard model, the predicted value (within a small range) of the Higgs mass agrees with later observation.
\item Depending on the precise short distance theory the deviation from the vacuum electroweak phase transition (mass term for the Higgs scalar) may be relevant or not. If relevant, the ratio between Fermi scale and Planck scale $\varphi_{0}/M$ is a free parameter that cannot be predicted. If the mass term is an irrelevant parameter, the ratio $\varphi_{0}/M$ becomes, in principle, predictable. Gravitational fluctuations lead in this case to self-organized criticality. At short distances the scalar mass term is driven precisely to the critical surface of the vacuum electroweak phase transition. Any nonzero Fermi scale $\varphi_{0}$ is then due to violations of the particle scale symmetry by the running of couplings in the effective low energy theory. It is possible that the running of the standard model couplings is too slow for producing a Fermi scale above $100$MeV. In this case additional particles beyond the standard model with masses not too far from the Fermi scale would be needed.
\item These considerations extend to scalar fields beyond the standard model, as for grand unification. All quartic scalar couplings are irrelevant and predicted to be very small at the Planck scale. This puts severe restrictions on model building. If the gravity induced anomalous dimension is large enough to ensure self-organized criticality for the electroweak phase transition, the same anomalous dimension renders all scalar mass terms irrelevant. The potential at the Planck scale is predicted to be almost flat in this case.
\end{enumerate}

The predictions of approximate scale symmetry for cosmology are the following:
\begin{enumerate}
\item The almost scale invariant primordial fluctuation spectrum finds a simple explanation if the ``beginning epoch'' of the universe can be associated with the vicinity of an UV-fixed point. Instead of the tuning of parameters in the inflaton potential, usually necessary to obtain sufficient inflation and an almost flat spectrum, the approximate quantum scale symmetry provides for a simple reason for the approximate scale invariance of the fluctuation spectrum. We have discussed Starobinski inflation, cosmon inflation and Higgs inflation in the light of approximate quantum scale symmetry. The end of inflation is triggered by the crossover away from the UV-fixed point. 
\item A small amplitude of the primordial fluctuations can find a natural explanation if the crossover away from the UV-fixed point involves more than one relevant parameter.
\item Approximate scale symmetry can give a simple reason for the presence of dynamical dark energy in the present epoch. If the Planck mass is given by a scalar field $\chi$ (in the scaling frame for the metric), the value of $\chi$ spontaneously breaks scale symmetry. In the absence of an intrinsic scale $\bar{\mu}$ this spontaneous breaking implies the presence of a massless Goldstone boson. In the presence of nonzero $\bar{\mu}$ the mass of the pseudo-Goldstone boson or cosmon, $m_{c}\approx \bar{\mu}^{2}/\chi$, decreases as $\chi$ increases. If $\chi/\bar{\mu}$ increases to very large values during the long cosmological evolution, the cosmon mass is tiny. The cosmon can play the role of the dynamical scalar field needed for quintessence. No tuning of parameters is needed for a small cosmon mass. This small mass finds a natural explanation in terms of symmetry, namely the spontaneously broken scale symmetry.
\item The tiny dark energy density in units of the Planck mass finds a simple possible explanation in the existence of scaling solutions for the radiation and matter dominated epochs. For these scaling solutions the potential energy of the scalar field $V_{E}(\chi)$, which is the source of dark energy, decreases at the same rate as the dominant radiation of matter component of the universe. The tiny ratio $V_{E}/M^{4}\approx 10^{-120}$ is then explained by the huge age of the universe, similar to the tiny ratio $\rho_{M}/M^{4}$ for the dark matter density $\rho_{M}$. The scaling solution implies a fraction of early dark energy, realistically around a percent.
\item Scale symmetry explains the absence of a fifth force leading to an apparent violation of the equivalence principle, and the absence of time variation of the fundamental constants in late cosmology. This holds even for a very small mass of the cosmon and even for cosmologies for which $\chi$ varies in time. The reason is that dimensionless couplings and mass ratios are independent of $\chi$ in the scale invariant standard model. Tiny residual variations of couplings and apparent violations of the equivalence principle may arise from violations of scale symmetry, typically in the beyond standard model sector.
\item The cosmological constant problem can find a solution by the properties of an infrared fixed point of quantum gravity. For the scale invariant effective action of an infrared fixed point, the Planck mass is given by $\chi$ (in the scaling frame of the metric and for an appropriate normalization of $\chi$). The value of the coupling $\tilde{\lambda}=U(\chi)/\chi^{4}$ has to approach the fixed point value $\tilde{\lambda}_{*}=0$ for $\chi\raw\infty$. Any positive value for $\tilde{\lambda}_{*}$ would violate the graviton barrier and lead to an instability in the graviton propagator. On the other hand, negative $\tilde{\lambda}_{*}$ would lead to an instability in the scalar sector. Both instabilities are avoided by the functional renormalization flow, as we have seen explicitly in the investigation of the scaling effective potential in dilaton quantum gravity. The graviton barrier implies that a positive $U$ can grow at most $\sim\bar{\mu}^{2}\chi^{2}$ for large $\chi$. For runaway cosmologies, where $\chi$ moves to infinity in the infinite future, the observable ratio $U/M^{4}$ vanishes asymptotically. In other words, the cosmological ``constant'' vanishes dynamically in the infinite future.
\end{enumerate}

The conceptual framework for our investigation is the Wilsonian approach to renormalization, in close analogy to the physics of critical phenomena in statistical physics. Practical computations for an inclusion of gravitational fluctuation effects become possible by the modern form of functional renormalization based on the effective average action. We hope that this approach can help to settle some of the remaining big issues and questions. Indeed, for every one of the three fixed points discussed in this review several key questions have not yet found a definite answer.

For the UV-fixed point a quantitatively precise and reliable computation of the running Planck mass, or the scaling function $w(\chi)$, is not yet available, even though the indications for the existence of the scaling solution are strong. This scaling behavior is needed in dependence on the precise particle content of a model. It is not settled what is the precise form of the effective action for quantum gravity in the UV-limit. In particular, is the coefficient $C$ multiplying the term $C R^{2}$ relevant or irrelevant? (If irrelevant, the large value of $C$ needed for Starobinski inflation is unlikely to be realized.) More generally, one has to establish the momentum and field dependence of the coefficients of the higher curvature terms (e.g. the functions $C$, $D$ etc. in eq.~\eqref{eq:129}). This will decide on the issue of stability, i.e. the presence or not of ghosts. 

For the interactions between particles and gravity at the UV-fixed point a key question asks: Is the mass term of the Higgs scalar an irrelevant parameter at the UV-fixed point defining quantum gravity? If yes -- for which microscopic models? Self-organized criticality of the vacuum electroweak phase transition would amount to a profound change for our view of the gauge hierarchy. Furthermore, it will be important to find out for which particle physics models the gauge and Yukawa couplings are irrelevant parameters at the UV-fixed point. In this case their values at the Planck scale become predictable, and gravity scale symmetry can be realized in the quantum-scale invariant standard model.

For the SM-fixed point of the standard model as an effective low energy theory the running gauge and Yukawa couplings are (marginally) relevant parameters. They induce mass scales violating particle scale symmetry. The question if these mass scales are intrinsic scales $\bar{\mu}$, or reflect spontaneous symmetry breaking of gravity scale symmetry by a scalar field $\chi$, depends on the issue of the realization of gravity scale symmetry in the standard model. An intrinsic or spontaneous obvious scale induced by the running of the strong gauge coupling is the confinement scale of QCD. It is important to settle if this is the largest scale reflecting the breaking of particle scale symmetry, or if there could be a larger scale in the Higgs sector. The issue is crucial for the case of self-organized criticality of the vacuum electroweak phase transition.

For the IR-fixed point and its connection to the UV-fixed point a first central question asks: What is the field dependence of the coefficient of the curvature scalar and the scalar kinetic term? Both issues are crucial for a computation of the properties of inflation and dynamical dark energy. Furthermore, one has to understand more precisely the issue of relevant parameters and intrinsic scales in the scalar potential.

Once these questions are settled, a new epoch for model building based on basic properties of complete renormalizable quantum field theories may start. Many prejudices about what is judged to be natural or not will look rather different in this perspective.

Already now a rather consistent overall picture for fundamental physics seems to emerge, as depicted in Fig.~\ref{fig:AA}. Quantum gravity coupled to particle physics can be realised as a renormalizable quantum field theory defined at the UV-fixed point. The properties of the standard model are determined by the SM-fixed point, to which the effective action flows after decoupling of the gravitational degrees of freedom. Finally, the physics at the largest length scales, associated to the size of our observable universe, reflects the presence of an IR-fixed point. 

Such a consistent picture does not imply that one has found the most fundamental theory. It is well conceivable that a fundamental theory is formulated in terms of other degrees of freedom, say only based on fermions. In such a theory the metric or gauge fields may emerge only as collective or composite fields. If an effective continuum quantum field theory with a metric field, gauge fields, scalars and fermions is valid for a sufficient range of length scales smaller than the Planck length, it has to be described by the UV-fixed point discussed in the present work. Some of the relevant parameters at the UV-fixed point could become predictable in a more fundamental theory.

\medskip\noindent
{\em Acknowledgment:} The author would like to thank L. Amendola, J. Donoghue, A. Eichhorn, R. Percacci, J. Pawlowski, E. Rabinovici, M. Reichert, G. Ross, J. Rubio, A. Salvio, F. Saueressig, M. Shaposhnikov, M. Yamada for useful discussion and suggestions. He thanks A. Schachner for typing the manuscript. This work is supported by the DFG-SFB 1225 ``ISOQUANT''.

%\begin{appendices}

\appendix

\section{Quantum dilatation current}\label{app:A} 

In this appendix we discuss the quantum dilatation current and its conservation for a quantum scale invariant theory. We employ the standard Noether construction, applied to the quantum effective action. Our procedure is closely related to refs.~\cite{PSW,CWQ,FHR2}. We present an example for an arbitrary number of scalar fields coupled to gravity, with arbitrary field dependence of the effective potential and coefficients of scalar kinetic terms and curvature scalar. We emphasize that all quantum effects are already incorporated in the quantum effective action. No further quantum corrections are present.

For the Noether construction of the quantum dilatation current $J^{\mu}_{D}$ we start with an effective action $\Gamma[\phi_{i}]$, with $\phi_{i}$ standing for arbitrary renormalized fields, i.e. metric, scalar fields, gauge fields and fermions. Canonical infinitesimal scale transformations are ($\alpha=1+\varepsilon$)
\begin{equation}\label{eq:DC1} 
\delta\phi_{i}=d_{i}\varepsilon\phi_{i}\, ,
\end{equation}
with $d_{i}$ the scaling dimensions of the respective fields. We take $\varepsilon(x)$ local and define
\begin{equation}\label{eq:DC2} 
\Gamma[\phi_{i},\varepsilon]=\Gamma[\phi_{i}+d_{i}\varepsilon(x)\phi_{i}]\, .
\end{equation}
With $\Gamma$ depending on $\phi_{i}$ and derivatives $\p_{\mu}\phi_{i}$, one has $\Gamma[\phi_{i},\varepsilon]$ depending on $\varepsilon(x)$ and derivatives $\p_{\mu}\varepsilon(x)$. The quantum dilatation current density is defined as
\begin{equation}\label{eq:DC3} 
\tilde{J}^{\mu}_{D}=\dfrac{\p\Gamma[\phi_{i},\varepsilon]}{\p(\p_{\mu}\varepsilon)}=\sum_{i}\, d_{i}\phi_{i}\,\dfrac{\p\Gamma}{\p(\p_{\mu}\phi_{i})}\, .
\end{equation}
This construction can be generalized if the effective action contains terms with second derivatives $\p_{\mu}\p_{\nu}\phi_{i}$ or even higher derivatives \cite{CWQ}.

For a scale invariant effective action the current density is conserved for all solutions of the field equations in absence of sources,
\begin{equation}\label{eq:DC4} 
\p_{\mu}\tilde{J}^{\mu}_{D}=\sum_{i}\, d_{i}\left [\phi_{i}\, \p_{\mu}\dfrac{\p\Gamma}{\p(\p_{\mu}\phi_{i})}+(\p_{\mu}\phi_{i})\, \dfrac{\p\Gamma}{\p(\p_{\mu}\phi_{i})}\right ]=0\, .
\end{equation}
This can be seen by employing the exact quantum field equations in absence of sources,
\begin{equation}\label{eq:DC5} 
\dfrac{\p\Gamma}{\p\phi_{i}}-\p_{\mu}\dfrac{\p\Gamma}{\p(\p_{\mu}\phi_{i})}=0\, ,
\end{equation}
yielding
\begin{equation}\label{eq:DC6} 
\p_{\mu}\tilde{J}^{\mu}_{D}=\sum_{i}\, d_{i}\left [\phi_{i}\, \dfrac{\p\Gamma}{\p \phi_{i}}+(\p_{\mu}\phi_{i})\, \dfrac{\p\Gamma}{\p(\p_{\mu}\phi_{i})}\right ]\, .
\end{equation}
The r.h.s of eq.~\eqref{eq:DC6} vanishes if we specialize to $\Gamma$ being invariant under the global scale transformation with $\varepsilon$ independent of $x$. (Local gauge symmetries contain the global symmetry such that $\tilde{J}^{\mu}_{D}$ is conserved in this case as well.) 

We define the dilatation current as
\begin{equation}\label{eq:DC7} 
J_{D}^{\mu}=g^{-\frac{1}{2}}\,\tilde{J}^{\mu}_{D}\, .
\end{equation}
For a scale invariant effective action it is covariantly conserved,
\begin{equation}\label{eq:DC8} 
D_{\mu}J^{\mu}_{D}=0\, .
\end{equation}
We emphasize that there are no quantum corrections to these statements. The conservation of the quantum dilatation current \eqref{eq:DC8} is exact if the quantum effective action is invariant under the transformation \eqref{eq:DC1}. The crucial ingredient is the exactness of the quantum field equation \eqref{eq:DC5}. Similar statements are, in general, not true for a scale invariant classical action.

As an example, we consider gravity coupled to a number of scalar fields $\chi_{a}$, with field dependent coefficients of the curvature scalar and scalar kinetic terms and effective potential,
\begin{equation}\label{eq:DC9} 
\Gamma=\int_{x}\,\sqrt{g}\left \{-\dfrac{F(\chi)}{2}R+\dfrac{1}{2}K_{ab}(\chi)\p^{\mu}\chi_{a}\p_{\mu}\chi_{b}+U(\chi)\right \}\, .
\end{equation}
The dilatation current receives contributions from the first two terms. For the first term $\sim R$ one has ($D^{2}=D_{\mu}D^{\mu}$)
\begin{equation}\label{eq:DC10} 
\Gamma_{R}[\phi,\varepsilon]=\Gamma_{R}[\phi]+\int_{x}\,\sqrt{g}\left \{\varepsilon R\left (F-\dfrac{1}{2}\chi_{a}\dfrac{\p F}{\p\chi_{a}}\right )-3FD^{2}\varepsilon\right \}\, .
\end{equation}
After partial integration ($FD^{2}\varepsilon\raw -\p^{\mu}F\p_{\mu}\varepsilon$) the contribution \eqref{eq:DC3} to the dilatation current density is
\begin{equation}\label{eq:DC11} 
\tilde{J}^{D}_{\mu(R)}=3\sqrt{g}\p^{\mu}F\, .
\end{equation}
Combining with the contribution of the scalar kinetic term the quantum dilatation current takes the simple form
\begin{equation}\label{eq:SD1} 
J^{\mu}_{D}=K_{ab}\p^{\mu}\chi_{a}\chi_{b}+3\p^{\mu}F\, .
\end{equation}

If a function $L(\chi)$ exists such that one can write
\begin{equation}\label{eq:SD2} 
K_{ab}\chi_{b}=\dfrac{\p L}{\p\chi_{a}}\, ,
\end{equation}
the quantum dilatation current can be written as a derivative of a kernel $L+3F$,
\begin{equation}\label{eq:SD3} 
J^{\mu}_{D}=\p^{\mu}\left (L+3F\right )\, .
\end{equation}
For the particular case of field-independent $K_{ab}$ one has
\begin{equation}\label{eq:SD4} 
L=\dfrac{1}{2}K_{ab}\chi_{a}\chi_{b}\, .
\end{equation}

The covariant derivative of the quantum dilatation current is given by
\begin{align}\label{eq:SD5} 
D_{\mu}J^{\mu}_{D}&=\chi_{a}\left (K_{ab}D^{2}\chi_{b}+\p_{\mu}K_{ab}\p^{\mu}\chi_{b}\right )\nn\\
&\quad +K_{ab}\p^{\mu}\chi_{a}\p_{\mu}\chi_{b}+3D^{2}F\, .
\end{align}
We next insert the quantum field equations. The scalar quantum field equations derived from the variation of the effective action \eqref{eq:DC9} read \cite{CWVG}
\begin{align}\label{eq:SD6} 
&K_{ab}D^{2}\chi_{b}+\p_{\mu}K_{ab}\p^{\mu}\chi_{b}-\dfrac{1}{2}\dfrac{\p K_{bc}}{\p\chi_{a}}\p^{\mu}\chi_{b}\p_{\mu}\chi_{c}\nn\\
&=\dfrac{\p U}{\p\chi_{a}}-\dfrac{1}{2}\dfrac{\p F}{\p\chi_{a}}R-q_{a}\, ,
\end{align}
with $q_{a}$ a possible source term, connected to degrees of freedom not contained in eq.~\eqref{eq:DC9}. The quantum gravitational field equation is given by
\begin{align}\label{eq:SD7} 
&F\left (R_{\mn}-\dfrac{1}{2}Rg_{\mn}\right )+D^{2}Fg_{\mn}-D_{\mu}D_{\nu}F\nn\\
&+\dfrac{1}{2}K_{ab}\p^{\rho}\chi_{a}\p_{\rho}\chi_{b}g_{\mn}-K_{ab}\p_{\mu}\chi_{a}\p_{\nu}\chi_{b}+Ug_{\mn}=T_{\mn}\, ,
\end{align}
where the energy momentum tensor $T_{\mn}$ refers again to a possible source related to additional degrees of freedom. Contracting eq.~\eqref{eq:SD7} with $g^{\mn}$,
\begin{equation}\label{eq:SD8} 
FR=3D^{2}F+K_{ab}\p^{\mu}\chi_{a}\p_{\mu}\chi_{b}+4U-T_{\mu}^{\mu}\, ,
\end{equation}
we can express $R$ in eq.~\eqref{eq:SD6} by a function of scalar fields and derivatives. Inserting these field equations into eq.~\eqref{eq:SD5} one finds
\begin{align}\label{eq:SD9} 
D_{\mu}J^{\mu}_{D}&=\chi_{a}\dfrac{\p U}{\p\chi_{a}}-4U+4(1-f)U+3(1-f)D^{2}F\nn\\
&\quad+(1-f)K_{ab}\p^{\mu}\chi_{a}\p_{\mu}\chi_{b}+\dfrac{1}{2}\dfrac{\p K_{ab}}{\p\chi_{c}}\chi_{c}\p^{\mu}\chi_{a}\p_{\mu}\chi_{b}\nn\\
&\quad +fT^{\mu}_{\mu}-\chi_{a}q_{a}
\end{align}
where
\begin{equation}\label{eq:SD10} 
f=\dfrac{1}{2}\chi_{a}\dfrac{\p\ln(F)}{\p\chi_{a}}\, .
\end{equation}

For a scale invariant effective action one has
\begin{align}\label{eq:SD11} 
\chi_{a}\dfrac{\p U}{\p\chi_{a}}&=4U\com \chi_{a}\dfrac{\p F}{\p\chi_{a}}=2F\, ,\nn\\
\chi_{c}\dfrac{\p K_{ab}}{\p\chi_{c}}&=0\, .
\end{align}
This implies $f=1$ and
\begin{equation}\label{eq:SD12} 
D_{\mu}J^{\mu}_{D}=T^{\mu}_{\mu}-\chi_{a}q_{a}\, .
\end{equation}
In the absence of sources (or for scale invariant sources) the dilatation current is indeed covariantly conserved, as required by eq.~\eqref{eq:DC8}. The conservation of the dilatation current can be viewed as another facet of our argument in sect.~\ref{sec:QScaleSymAndQuqnatumEffAction} that the absence of intrinsic mass or length scales in the effective action implies quantum scale symmetry. The relations \eqref{eq:SD11} are a direct consequence of the absence of intrinsic scales.

For $K_{ab}$ obeying the relation \eqref{eq:SD2} a conserved dilatation current implies
\begin{equation}\label{eq:SD13} 
D^{2}(L+3F)=0\, .
\end{equation}
In a homogeneous isotropic expanding universe this reads, with $a$ the scale factor in the Robertson-Walker metric and $H=\dot{a}/a$ the Hubble parameter,
\begin{equation}\label{eq:SD14} 
\left (\p_{t}^{2}+3H\p_{t}\right )(L+3F)=0\, ,
\end{equation}
or
\begin{equation}\label{eq:SD15} 
\p_{t}(L+3F)=c_{LF}a^{-3}\, .
\end{equation}
Due to Hubble damping $L+3F$ quickly settles to a constant value. For a single field $\chi$ this corresponds to the metron reaching a constant value early in cosmology. Correspondingly, the dilaton is frozen early in cosmology and plays no role subsequently, as discussed in ref.~\cite{CWQ}. The generalization to several scalar fields is investigated in ref.~\cite{FHR2}.

If intrinsic mass or length scales are present in the effective action the relations \eqref{eq:SD11} no longer hold. The corresponding non-vanishing contributions on the r.h.s. of eq.~\eqref{eq:SD9} are the dilatation anomaly. For a non-vanishing dilatation anomaly the combination $L+3F$ no longer settles to a constant value. For example, one may find runaway cosmologies where the cosmon, the pseudo-Goldstone boson of spontaneously broken scale symmetry, increases continuously without setting to a constant value. The mass of the cosmon vanishes in the limit of vanishing dilatation anomaly.

%\end{appendices}

%%%NOTE BIB: FRATSE2->FRATSE

\bigskip\noindent

\nocite{*}
%Noch nicht zitierte Referenzen: %\cite{GIL,CWG,WEICE,DEOB,WEIAS,CWFR,MRCW,MRQG,SOU,DP,RS,CWIQ,FU1,COL1,GW,POL,COLW,PSW,WIL,CWGF,BORW,BZA,LS,BS1,CWGFE,SW,CAS}

%\cite{CWM,IMCWF}

%%%%%%%%%%%%%%
\vspace{2.0cm}\noindent

\bibliographystyle{utphys}
\bibliography{Quantum_scale_symmetry}

\end{document}